\documentclass[
  aps,pra,
  twocolumn,
  superscriptaddress,
  floatfix
]{revtex4-1}
\usepackage[dvipdfmx]{graphicx}
\usepackage{dcolumn}
\usepackage{bm}
\usepackage{ulem}
\usepackage{amsmath}
\usepackage{amssymb}
\usepackage{txfonts}
\usepackage{hyperref}
\usepackage{color} 
\newcommand{\bs}   {\boldsymbol}
\newcommand{\mb}   {\mathbf}
\newcommand{\mr}   {\mathrm}

\newcommand{\mcal} {\mathcal}
\newcommand{\imag} {\mathrm{i}}
\newcommand{\dd}   {\mathrm{d}}
\newcommand{\e}    {\mathrm{e}}
\newcommand{\bra}  {\langle}
\newcommand{\ket}  {\rangle}
\newcommand{\up}   {\uparrow}
\newcommand{\dn}   {\downarrow}
\newcommand{\s}    {\sigma}
\newcommand{\w}    {\omega}

\newcommand{\eps}  {\epsilon}
\newcommand{\Omegapersite} {\Omega}
\newcommand{\Sact} {{\mathcal S}}
\renewcommand{\star} {\dag}
\hypersetup{
  colorlinks=true,
  linkcolor=[rgb]{0.60,0.00,0.0},
  citecolor=[rgb]{0.0,0.0,0.6},
  urlcolor=[rgb]{0.0,0.0,0.6},
  setpagesize=false
}
\begin{document}

\title{
  Variational cluster approach to thermodynamic properties of interacting fermions at finite temperatures: 
  A case study of the two-dimensional single-band Hubbard model at half filling
}

\author{Kazuhiro~Seki}
\affiliation{International School for Advanced Studies (SISSA), Via Bonomea 265, 34136, Trieste, Italy}
\affiliation{Computational Condensed Matter Physics Laboratory, RIKEN Cluster for Pioneering Research (CPR), Saitama 351-0198, Japan}
\affiliation{Computational Materials Science Research Team, RIKEN Center for Computational Science (R-CCS),  Hyogo 650-0047,  Japan}

\author{Tomonori~Shirakawa}
\affiliation{International School for Advanced Studies (SISSA), Via Bonomea 265, 34136, Trieste, Italy}
\affiliation{Computational Condensed Matter Physics Laboratory, RIKEN Cluster for Pioneering Research (CPR), Saitama 351-0198, Japan}
\affiliation{Computational Materials Science Research Team, RIKEN Center for Computational Science (R-CCS),  Hyogo 650-0047,  Japan}
\affiliation{Computational Quantum Matter Research Team, RIKEN, Center for Emergent Matter Science (CEMS), Saitama 351-0198, Japan}

\author{Seiji~Yunoki}
\affiliation{Computational Condensed Matter Physics Laboratory, RIKEN Cluster for Pioneering Research (CPR), Saitama 351-0198, Japan}
\affiliation{Computational Materials Science Research Team, RIKEN Center for Computational Science (R-CCS),  Hyogo 650-0047,  Japan}
\affiliation{Computational Quantum Matter Research Team, RIKEN, Center for Emergent Matter Science (CEMS), Saitama 351-0198, Japan}

\begin{abstract}
  We formulate a finite-temperature scheme of the variational cluster approximation (VCA) 
  particularly suitable for an exact-diagonalization cluster solver. 
  Based on the analytical properties of the single-particle Green's function matrices, 
  we explicitly show the branch-cut structure of logarithm of the complex determinant functions 
  appearing in the self-energy-functional theory (SFT) 
  and whereby construct an efficient scheme for the finite-temperature VCA. 
  We also derive the explicit formulas for entropy and specific heat within the framework of the SFT. 
  We first apply the method to explore the antiferromagnetic order in a half-filled Hubbard 
  model by calculating the entropy, specific heat, and single-particle excitation spectrum for different 
  values of on-site Coulomb repulsion $U$ and temperature $T$. We also calculate the $T$ dependence 
  of the single-particle excitation spectrum in the strong coupling region, and discuss the overall similarities to 
  and the fine differences from the spectrum obtained by the spin-density-wave mean-field theory at low temperatures 
  and the Hubbard-I approximation at high temperatures. 
  Moreover, we show a necessary and sufficient condition for the third law of thermodynamics in the SFT. 
  On the basis of the thermodynamic properties, such as the entropy and the double occupancy, 
  calculated via the $T$ and/or $U$ derivative of the grand potential, we obtain a crossover diagram 
  in the $(U,T)$-plane which separates a Slater-type insulator and a Mott-type insulator. 
  Next, we demonstrate the finite-temperature scheme in the cluster-dynamical-impurity approximation (CDIA), 
  i.e., the VCA with noninteracting bath orbitals attached to each cluster, and 
  study the paramagnetic Mott metal-insulator transition in the half-filled Hubbard model. 
  Formulating the finite-temperature CDIA, we first address 
  a subtle issue regarding the treatment of the artificially introduced bath degrees of freedom  
  which are absent in the originally considered Hubbard model. 
  We then apply the finite-temperature CDIA to calculate the finite-temperature 
  phase diagram in the $(U,T)$-plane. 
  Metallic, insulating, coexistence, and crossover regions are distinguished from 
  the bath-cluster hybridization-variational-parameter dependence of the grand-potential functional. 
  We find that the Mott transition at low temperatures is discontinuous, 
  and the coexistence region of the metallic and insulating states persists down to zero temperature.  
  The result obtained here by the finite-temperature CDIA 
  is complementary to 
  the previously reported zero-temperature CDIA phase diagram. 
\end{abstract}

\date{\today}
\maketitle

\section{Introduction}
The first successful foundation of the perturbative treatment for finite-temperature many-particle 
quantum systems was constructed in 1950s by Matsubara, who introduced the imaginary-time 
Green's function to formulate the many-body perturbation theory~\cite{Matsubara1955}. 
Soon after this proposal, the physical and mathematical aspects of the formulation,  
including the Fourier expansion of the imaginary-time Green's function with 
discrete (Matsubara) frequencies~\cite{Ezawa1957}, 
have been quickly developed and these are summarized in the classic textbooks~\cite{AGD,FW}. 
Although there have been continuous efforts in developing many-body techniques 
along this line, 
the application to strongly correlated systems beyond the perturbative treatment 
is still one of the most challenging issues in many-particle quantum physics~\cite{text}.

Recently, a novel variational principle for many-fermion systems, 
based on the Luttinger-Ward formalism for the grand potential~\cite{Luttinger1960} 
in a nonperturbative way~\cite{Potthoff2006} using a functional integral form~\cite{NO}, 
has been formulated. 
This formalism, called self-energy-functional theory (SFT)~\cite{Potthoff2003a,Potthoff2003b,Potthoff2012},  
provides a unified perspective for constructing different quantum cluster approximations~\cite{Maier2005_RMP,Senechal2008} 
such as the dynamical mean-field theory (DMFT)~\cite{Georges1996} and 
its cluster extension (CDMFT)~\cite{Lichtenstein2000,Kotliar2001},  
the cluster perturbation theory (CPT)~\cite{Gros1993,Senechal2000,Senechal2002,Senechal2012},  
the variational cluster approximation (VCA)~\cite{Potthoff2003_PRL},  
and the cluster dynamical impurity approximation (CDIA)~\cite{Balzer2009,Balzer2010cdia,Senechal2010}. 
In particular, the VCA and the CDIA calculate the grand potential 
and thus the thermodynamic quantities can be readily derived. 
Furthermore, these methods allow one to calculate the translationally 
invariant single-particle Green's function by combining with, for example, the CPT~\cite{Dahnken2004}. 
Until now, the SFT and related methods have been extended to 
fermion systems with long-range interactions~\cite{Tong2005}, 
interacting boson systems~\cite{Koller2006,Knap2010,Knap2011,Arrigoni2011,Ejima2012}, 
localized quantum spin systems~\cite{Filor2010, Filor2014}, 
nonequilibrium fermion systems~\cite{Hofmann2013,Hofmann2016,Asadzadeh2016},  
and quantum chemistry calculations~\cite{Kosugi2018}.

Although the SFT is formulated at finite temperatures 
by its nature~\cite{Luttinger1960,Potthoff2006,NO,Potthoff2003a,Potthoff2003b}, 
the VCA and CDIA are applied mostly at zero temperature, 
with some exceptions~\cite{Potthoff2003b,Filor2010,Filor2014,Hofmann2016,Pozgajcic2004,Inaba2005_PRB,Inaba2005_JPSJ,Eckstein2007,
Eder2007,Eder2008,Li2009,Seki2014,Eder2015,Seki2016}. 
One of the reasons is because the main interest lies in the ground state where 
quantum fluctuations are usually strongest. 
Another reason is because of lack of a systematic description of an efficient algorithm 
for finite-temperature calculations, 
especially when either the full diagonalization of the Hamiltonian or 
the Lanczos-type method~\cite{Weisse,Prelovsek} is employed as a cluster solver. 
In fact, recent developments of experimental techniques have revealed many intriguing aspects of 
temperature dependent properties of strongly correlated systems, some of which will be 
described below. The development of an efficient algorithm at finite temperatures 
is thus highly desired.

A series of 5$d$ transition metal oxides has attracted much attention because of its rich physics 
induced by the inherently strong relativistic spin-orbit coupling which entangles spin and orbital degrees 
of freedom. 
For example, a novel effective total angular momentum $J_{\mr{eff}} = \frac{1}{2}$ antiferromagnetic 
insulating (AFI) state has been observed in 
Sr$_2$IrO$_4$~\cite{Kim2008,Kim2012,Jackeli2009,Jin2009,Watanabe2010,Shirakawa2011,Watanabe2011,Sato2015}.  
Furthermore, because of its similarity to the cuprate superconductors, 
this material is expected to show a pseudospin singlet 
$d$-wave superconductivity if mobile carriers are introduced~\cite{Wang2011,Watanabe2013,Yang2014,Meng2014}. 
Although no direct evidence for superconductivity has been observed, there are several experiments 
showing precursors of $d$-wave superconductivity or electronic structures similar to 
cuprates~\cite{Kim2014,Torre2015,Yan2015,Liu2015,Zhao16,Kim2016,Battisti17,Seo17,Terashima17,Chen18}.

Recently, it has been under the debate whether the AFI state in Sr$_2$IrO$_4$ 
is a weak-coupling Slater-type insulator or a strong-coupling Mott-type insulator~\cite{Arita2012,Watanabe2014}. 
Since 5$d$ electrons are less localized than 3$d$ electrons, the effects of electron correlations 
in 5$d$ transition metal oxides are expected smaller than those in 3$d$ transition metal oxides. 
Therefore, it is more likely that the Slater-type insulator might occur in 5d electron systems. 
Indeed, $5d$ transition metal oxides NaOsO$_3$ and SrIr$_{1-x}$Sn$_x$O$_3$ have been the only 
accepted Slater-type insulators so far~\cite{Calder2012,Cui2016}. 
The resistivity measurements for Sr$_2$IrO$_4$ have found no indication of change 
through the N\'{e}el temperature $T_{\rm N}$~\cite{Chikara2009}. 
Furthermore, the temperature dependent scanning tunneling microscopy/spectroscopy measurements 
for Sr$_2$IrO$_4$ have revealed a pseudogap behavior even above $T_{\rm N}$~\cite{Li2013}. 
These behaviors can not be reproduced by DMFT calculations~\cite{Li2013} because the DMFT does not 
take into account spatial magnetic fluctuations which develop near the phase transition. 
It is also remarkable that the recent angle-resolved photoemmision spectroscopy experiment has observed 
the Slater to Mott crossover with decreasing temperature in the metal-insulator transition of another $5d$ 
electron system Nd$_2$Ir$_2$O$_7$~\cite{Nakayama2016}.  
It is therefore highly desirable to develop theoretical methods which can treat spatial magnetic 
fluctuations and allow one to calculate finite-temperature quantities, 
including single-particle excitation spectra, 
down to sufficiently low temperature $T$, 
typically in a range of $0 < T  \lesssim t^2/U$ for a Hubbard model. 

The major difficulty of the conventional finite-temperature VCA is 
the increase of the number $N_{\rm pole}$ of the single-particle excitation energies 
which must be summed up to calculate the grand-potential functional~\cite{Potthoff2003b,Aichhorn2006_Qmat}. 
The rapid increase of $N_{\rm pole}$ at finite temperatures compared to zero temperature 
is simply because one has to consider the single-particle-excited states not only from 
the cluster's ground state but also from the several lowest or all cluster's excited states.  
Since the $N_{\rm pole} \times N_{\rm pole}$ Hermitian matrix has to be diagonalized 
at each momentum to obtain the single-particle excitation energies~\cite{Aichhorn2006_Qmat}, 
the large $N_{\rm pole}$ severely limits the finite-temperature VCA calculations, 
even if the full diagonalization of the cluster's Hamiltonian can be performed 
without any difficulty. 

To overcome this difficulty, here 
we provide an efficient scheme of the finite-temperature VCA with 
the exact-diagonalization method as a cluster solver. 
We carefully analyze the analytical properties of logarithm 
of the complex determinant functions, which appear when the grand-potential functional is 
calculated in the SFT, and treat the exponentially increasing number of poles without actually 
summing them.
Our scheme is based on the same idea proposed earlier in Ref.~\cite{Eder2008}, 
but simplifies the integrand of the grand-potential functional as 
in the zero-temperature scheme described in Ref.~\cite{Senechal2008}. 
We also derive the analytic formulas for entropy and specific heat within the framework  
of the SFT for which the exact-diagonalization method is easily applied. 

For demonstration, we apply this method to the single-band Hubbard model 
on the square lattice at half filling and calculate various thermodynamic quantities as well as 
the single-particle excitation spectra at finite temperatures. 
Based on the temperature and the interaction dependence of the 
thermodynamic quantities,
we discuss the crossover from a Slater-type insulator to a Mott-type insulator in the paramagnetic 
state. 
We also apply this method to the finite-temperature CDIA, i.e., the finite-temperature 
VCA with bath orbitals attached to each cluster, and examine the Mott metal-insulator 
transition in the paramagnetic state at half filling. 
We construct the finite-temperature phase diagram 
from the analysis of the grand-potential functional, and also investigate 
the single-particle excitations 
in the finite-temperature CDIA.

The rest of this paper is organized as follows. 
After introducing the single-band Hubbard model in Sec.~\ref{model}, 
a finite-temperature VCA 
with the exact-diagonalization cluster solver is described in depth in Sec.~\ref{method}. 
The block-Lanczos method for cluster single-particle Green's functions 
is described in Sec.~\ref{sec.BL}. 
The method is applied in Sec.~\ref{results} to 
the single-band Hubbard model and calculate various quantities at finite temperatures, 
including grand potential, entropy, specific heat, 
and single-particle excitation spectra within the VCA. 
The paramagnetic Mott metal-insulator transition at half filling is 
also investigated within the CDIA in Sec.~\ref{cdia_results}. 
In deriving the formalism of the CDIA, we address an issue of how to 
appropriately treat the contribution of the bath degrees of freedom to the grand-potential functional. 
Section~\ref{summary} is devoted to the summary of this paper 
and the discussion on other applications and further extensions.
More technical details are provided in Appendixes~\ref{app.anotherOmega},~\ref{app.ST},~\ref{app.dG} and
\ref{app.energy}.

\section{Model}\label{model}

We consider the two-dimensional single-band Hubbard model on the square lattice defined as 
\begin{eqnarray}\label{eq.ham}
  \hat{H} &=&
  - \sum_{\bra i,j \ket} \sum_\sigma t_{ij} \left(\hat{c}_{i\s}^{\dag} \hat{c}_{j\s} + \mathrm{H.c.}\right) \nonumber \\
  &+& U   \sum_{i} \hat{n}_{i \up} \hat{n}_{i \dn} 
  - \mu \sum_{i}\sum_\sigma \hat{n}_{i \s},
\end{eqnarray}
where $\hat{c}_{i\s}$ ($\hat{c}_{i\s}^\dag$) denotes the annihilation (creation) operator of an electron with spin 
$\s\ (=\uparrow,\downarrow)$ at site $i$ and 
$\hat{n}_{i\s} = \hat{c}_{i\s}^\dag \hat{c}_{i\s}$. 
Notice that operators are indicated with hat. 
The hopping integral $t_{ij}=t$ is between the nearest neighbor sites
$i$ and $j$ on the square lattice and  
the sum in the first term denoted as $\langle i,j\rangle$ runs 
over all independent pairs of sites $i$ and $j$. 
The on-site Coulomb repulsion between electrons is represented by $U$ and  
the chemical potential $\mu$ is determined so as to keep the average electron density $n$ 
at half filling, i.e., $n=1$. 
Hereafter, we set $\hbar = k_\mr{B} = 1$ and we use $t$ as the energy unit unless otherwise 
stated. We also set the lattice constant to be one. 
We refer to $z$ and $\w$ as complex and real number, respectively. 
Although here we choose this particular model, the formulation described below is readily applied 
for any fermion systems with intrasite interactions, including multi-band Hubbard models.

\section{Variational cluster approximation at finite temperatures}\label{method}

In this section, we describe a formalism of the finite-temperature VCA 
with the exact-diagonalization cluster solver. 
The VCA is one of the self-consistent quantum-cluster methods based 
on the SFT, 
which by its nature is formulated at finite temperatures~\cite{Potthoff2003a, Potthoff2003b, Potthoff2012}.

\subsection{Self-energy-functional theory}\label{method.sft}

In the SFT, the grand-potential functional $\Omega[\bs{\Sigma}]$ as 
a functional of the self-energy $\bs{\Sigma}$ is given as
\begin{equation}\label{eq.LW-SFT}
  \Omega[\bs{\Sigma}] = {\cal F} [ \bs{\Sigma} ] - \frac{1}{\beta} \mr{Tr} \ln \left( -\bs{G}_{0}^{-1} + \bs{\Sigma} \right),
\end{equation} 
where 
\begin{equation}
{\cal F} [ \bs{\Sigma} ] = \Phi \bigl[ \bs{G}[\bs{\Sigma}] \bigr]  - \frac{1}{\beta} \mr{Tr}\bigl( \bs{G}[\bs{\Sigma}]\, \bs{\Sigma} \bigr)
\end{equation} 
is the Legendre transform of the Luttinger-Ward potential $\Phi[\bs{G}]$ and 
the single-particle Green's function $\bs{G}[\bs{\Sigma}]$ is given 
as the functional of $\bs{\Sigma}$~\cite{Luttinger1960, Potthoff2003a}. 
$\beta=1/T$ is the inverse temperature and 
$\bs{G}_{0}$ is the noninteracting single-particle Green's function. 
$\mr{Tr}$ represents the functional trace which runs over 
all (both spatial and temporal, either discrete or continuous) variables of the summand.  
For example, when the system is at equilibrium, the summand 
becomes diagonal with respect to the Matsubara frequency 
\begin{equation}
  \imag \w_\nu = (2\nu + 1 )\pi \imag T, 
\end{equation}
where $\imag=\sqrt{-1}$ and $\nu=0,\pm 1, \pm 2, \cdots$~\cite{Ezawa1957},   
as explicitly shown below in Eq.~(\ref{eq.Tr}). 
The stationary condition 
\begin{equation}\label{eq.VP}
\left. \frac{\delta \Omega[\mb{\Sigma}]}{\delta \bs{\Sigma}}\right|_{{\mb\Sigma}={\mb\Sigma^*}} = 0
\end{equation}
gives the Dyson's equation 
\begin{equation}
{\bs G}^{-1}[\bs\Sigma^*] = {\bs G}_0^{-1} - {\bs\Sigma}^*,
\end{equation}
and the functionals $\Omega[\bs{\Sigma^*}]$ and $\bs{G}[\bs{\Sigma^*}]$ 
at the stationary point are the grand potential and the single-particle Green's function 
of the system, respectively~\cite{Potthoff2003a, Luttinger1960}. 
Therefore, the self-energy $\bs{\Sigma}$ is considered as a trial function for the variational 
calculation.

The VCA is an approximate but nonperturbative method to calculate the grand 
potential~\cite{Potthoff2006}, and is based on the fact that the functional form of ${\cal F} [\bs{\Sigma}]$ 
depends only on the interaction terms, but not the one-body terms, of the Hamiltonian $\hat{H}$.
In the VCA, the lattice on which the Hamiltonian $\hat{H}$ is defined is divided into disconnected 
finite-size clusters with no inter-cluster terms, and each cluster is described by Hamiltonian $\hat{H}'$. 
Although the clusters are not necessarily identical with each other, here we assume for simplicity 
that they are identical.  
The reference system is introduced as a collection of these disconnected clusters 
forming a superlattice. The cluster Hamiltonian $\hat{H}'$ must have the same interaction terms as the 
original Hamiltonian $\hat{H}$ but the one-body terms can be different. 
Therefore, the functional form of ${\cal F}[\bs{\Sigma}]$ for the reference system is exactly the 
same as that for the original system.

The exact grand potential of the reference system is 
\begin{equation}
  \Omega_{\rm r}[\bs{\Sigma}_{\rm r}] = {\cal F} [\bs{\Sigma}_{\rm r}] 
  - \frac{1}{\beta}\mr{Tr} \ln \left(-\bs{G}_{{\rm r}0}^{-1} + \bs{\Sigma}_{\rm r}\right),
\end{equation} 
where $\bs{\Sigma}_{\rm r}$ and $\bs{G}_{{\rm r}0}$ are the exact self-energy and the  
noninteracting single-particle Green's function of the reference system, respectively. 
Since ${\cal F}[\bs{\Sigma}]$ shares the same functional form 
in the original and reference systems, 
we can eliminate the unknown ${\cal F} [\bs{\Sigma}]$ from Eq.~(\ref{eq.LW-SFT}) by 
assuming that the trial self-energy $\bs{\Sigma}$ space of the original system 
is restricted within the self-energy $\bs{\Sigma}_{\rm r}$ space of the reference system, which is parametrized with 
a set of one-particle parameters $\bs{\lambda}$, appearing 
as the one-body terms in the Hamiltonian for the reference system. 
The resulting approximate grand-potential functional 
for the original system is thus 
\begin{equation}\label{eq.Omega}
  \Omega [\bs{\Sigma}_{\rm r}] = \Omega_{\rm r}[\bs{\Sigma}_{\rm r}]
  - \frac{1}{\beta} \mr{Tr} \ln \left(\bs{I}-\bs{V}\bs{G}_{\rm r}[\bs\Sigma_{\rm r}]\right), 
\end{equation}
where $\bs{I}$ is a unit matrix,
\begin{equation}\label{V_def}
  \bs{V}  =  \bs{G}_{{\rm r}0}^{-1} - \bs{G}_{0}^{-1}
\end{equation}
represents the difference of the one-body terms between the original and reference systems, and 
\begin{equation}
  \bs{G}_{\rm r}[\bs\Sigma_{\rm r}]=\left(\bs{G}_{{\rm r}0}^{-1} - \bs{\Sigma}_{\rm r}\right)^{-1}
\end{equation} 
is the exact Green's function of the reference system~\cite{Senechal2008}. 
Because a set of one-particle parameters $\bs{\lambda}$ is considered as the variational 
parameter~\cite{Potthoff2012},  
the variational principle in Eq.~(\ref{eq.VP}) 
is now regarded as the stationary condition for these variational parameters, i.e., 
\begin{equation}
  \left. \frac{\partial \Omega \left[ \bs{\Sigma}_{{\rm r},\bs{\lambda} } \right] }{\partial \bs{\lambda}} \right|_{\bs{\lambda}=\bs{\lambda}^{*}} = \bs{0},
  \label{eq:gp}
\end{equation}
where $\bs{\lambda}^{*}$ is a set of optimal variational parameters.

Since the reference system is composed of the disconnected clusters 
on the superlattice, $\mr{Tr}$ in Eq.~(\ref{eq.Omega}) is now explicitly given as 
\begin{equation}\label{eq.Tr}
  \mr{Tr} [\cdots] = \sum_{\nu=-\infty}^\infty \sum_{\tilde{\mb{k}}} \e^{\imag \w_\nu 0^+} \mr{tr} [\cdots],
\end{equation}
where $\tilde{\mb{k}}$ is a wave vector belonging to the Brillouin zone of the superlattice 
(i.e., the reduced Brillouin zone) 
and $\mr{tr}$ in the right-hand side represents trace over the remaining indices such as 
spins, orbitals, and sites within the cluster. 
The convergence factor $\e^{\imag \w_\nu 0^+}$ with $0^+$ being infinitesimally small positive 
is due to the causality at an equal imaginary time~\cite{Luttinger1960} and  
allows us to convert the Matsubara sum into the contour integral involving the Fermi-distribution function 
in the complex $z$ plane. In particular, the convergence factor plays a role 
if the integrand decays slowly as $1/z$ for large $|z|$ 
and the path of the contour integral reaches to the infinity in the left-half plane. 
However, as shown in the following, the contour proposed here for the VCA at finite temperatures 
is within a finite range. Therefore, we omit the convergence factor hereafter.

\subsection{Grand-potential functional}\label{method.eq}

Using the relation $\mr{tr} \ln [\cdots] = \ln \det [\cdots]$, 
the grand-potential functional $\Omegapersite \ (:=\Omega[\bs{\Sigma}_{\rm r}]/NL_{\rm c})$ per site is now given as
\begin{eqnarray}\label{eq.Omega-num}
  \Omegapersite = 
  \frac{1}{L_{\rm c}}\Omega' 
  - \frac{T}{NL_{\mr{c}}}  \sum_{\nu=-\infty}^\infty \sum_{\tilde{\mb{k}}} 
  \ln \det 
  \left[ 
    \bs{I}-\bs{V}(\tilde{\mb{k}}) \bs{G}'(\imag \w_\nu)
  \right], 
\end{eqnarray}
where $\Omega'$ and $\bs{G}'(\imag \w_\nu)$ are respectively the grand potential 
(i.e., $\Omega_{\rm r}[\bs{\Sigma}_{\rm r}]/N$) and the 
single-particle Green's function of the single cluster containing $L_{\rm c}$ sites, and 
$N$ is the number of clusters. $\bs{V}(\tilde{\mb{k}})$ is the Fourier transform of 
Eq.~(\ref{V_def}) with respect to the superlattice. 
For simplicity, the functional dependence on $\bs{\Sigma}_{\rm r}$ is omitted 
in Eq.~(\ref{eq.Omega-num}). 
The exact grand potential $\Omega'$ of the single cluster is evaluated as 
\begin{equation}\label{eq:omega}
  \Omega' = -\frac{1}{\beta} \ln \sum_{s=0}^{ s_{\mr{max}}} \exp(-\beta E_{s}),
\end{equation}
where $E_s$ is the $s$th eigenvalue of $\hat{H}'$ with 
$E_0\leqslant E_1\leqslant E_2 \leqslant \cdots \leqslant E_{s_{\rm max}}$. 
Note that the chemical-potential term is also included in the Hamiltonian 
$\hat{H}'$ [see Eq.~(\ref{eq.ham})]. 
In practical calculations, 
the sum in Eq.~(\ref{eq:omega}) is terminated at 
$s_\mr{max}$ for a given temperature $T$ in order to save the computational cost. 
This is a legitimate approximation 
because the contribution from excited states with larger $E_s$ becomes exponentially smaller. 
We choose $s_\mr{max}$ to satisfy 
\begin{equation}
  \exp{(-\beta E_{s_{\mr{max}}})}/\exp{(-\beta E_{0})} \geqslant \eps,
  \label{truncation}
\end{equation}
where $E_0$ is the ground state energy of $\hat{H}'$ and $\eps$ is a threshold for thermal fluctuations~\cite{Aichhorn2003}. 
We typically set 
$\eps = 1 \times 10^{-6}$ for
all the clusters (see Fig.~\ref{fig.clusters}).    
This $\eps$ value small enough to safely ignore the contribution from 
high-energy excited states in all quantities studied here.

The single-particle Green's function $\bs{G}{'}(z)$ of the cluster is given as
\begin{equation}\label{eq.G}
  G{'}_{ij}^{\s\s'} (z) = 
  \sum_{s=0}^{ s_\mr{max}} \e^{\beta (\Omega' -  E_s)} 
  \left( G^{\s\s',+}_{ij,s}  (z) + G^{\s\s',-}_{ij,s} (z) \right), 
\end{equation}
where 
\begin{eqnarray}
  G^{\s\s',+}_{ij,s} (z) &=& \left\bra \Psi_s \left| \hat{c}_{i\s}        \left[z -\left(\hat{H}' - E_s\right)\right]^{-1} \hat{c}_{j\s'}^\dag \right| \Psi_s \right\ket, \label{eq.Ge}\\
  G^{\s\s',-}_{ij,s} (z) &=& \left\bra \Psi_s \left| \hat{c}_{j\s'}^\dag  \left[z +\left(\hat{H}' - E_s\right)\right]^{-1} \hat{c}_{i\s}       \right| \Psi_s \right\ket, \label{eq.Gh}
\end{eqnarray}
and $|\Psi_s \ket$ is the $s$th eigenstate of $\hat{H}'$. 
Notice that i) the sum in Eq.~(\ref{eq.G}) is terminated at $ s_\mr{max}$ and ii) the same expression 
for the single-particle Green's functions of a cluster is employed in the CDMFT
with exact-diagonalization impurity solvers~\cite{Perroni2007, Capone2007, Liebsch2012}.  
It is apparent in Eq.~(\ref{eq.G}) that $\bs{G}'(\imag \w_\nu) \in \mathbb{C}^{L \times L}$ 
and thus $\bs{V}( \tilde{\mb{k}} ) \in \mathbb{C}^{L \times L}$ in Eq.~(\ref{eq.Omega-num}),  
where $\mathbb{C}^{m \times n}$ represents a set of $m \times n$ complex matrices and 
$L=2L_{\rm c}$ denotes the number of the single-particle labels in the cluster, including 
the spin degrees of freedom for the single-band Hubbard model in Eq.~(\ref{eq.ham}). 
Equations ~(\ref{eq.Ge}) and (\ref{eq.Gh}) are calculated efficiently 
by employing the block-Lanczos method. 
Since the efficient calculation of the cluster's single-particle Green's function is 
crucial for the efficient calculations of VCA in particular at finite temperatures, 
the block-Lanczos method for the single-particle Green's function will be described separately in Sec.~\ref{sec.BL}.

\begin{figure}
  \begin{center}
    \includegraphics[width=1.0\columnwidth]{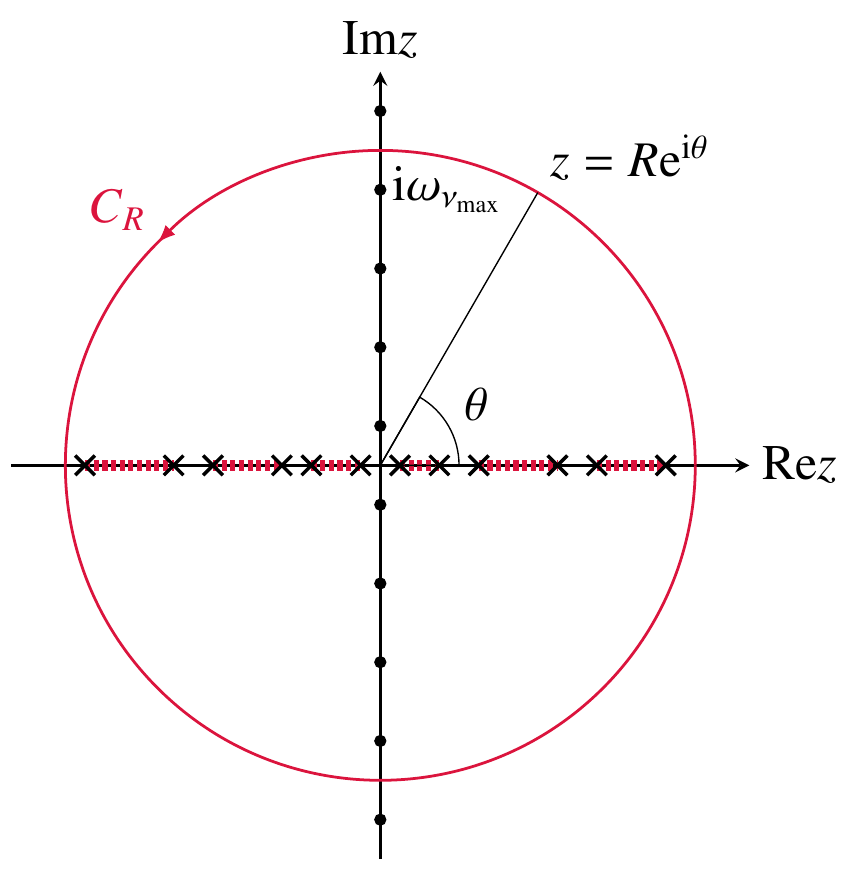}
    \caption{
      Complex $z$ plane for the 
      contour integral appearing in Eqs.~(\ref{eq.Matsubara_sum}) and~(\ref{eq.Matsubara_sum2}).  
      The fermionic Matsubara frequencies are denoted by solid dots on the imaginary axis, where 
      the Fermi-distribution function $n_{\rm F}(z)$ displays singularities.  
      The contour $C_R$, on which the complex frequency is represented as 
      $z=R\e^{\imag \theta}$, 
      is shown by a red solid line with arrow.  
      The branch cuts and branch points of $\ln \det [\bs{I} - \bs{V}(\tilde{\mb{k}})\bs{G}'(z)]$ 
      (see Sec.~\ref{sec.cuts}) are 
      also indicated by red dotted lines and crosses on the real axis, respectively. 
      \label{contour}}
  \end{center}
\end{figure}

In the calculation of the grand-potential functional at finite temperatures, 
there appears the infinite sum over the Matsubara frequencies, which cannot be performed directly. 
In addition, the contribution from the high-frequency part is not negligible because
the integrand decays in frequency as $\sim -{\rm tr}[{\bs{V}(\tilde{\mb{k}})}]/z$~\cite{Senechal2008}.
Therefore, the sum over Matsubara frequencies in Eq.~(\ref{eq.Omega-num}) is evaluated by the combination 
of the direct summation and a contour integral~\cite{Eder2008, Wildberger1995, Lu2009}. 
The low-frequency part $(|\w_\nu| \leqslant \w_{\nu_{\mr{max}}})$ is summed explicitly, 
while the high-frequency part of the sum is replaced by the contour integral along 
the closed path $C_R$ as shown in Fig.~\ref{contour}, i.e., 
\begin{eqnarray}\label{eq.Matsubara_sum}
  T \sum_{\nu= -\infty}^{\infty} [\cdots] 
  = T \sum_{\nu = -\nu_{\rm max} -1}^{\nu_{\rm max}} [\cdots] 
  + \oint_{C_R} \frac{\dd z}{2 \pi \imag } n_{\rm F}(z) [\cdots],  
\end{eqnarray}
where 
\begin{equation}
  n_{\rm F}(z) = \left[\exp{(\beta z)} + 1\right]^{-1}
\end{equation}
is the Fermi-distribution function. 
On the path $C_R$, the complex frequency $z$ is represented as $z = R \exp(\imag \theta)$ where 
$R > 0$ is a fixed radius and $\theta$ is a variable angle. 
The radius $R$ must be larger than the cutoff Matsubara frequency $\w_{\nu_\mr{max}}$
and smaller than the next-higher one $\w_{\nu_\mr{max}+1}$, i.e., 
$\w_{\nu_\mr{max}} < R < \w_{\nu_\mr{max}+1}$.  
In addition, since $n_{\rm F}(z)$ in the integrand exhibits poles 
at the fermionic Matsubara frequencies, 
it is better to choose $R$ to be a bosonic Matsubara frequency, 
which is the midpoint of the two successive fermionic Matsubara frequencies. 
Therefore, we choose
\begin{equation}
  \label{eq.radius}
  R = \w_{\nu_{\mr{max}}} + \pi T = 2 (\nu_{\rm max}+1) \pi T. 
\end{equation}
Since the second term on the right-hand side of Eq.~(\ref{eq.Matsubara_sum}) 
is the contour integral of the complex logarithmic function, 
the location of branch cuts of the integrand must be examined carefully. 
It is shown in Sec.~\ref{sec.cuts} that 
the contour integral in Eq.~(\ref{eq.Matsubara_sum}) can be safely performed 
as long as the radius $R$ of the contour $C_R$ is large enough to enclose 
the poles of $\det \bs{G}'(z)$ and $\det{\tilde{\bs{G}}(\tilde{\mb{k}}, z)}$, where  
\begin{eqnarray}\label{G_CPT}
  \tilde{\bs{G}}(\tilde{\mb{k}}, z) 
  &=& \left[\bs{G}_0(\tilde{\mb{k}},z)^{-1} -\bs{\Sigma}_{\rm r}(z) \right]^{-1} \notag \\
  &=& \left[\bs{G}'(z)^{-1} -\bs{V}(\tilde{\mb{k}}) \right]^{-1}   
\end{eqnarray} 
is the approximate single-particle Green's function of the original system $\hat H$ 
within the CPT, as discussed in Sec.~\ref{cpt}, and 
$\bs{G}_0(\tilde{\mb{k}},z)$ is the Fourier transform of $\bs{G}_0(z)$, 
i.e., the noninteracting single-particle Green's function of the original system $\hat H$, 
with respect to the superlattice of the clusters. Note also that Eq.~(\ref{V_def}) is used for the second equality 
in Eq.~(\ref{G_CPT}).

The cutoff Matsubara 
frequency $\w_{\nu_{\max}}$ can be estimated as follows. 
Since $\tilde{\bs{G}}(\tilde{\mb{k}}, z)$ is the approximate 
single-particle Green's function of the original system $\hat H$ 
[see Eq.~(\ref{eq.periodization}) in Sec.~\ref{cpt}], 
we can assume that  
the largest pole is approximately given as $a(\w'_{\rm max} + W)$, where $a$ is 
a dimensionless constant of the order of $1$, $\w'_{\rm max}$ is the largest (in absolute value) 
single-particle excitation energy of the cluster,  
and $W$ is the noninteracting bandwidth of $\hat H$. 
Therefore, we can safely chose 
  $\w_{\nu_\mr{max}}$ as the minimal fermionic Matsubara frequency which satisfies 
$\w_{\nu_{\mr{max}}} > a (\w'_{\rm max} + W)$. 
We typically set $a=2$ and find that this performs efficiently. 
Note also that the largest single-particle excitation energy $\w'_{\rm max}$ 
of the cluster 
can be readily calculated by the Lanczos method for the single-particle Green's function.

Equation~(\ref{eq.Matsubara_sum}) now reduces to  
\begin{eqnarray}\label{eq.Matsubara_sum2}
  T \sum_{\nu= -\infty}^{\infty} [\cdots] 
  &=& 2T \sum_{\nu =0}^{\nu_{\rm max}} {\rm Re} [\cdots] \notag \\ 
  &+& {\rm Re} \int_{0}^  {\pi} \frac{\dd \theta R \e^{\imag \theta}}{  \pi} n_{\rm F}(R \e^{\imag \theta}) [\cdots]. 
\end{eqnarray}
Here the symmetry of the integrand with respect to the real axis is employed 
to halve the range of the sum and the integral. 
The justification for this is essentially for the same reason 
in the zero-temperature calculation~\cite{Senechal2008} [also see Eq.~(\ref{eq.antisymmetry})]. 
The integral in Eq.~(\ref{eq.Matsubara_sum2}) can be readily evaluated 
because the exact-diagonalization method 
allows one to compute the single-particle Green's function $\bs{G}'(z)$ 
for an arbitrary complex frequency $z$ 
(see Sec.~\ref{sec.BL} and Appendix~\ref{app.dG}).

Finally, we leave a note on the summation over $\tilde{\mb{k}}$ in Eq.~(\ref{eq.Omega-num}). 
In order to achieve a desired accuracy, 
the coarser $\tilde{\mb{k}}$ grid is adapted for the larger frequencies (in absolute value)
because the integrand, i.e., $\bs{G}'(z)$, becomes smoother for the frequency away from the real axis. 
Therefore, the summation over $\tilde{\mb{k}}$ should be performed  
with the different number of $\tilde{\mb{k}}$ points adapted separately for each frequency.  
This can be applied not only for the calculation of the grand-potential functional in Eq.~(\ref{eq.Omega-num}) 
but also for the calculation of other thermodynamic quantities such as entropy and specific heat as well as 
for the expectation value of single-particle operators.

\subsection{Remarks on branch cuts}\label{sec.cuts}

In order to justify Eq.~(\ref{eq.Matsubara_sum}) 
with the properly chosen cutoff Matsubara frequency in Eq.~(\ref{eq.radius}), 
we examine the branch-cut structure of $\ln \det \left [\bs{I}-\bs{V}(\tilde{\mb{k}})\bs{G}'(z) \right]$.  
For this purpose, it is useful to rewrite $\ln \det \left [\bs{I}-\bs{V}(\tilde{\mb{k}})\bs{G}'(z) \right]$ as 
\begin{eqnarray}
  \ln \det \left[\bs{I} - \bs{V}(\tilde{\mb{k}}) \bs{G}'(z) \right] 
  &=& \ln \frac{\det \bs{G}'(z)}{\det \tilde{\bs{G}}(\tilde{\mb{k}},z)} \label{eq.detratio} \label{eq.branch0}\\ 
  &=&\sum_{p=1}^{N_{\rm pole}} \ln \left( \frac{z - \w_{\tilde{\mb{k}},p}}{z - \w_{p}} \right),  
  \label{eq.branch}
\end{eqnarray}
where $\w_p$  and $\w_{\tilde{\mb{k}},p}$ are poles of
$\det \bs{G}'(z)$ and $\det \tilde{\bs{G}}(\tilde{\mb{k}},z)$, respectively, and 
$N_{\rm pole}$ is the number of poles of the determinants. 
The second equality follows from the fact that the entries of each matrix  
are the rational function of $z$ 
and thus the determinant can be written as a fraction of polynomials~\cite{Seki2017}, i.e.,  
\begin{eqnarray}
  \det{\bs{G}'}(z)
  &=& \frac{\prod_{r=1}^{N_{\rm zero}} (z - \zeta_r)}{\prod_{p=1}^{N_{\rm pole}} (z - \w_p)} \label{eq.detG'} 
\end{eqnarray}
and 
\begin{eqnarray}
  \det{\tilde{\bs{G}}}(\tilde{\mb{k}},z) 
  &=& \frac{\prod_{r=1}^{N_{\rm zero}} (z - \zeta_r)}{\prod_{p=1}^{N_{\rm pole}} (z - \w_{p,\tilde{\mb{k}}})}. \label{eq.detGcpt} 
\end{eqnarray} 
Here $\zeta_r$ is the real frequency at which the determinants become zero, e.g., $\det{\bs{G}'}(z=\zeta_r)=0$, and  
$N_{\rm zero}$ is the number of zeros of the determinants in the complex $z$ plane. 
Recalling that $\bs{G}'(z) \in \mathbb{C}^{L \times L}$ and 
$ \tilde{\bs{G}}(\tilde{\mb{k}},z) \in \mathbb{C}^{L \times L}$, 
$N_{\rm pole}$ and $N_{\rm zero}$ must be related with 
\begin{equation}
  N_{\rm pole} - N_{\rm zero} = L \label{eq.NzeroNpole} 
\end{equation}
because the diagonal elements of the Green's function decay in 
frequency as $1/z$  and the offdiagonal elements decay faster than $1/z$ for large $|z|$ 
to satisfy the anti-commutation relation of the fermion operators [see Eq.~(\ref{eq.sumrule})].  
Further analytical properties of the single-particle Green's function matrix 
can be found, 
for example, in Refs.~\cite{Eder2008,Seki2017,Dzyaloshinskii2003,Seki2016_tetra}.

\begin{figure}
  \begin{center}
    \includegraphics[width=1.0\columnwidth]{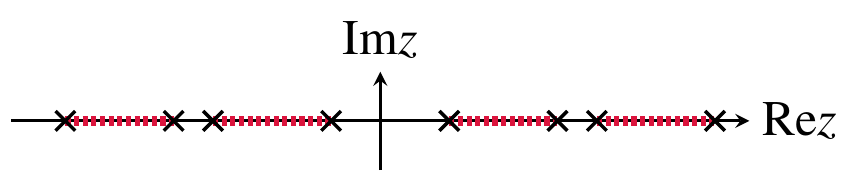}
    \caption{
      The branch cuts of $\ln \det \left[\bs{I}-\bs{V}(\tilde{\mb{k}})\bs{G}'(z)\right]$       
      with $N_{\rm pole} = 4$ in the complex $z$ plane [see Eq.~(\ref{eq.branch})].  
      The crosses represent the poles of $\det {\bs{G}'(z)}$ and $\det {\tilde{\bs{G}}(\tilde{\mb{k}},z)}$.  
      The thick red dotted lines represent the branch cuts across which  
      ${\rm Im}\ln \det \left[\bs{I}-\bs{V}(\tilde{\mb{k}})\bs{G}'(z)\right]$ 
      changes discontinuously by $\pm 2 \pi$. 
      The branch cuts are all on the real-frequency axis in a finite range 
      bounded by the largest and smallest poles of $\det \bs{G}'(z)$ and $\det \tilde{\bs{G}}(\tilde{\mb{k}},z)$.  
      \label{fig.cuts}}
  \end{center}
\end{figure}

Notice in Eqs.~(\ref{eq.detG'}) and (\ref{eq.detGcpt})  
that $\det \bs{G}'(z)$ and $\det \tilde{\bs{G}}(\tilde{\mb{k}},z)$ become 
zero at the same frequencies $z$ 
because they share the same self-energy $\bs{\Sigma}_{\rm r}(z)$ of the cluster [see Eq.~(\ref{G_CPT})].  
Therefore, the contributions of $N_{\rm zero}$ zeros in Eq.~(\ref{eq.branch0}) cancel out  
and only the contributions of $N_{\rm pole}$ poles remain in Eq.~(\ref{eq.branch}).  
It is thus clear from Eq.~(\ref{eq.branch}) that $\ln \det \left [\bs{I}-\bs{V}(\tilde{\mb{k}})\bs{G}'(z) \right]$ has 
$N_{\rm pole}$ branch cuts on the real-frequency axis with finite intervals, as schematically shown 
in Fig.~\ref{fig.cuts}. 
Therefore, as long as $C_R$ in Eq.~(\ref{eq.Matsubara_sum}) is chosen to enclose 
all these poles of $\det{\bs{G}}'(z)$ and $\det{\tilde{\bs{G}}}(\tilde{\mb{k}},z)$, 
i.e., $R > \max(|\w_p|,|\w_{\tilde{\mb{k}},p}|)$, 
the branch cuts of $\ln \det \left [\bs{I}-\bs{V}(\tilde{\mb{k}})\bs{G}'(z) \right]$ are all included inside the 
contour path and hence do not influence the calculation of  
the grand-potential functional $\Omegapersite$.

This preferable analytical property of 
$\ln \det \left[\bs{I}-\bs{V}(\tilde{\mb{k}})\bs{G}'(z)\right]$ for the contour integral 
in Eq.~(\ref{eq.Matsubara_sum}) results from 
the cancellation of the zeros of $\det \bs{G}'(z)$ and $\det \tilde{\bs{G}}(\tilde{\mb{k}},z)$. 
The cancellation occurs because the exact self-energy $\bs{\Sigma}_{\rm r}(z)$ 
in $\bs{G}'(z)$ for the cluster Hamiltonian ${\hat H}'$ is used in 
$\tilde{\bs{G}}(\tilde{\mb{k}},z)$ for the original system $\hat H$, 
which is the essential point of the SFT for deriving 
the practical quantum cluster approaches~\cite{Potthoff2003_PRL}.  
Basically the same argument is applied for the cancellation of ``$R_{\Sigma}$'' 
in Ref.~\cite{Potthoff2003b}. 
However, it should be reminded that, according to the SFT, 
the sharing of the same self-energy is not sufficient 
to eliminate the Legendre transform of 
the Luttinger-Ward potential, i.e., $\mcal{F}[{\bs\Sigma}]$. 
In order to do so, the original system of interest and the reference system must share the same ``interaction term'' 
and the self-energy.

On the other hand, Eqs.~(\ref{eq.detG'})--(\ref{eq.NzeroNpole}) 
indicate that the branch-cut structure of $\ln \det \bs{G}'(z)$ and $\ln \det \tilde{\bs{G}}(\tilde{\mb{k}},z)$ 
is different from that of $\ln \det \left[\bs{I}-\bs{V}(\tilde{\mb{k}})\bs{G}'(z)\right]$ 
because $N_{\rm pole} \not = N_{\rm zero}$. 
The branch-cut structure of these two functions is better understood 
in the extended complex plane or the Riemann sphere, 
consisting of the complex number $\mathbb{C}$ and the point at infinity $\infty$. 
In the extended complex plane, 
the number of poles ($N_{\rm pole}^{\rm ext.}$) must be the same as 
that of the zeros ($N_{\rm zero}^{\rm ext.}$) because the infinity is included~\cite{Ahlfors}.  
In the present case, the multiple $L$ zeros of 
$\det \bs{G}'(z)$ and $\det \tilde{\bs{G}}(\tilde{\mb{k}},z)$ locate at $\infty$, and thereby 
$N_{\rm zero}^{\rm ext.}=N_{\rm zero} + L=N_{\rm pole}=N_{\rm pole}^{\rm ext.}$.  
Thus, $L$ branch cuts must lie between some points on the real axis and $\infty$ for 
$\ln \det \bs{G}'(z)$ and $\ln \det \tilde{\bs{G}}(\tilde{\mb{k}},z)$.  
Therefore, the contour integrals of 
$\ln \det \bs{G}'(z)$ and $\ln \det \tilde{\bs{G}}(\tilde{\mb{k}},z)$ 
along $C_R$ should {\it not} be performed {\it separately}  
because the integral variable $z$ may cross the different branch cuts. 
Instead, the contour integral should be performed for 
the logarithm of the ratio of these two functions as in Eq.~(\ref{eq.detratio}), 
because the integrand remains on the principal branch 
and thus it is single valued through the contour integral along $C_R$ 
for sufficiently large $R$ [Eq.~(\ref{eq.radius})].

\begin{figure*}
  \begin{center}
    \includegraphics[width=16.0cm]{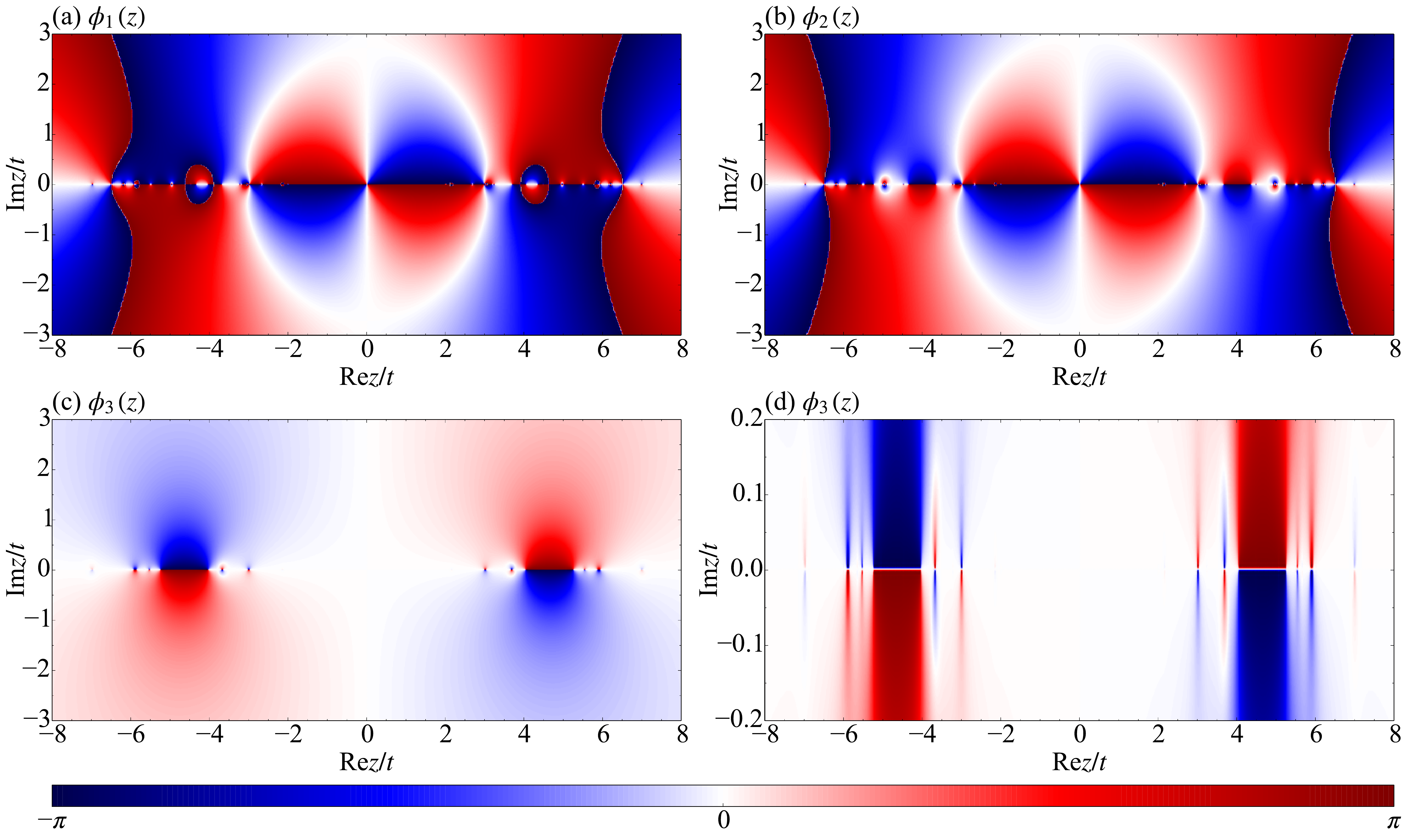}
    \caption{
      Intensity plots of 
      (a) $\phi_1 (z) = {\rm Im} \ln \det \bs{G}'_{\s}(z)$, 
      (b) $\phi_2 (z) = {\rm Im} \ln \det \tilde{\bs{G}}_{\s}(\tilde{\mb{k}},z)$, and 
      (c) $\phi_3 (z) = {\rm Im} \ln \det \left[\bs{I}-\bs{V}_{\s}(\tilde{\mb{k}})\bs{G}'_{\s}(z)\right]$     
      in the complex $z$ plane for the Hubbard model on the square lattice with $L_{\rm c} = 2 \times 2$ and 
      $\tilde{\mb{k}}=(0,0)$. The other parameters are $U/t=8$, $\mu/t=4$, and $T/t=0.1$. 
      (d) Enlarged figure of (c) near the real axis. 
      The phases $\phi_1(z)$, $\phi_2(z)$, and $\phi_3(z)$ are plotted in the range of 
      $-\pi < \phi_i \leqslant \pi$, as indicated by the color bar. 
      \label{fig.cuts2}}
  \end{center}
\end{figure*}

To better understand the analytical properties of these logarithm-determinant functions 
appearing in the SFT, Fig.~\ref{fig.cuts2} shows
the imaginary parts of these functions, i.e., 
\begin{eqnarray}
  \phi_1(z)&=& {\rm Im} \ln \det \bs{G}'_{\s}(z), \\
  \phi_2(z)&=& {\rm Im} \ln \det \tilde{\bs{G}}_{\s}(\tilde{\mb{k}},z), 
\end{eqnarray}
and
\begin{eqnarray}
  \phi_3(z)&=& {\rm Im} \ln \det \left[\bs{I}-\bs{V}_{\s}(\tilde{\mb{k}})\bs{G}_{\s}'(z)\right], 
\end{eqnarray}  
numerically calculated for the single-band Hubbard model $\hat H$ on the square lattice 
with a reference system of $L_{\rm c} = 2 \times 2$ site cluster (see Fig.~\ref{fig.clusters}) 
at $U/t=8$, $\mu/t=4$, $T/t=0.1$, and $\tilde{\mb{k}}=(0,0)$, assuming 
the same one-body terms as in $\hat H$ (i.e., no variational parameters) for the reference system. 
Here, $\bs{G}'_{\s}(z)$, $\tilde{\bs{G}}_{\s}(\tilde{\mb{k}},z)$, and $\bs{V}_{\s}(\tilde{\mb{k}})$ denote 
the block-diagonal elements of $\bs{G}'(z)$, $\tilde{\bs{G}}(\tilde{\mb{k}},z)$, and $\bs{V}(\tilde{\mb{k}})$ 
with respect to the spin index $\sigma$, respectively, e.g., $\bs{G}'(z)=\bs{G}'_{\up}(z) \oplus \bs{G}'_{\dn}(z)$. 
Thus, their matrix dimension is $L_{\rm c} \times L_{\rm c}$. 
The range of phases is $-\pi < \phi_i(z) \leqslant \pi$ for $i=1,2$, and $3$, as indicated in Fig.~\ref{fig.cuts2}.  
The branch cuts are therefore located at the boundaries where a sudden change  
of the color from blue to red (from $-\pi$ to $\pi$) and vice versa occurs in Fig.~\ref{fig.cuts2}.

It is first noticed in Fig.~\ref{fig.cuts2} that the phases $\phi_1(z)$, $\phi_2(z)$, and $\phi_3(z)$ are 
antisymmetric in the complex $z$ plane with respect to the real axis, i.e., 
\begin{equation}
  \phi_i(z^*)=-\phi_i(z). 
  \label{eq.antisymmetry}
\end{equation}
This is readily shown from the fact that 
$\bs{G}'(z^*) = \bs{G}'(z)^\dag$, 
$\bs{V}(\tilde{\mb{k}}) = \bs{V}(\tilde{\mb{k}})^\dag$, and 
$\det\bs{A}^\dag = \left(\det \bs{A} \right)^*$ for a regular matrix $\bs{A}$.  
The antisymmetry with respect to the imaginary axis, i.e., $\phi_i(-z)=-\phi_i(z)$, found in 
Fig.~\ref{fig.cuts2} is due to 
the particle-hole symmetry for this example.

More interestingly, Figs.~\ref{fig.cuts2}(a) and \ref{fig.cuts2}(b) show clearly that 
both $\ln \det \bs{G}_{\s}(z)$ and 
$\ln \det \tilde{\bs{G}}_{\s}(\tilde{\mb{k}},z)$ have 
branch cuts located in the complex $z$ plane off the real axis, in addition to branch cuts on the real axis. 
In particular, we can find the four ($=L_{\rm c}$) branch cuts 
connecting branch points on the real axis and the infinity.  
On the other hand, 
the branch cuts of $\ln \det \left[\bs{I}-\bs{V}_{\s}(\tilde{\mb{k}})\bs{G}'_{\s}(z)\right]$ 
are all on the real axis, as shown in Figs.~\ref{fig.cuts2}(c) and \ref{fig.cuts2}(d). 
Therefore, the contour integral of $\ln \det \left[\bs{I}-\bs{V}_{\s}(\tilde{\mb{k}})\bs{G}'_{\s}(z)\right]$ 
is well defined as long as the radius of the path $C_R$ is large enough,  
while the contour integrals of $\ln \det \bs{G}_{\s}'(z)$ and $\ln \det \tilde{\bs{G}}_{\s}(\tilde{\mb{k}},z)$ are 
not well defined in general.

The analytical properties of the logarithm-determinant functions examined here are 
also essential for the analysis of the grand-potential functional $\Omega$ 
in Appendix~\ref{app.anotherOmega}, where the application of the kernel-polynomial method 
(KPM)~\cite{Weisse2006} for the VCA is also discussed.

\subsection{Entropy and specific heat}\label{thermodynamic}
Thermodynamic quantities such as entropy $S$ and specific heat $C$ are derived 
from temperature derivatives of the grand potential.  
It should be noted however that the grand potential depends on the temperature both explicitly 
and implicitly.
The explicit dependence is from the Boltzmann factor in the grand potential and 
the single-particle Green's function of the reference system [see Eqs.~(\ref{eq.Omega-num}), 
(\ref{eq:omega}), and (\ref{eq.G})]. 
The implicit dependence is due to the fact that the optimal variational parameters 
$\bs{\lambda}^{*}(T)$ depend on the temperature. 
This is because the stationary condition
\begin{equation}
  \left. \frac{\partial \Omega(T,\bs{\lambda})}{\partial \bs{\lambda}} \right|_{\bs{\lambda}=\bs{\lambda}^{*}} = \bs{0}
\end{equation}
gives the temperature dependent optimal variational parameters $\bs{\lambda}^{*}(T)$ 
(for example, see Fig.~\ref{grandpotential}), despite the fact that the variational parameters 
$\bs{\lambda}$ themselves are independent of the temperature.
Therefore, the temperature dependence of the grand potential should be considered as 
$\Omega = \Omega(T, \bs{\lambda}^{*}(T))$. 
The implicit dependence on the external magnetic field of the grand-potential functional
have already been pointed out in Refs.~\cite{Eder2010chi} and \cite{Balzer2010}.

The entropy $S$ is the first derivative of the grand potential 
with respect to the temperature and is given as   
\begin{eqnarray}
  S 
  = -\frac{\dd \Omegapersite}{\dd T} 
  =
  -\frac{\partial \Omegapersite}{\partial T} 
  - \frac{\dd \bs{\lambda}^{*}}{\dd T} \cdot
  \frac{\partial \Omegapersite}{\partial\bs{\lambda}^{*}} .
  \label{eq.ST}
\end{eqnarray}
The second term of the right hand side of Eq.~(\ref{eq.ST}) is zero because near the stationary point 
$\bs{\lambda}^{*}$ at a fixed $T$ the grand potential has a quadratic form
\begin{equation}
  \Omegapersite(\bs{\lambda}^{*}+\bs{h}) \approx \Omegapersite(\bs{\lambda}^{*}) 
  + \frac{1}{2} \sum_{i,j} \left. \frac{\partial^2 \Omegapersite}{\partial \lambda_i \partial \lambda_j}\right|_{\bs{\lambda}=\bs{\lambda}^{*}} h_i h_j,
\end{equation}
where $\bs{h}=\bs{\lambda} - \bs{\lambda}^{*}$, and 
therefore $\partial \Omegapersite(T,\bs{\lambda}^{*}) / \partial \bs{\lambda}^{*} = \bs{0}$.
The entropy per site is thus 
\begin{eqnarray}
  S 
  &=& \frac{1}{L_{\rm c}} S'  \nonumber \\
  &+& \frac{1}{NL_{\mr{c}}} \sum_{\nu=-\infty}^\infty \sum_{\tilde{\mb{k}}} \left(1 + \frac{1}{\beta} {\rm D}_T \right)  
  \ln \det 
  \left[ 
    \bs{I}-\bs{V}(\tilde{\mb{k}}) \bs{G}'(\imag \w_\nu)
  \right], 
  \label{SFTentropy}
\end{eqnarray}
where $S' = -\partial \Omega' /\partial T$ is the exactly calculated entropy of the cluster.  
Equation~(\ref{SFTentropy}) can be derived 
from the $T$-derivative of Eq.~(\ref{eq.Omega-num}) 
by taking into account the $T$ dependence of the Matsubara frequencies. 
In Appendix~\ref{app.ST}, we also show that Eq.~(\ref{SFTentropy}) 
can be derived by converting the sum over Matsubara frequencies 
into the contour integral involving the Fermi-distribution function. 
In the above equation, we have introduced the following temperature derivative operator 
[also see Eqs.~(\ref{T-derivative1}) and (\ref{T-derivative2})] 
\begin{equation}
  {\rm D}_T := \frac{\partial}{\partial T} + \imag \w_\nu \beta \frac{\partial }{\partial (\imag \w_\nu)}. 
\end{equation}
The last term of Eq.~(\ref{SFTentropy}) is then given as
\begin{equation}\label{SS}
  {\rm D}_T \ln \det (\bs{I}-\bs{VG}')  
  = -\mr{tr} \left[ (\bs{I}-\bs{VG}')^{-1} \bs{V} ({\rm D}_T \bs{G}') \right],
\end{equation}
where $\partial_T \ln \det \bs{A} (T) = \mr{tr} \left[ \bs{A}(T)^{-1} \partial_T \bs{A}(T) \right]$ is 
used for any regular and differentiable matrix $\bs{A}(T)$.
The infinite sum of Matsubara frequencies in the right-hand side of Eq.~(\ref{SFTentropy}) can be 
decomposed into the finite sum of Matsubara frequencies and the contour integral, as in 
Eq.~(\ref{eq.Matsubara_sum}), 
because the frequency derivative of 
$\ln \det  \left[  \bs{I}-\bs{V}(\tilde{\mb{k}}) \bs{G}'(\imag \w_\nu) \right]$ 
simply results in the sum of discrete poles distributed within a finite range on the real-frequency axis 
[see Eq.~(\ref{eq.branch})].

The specific heat $C$ is obtained by the second derivative of the grand potential 
with respect to the temperature and is given as
\begin{eqnarray}
  C 
  = - T \frac{\dd^2\Omegapersite}{\dd T^2} 
  = - T \frac{\partial^2 \Omegapersite}{\partial T^2} 
  - T \frac{\dd \bs{\lambda}^{*}}{\dd T} \cdot \frac{\partial^2 \Omegapersite}{\partial \bs{\lambda}^{*} \partial T}.
  \label{eq.CT}
\end{eqnarray}
The first term in the right-hand side of Eq.~(\ref{eq.CT}) is expressed as 
\begin{eqnarray}
  \label{C-matrix}
  &-& T 
  \frac{\partial^2 \Omegapersite}{\partial T^2}
  =  \frac{1}{L_{\rm c}} C' \notag \\
  &+& \frac{1}{NL_{\mr{c}}} \sum_{\nu=-\infty}^\infty \sum_{\tilde{\mb{k}}} 
  \left( \frac{2}{\beta} {\rm D}_T + \frac{1}{\beta^2} {\rm D}_T^2 \right) 
  \ln \det 
  \left[ 
    \bs{I}-\bs{V}(\tilde{\mb{k}}) \bs{G}'(\imag \w_\nu)
  \right],
\end{eqnarray}
where $C' = -T \partial^2 \Omega' / \partial T^2$ is the exactly calculated specific heat of the cluster.
The last term in the right hand side of Eq.~(\ref{C-matrix}) is given as 
\begin{eqnarray} \label{CC}
  {\rm D}_T^2 \ln \det (\bs{I}-\bs{VG}') = 
  &-& 
  {\rm tr} \left[ \left\{ (\bs{I}-\bs{VG}')^{-1} \bs{V} ({\rm D}_T   \bs{G}') \right\}^2 \right] \notag \\ 
  &-&
  {\rm tr} \left[ (\bs{I}-\bs{VG}')^{-1} \bs{V} ({\rm D}_T^2 \bs{G}') \right],
\end{eqnarray}
where $\partial_T \bs{A}(T)^{-1} = - \bs{A}(T)^{-1} \bigl[\partial_T \bs{A}(T)\bigr] \bs{A}(T)^{-1}$ 
is used. 
Note that, in contrast to the entropy, the second term in the right hand side of Eq.~(\ref{eq.CT}) 
does not vanish in general. 
Since the variational parameter dependence of the grand potential is not analytically known, 
the specific heat can be calculated much easier by numerically 
differentiating the entropy or the grand potential with respect to $T$.

Before ending this subsection, three remarks are in order. 
First, derivatives of $\bs{G}'(z)$ with respect to $T$ and $z$ are required for 
${\rm D}_T \bs{G}'(\imag \w_\nu)$ and ${\rm D}_T^2 \bs{G}'(\imag \w_\nu)$.  
Since $\bs{G}'(z)$ depends on $T$  
only through the Boltzmann factor [see Eq.~(\ref{eq.G})], 
$\partial_T \bs{G}'(z)$ and $\partial_T^2 \bs{G}'(z)$ are easily obtained.  
More specifically, $\partial_T \bs{G}'(z)$ and $\partial_T^2 \bs{G}'(z)$ are
obtained by replacing the factor $\e^{\beta (\Omega' - E_s)}$ in Eq.~(\ref{eq.G}) with
\begin{eqnarray}
\partial_T   \e^{\beta (\Omega' - E_s)} &=& \e^{\beta (\Omega' - E_s)} (E_s - E') \beta^2 
\end{eqnarray}
and 
\begin{eqnarray}
\partial_T^2 \e^{\beta (\Omega' - E_s)} &=& \e^{\beta (\Omega' - E_s)} 
\left\{
(E_s - E')^2 \beta^4 \right. \notag \\
&-&
\left. C' \beta^2 -2(E_s - E') \beta^3 \right\},
\end{eqnarray}
respectively, where 
\begin{equation}
  E'= \Omega' + T S'
\end{equation}
is the internal energy of the cluster. 
Note that from our definition of the Hamiltonian in Eq.~(\ref{eq.ham}) 
the internal energy $E'$ includes the chemical-potential term. 
$\partial_z \bs{G}'(z)$ and $\partial_z^2 \bs{G}'(z)$  
can be easily evaluated when $\bs{G}'(z)$ is given in the Lehmann representation (see Sec.~\ref{sec.Lehmann}). 
However, if the single-particle Green's function $\bs{G}'(z)$ is evaluated by the continued-fraction expansion, 
the evaluation of $\partial_z \bs{G}'(z)$ and $\partial_z^2 \bs{G}'(z)$ is slightly involved 
and the detail is summarized in Appendix~\ref{app.dG}. 

Second, one may tempt to rewrite 
$ \left[ \bs{I}-\bs{V}(\tilde{\mb{k}}) \bs{G}'(\imag \w_\nu)  \right]^{-1} \bs{V}(\tilde{\mb{k}}) 
=  \left[ \bs{V}^{-1}(\tilde{\mb{k}}) - \bs{G}'(\imag \w_\nu) \right]^{-1} $ 
in Eqs.~(\ref{SS}) and (\ref{CC}), assuming that $\bs{V}(\tilde{\mb{k}})$ is a regular (invertible) 
matrix for arbitrary $\tilde{\mb{k}}$. 
However, this is often not the case 
because several eigenvalues of the Hermitian matrix $\bs{V}(\tilde{\mb{k}})$, which 
describes the inter-cluster hopping terms as defined in Eq.~(\ref{V_def}), 
are often zero and thus $\bs{V}^{-1}(\tilde{\mb{k}})$ does not always exist.

Third, we have regarded that the chemical potential $\mu$ is independent of temperature $T$ 
as in the grand canonical ensemble, i.e., $\Omega = \Omega\bigl(T, \mu, \bs{\lambda}^{*}(T)\bigr)$. 
However, generally one would fix the particle density $n$ 
by tuning the chemical potential at a given $T$. In this case, 
the implicit dependence on the temperature of the grand potential through the 
chemical potential should also be considered, i.e., $\Omega = \Omega\bigl(T, \mu(T), \bs{\lambda}^{*}(T)\bigr)$.

\subsection{Reference system}\label{reference}

As described in Sec.~\ref{method.sft}, the reference system is composed of disconnected clusters 
and the Hamiltonian in each cluster is described as  
\begin{equation}
  \label{eq.Href}
  \hat{H}'       = \hat{H} +  \hat{H}_{h'}, 
\end{equation}
where $\hat{H}$ is the same Hamiltonian as in Eq.~(\ref{eq.ham}) but is defined only within the cluster with 
open boundary conditions, and 
\begin{equation}
  \label{eq.v-parm}
  \hat{H}_{h'}   = h'    \sum_{i} \e^{\imag \mb{Q} \cdot \bs{r}_i} \left( \hat{n}_{i \up} - \hat{n}_{i \dn} \right). 
\end{equation}
The second term $\hat{H}_{h'}$ introduces the variational 
magnetic field $h'$ 
in order to investigate the antiferromagnetism in the Hubbard model. 
Here, $\mb{Q} = (\pi,\pi)$ 
and ${\bs r}_i$ represents the location of site $i$ in a cluster. 
Since we consider the particle-hole symmetric case, 
the particle density can be kept at half filled ($n=1$) without 
introducing the variational site-independent energy~\cite{Aichhorn2006_muvar}.

Although the variational magnetic field is applied along the $z$ direction in Eq.~(\ref{eq.v-parm}), 
the solutions for in-plane and out-of-plane antiferromagnetism are 
degenerated. 
We consider the out-of-plane antiferromagnetism because 
the $z$-component of spin is conserved in $\hat{H}_{h'}$. 

\begin{figure}
  \begin{center}
    \includegraphics[width=1.0\columnwidth]{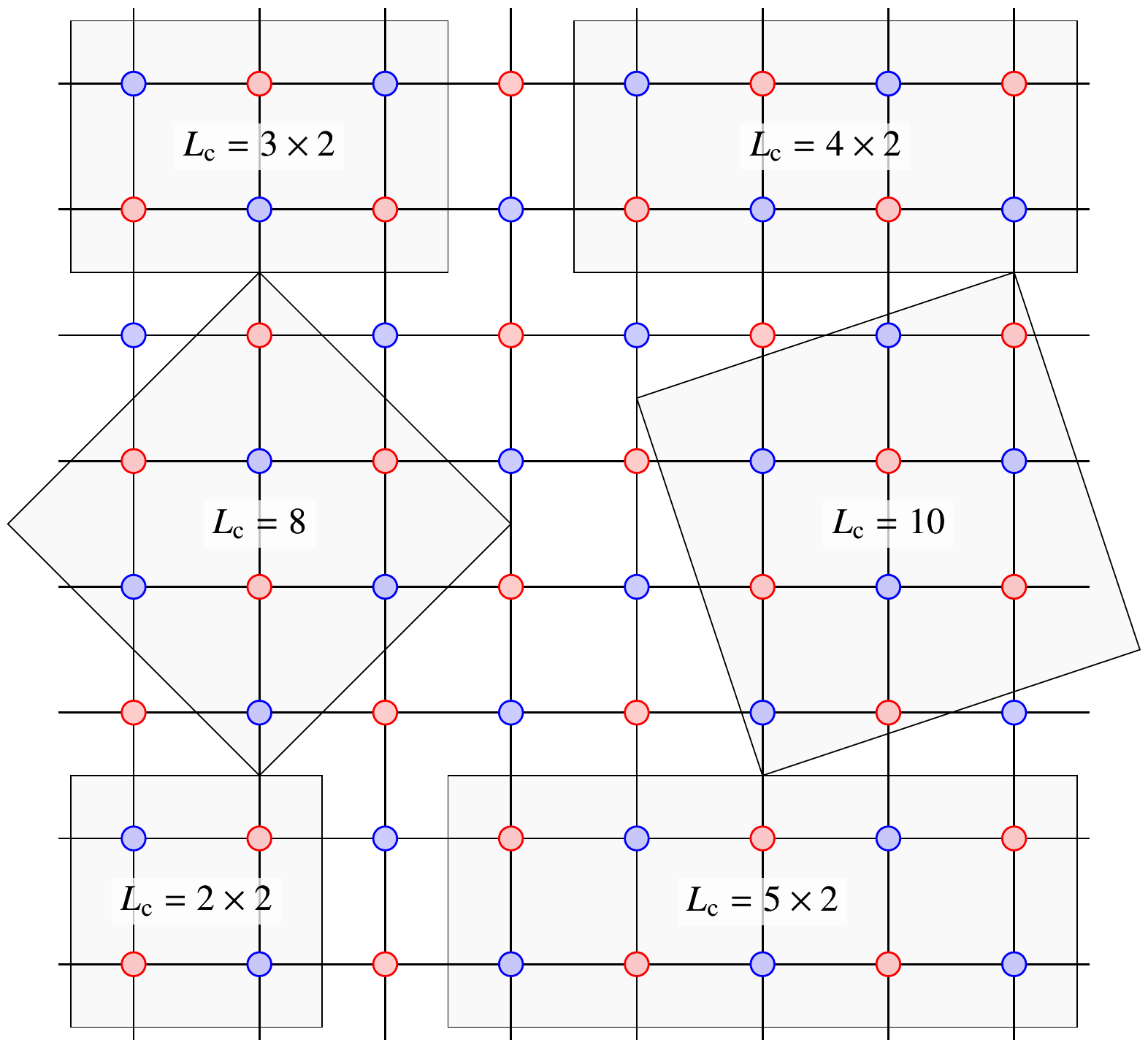}
    \caption{
      Clusters used in this study are indicated by solid lines. The size of each cluster is also indicated. 
      The primitive translational vectors of each cluster is summarized in Table~\ref{table.R1R2}. 
      The red (blue) circles represent sites on sublattice $A$ ($B$) of the square lattice. 
      \label{fig.clusters}}
  \end{center}
\end{figure}

The optimal variational parameter $h'^*$ is 
determined so as to satisfy the stationary condition 
\begin{equation}
\left. \frac{\partial \Omega}{\partial h'} \right|_{h' = h'^{*}} 
= 0. 
\label{eq:sc0}
\end{equation} 
A solution with $h'^{*} \not= 0$ corresponds to an antiferromagnetic state.
The clusters used here are shown in Fig.~\ref{fig.clusters}.  
The corresponding primitive translational vectors $\mb{R}_1$ and $\mb{R}_2$
for each cluster are given in Table~\ref{table.R1R2}.

\begin{table}
  \caption{
    \label{table.R1R2}    
    The primitive translational vectors $\mb{R}_1$ and $\mb{R}_2$ 
    of the clusters shown in Fig.~\ref{fig.clusters}. 
  }
  \begin{tabular}{ccc}
    \hline \hline
    Cluster  & $\mb{R}_1$ & $\mb{R}_2$  \\
    \hline
    $2 \times 2$ & $(2,0)$ & $(0,2)$  \\
    $3 \times 2$ & $(3,1)$ & $(0,2)$  \\
    $4 \times 2$ & $(4,0)$ & $(0,2)$  \\
    $ 8 $        & $(2,2)$ & $(-2,2)$ \\
    $5 \times 2$ & $(5,1)$ & $(0,2)$ \\
    $10$         & $(3,1)$ & $(-1,3)$  \\
    \hline \hline
  \end{tabular}
\end{table}

\subsection{Cluster perturbation theory}\label{cpt}

The quantum-cluster methods including the VCA break the translational symmetry and 
thus an appropriate prescription is necessary to obtain the translationally invariant single-particle 
Green's functions~\cite{Senechal2012,Sakai2012}. 
For this purpose, here, we employ the CPT~\cite{Senechal2000,Senechal2002,Senechal2012},  
in which the single-particle Green's function is given as
\begin{equation}
  \mcal{G}^{\s \s'} (\mb{k}, z) = 
  \frac{1}{L_{\mr{c}}}  
  \sum_{i,j}^{L_{\mr{c}}} 
  \left[ \tilde{\bs{G}}^{\s \s'} (\mb{k}, z)\right]_{ij} 
  \e^{-\imag \mb{k} \cdot (\mb{r}_i - \mb{r}_j )}, 
  \label{eq.periodization}
\end{equation}
where $\mb{r}_{i}$ is the position of site $i$ within a cluster and 
$\tilde{\bs{G}}^{\s \s'} (\mb{k}, z)$ is the $(\s,\s')$ element of $\tilde{\bs{G}}(\mb{k}, z)$ 
defined in Eq.~(\ref{G_CPT}). 
The CPT is readily extended to finite temperatures by 
using the single-particle Green's function of a cluster at finite temperatures given in 
Eq.~(\ref{eq.G})~\cite{Aichhorn2003,Kawasugi2016}.

The CPT is exact both in the noninteracting limit $(U=0)$ and in the atomic limit $(t=0)$, 
and is expected to be a good approximation in strongly interacting regime 
since it is derived originally from the strong coupling expansion for the single-particle Green's functions. 
The CPT approximation is practically improved with increasing the cluster size $L_{\rm c}$ and 
becomes exact for $L_{\rm c}\to\infty$, independently of $U/t$~\cite{Senechal2000}.  
When the exact-diagonalization cluster solver is employed, 
the size of clusters which can be treated is rather limited, 
typical clusters being shown in Fig.~\ref{fig.clusters}, especially at finite temperatures 
where higher excited states are required. However, the CPT can treat 
spatial fluctuations exactly within a cluster and is expected to be 
a better approximation at high temperatures (e.g., $T \gg t^2/U$ for the Hubbard model) 
for a given finite-size cluster because the spatial fluctuations generally become short-ranged at high temperatures. 
Indeed, quantum Monte Carlo studies for relatively large system sizes~\cite{Preuss1995,Preuss1997,Groeber2000} 
has shown 
that at high temperatures the dispersion relation which can be identified 
in the single-particle excitation spectrum of the single-band Hubbard model 
resemble those obtained by the Hubbard-I approximation~\cite{Groeber2000,Hubbard1963,Gebhard1997}, 
which neglects the spatial correlations and corresponds to the CPT approximation with 
$L_{\rm c}=1$.

\subsection{Comparison with previous formalism}

Here we briefly summarize the previous VCA studies at finite temperatures 
and compare those finite temperature schemes with our formalism developed here. 

The VCA with a single bath-impurity cluster was first applied for the single-band Hubbard model 
to study metal-insulator transitions at finite temperatures~\cite{Potthoff2003b} 
and later for a particle-hole asymmetric Hubbard model away from the half filling~\cite{Eckstein2007}. 
The extension to multi-band Hubbard models has also been reported~\cite{Inaba2005_PRB, Inaba2005_JPSJ}. 
The thermodynamics and the single-particle excitations at finite temperatures 
for a periodic Anderson model~\cite{Eder2007} and the multi-band Hubbard models for  
3$d$ transition-metal oxides combined with the realistic band-structure calculation 
have also been reported~\cite{Eder2008, Eder2015}. 
Moreover, a finite-temperature VCA algorithm for Hubbard-like models with a  
continuous-time quantum Monte Carlo (CTQMC) 
cluster solver has been proposed to examine the temperature dependence of 
thermodynamic quantities for the single-band Hubbard model~\cite{Li2009}.

From the technical point of view, these previous finite-temperature VCA methods except for 
Ref.~\cite{Li2009} are based on 
the analytical expression of the grand-potential functional at finite temperatures~\cite{Potthoff2003b}, 
which requires the explicit evaluation of the poles of $\bs{G}'(z)$ and $\tilde{\bs{G}}(\tilde{\mb{k}},z)$ 
(aslo see Appendix~\ref{app.anotherOmega}). 
The poles of the single-particle Green's functions 
can be obtained either by numerically solving the 
nonlinear equations of $\det{\bs{G}'(z)}^{-1}=0$ and
$\det{\bs{G}(\tilde{\mb{k}},z)}^{-1}=0$~\cite{Inaba2005_PRB, Inaba2005_JPSJ, Eder2007} 
or employing the $Q$-matrix method~\cite{Aichhorn2006_Qmat}. 
The $Q$-matrix method gives the poles of the single-particle Green's functions 
as eigenvalues of a momentum-dependent Hermitian matrix. 
Since solving the nonlinear equation is in general less stable than the eigenvalue problem, 
the $Q$-matrix method can be considered as a preferable method to solve 
$\det{\bs{G}'(z)}^{-1}=0$ and $\det{\bs{G}(\tilde{\mb{k}},z)}^{-1}=0$. 
Although the $Q$-matrix method gives accurate results, 
the dimension of the Hermitian matrix is as large as the number of 
the \textit{pair} of the excited states in the cluster 
and thus the method rapidly becomes unfeasible at finite temperatures~\cite{note1}. 
For example, the number of poles of the single-particle Green's function 
for the single-band Hubbard model on an eight-site cluster at half filling, 
which can be fully diagonalized without difficulties, 
exceeds $\mcal{O}(10^5)$  at $T/t=0.35$ even if the truncation scheme in Eq.~(\ref{truncation}) is employed. 
Since the number of poles corresponds to the dimension of the momentum-dependent Hermitian matrix 
to be diagonalized in the $Q$-matrix method, the diagonalization of the Hermitian matrix is difficult 
to be performed in the realistic computational time. 
This is the main reason why the previous finite-temperature VCA studies have been limited for relatively 
small clusters, especially, when the exact-diagonalization cluster solver is employed.

In this paper, we propose another scheme for 
the finite-temperature VCA with the exact-diagonalization cluster solver, 
which is a natural extension of the scheme at zero temperature~\cite{Senechal2008}. 
The main advantage of our method is that it requires neither 
the explicit evaluation of the poles of $\bs{G}(z)$ nor $\tilde{\bs{G}}(\tilde{\mb{k}},z)$. 
Instead, the grand-potential functional is calculated with 
the simple matrix operations of $\bs{G}'(z)$ and $\bs{V}(\tilde{\mb{k}})$ and 
the simple numerical line integrals in the complex plane, by taking full account of 
the analytical properties of the finite-temperature single-particle Green's functions. 
This has a significant advantage in saving computational time, which thus allows one to 
treat the larger clusters as compared with the previous studies. 
Our scheme is based on the same idea which has been proposed earlier in Ref.~\cite{Eder2008} 
but the integrand of the grand-potential functional in our scheme is as simple as 
that in the zero-temperature calculation~\cite{Senechal2008}. 
This simplification is indeed justified 
by analyzing the analytical properties of the integrand in Sec.~\ref{sec.cuts}. 
Our method should be considered to be complementary to the finite-temperature VCA with 
the CTQMC cluster solver, which often encounters difficulties at low temperatures~\cite{Li2009}.

Recently, a method for the finite-temperature VCA 
on quantum computers has been reported~\cite{Dallaire-Demers2016}. 
They have considered a two-site Hubbard cluster as an example and 
shown that the grand-potential functional varies in a large energy scale of $\sim 20t$ 
with the change of the variational parameters (within the energy scale of $\sim t$)  
even for the noninteracting case~\cite{Dallaire-Demers2016}. 
However, the self-energy should vanish in the noninteracting limit and 
therefore no variational-parameter dependence 
of the grand-potential functional is expected. 
Although there may be some issues to be solved, the finite-temperature VCA 
on quantum computers is certainly an interesting direction for the future research.

\section{Block Lanczos method for a single-particle Green's function}
\label{sec.BL}

Since the single-particle Green's functions of the cluster have to be calculated repeatedly in Eqs.~(\ref{eq.Ge}) 
and (\ref{eq.Gh}), 
the finite-temperature VCA is computationally $s_{\rm max} + 1$ times more demanding 
than the zero-temperature VCA. 
Therefore, an efficient evaluation of the single-particle Green's functions of the cluster is crucial. 
This section is devoted to describe the block-Lanczos method to 
evaluate the single-particle Green's functions in the Lehmann representation. 
First, we summarize the following three points (i), (ii), and (iii) to explain 
why the block-Lanczos method is preferable to the finite-temperature VCA

(i) As described in details in this section, the block-Lanczos method  
can be faster than the standard Lanczos method 
to calculate the single-particle Green's functions of the cluster 
at the expense of additional memory storage for the block-Lanczos vectors. 
  
(ii) The block-Lanczos method is robust against 
the loss of orthogonality of Lanczos vectors as compered to the standard Lanczos method. 
This is because the Lanczos vectors are explicitly orthonormalized 
within the block size $L$ at each block-Lanczos step [see Eq.~(\ref{BL3})]. 
Therefore, the block-Lanczos method can describe the excited states and hence
compute the excitation spectrum more accurately than the standard Lanczos method. 
This advantage of block-Lanczos method holds also 
for solving the eigenvalue problem of the cluster Hamiltonian. 
We employ the block-Lanczos method to compute low-lying eigenvalues and eigenstates 
$\{E_s, |\Psi_s \rangle \}$ of the cluster Hamiltonian when the dimension $N_s$ of 
the Hilbert space of the cluster for a given subspace 
(labeled by, e.g., particle number, $z$-component of total spin, and 
point-group symmetry) is large (typically when $N_s \geqslant 10000$). 
Otherwise, we use the LAPACK routines~\cite{lapack} to find all or selected $\{E_s, |\Psi_s \rangle \}$, 
according to the truncation scheme in Eq.~(\ref{truncation}). 

(iii) The block-Lanczos method can be even more efficient 
than the band-Lanczos method~\cite{Freund} in computational time.   
This is because the block-Lanczos method 
can be implemented on the basis of the level-3 BLAS and LAPACK routines~\cite{lapack} 
due to the block-wise extension of the Krylov space. 
For example, the block-diagonal entries $\bs{A}_j$ and the block-subdiagonal entries $\bs{B}_j$ 
of the Hamiltonian matrix $\bs{T}_k$ can be constructed by 
a matrix-matrix multiplication and a QR factorization, respectively 
[see Eqs.~(\ref{BL1})--(\ref{BL3}) and (\ref{Bandmatrix})]. 
The matrix-vector multiplication required for the block-Lanczos method in Eq.~(\ref{BL1}) 
can also be implemented efficiently as the sparse-matrix by tall-skinny-matrix multiplication,
where the sparse matrix is the Hamiltonian matrix $\bs{H}'$ and 
the tall-skinny matrix is the set of the block-Lanczos vectors $\bs{Q}_k$.   

The block-Lanczos method coincides with 
the band-Lanczos method if the deflation 
(i.e., deletion of almost linearly dependent vectors 
during the process of extending the Krylov space) 
does not occur~\cite{Freund}. 
In our experience, the deflation may occur when noninteracting orbitals are introduced 
as in the CDIA. 
However, in the VCA, we have not met the necessity of the deflation so far. 
Therefore, the block-Lanczos method is still useful for the VCA. 

In the following, we describe the block-Lanczos method. 
Sections~\ref{sec.Lehmann} and \ref{sec.basis} are devoted to preliminaries, while 
Secs.~\ref{sec.inivec}, \ref{sec.blm}, and \ref{sec.sumrule}
are devoted to technicalities for a practical implementation of the block-Lanczos method.

\subsection{Lehmann representation}\label{sec.Lehmann}

Inserting the identity operator $\hat{1}=\sum_{r=1}^{N_{\rm st}^\pm} |\Psi_r^\pm \ket \bra \Psi_r^\pm|$ 
into Eqs.~(\ref{eq.Ge}) and (\ref{eq.Gh}) yields the Lehmann representation of the single-particle 
Green's function 
\begin{eqnarray}
  G_{ij,s}^{+}(z)   
  &=& \sum_{r=1}^{N_{\rm st}^+} \frac{
    \bra \Psi_s |  \hat{c}_i      | \Psi_r^+ \ket
    \bra \Psi_r^+ |  \hat{c}_j^\dag | \Psi_s \ket}
  {z - (E_r^+ - E_s)} 
  \label{eq.GeLehmann}
\end{eqnarray}
for the particle-addition part and 
\begin{eqnarray}
  G_{ij,s}^{-}(z)   
  &=& \sum_{r=1}^{N_{\rm st}^-} \frac{
    \bra \Psi_s |  \hat{c}_j^\dag | \Psi_r^- \ket
    \bra \Psi_r^- |  \hat{c}_i      | \Psi_s \ket}
  {z + (E_r^- - E_s)} 
  \label{eq.GhLehmann}
\end{eqnarray} 
for the particle-removal part, where  
$i\,(=1,2,\cdots,L)$ represents the generalized single-particle index, including 
the site and spin indices, and $| \Psi_s \ket$ ($ | \Psi_r^\pm \ket$) is the eigenstate of the cluster Hamiltonian 
$\hat H'$ in the $N$ ($N\pm1$) electron subspace with its eigenvalue $E_s$ ($E_r^\pm$)~\cite{Note_pointgroup}. 
The dimension of the Hilbert space for $\hat H'$ in the ($N\pm1$)-electron subspace is denoted as 
$N_{\rm st}^\pm$. 
The exact single-particle Green's functions in the Lehmann representation given in 
Eqs.~(\ref{eq.GeLehmann}) and ~(\ref{eq.GhLehmann}) are 
evaluated when the full diagonalization of the Hamiltonian matrix is possible with a reasonable amount of 
computational time.

However, the exponential growth of the dimension of the Hilbert space for $\hat H'$ 
restricts the full diagonalization to, e.g., $L_{\rm c} \leqslant 8$ 
for the half-filled single-band Hubbard model in practice.  
Therefore, the Lanczos method is often applied to calculate the single-particle Green's functions 
of the cluster with $L_{\rm c}\geqslant 10$ by taking advantage of the sparsity of the Hamiltonian 
matrix~\cite{Prelovsek,Fulde1995,Dagotto1994}. 
Since the CPT and the VCA prefer the open-boundary clusters to better approximate 
the infinite system~\cite{Senechal2002,Potthoff2003_PRL}, 
the momentum of the cluster is not a good quantum number. 
Moreover, in the VCA, variational parameters 
which break point-group, time-reversal, or gauge symmetry of the cluster Hamiltonian 
are often introduced to examine possible symmetry-breaking states.  
Thus, in the standard Lanczos method with a single Lanczos vector, 
at most $O(L^2)$ Lanczos procedures 
are required to obtain all elements of $\bs{G}^+_s(z)$ and $\bs{G}^-_s(z)$.

On the other hand, in the block-Lanczos method, 
two block-Lanczos procedures are sufficient, each for $\bs{G}^+_s(z)$ and $\bs{G}^-_s(z)$, 
to calculate the single-particle Green's functions. 
The number of matrix-vector multiplications, which are the most numerically demanding, 
is then reduced by a factor of $L$ in the block-Lanczos method, as compared with the standard Lanczos 
method, at a cost of the memory workspace for keeping two sets of $L$ Lanczos vectors.

As in the standard Lanczos method for the single-particle Green's function~\cite{Balzer2011}, 
Eqs.~(\ref{eq.GeLehmann}) and (\ref{eq.GhLehmann}) are approximately computed 
in the block-Lanczos method by truncating the intermediate (single-particle excitated) states as 
\begin{eqnarray}
  G_{ij,s}^{+}(z)   
  &\approx& \sum_{l=1}^{M^+} 
  \frac{
    \bra \Psi_s   |  \hat{c}_i      | \psi_l^+ \ket
    \bra \psi_l^+ |  \hat{c}_j^\dag | \Psi_s \ket}
  {z - (\eps_l^+ - E_s)} 
  \label{eq.GeLehmannBL}
\end{eqnarray}
and 
\begin{eqnarray}
  G_{ij,s}^{-}(z)   
  &\approx& \sum_{l=1}^{M^-} 
  \frac{
    \bra \Psi_s   |  \hat{c}_j^\dag | \psi_l^- \ket
    \bra \psi_l^- |  \hat{c}_i      | \Psi_s \ket}
  {z + (\eps_l^- - E_s)},
  \label{eq.GhLehmannBL}
\end{eqnarray} 
where $\eps_l^\pm$ and $|\psi_l^\pm \ket$ are approximate (Ritz) eigenvalue and eigenstate of $\hat{H}'$ 
in the $(N\pm 1)$-electron subspace obtained by the block-Lanczos method 
and $M^\pm$ is the number of the excited states calculated 
for the particle-addition/removal spectrum.

This approximation can be considered as an approximation 
for the Hamiltonian in the $(N \pm 1)$-electron subspace. 
The exact spectral representation of the Hamiltonian is given as 
\begin{equation} 
  \hat{H}' = \hat{P}_{\rm Eig} \hat{H}' \hat{P}_{\rm Eig} = \sum_{r=1}^{N_{\rm st}^{\pm}} E_r^\pm | \Psi_r^\pm \ket \bra \Psi_r^\pm|, 
\end{equation} 
where $\hat{P}_{\rm Eig} = \sum_{r=1}^{N_{\rm st}^{\pm}} |\Psi_r^\pm \ket \bra \Psi_r^\pm| \, (=\hat{1})$ is the 
projection operator with the exact eigenstates $|\Psi_r^\pm \ket$. 
Accordingly, the resolvent is given as 
\begin{equation}
  \left[z \mp \left(\hat{H}' - E_s\right) \right]^{-1} = \sum_{r=1}^{N_{\rm st}^{\pm}} \frac{|\Psi_r^\pm \ket \bra \Psi_r^\pm|}{z \mp (E_r^\pm - E_s)},  
\end{equation}
On the other hand, the spectral representation of the Hamiltonian is approximated in 
Eqs.~(\ref{eq.GeLehmannBL}) and (\ref{eq.GhLehmannBL}) as  
\begin{equation}
  \hat{H}' \approx \hat{P}_{\rm Ritz} \hat{H}' \hat{P}_{\rm Ritz} = \sum_{l=1}^{M^{\pm}} \eps_{l}^{\pm} | \psi_l^{\pm} \ket \bra \psi_l^{\pm}|, 
\end{equation} 
where $\hat{P}_{\rm Ritz} = \sum_{l=1}^{M^{\pm}} | \psi_l^{\pm} \ket \bra \psi_l^{\pm} |$ is the projection 
operator with the Ritz states $| \psi_l^{\pm} \ket$~\cite{Prelovsek,Jaklic2000}. 
Accordingly, the resolvent is approximated as 
\begin{equation}
  \left[z \mp \left(\hat{H}' - E_s\right) \right]^{-1} \approx \sum_{l=1}^{M^{\pm}} \frac{|\psi_l^{\pm} \ket \bra \psi_l^{\pm}|}{z \mp (\eps_l^{\pm} - E_s)}.   
\end{equation}
As described below, the Ritz states should be obtained from the block-Lanczos procedure 
starting with appropriate initial states as in Eqs.~(\ref{initialstates}) and (\ref{initialstates2}).  

For simplicity, we shall focus on 
the particle-addition part of the single-particle Green's functions $G_{ij,s}^+(z)$ in Eq.~(\ref{eq.GeLehmannBL}) 
and describe how the block-Lanczos method can be applied to accelerate the calculation. 
However, the following argument is applied 
straightforwardly to 
the particle-removal part of the single-particle Green's functions $G_{ij,s}^-(z)$ in Eq.~(\ref{eq.GhLehmannBL}).

\subsection{Numerical representation of operators and states}\label{sec.basis}

In the exact diagonalization method, 
the second-quantized operators and many-body states 
are represented in the many-body configuration basis $| x \ket$, 
e.g., direct products of local electron configurations~\cite{Jafari2008}, 
which form the complete orthonormal system, i.e.,  
\begin{eqnarray} 
  \sum_{x} | x \ket \bra x | &=&  \hat{1} \label{complete} 
\end{eqnarray}
and 
\begin{eqnarray}
  \bra x | x' \ket &=& \delta_{x,x'}. \label{orthonormal} 
\end{eqnarray}
For example, an operator $\hat{O} = \sum_{x,x'} |x \ket \bra x | \hat{O} | x' \ket \bra x'|$ is represented 
as a matrix $\bs{O}$ with the matrix element 
\begin{equation}
  [{\bs{O}}]_{xx'} = \bra x | \hat{O} | x' \ket,  
\end{equation}
and a many-body state $|\phi \ket = \sum_{x} \bra x | \phi \ket | x \ket $ as a vector $\bs{\phi}$ 
with the vector component 
\begin{equation}
  [\bs{\phi}]_x = \bra x | \phi \ket. 
\end{equation}

\subsection{Initial vectors for block-Lanczos method}\label{sec.inivec}

On the analogy of the standard Lanczos method 
for dynamical correlation functions~\cite{Prelovsek,Fulde1995,Dagotto1994}, 
we consider a set of one-electron added states 
\begin{equation}
  \hat{c}_1^\dag \left|\Psi_s \right\ket, \ \hat{c}_2^\dag \left|\Psi_s \right\ket, \quad \cdots, \quad \hat{c}_L^\dag \left| \Psi_s \right \ket, 
  \label{initialstates}
\end{equation}
and represent them as a single rectangular matrix $\bs{S} \in \mathbb{C}^{N_{\rm st}^{+} \times L}$ with the matrix element 
\begin{equation} 
  \label{Sdef} 
  \left[\bs{S}\right]_{xi}  = \bra x | \hat{c}_i^\dag | \Psi_s \ket. 
\end{equation}
Note that the $L$ column vectors contained in $\bs{S}$ are not orthonormalized in general. 
Since the block-Lanczos algorithm requires the initial vectors to be orthonormalized~\cite{Chatelin}, 
we apply the QR factorization to obtain the orthonormal vectors, i.e., 
\begin{equation}
  \label{SQR}
  \bs{S} = \bs{Q}_1 \bs{B}_0,
\end{equation}
where $\bs{Q}_1 \in \mathbb{C}^{N_{\rm st}^{+} \times L}$ 
is composed of $L$ orthonormal column vectors, 
\begin{equation}
  \bs{Q}_1^\dag \bs{Q}_1 = \bs{I}_{L} 
  \label{eq.Qorthonormal}
\end{equation}
with $\bs{I}_{L} $ being the $(L\times L)$ unit matrix, 
and $\bs{B}_0 \in \mathbb{C}^{L \times L}$ is an upper-triangular matrix. 
For the QR factorization in Eq.~(\ref{SQR}) and 
also later in Eq.~(\ref{BL3}), we employ 
the Cholesky QR2 algorithm~\cite{Yamamoto2014,Fukuya2014}, 
which is found faster than the Householder QR or 
the modified Gram-Schmidt methods for most cases studied here. 

The static correlation function can be calculated as 
\begin{eqnarray}
  \bra \Psi_s | \hat{c}_i \hat{c}_j^\dag | \Psi_s \ket 
  = \sum_{x} \bra \Psi_s | \hat{c}_i | x \ket \bra x | \hat{c}_j^\dag | \Psi_s \ket 
  &=& [\bs{S}^\dag \bs{S}]_{ij} \notag \\
  &=& [\bs{B}_0^\dag \bs{B}_0]_{ij}. 
  \label{B0B0}
\end{eqnarray}
This is analogous to the standard Lanczos method [see Eq.~(\ref{b02})].

\subsection{Block Lanczos method}\label{sec.blm}

The block-Lanczos method first prepares
the $L$ column vectors $\bs{Q}_1$ defined in Eq.~(\ref{SQR}) for the initial block-Lanczos vector 
and constructs successively the block-Lanczos vectors $\bs{Q}_2, \bs{Q}_3, \cdots, \bs{Q}_{k_{\rm max}^+}$ 
by iterating the following procedures: 
\begin{eqnarray}
  \label{BL1}
  \bs{A}_k &:=& \bs{Q}_{k}^\dag \bs{H}' \bs{Q}_{k} \\
  \label{BL2}
  \bs{X}_k &:=& \bs{H}' \bs{Q}_{k} - \bs{Q}_{k} \bs{A}_k  - \bs{Q}_{k-1} \bs{B}_{k-1}^\dag  \\
  \label{BL3}
  \bs{X}_k &=:& \bs{Q}_{k+1}\bs{B}_k 
\end{eqnarray} 
for $k=1$ to  $k_{\rm max}^+$~\cite{Chatelin}. 
Here $\bs{Q}_0 := \bs{0}$ and  
$\left[ \bs{H}' \right]_{xx'} = \bra x | \hat{H}' | x' \ket$ is the matrix representation of 
the cluster Hamiltonian $\hat{H}'$ given in Eq.~(\ref{eq.Href}). 
The procedure in Eq.~(\ref{BL3}) should be read as the QR factorization 
of $\bs{X}_k \in \mathbb{C}^{N_{\rm st}^{+} \times L}$ yielding the $(k+1)$st block-Lanczos vector 
$\bs{Q}_{k+1}\in \mathbb{C}^{N_{\rm st}^{+} \times L}$ and an upper-triangular matrix 
$\bs{B}_k \in \mathbb{C}^{L \times L}$. 
The procedure in Eq.~(\ref{BL1}) requires 
$L$ matrix-vector multiplications to construct $\bs{H}' \bs{Q}_k$.  
Note also that $\bs{A}_k \in \mathbb{C}^{L \times L}$ is Hermitian 
since $\bs{H}' \in \mathbb{C}^{N_{\rm st}^+ \times N_{\rm st}^+}$ is Hermitian. 
As shown in the following, $M^+ = k_{\rm max}^+ L$ is the number of poles in 
the particle-addition part of the single-particle Green's function for $|\Psi_s\ket$ [see Eq.~(\ref{eq.GeLehmannBL})].  
We typically take $M^+ \lesssim~ 300$   
as in the zero-temperature calculations~\cite{Senechal2008}.

Let us define 
$\bs{Q}_{\rm L} 
:= \left[ \bs{Q}_1, \cdots, \bs{Q}_{k_{\rm max}} \right] 
\in \mathbb{C}^{N_{\rm st}^{+} \times M^+}$ 
in which $M^+$ Lanczos vectors are contained. 
The Lanczos vectors are orthonormalized, i.e., 
\begin{equation}
  \bs{Q}_{\rm L}^\dag \bs{Q}_{\rm L}=\bs{I}_{M^+}.
  \label{QQI}
\end{equation} 
Defining the Lanczos state $|q_m\ket$ by 
\begin{equation}
  \bra x | q_m\ket = [\bs{Q}_{\rm L}]_{xm},   
  \label{def_lanczosstate}
\end{equation}
Eq.~(\ref{QQI}) is simply rewritten as  
\begin{equation} 
  \bra q_m | q_n \ket = \delta_{m,n}.     
  \label{qq}
\end{equation}
Thus the Lanczos states are orthonormalized. 
However, since $M^+ \ll N_{\rm st}^{+}$ in practice, 
the Lanczos states may not form a complete set for the 
$(N+1)$-electron Hilbert space. 
In other words, the Lanczos method allows one to 
approximate many-body states within the limited number $M^+$ 
of the orthonormalized basis states $|q_m\ket$.

After the procedure (\ref{BL1}) of the $k$th block-Lanczos iteration, 
a matrix representation $\bs{T}_k$ of the cluster Hamiltonian $\hat{H}'$ 
in the Lanczos basis  
\begin{equation}
  [\bs{T}_k]_{mn} = \bra q_m | \hat{H}' | q_n \ket  
  \label{eq.Tk}
\end{equation}
can be constructed. It is readily found from Eqs.~(\ref{BL1})-(\ref{BL3}) that 
$\bs{Q}_{j'}^\dag \bs{H}' \bs{Q}_{j} = \bs{A}_j \delta_{j',j} + \bs{B}_j \delta_{j',j+1} + \bs{B}_{j'}^\dag \delta_{j',j-1}$ 
and thus the reduced Hamiltonian matrix 
$\bs{T}_k \in \mathbb{C}^{kL \times kL}$ is a Hermitian-band matrix with a bandwidth $L$ containing 
$\bs{A}_j$ with $j=1,2,\cdots,k$ in the diagonal and 
$\bs{B}_j$ ($\bs{B}_j^\dag$) with $j=1,2,\cdots,k-1$ in the subdiagonal (superdiagonal) blocks, i.e.,  
\begin{equation}
  \label{Bandmatrix}
  \bs{T}_k =
  \left[
    \begin{array}{ccccc}
      \bs{A}_1 &  \bs{B}^\dag_1 & 0 & \cdots & 0      \\
      \bs{B}_1 &  \bs{A}_2      & \bs{B}^\dag_2 & \ddots & \vdots \\
      0 & \ddots        & \ddots & \ddots & 0 \\
      \vdots & \ddots & \bs{B}_{k-2} & \bs{A}_{k-1} & \bs{B}^\dag_{k-1} \\
      0 & \cdots & 0 & \bs{B}_{k-1} & \bs{A}_k 
    \end{array}
  \right]. 
\end{equation}

The Ritz state can be obtained as follows. Let us define $\bs{T} := \bs{T}_{k_{\rm max}^+}$.   
Since $\bs{T}$ is Hermitian, there exist a unitary matrix $\bs{U}$ and a diagonal matrix $\bs{D}$ such that 
\begin{equation}
  \bs{D} = \bs{U}^\dag \bs{T} \bs{U} = {\rm diag}(\eps_1^+, \cdots, \eps_{M^+}^+).  
\end{equation} 
Recalling in Eq.~(\ref{eq.Tk}) that $\bs{T}$ is a matrix representation of $\hat{H}'$ in the Lanczos states, 
i.e.,  $[\bs{T}]_{mn} = \sum_{x x'} \bra q_m | x \ket \bra x | \hat{H}' | x' \ket \bra x' | q_n \ket $, or 
equivalently  
\begin{equation}
  \bs{T} = \bs{Q}_{\rm L}^\dag \bs{H}' \bs{Q}_{\rm L},  
\end{equation}
we find that 
\begin{eqnarray}
  \bs{D}&=& \left(\bs{Q}_{\rm L} \bs{U} \right)^\dag \bs{H}' \left(\bs{Q}_{\rm L} \bs{U} \right). 
\end{eqnarray}
Therefore, the Ritz state $|\psi_l^+\ket$ which satisfies 
$\hat{H}'|\psi_l^+ \ket = \eps_l^+ |\psi_l^+ \ket$ is given by 
\begin{equation} 
 \bra x | \psi_l^+ \ket = [\bs{Q}_{\rm L}\bs{U}]_{xl}. 
 \label{Ritzvector}
\end{equation}
In terms of the Lanczos states $| q_m \ket$, the Ritz state $|\psi_l^+\ket$ can be represented as 
\begin{equation}
  | \psi_l^+ \ket = \sum_{x} | x \ket \bra x | \psi_l^+ \ket = \sum_{m = 1}^{M^+} [\bs{U}]_{ml} | q_m \ket, 
  \label{Ritzstate}
\end{equation}
i.e., 
the linear combination of the $M^+ (\ll N_{\rm st}^+)$ Lanczos states $|q_m\ket$ 
with the coefficients being the eigenvectors $[\bs{U}]_{ml}$ 
of the reduced Hamiltonian matrix $\bs{T}$.    
It is readily found from the orthonormality of the Lanczos states in Eq.~(\ref{qq}) that 
\begin{equation}
  \bra \psi_l^+ | \psi_m^+ \ket = \delta_{lm}.  
\end{equation} 
Since the Ritz states are orthonormalized, we can define a projection operator 
\begin{equation}
  \hat{P}_{\rm Ritz} = \sum_{l=1}^{M^+} | \psi_l^+ \ket \bra \psi_l^+ |, 
  \label{eq.P_Ritz}
\end{equation}
which satisfies $\hat{P}_{\rm Ritz}^2=\hat{P}_{\rm Ritz}$ 
and acts as an identity operator for linear combinations of the Ritz states, e.g., 
$\hat{P}_{\rm Ritz} \left(\sum_l a_l |\psi_l^+ \ket \right) = \sum_l a_l |\psi_l^+ \ket$ 
with $a_l$ being complex number. 
Inserting this projection operator into Eq.~(\ref{eq.Ge}), we finally obtain 
the approximated single-particle Green's function given in Eq.~(\ref{eq.GeLehmannBL}). 
The Ritz values $\eps_1^+, \eps_2^+,\cdots, \eps_{M^+}^+$ of $\bs{T}$ thus 
correspond to the poles of the single-particle Green's function in Eq.~(\ref{eq.GeLehmannBL}).

\subsection{Spectral-weight sum rule and high-frequency expansion of single-particle Green's function}\label{sec.sumrule}

Now we consider the spectral weight of the single-particle Green's function 
which appears in the numerator of Eq.~({\ref{eq.GeLehmannBL}}).
From Eqs.~(\ref{complete}), (\ref{Sdef}) and (\ref{Ritzvector}), we find that 
\begin{eqnarray}
  \bra \psi_l^+ | \hat{c}_j^\dag | \Psi_s \ket 
  &=&\left[\bs{U}^\dag \bs{Q}_{\rm L}^\dag \bs{S} \right]_{lj} 
  = \left[\bs{U}^\dag \bs{Q}_{\rm L}^\dag \bs{Q}_1 \bs{B}_0 \right]_{lj} \notag \\ 
  &=&\sum_{n=1}^{L} \left[\bs{U}^\dag\right]_{ln} \left[\bs{B}_{0}\right]_{nj}  
  \label{matelm1} 
\end{eqnarray} 
Therefore, the spectral weight does not require 
the set of Lanczos vectors $\bs{Q}_{\rm L}$  to be stored 
but instead only rather smaller matrices $\bs{U}$ and $\bs{B}_0$.  
The upper bound $L$ of the sum over $n$ in Eq.~(\ref{matelm1}) 
can be replaced by $j$ because $\bs{B}_0$ is the upper-triangular matrix.

Here we show that the spectral-weight sum rule 
is satisfied for the single-particle Green's function~\cite{Luttinger1961,NO} represented with 
the block-Lanczos basis in Eqs.~(\ref{eq.GeLehmannBL}) and (\ref{eq.GhLehmannBL}). 
For the numerator of the particle-addition part of the single-particle Green's function in Eq.~(\ref{eq.GeLehmannBL}), 
we find from Eqs.~(\ref{B0B0}) and (\ref{matelm1}) that 
\begin{eqnarray}
  \sum_{l=1}^{M^+} \bra \Psi_s | \hat{c}_i | \psi_l^+ \ket \bra \psi_l^+ |\hat{c}_j^\dag |\Psi_s \ket 
  &=& \bra \Psi_s | \hat{c}_i \hat{c}_j^\dag |\Psi_s \ket. 
\end{eqnarray}
Similarly, for the particle-removal part of the single-particle Green's function in Eq.~(\ref{eq.GhLehmannBL}), 
we can find that  
\begin{eqnarray}
  \sum_{l=1}^{M^-} \bra \Psi_s | \hat{c}_j^\dag | \psi_{l}^- \ket \bra \psi_{l}^- | \hat{c}_i |\Psi_s \ket 
  &=& \bra \Psi_s | \hat{c}_j^\dag \hat{c}_i |\Psi_s \ket, 
\end{eqnarray}
provided that the initial block-Lanczos states are chosen as 
\begin{equation}
  \hat{c}_1 \left|\Psi_s \right\ket, \  \hat{c}_1 \left|\Psi_s \right\ket, \quad \cdots, \quad \hat{c}_L \left| \Psi_s \right \ket, 
  \label{initialstates2}
\end{equation}
instead of those given in Eq.~(\ref{initialstates}). 
We thus find for a high frequency $|z|\to\infty$ that the single-particle Green's function in 
Eq.~(\ref{eq.G}) is 
\begin{eqnarray}
  G'_{ij}(z) 
  &=& \frac{1}{z} 
  \sum_{s=0}^{s_{\rm max}} \e^{\beta(\Omega' - E_s)} 
  \bra \Psi_{s} | \left\{\hat{c}_i, \hat{c}_j^\dag \right\} |\Psi_s \ket + \mcal{O}\left(\frac{1}{z^2} \right)\notag \\
  &=& \frac{\delta_{ij}}{z} + \mcal{O} \left(\frac{1}{z^2} \right),    
  \label{eq.sumrule}
\end{eqnarray}
where $\{\hat{c}_i, \hat{c}_j^\dag \}=\hat{c}_i \hat{c}_j^\dag+\hat{c}_j^\dag\hat{c}_i = \delta_{ij}$ 
is used. Therefore, the block-Lanczos method respects the spectral-weight sum rule 
of the single-particle Green's function.

Next, we show that the high-frequency expansion 
of the single-particle Green's function can also be easily obtained when the 
Lehmann representation of the single-particle Green's function is available. 
The high-frequency expansion of the single-particle Green's function can be written as 
\begin{equation}
  G'_{ij}(z) = \sum_{k=0}^{\infty} \frac{\mcal{M}_{ij}^{(k)}} {z^{k+1}}, \label{eq.highw}
\end{equation}
where 
\begin{equation}
  \mcal{M}_{ij}^{(k)} = \oint \frac{\dd z}{2 \pi \imag} z^k G'_{ij}(z)
  \label{Mij}
\end{equation}
is the $k$th moment of the single-particle Green's function~\cite{Harris1967,Senechal2002,Seki2011,Sakai2012}. 
The contour in Eq.~(\ref{Mij}) should enclose in a counter-clockwise manner 
all poles of the single-particle Green's function, which are 
on the real frequency axis distributed within a limited range of frequency. 
Since $G'_{ij}(z)$ is given by Eqs.~(\ref{eq.G}), (\ref{eq.GeLehmannBL}) and (\ref{eq.GhLehmannBL}), 
the contour integral in Eq.~(\ref{Mij}) can be performed readily as 
\begin{eqnarray}
  \mcal{M}_{ij}^{(k)}
  &=& \sum_{s=0}^{s_{\rm max}} \e^{\beta(\Omega'-E_s)} \notag \\
  &\times&
  \left[\sum_{l=1}^{M^+} \left(\eps_l^+ - E_s \right)^k \bra \Psi_s | \hat{c}_i | \psi_l^+ \ket \bra \psi_l^+ | \hat{c}_j^\dag |\Psi_s \ket  \right. \notag \\
  &+&
  \left. \sum_{l=1}^{M^-} \left(E_s - \eps_{l}^- \right)^k \bra \Psi_s | \hat{c}_j^\dag | \psi_{l}^- \ket \bra \psi_{l}^- | \hat{c}_i |\Psi_s \ket \right]. 
  \label{Mij2}
\end{eqnarray}  
Equation~(\ref{Mij2}) thus shows that matrix $\mcal{M}_{ij}^{(k)}$ is Hermitian, i.e.,  
\begin{equation}
  \mcal{M}_{ji}^{(k)} = \left(\mcal{M}_{ij}^{(k)} \right)^*,
\end{equation}  
and the single-particle Green's function satisfies 
\begin{equation}
  G'_{ji}(z)=\left(G'_{ij}(z^*)\right)^*.
\end{equation} 
Note that $\mcal{M}_{ij}^{(0)} = \delta_{ij}$ due to the anti-commutation relation 
of the fermion operators as shown in Eq~(\ref{eq.sumrule}).

The high-frequency expansion in Eq.~(\ref{eq.highw}) can significantly 
reduce the computational cost for $G'_{ij} (z)$ 
especially at high temperatures.  
This is because $\mcal{M}_{ij}^{(k)}$ is independent of the frequency $z$ 
and therefore the ``once and for all'' calculation of $\mcal{M}_{ij}^{(k)}$ is sufficient, 
while the calculation from Eqs.~(\ref{eq.G}),~(\ref{eq.GeLehmannBL}) and (\ref{eq.GhLehmannBL}) 
requires $\mcal{O}(N_{\rm pole})$ operations 
for each complex frequency $z$. 
For example, even for the $L_{\rm c}=8$ cluster, the number 
$N_{\rm pole} = (s_{\rm max}+1) \times (M^+ + M^-)$ of poles with nonzero spectral weight reaches $ \sim \mcal{O}(10^7)$ 
if all the excited states ($s_{\rm max}+1=4^8$) are necessary, e.g., at high temperatures. 
The high-frequency expansion of $G'_{ij}(z)$ is useful to evaluate 
the contour-integral part, i.e., 
the second term of the right-hand side in Eq.~(\ref{eq.Matsubara_sum}), 
of the grand-potential functional.   
In general, 
the value of the integral evaluated using the high-frequency expansion of $G'_{ij}(z)$ up to the 15th-order 
in Eq.~(\ref{eq.highw})
agrees with 
that calculated using the full Lehmann-represented $G'_{ij}(z)$ in
Eqs.~(\ref{eq.G}),~(\ref{eq.GeLehmannBL}), and (\ref{eq.GhLehmannBL})   
within the accuracy of approximately ten digits.

Finally, it should be emphasized that the selection of the initial block-Lanczos vectors 
in Eqs.~(\ref{initialstates}) and (\ref{initialstates2}) for the particle-addition and 
particle-removal spectra, respectively, is crucial to justify 
the approximations in Eqs.~(\ref{eq.GeLehmannBL}) and (\ref{eq.GhLehmannBL}), 
as in the Lanczos method for dynamical correlation functions 
with a single initial vector~\cite{Dagotto1994,Prelovsek}. 
The importance of the selection of the initial block-Lanczos vectors also 
resembles the recently proposed 
block-Lanczos density-matrix-renormalization-group method, where the block-Lanczos transformation 
maps general multi-orbital multi-impurity Anderson models 
to quasi-one-dimensional models with keeping the two-body interactions local if the initial block-Lanczos 
states are properly chosen~\cite{Shirakawa2014}.

\section{Application of finite-temperature VCA}\label{results}

In this section, we demonstrate the finite-temperature scheme of the VCA proposed here 
by exploring the finite-temperature properties 
of the two-dimensional single-band Hubbard model on the square lattice described by the 
Hamiltonian in Eq.~(\ref{eq.ham}) at half filling with considering the antiferromagnetic order 
in the reference system 
[Eqs.~(\ref{eq.Href})--(\ref{eq.v-parm})].

\subsection{N\'{e}el Temperature}

\begin{figure}
  \begin{center}
    \includegraphics[width=7.5cm]{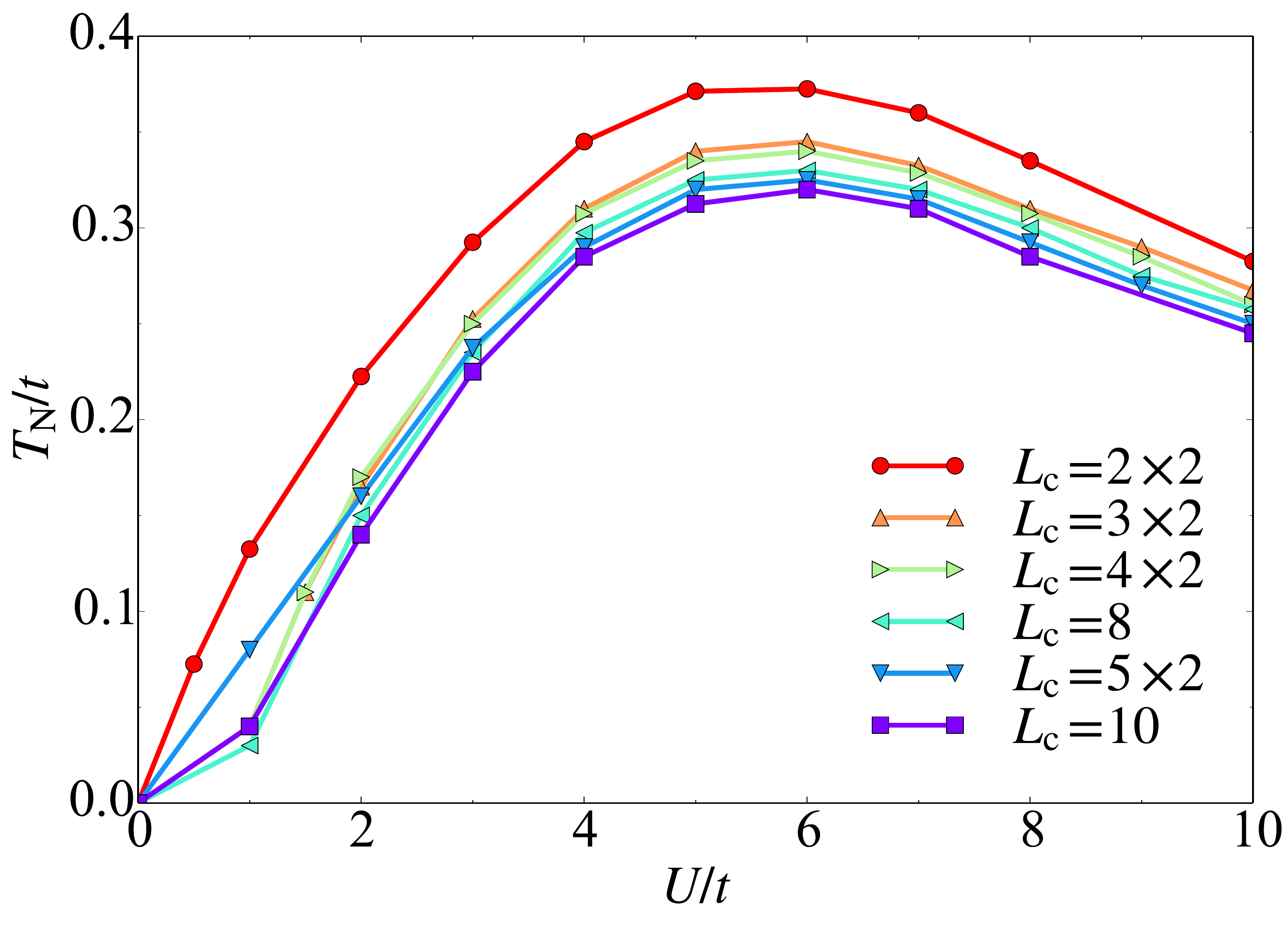}
    \caption{
      The N\'{e}el temperature $T_{\rm N}$ 
      obtained by the VCA 
      with various clusters indicated in the figure 
      (also see Fig.~\ref{fig.clusters} and Table~\ref{table.R1R2}). 
      \label{phasediagram}}
  \end{center}
\end{figure}

The $U$-dependence of the N\'{e}el temperature $T_{\rm N}$  
for various clusters is shown in Fig.~\ref{phasediagram}. 
Although the Mermin-Wagner theorem prohibits any continuous symmetry breaking at finite temperatures 
in two dimensions~\cite{Mermin1966}, 
the VCA finds $T_{\rm N}>0$ for the clusters studied here. 
This is because the VCA neglects the longer-range correlations beyond the cluster size. 
The VCA can describe the quantum fluctuations exactly within a cluster, 
while the antiferromagnetic correlations beyond the cluster are treated 
in a mean-field level by introducing the variational parameter which explicitly 
breaks the symmetry as in Eq.~(\ref{eq.v-parm}). 
Indeed, a systematic study for the finite-size scaling of $T_{\rm N}$ 
in the dynamical-cluster approximation with a QMC solver at $U/t=8$ has shown 
that $T_{\rm N}$ approaches to zero logarithmically with increasing the size of clusters~\cite{Maier2005}. 
We also note that the magnitude of $T_{\rm N}$ reasonably agrees with
a CDMFT study for the Hubbard model on the square lattice~\cite{Sato2016}. 
As expected in Fig.~\ref{phasediagram}, the larger clusters tend to show smaller $T_{\rm N}$, 
although this is not the case for $U/t \lesssim 2$ where the finite-size effect 
on $T_{\rm N}$ seems significant.

Nevertheless, as shown in Fig.~\ref{phasediagram}, 
$T_{\rm N}$ shows a maximum around $U/t \sim 6$, 
independently of the size of clusters in the reference system, 
and decreases as $T_{\rm N} \propto J=4t^2/U$, expected in the large $U$ regime 
where the half-filled Hubbard model is approximated by the spin $1\over 2$ antiferromagnetic Heisenberg model 
with the exchange interaction $J=4t^2/U$. 
We also find 
that the $U$ dependence of $T_{\rm N}$ is rather similar to 
that of the optimal variational parameter $h'^*$ at $T=0$ (see Fig. 6 in Ref.~\cite{Senechal2008})  
than the order parameter $m$ at $T=0$ times $U$, 
the latter being expected in the 
spin-density-wave (SDW) mean-field theory.

\subsection{Grand-potential functional}

\begin{figure*}
  \begin{center}
    \includegraphics[width=1.0\textwidth]{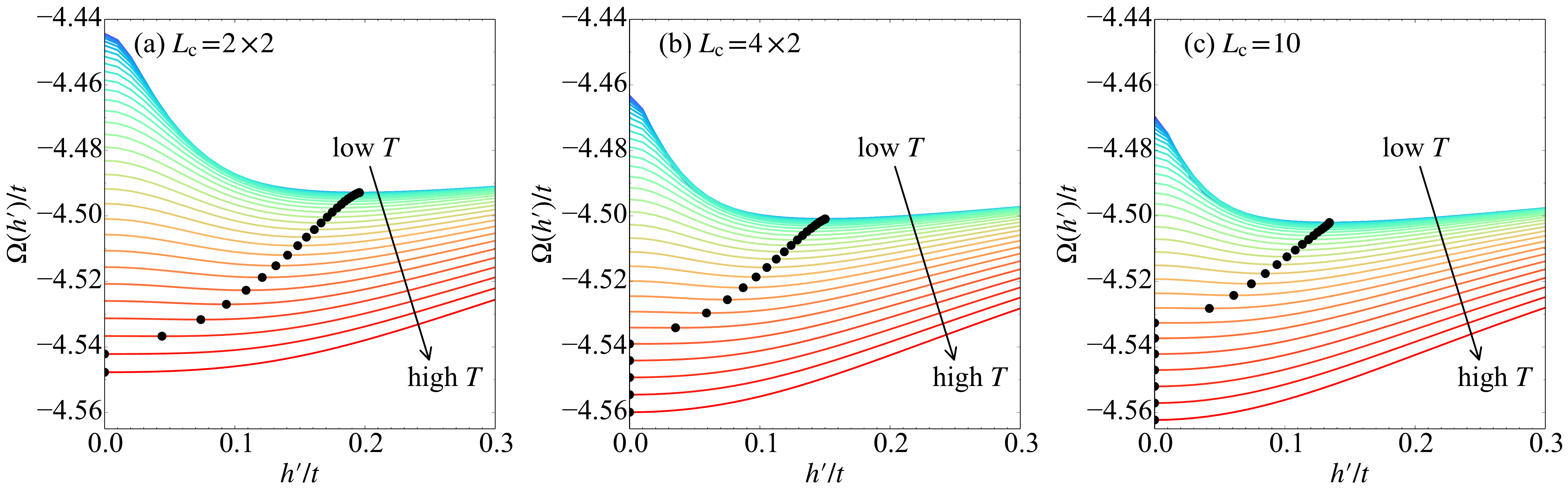}
    \caption{
      The grand-potential functional $\Omega$ 
      as a function of variational parameter $h'$ for $U/t=8$ at temperatures  
      $T/t=0.01, 0.02, \cdots, 0.34$, and $0.35$ (from violet to red lines). The chemical potential 
      is set at $\mu=U/2$ . 
      The clusters used are 
      (a) $2 \times 2$, 
      (b) $4 \times 2$, and 
      (c) $10$ sites. 
      Each dot indicates the variational parameter $h'^*$ where the stationary condition is satisfied 
      with the lowest grand potential for a given temperature. 
      The solution with $h'^* \not=0$ indicates the antiferromagnetic state, 
      while $h'^* = 0$ corresponds to the paramagnetic state. 
      Notice that the grand-potential functionals for $T/t \leqslant 0.04$ 
      are almost degenerate in the cases studied here. 
      \label{grandpotential}}
  \end{center}
\end{figure*}

Figure~\ref{grandpotential} shows the results of the grand-potential 
functional as a function of variational parameter 
$h'$ at $U/t=8$ with $\mu=U/2$ for various temperatures 
using $L_{\rm c} = 2 \times 2$, $4 \times 2$, and $10$ site clusters. 
Each dot indicates the optimal variational parameter $h'^*$ for a given temperature, 
which satisfies the stationary condition [see Eq.~(\ref{eq:sc0})] with the lowest grand potential. 
Therefore, for example, from Fig.~\ref{grandpotential}(c), we can estimate that 
$T_{\rm N}/t \approx 0.285$ for the $L_{\rm c} = 10 $ site cluster at $U/t=8$. 
Similarly, we can estimate $T_{\rm N}$ for other clusters with varying 
$U/t$ to eventually obtain the results 
shown in Fig.~\ref{phasediagram}.

We notice in Fig.~\ref{grandpotential} that the larger cluster tends to show 
the shallower minimum of the grand potential 
[i.e., the smaller $\Omega(0)-\Omega(h'^*)$]  
for the antiferromagnetic solution with $h'^* \not =0$. 
We also find in Fig.~\ref{grandpotential} that $h'^{*}$ at the lowest temperature becomes smaller 
for the larger cluster, indicating that the smaller magnetic field can stabilize the symmetry-broken 
state for the larger cluster. 
It is expected that, 
with increasing the cluster size, $h'^*$ would approach to the ``true'' Weiss field, 
i.e., an infinitesimally small field, 
which induces the symmetry-broken state at $T=0$ in the thermodynamic limit, 
as already shown in Ref.~\cite{Dahnken2004}.

\subsection{Entropy and specific heat}

\begin{figure}
  \begin{center}
      \includegraphics[width=6.0cm]{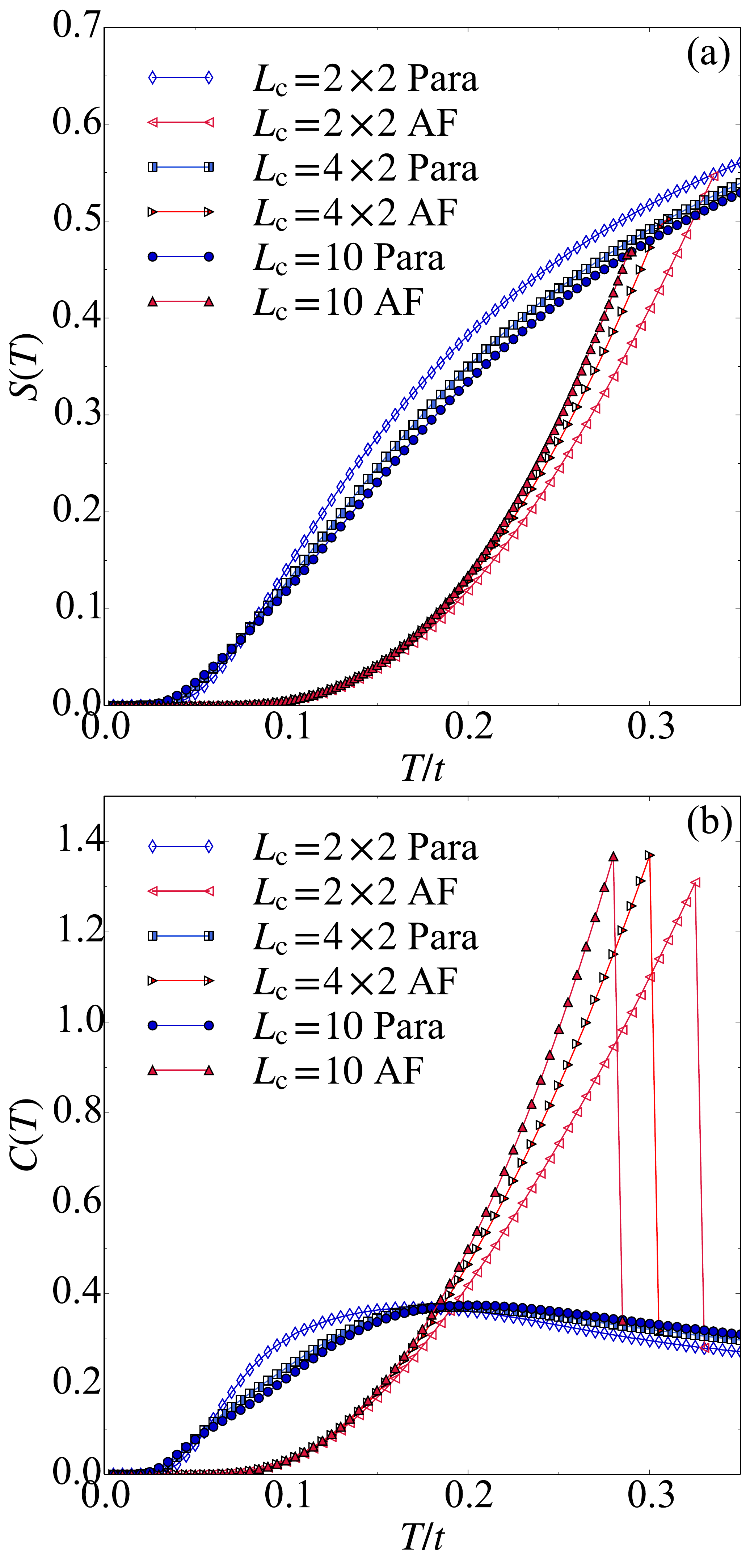}
      \caption{
        Temperature dependence of 
        (a) entropy $S(T)$ and 
        (b) specific heat $C(T)$ at $U/t=8$ 
        for the paramagnetic solution and 
        the antiferromagnetic solution with 
        the clusters of $L_{\rm c} = 2 \times 2$,  $4 \times 2$, and $10$ sites. 
        In (b), 
        $C(T)$ for the antiferromagnetic solution just below $T_{\rm N}$ is connected to
        that for the paramagnetic solution just above $T_{\rm N}$.
        All the results are obtained for the insulating phase in the sense
        that the single-particle gap at the Fermi level is finite (see also Fig.~\ref{spectra})
      }\label{ctst}
  \end{center}
\end{figure}

The temperature dependence of the grand potential is weaker 
for the antiferromagnetic solutions with $h'^* \not = 0$  
than for the paramagnetic solutions with $h'^*=0$, 
irrespectively of the size of clusters. 
Since $S(T)=-\partial_T \Omega$, 
the weaker dependence on the temperature of the grand potential 
indicates the smaller entropy in antiferromagnetic phase 
compared to the paramagnetic phase. 
As shown in Fig.~\ref{ctst}(a), this is indeed the case. 
Figure~\ref{ctst} shows the temperature dependence of 
the entropy $S(T)$ and the specific heat $C(T)$ for $U/t=8$ 
calculated using the clusters of
$L_{\rm c} = 2 \times 2$, $4\times 2$ and $10$ sites. 
The results are obtained both for the paramagnetic and antiferromagnetic solutions.

The entropy in Fig.~\ref{ctst}(a) is calculated from Eq.~(\ref{SFTentropy}) and 
it is confirmed that the results agree with those obtained by numerically differentiating the grand potential 
with respect to $T$. 
The specific heat shown in Fig.~\ref{ctst}(b) is calculated from the numerical differentiation of the entropy 
with respect to $T$ and it is confirmed that the results for the paramagnetic states  
agree with those obtained from Eq.~(\ref{C-matrix}). 
The reason is simply because the optimal variational parameters are $(h'^{*},\eps^{*}) = (0,0)$ 
for the paramagnetic states and therefore $\dd \bs{\lambda}^{*}(T) / \dd T=0$
in the second term of the right-hand side in Eq.~(\ref{eq.CT}), while $\dd \bs{\lambda}^{*}(T) / \dd T\ne0$ for the 
antiferromagnetic states in general.

The entropy 
shows a kink and correspondingly the specific heat exhibits a jump at 
$T_{\rm N}$, indicating that the phase transition is of the second order. 
The entropy for the antiferromagnetic solution is suppressed below $T_{\rm N}$ as compared to that for the 
paramagnetic solution because the spin fluctuations are 
reduced in the ordered phase. 
Both the entropy and the specific heat decay exponentially at low temperatures 
even in the antiferromagnetic phase, where a gapless magnon excitation is expected. 
The gapful behavior found here 
is due to the finite-size effect where the VCA fails to incorporate 
the long-range spin fluctuations and thus to describe the gapless magnon excitations. 
Indeed, as shown in Appendix~\ref{app.anotherOmega}, the thermodynamic quantities in the SFT 
are expressed only in terms of
the exact quantities of the (small) cluster and 
approximate single-particle excitation energies of the infinitely large system.  

The temperature dependence of the entropy and the specific heat in a high temperature region is 
further discussed in Sec.~\ref{sec:SM}.

\subsection{The third law of thermodynamics in SFT}\label{thirdlaw} 

The entropy $S(T)$ shown in Fig.~\ref{ctst}(a) becomes 
zero in the zero-temperature limit, implying that the third law 
of the thermodynamics, 
\begin{equation}
  \lim_{T \rightarrow 0} S(T) = 0,  
\end{equation}
is satisfied. Here, 
we show that the third law of the thermodynamics 
is fulfilled in the SFT if and only if the entropy $S'(T)$ of the cluster becomes zero 
in the zero temperature limit.

Let us consider the internal energy $E$ per site defined as 
\begin{equation}
  E = \Omega + TS.
  \label{internalE}
\end{equation} 
It should be noted again that 
the internal energy $E$ includes the chemical-potential term because of the definition of the Hamiltonian 
in Eq.~(\ref{eq.ham}). 
From $\Omega$ in Eq.~(\ref{eq.Omega-num}) and $S$ in Eq.~(\ref{SFTentropy}), 
we obtain that 
\begin{equation}\label{eq.internalE}
  E = \frac{1}{L_{\rm c}} E' + \frac{1}{N L_{\rm c}\beta^2}  \sum_{\nu=-\infty}^\infty \sum_{\tilde{\mb{k}}} 
  {\rm D}_T
  \ln \det 
  \left[ 
    \bs{I}-\bs{V}(\tilde{\mb{k}}) \bs{G}'(\imag \w_\nu)
  \right], 
\end{equation}
where $E' = \Omega' + TS'$ is the internal energy of the cluster. 
Since $\lim_{T\to 0} E = \lim_{T \to 0} \Omega$ as in Eq.~(\ref{internalE}), 
the comparison between 
the internal energy $E$ in Eq.~(\ref{eq.internalE}) and 
the grand potential $\Omega$ in Eq.~(\ref{eq.Omega-num}) in the zero-temperature limit leads 
\begin{equation}
  -\lim_{T \rightarrow 0} \mr{Tr} \ln (\bs{I}-\bs{VG}') 
  = \lim_{T \rightarrow 0} \beta^{-1} \mr{Tr} \left[{\rm D}_{T} \ln (\bs{I}-\bs{VG}')\right].
\end{equation} 
Substituting this into Eq.~(\ref{SFTentropy}) in the zero-temperature limit yields to 
\begin{equation}
  \lim_{T \rightarrow 0} S(T) = \frac{1}{L_{\rm c}} \lim_{T \rightarrow 0} S'(T),  
\end{equation}
where $\lim_{T\to 0} E' = \lim_{T \to 0} \Omega'$ is also used. 
Therefore, the third law of the thermodynamics is fulfilled if and only if 
the entropy $S'(T)$ of the cluster becomes zero in the zero temperature limit, i.e., 
\begin{equation}
  \lim_{T \rightarrow 0} S'(T) = 0.  
  \label{eq:s0}
\end{equation}

Two remarks are in order. First, Eq.~(\ref{eq:s0}) is satisfied whenever the ground state of the cluster is unique 
even for a paramagnetic insulating state. 
This is the reason why $\lim_{T \to 0} S(T) = 0$ for the paramagnetic state in 
Fig.~\ref{ctst}(a), instead of $\lim_{T \to 0} S(T) = \ln 2$ found in the single-site DMFT~\cite{Rosenberg1994} 
and in the dynamical impurity approximation~\cite{Potthoff2003b}. 
Second, the condition to satisfy the third law of the thermodynamics in the SFT 
resembles the condition to guarantee the Luttinger theorem at zero temperature in the SFT~\cite{Ortloff2007}. 
Indeed, it has been shown that the Luttinger theorem 
is valid in the SFT if and only if the 
single-particle Green's function $\bs{G}'(\imag \w_\nu)$ of the cluster respects the Luttinger theorem, 
where the Luttinger theorem for a small and open-boundary cluster is defined 
in terms of the singularities of the single-particle Green's function $\bs{G}'(\imag \w_\nu)$~\cite{Ortloff2007}.

\subsection{Single-particle excitation spectrum}

\begin{figure*}
  \begin{center}
    \includegraphics[width=0.99\textwidth]{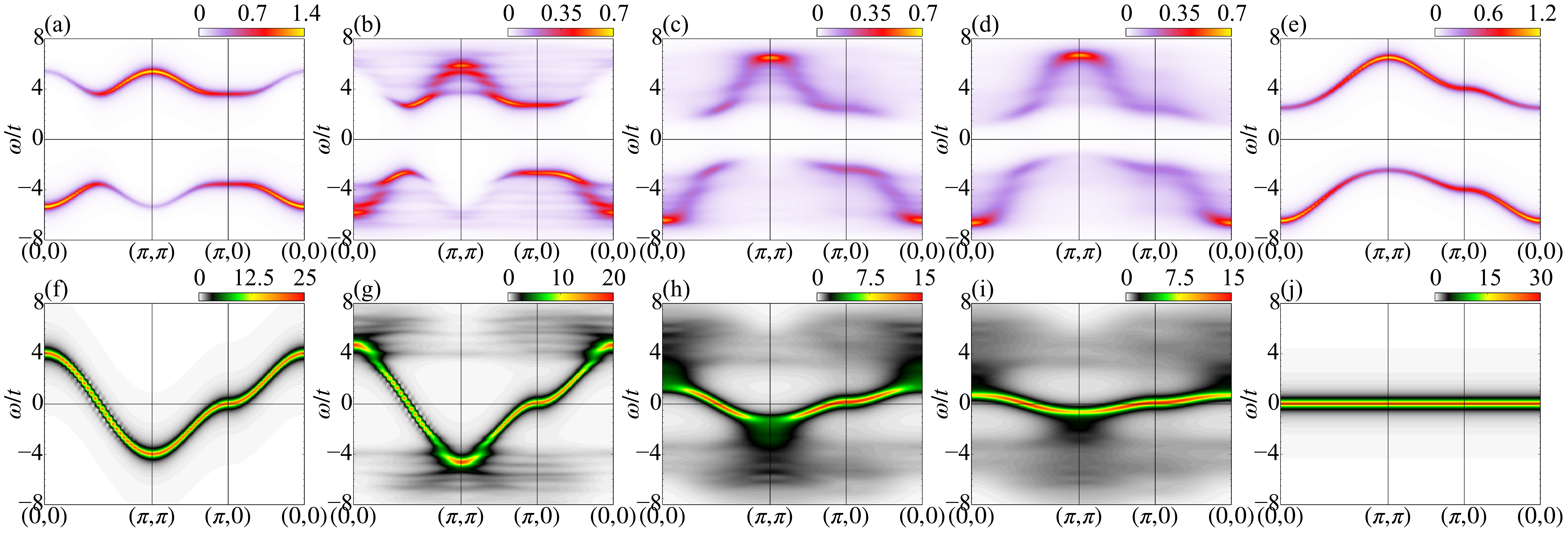}
    \caption{
      Single-particle excitation spectrum $\mcal{A}(\mb{k},\w)$ [(a)--(e)] and 
      the imaginary part ${\mcal S}(\mb{k},\w)$ of the self energy [(f)--(j)] 
      for the half-filled Hubbard model with $U/t=8$ calculated using 
      (a, f) the self-consistent SDW mean-field theory at $T/t=0$ in the antiferromagnetic state, 
      (b, g) the VCA at $T/t=0.001$ in the antiferromagnetic state, 
      (c, h) the VCA at $T/t=0.3$ in the paramagnetic state, 
      (d, i) the VCA at $T/t=0.5$ in the paramagnetic state, and 
      (e, j) the Hubbard-I approximation in the paramagnetic state. 
      The horizontal line at $\w=0$ denotes the Fermi level.
      The Lorentzian broadening of $\eta/t=0.2$ is used. 
      The $L_{\rm c } = 10$ cluster is used for the VCA in (b)--(d) and (g)--(i).  
      The results for the Hubbard-I approximation in (e) and (j) are independent of the temperature. 
      Note that different figures use different intensity scales as indicated in the color bars. 
      \label{spectra}}
  \end{center}
\end{figure*}

The single-particle excitation spectrum ${\mcal A}(\mb{k},\w)$ for the original system of interest is calculated 
from the single-particle Green's function ${\cal G}^{\s \s'} (\mb{k},z)$ in Eq.~(\ref{eq.periodization}) as 
\begin{equation}
  {\mcal A}(\mb{k},\w) = - \frac{1}{\pi} {\rm Im} {\cal G}^{\s \s} (\mb{k},\w+ \imag\eta), 
\end{equation}
where $\eta$ is real positive infinitesimal. The typical results for the Hubbard model with $U/t=8$ 
in the antiferromagnetic state at $T/t=0.001$ and in the paramagnetic states at $T/t=0.3$ and 
$0.5$ are shown in Figs.~\ref{spectra}(b)--\ref{spectra}(d) (also see Fig.~\ref{phasediagram}). 
Here, the single-particle Green's function is averaged over the 
two sublattices A and B within the cluster, and therefore $\mcal{A}(\mb{k},\w)$ does not depend on spin $\s$ 
even in the antiferromagnetic state.

To further analyse the single-particle excitations, 
we also show in Figs.~\ref{spectra}(g)--\ref{spectra}(i) the imaginary part of the self-energy
\begin{equation}
  {\mcal S}(\mb{k},\w) = - \frac{1}{\pi} {\rm Im}  {\Sigma}^{\s \s} (\mb{k},\w+ \imag\eta), 
\end{equation} 
where the self-energy ${\Sigma}^{\s \s} (\mb{k},z)$ for the original system is defined as 
\begin{equation}
\Sigma^{\s \s}(\mb{k},z)=  z - \eps_{\mb{k}} - \mcal{G}^{\s\s}(\mb{k},z)^{-1}
\end{equation}
with $\eps_{\mb{k}}=-2t (\cos{k_x} + \cos{k_y} )$ 
being the noninteracting band dispersion. 
Note that ${\mcal S}(\mb{k},\w) \geqslant 0$ because ${\mcal A}(\mb{k},\w) \geqslant 0$. 
The divergence of ${\mcal S}(\mb{k},\w)$ corresponds to the zero of ${\mcal A}(\mb{k},\w)$, 
thus implying the presence of the single-particle gap~\cite{Eder2011,Sakai2014,Seki2016_tetra,Sakai2016}.  
In practice, the divergence of ${\mcal S}(\mb{k},\w)$ appears as the peak due to the finite $\eta$.

Figure~\ref{spectra}(b) shows the single-particle excitation spectrum for the antiferromagnetic phase 
at $T/t=0.001$, where the temperature is low enough so that thermal excitations are negligible. 
Since the mean-field approximation is expected to be relevant 
in a symmetry-broken state, we compare the result with the SDW mean-field theory 
in which the single-particle Green's function can be given as 
\begin{equation}
  {\cal G_{\mr{SDW}}} (\mb{k},z) = \left[z - \eps_{\mb{k}} - \Sigma_{\rm SDW}(\mb{k},z)  \right]^{-1}
\end{equation} 
and 
\begin{equation}
  \Sigma_{\mr{SDW}} ( \mb{k},z) = \frac{U^2m^2}{z - \eps_{\mb{k+Q}}}  
  \label{eq:se_SDW}
\end{equation} 
with $m$ being the staggered magnetization per site 
and ${\mb Q}=(\pi,\pi)$~\cite{Fulde2012,Imada2011}. 
The single-particle excitation spectrum for the SDW mean-field theory is shown in Fig.~\ref{spectra}(a). 
Indeed, the SDW spectrum very much resembles the VCA result for the antiferromagnetic phase at 
$T/t=0.001$, including both the spectral weight and the dispersion. The most characteristic feature is 
the next-nearest-neighbor-hopping-like dispersion~\cite{Eder1990,Trugman1988}. 
Namely, the dispersion bends 
downward (upward) in the second (first) antiferromagnetic Brillouin 
zone along $(\pi/2,\pi/2) \to (\pi,\pi)$ and $(\pi,0) \to (\pi,\pi)$ 
[$(\pi/2,\pi/2) \to (0,0)$ and $(\pi,0) \to (0,0)$] for the occupied (unoccupied) states. 
It is tempting to conclude that the overall agreement of the VCA and the SDW results is due to 
the mean-field like treatment of the symmetry-broken state in the VCA. 
However, the QMC study has also found that the single-particle excitation spectrum is in good agreement 
with the SDW dispersion~\cite{Groeber2000}. 
Moreover, it should be noted that even when the antiferromagnetic 
long-range order 
is absent, 
the single-particle excitation spectrum shows the dispersion folding downward along 
$(\pi/2,\pi/2) \to (\pi,\pi)$ and $(\pi,0) \to (\pi,\pi)$ for the occupied states, 
similarly the upward-folding dispersion along $(\pi/2,\pi/2) \to (0,0)$ and $(\pi,0) \to (0,0)$ 
for the unoccupied states, in the presence of the short-range 
antiferromagnetic spin fluctuation at zero temperature~\cite{Eder2010,Eder2011}.

Although the overall features are similar, the details of the single-particle excitation spectra 
are different between the VCA and the SDW mean-field theory. 
The main characteristic feature of the single-particle excitations for the VCA is found in 
the low-energy dispersion. Since $\mcal{A}(\mb{k},\w\leqslant0) = \mcal{A}(\mb{k}+\mb{Q},-\w\geqslant 0)$ for the 
particle-hole symmetric case, we focus only on the occupied spectrum in the following. 
As shown in Fig.~\ref{spectra}(b), 
we can find a less-dispersive dispersion in a range of $-4t < \w < -3t$ around $\mb{k}=(0,0)$. 
This can be assigned to the single-particle excitations 
associated with the antiferromagnetic fluctuation of the energy scale of $J= 4t^2/U$. 
Such renormalized dispersion has also been observed in exact-diagonalization studies of 
the Hubbard model as well as the $t$-$J$ model and can be 
well described by the spin-bag picture~\cite{Eder1994,Schrieffer1989}. 
Therefore, the short-range antiferromagnetic fluctuations, 
which are absent in the SDW mean-field theory, make the fine but important difference in the low-energy 
excitations. 
We also note that the other less-dispersive dispersion around $\mb{k} = (0,0)$ and $\w \sim -6.5t$ 
found in the VCA for the antiferromagnetic phase at $T=0.001t$ [Fig.~\ref{spectra}(b)] 
is absent in the SDW mean-field theory [Fig.~\ref{spectra}(a)]. 
Since this less-dispersive dispersion remains even above $T_{\rm N}$ in the paramagnetic state, 
as shown in Figs.~\ref{spectra}(c) and \ref{spectra}(d),  
the origin can be assigned to localized holes.

The single-particle gap remains finite at high temperatures above $T_{\rm N}$ in the paramagnetic state. 
This is in sharp contrast to the SDW mean-field theory. 
The dispersion relation 
in the single-particle excitation spectrum is also quite different 
from that for the SDW mean-field theory, but rather resembles 
the dispersion relation for the Hubbard-I approximation, as shown in Figs.~\ref{spectra}(c)--\ref{spectra}(e). 
In particular, the characteristic feature of the dispersion found in the antiferromagnetic state, i.e., 
the dispersion bending downward (upward) in the second (first) antiferromagnetic Brillouin zone along 
$(\pi/2,\pi/2) \to (\pi,\pi)$ and $(\pi,0) \to (\pi,\pi)$ 
[$(\pi/2,\pi/2) \to (0,0)$ and $(\pi,0) \to (0,0)$] for the occupied (unoccupied) states, is now absent.  
The overall feature of the dispersion at high temperatures in the paramagnetic state 
is instead well reproduced by the Hubbard-I approximation.

The single-particle Green's function ${\cal G}_{\rm  H\text{-}I} (\mb{k},z)$ 
in the Hubbard-I approximation is given by 
\begin{equation}
  {\cal G}_{\rm  H\text{-}I} (\mb{k},z) = \left[z - \eps_{\mb{k}} - \Sigma_{\rm H\text{-}I} (z) \right]^{-1},
\end{equation}
where the self-energy 
\begin{equation}
  \Sigma_{\rm H\text{-}I} (z) = \frac{U^2 n_\s (1-n_{\s})}{z} 
  \label{eq.SigmaHI}
\end{equation}
corresponds to that of single-site Hubbard model 
and $n_{\s}$ is the electron density with spin $\s\,(=\uparrow,\downarrow)$~\cite{Hubbard1963,Gebhard1997}. 
At half filling in the paramagnetic phase, $n_\up = n_\dn = 1/2$. 
The self-energy $\Sigma_{\rm H\text{-}I} (z)$ of the Hubbard-I approximation is spatially local because 
the Hubbard-I approximation takes into account the local electron correlations at a single site 
but neglects the spatial correlations. Therefore, the self-energy is independent of the momentum 
and ${\mcal S}(\mb{k},\w)$ exhibits a flat dispersion, as shown in Fig.~\ref{spectra}(j). 
It is also interesting to observe in Figs.~\ref{spectra}(g)-\ref{spectra}(i) the gradual reduction of the bandwidth of 
the dispersion in ${\mcal S}(\mb{k},\w)$ 
with increasing $T$, 
implying 
the crossover from $\Sigma_{\rm SDW}(\mb{k},z)$-like self-energy to $\Sigma_{\rm H\text{-}I}(z)$-like one.  
The qualitative agreement between the single-particle excitation spectra for the Hubbard-I approximation and the VCA 
at high temperatures above $T_{\rm N}$ is understood because the thermal fluctuations are strong enough to 
destroy the spin 
correlations but not high enough to unfreeze the charge degrees of freedom for the temperatures shown 
in Figs.~\ref{spectra}(b)-(d). 
This is consistent with the entropy $S(T)$ at $T/t\sim0.5$, where $S(T)$ 
is comparable to $\ln{2} = 0.693$, not $\ln{4} = 1.386$, as shown in Fig.~\ref{ctst}(a).

We now remark on the substantial difference in the 
intensity of ${\mathcal S}(\mb{k},\w)$ 
between the VCA and the Hubbard-I approximation. 
It is noticed in Fig.~\ref{spectra} that the single-particle gap as well as the intensity of ${\mcal S}(\mb{k},\w)$ 
near the Fermi level in the VCA at high temperatures above $T_{\rm N}$ in the paramagnetic state 
is quite smaller than that in the Hubbard-I approximation. 
The difference of the single-particle gap can be understood 
by analyzing the moments of the single-particle Green's function for the Hubbard model 
up to the second order~\cite{Harris1967,Seki2011}, i.e., 
\begin{eqnarray}
  &&\int_{-\infty}^{\infty} \dd \w {\mcal A}(\mb{k},\w) = 1, \label{M0} \\ 
  &&\int_{-\infty}^{\infty} \dd \w \w {\mcal A}(\mb{k},\w) = \eps_{\mb{k}} -\mu + Un_\s, \label{M1} 
\end{eqnarray}
and 
\begin{eqnarray}
  &&\int_{-\infty}^{\infty} \dd \w \w^2 {\mcal A}(\mb{k},\w) -  \left[\int_{-\infty}^{\infty} \dd \w \w {\mcal A}(\mb{k},\w) \right]^2 \notag \\ 
  &&\quad\quad\quad\quad\quad\quad\quad = U^2n_\s(1-n_\s) \label{M2}. 
\end{eqnarray}
Note that $n_{\up}=n_{\dn}=1/2$ at half filling. 
Equation~(\ref{M0}) implies that the spectral function $\mcal{A}(\mb{k},\w)$ 
can be considered as a distribution function with respect to  $\w$. 
Equation~(\ref{M1}) indicates that the center of gravity of ${\mcal A}(\mb{k},\w)$
with respect to $\w$ is given by that in the noninteracting limit $\eps_{\bs{k}} - \mu$ 
with the correction of the Hartree potential $Un_\s$~\cite{Eder2011}, which 
cancels the chemical potential $\mu=U/2$ in the present case. 
Equation~(\ref{M2}) indicates that the variance of the spectral function is $U^2 n_\s(1-n_\s)$, 
and thus $\mcal{A}(\mb{k},\w)$ is distributed along the $\w$ axis 
with the standard deviation $U \sqrt{n_\s(1-n_\s)}$ around the center of gravity $\eps_{\bs{k}}$. 
It has been shown by the high-frequency expansion that Eq.~(\ref{M2}) can also be related 
to the spectral-weight sum rule for the self-energy~\cite{Koch2008,Seki2011} 
\begin{equation}
  \int_{-\infty}^{\infty} \dd \w  \mcal{S}(\mb{k},\w) = U^2 n_\s (1-n_\s) = \frac{U^2}{4},  
  \label{eq.Selfsumrule}
\end{equation} 
where we set $n_{\up}=n_{\dn}=1/2$ in the last equality.    
The total amount of the imaginary part of the self-energy 
is thus determined solely by $U$ and the electron density $n_\s$. 
From Eq.~(\ref{eq.SigmaHI}), we can show that the Hubbard-I approximation 
satisfies the sum rule but all the intensity is concentrated on 
the single ``band'' of ${\mcal S}(\mb{k},\w)$, as shown in Fig.~\ref{spectra}(j). 
Therefore, there exist only the upper and lower Hubbard bands 
with no incoherent spectra in the Hubbard-I approximation. 
On the other hand, ${\mcal S}(\mb{k},\w)$ in the VCA is distributed over the energy scale of 
$\approx U$ in the $\w$ axis to generate not only the Hubbard gap accross the Fermi level but also 
the incoherent single-particle excitations at the high energy. 
Therefore, the intensity of $\mcal{S}(\mb{k},\w)$ near the Fermi level 
is necessarily smaller in the VCA than in the Hubbard-I approximation.

Finally, we comment on the results for the single-particle excitation spectrum of the Hubbard model 
at half filling obtained by other methods such as the DMFT and the QMC. 
In the DMFT, a quasiparticle band with the 
narrow bandwidth appears near the Fermi level for $U/t=8$, 
even when the nonlocal correlations are included~\cite{Kusunose2006}. 
On the other hand, in the VCA, the single-particle excitation spectrum does not show 
such a coherent excitation near the Fermi level at any temperature, as shown in  Figs.~\ref{spectra}(b)--\ref{spectra}(d). 
This is because, unlike the DMFT, 
the VCA treats open-boundary clusters without bath orbitals and hence  
the Kondo-resonance-like peak and the coherent quasiparticle excitation 
near the Fermi level~\cite{Nozieres1998,Kotliar1999} 
may not be represented. 
According to the numerically exact QMC studies, 
the coherent quasiparticle excitations near the Fermi level are hardly observed 
for $U/t=8$~\cite{Groeber2000} and even for $U/t=4$~\cite{Rost2012} at half filling. 
In this sense, 
the VCA better agrees with the QMC than the DMFT for the single-particle excitations near the Fermi level 
at half filling.

\subsection{Slater to Mott crossover}\label{sec:SM}

It has been demonstrated recently that 
the weak-coupling Slater-type antiferromagnet and the strong-coupling Mott-type antiferromagnet 
can be well characterized by the energy-gain mechanism of the antiferromagnetic state, i.e., 
whether the antiferromagnetic ordered state gains the interaction energy or the kinetic energy 
relative to the paramagnetic state, 
for the three-orbital Hubbard model analyzed using the variational Monte Carlo method~\cite{Watanabe2014} and 
for the single-band Hubbard model using the variational Monte Carlo method~\cite{Tocchio2016} and the CDMFT method~\cite{Fratino2017}. 
In the CDMFT study, the evolution of the density of states as functions of $T$ 
and $U$ has also been studied~\cite{Fratino2017}.
The energy-gain mechanism of the antiferromagnetic phase of the double perovskite La$_2$NiTiO$_6$ 
has been studied based on the DMFT for an {\it ab initio}-derived multiorbital model~\cite{Karolak2015} 
with predicting the realization of a spin-$1$ strong-coupling antiferromagnet. 
These theoretical approaches of quantifying the energy-gain mechanism for the antiferromagnetic state 
over the paramagnetic state
at $T=0$ in two-dimensional systems or
at low temperatures in three-dimensional systems 
are quite valuable to distinguish the Slater-type antiferromagnet and the Mott-type antiferromagnet.

Here, we attempt to characterize the Slater-to-Mott crossover  
by calculating the thermodynamic quantities including the entropy, 
the specific heat, and the double occupancy in the paramagnetic state in the $(U,T)$ plane.  
We note that the crossover of the two-dimensional Hubbard model 
in the $(U,T)$ plane can also be explored experimentally, 
because the double occupancy and the entropy of the two-dimensional Hubbard model 
from a weak to a strong coupling region, $0 \lesssim U/t \lesssim 20$, has been measured 
recently in ultracold atoms in an optical lattice~\cite{Cocchi2016,Cocchi2017}. 
Therefore, the results obtained here can be tested 
by the ultracold-atom experiment.

\subsubsection{Entropy and specific heat}

\begin{figure*}
  \begin{center}
    \includegraphics[width=1.0\textwidth]{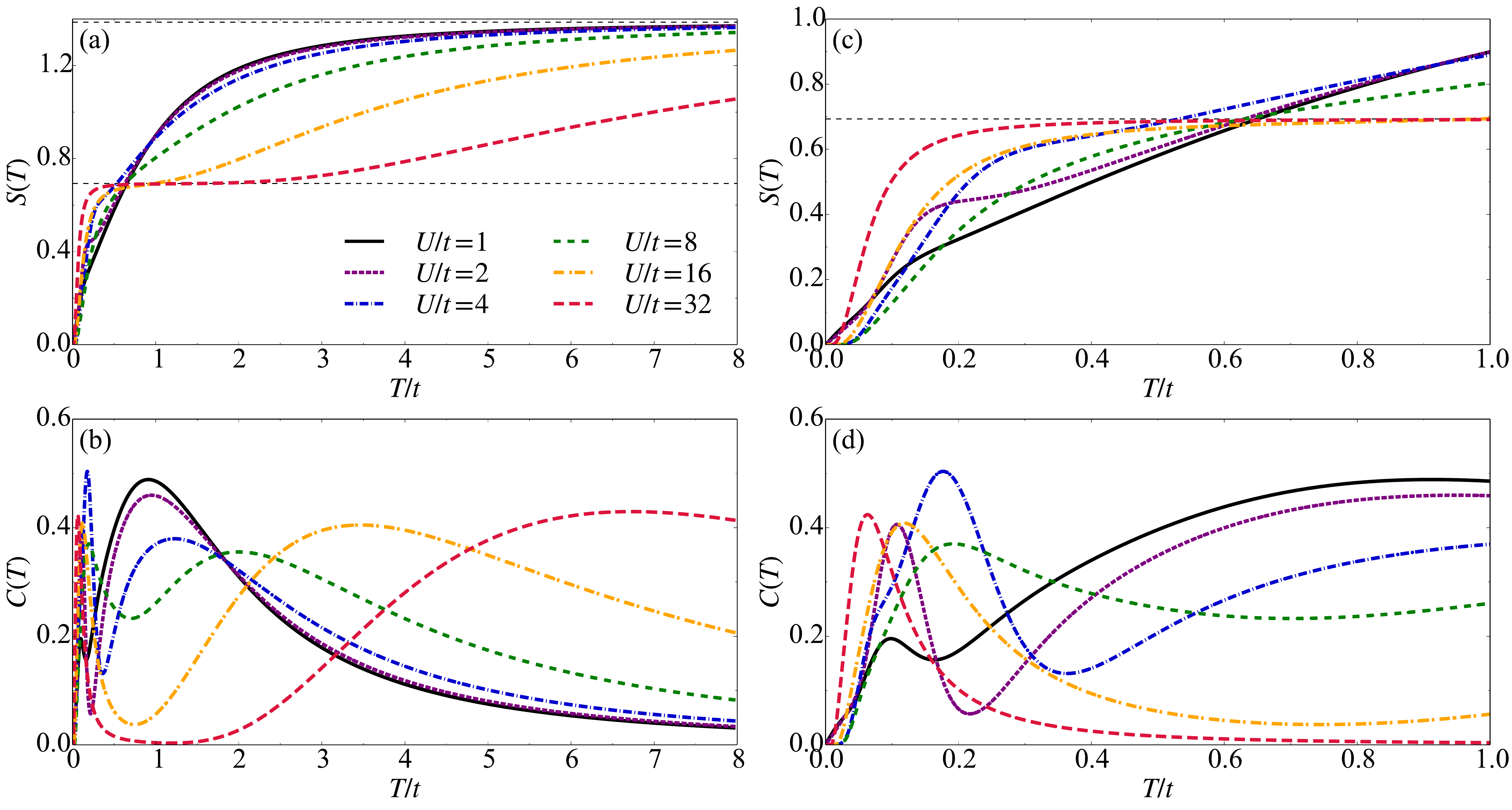}
    \caption{
      Temperature dependence of 
      (a) the entropy $S(T)$ and 
      (b) the specific heat $C(T)$ for $U/t=1,2,4,8,16$, and $32$.  
      The horizontal dashed lines in (a) represent 
      $S(T)=\ln 2 = 0.693$ and $S(T) = \ln 4 = 1.386$.
      (c) and (d) are enlarged plots of (a) and (c) for $0 \leqslant T/t \leqslant 1$, respectively. 
      The results are for the paramagnetic solution with the cluster of $L_{\rm c} = 4 \times 2$ sites. 
      }
      \label{stct_highT}
  \end{center}
\end{figure*}

Figure~\ref{stct_highT} shows the entropy $S(T)$ and the specific heat $C(T)$ 
in $0 \leqslant T/t \leqslant 8$ for $U/t=1,2,4,8,16$ and $32$.
The increment of $T/t$ is set to be 0.01.
The highest temperature $T/t=8$ is comparable to the band width $W/t=8$ of the square lattice, 
which might be too high for realistic materials to keep their 
lattice structures but we consider such high temperatures to be comparable with 
the previous study~\cite{Li2009}. 
For $U/t=16$ and $32$, a plateau-like temperature dependence of 
$S(T)\approx \ln 2$ can be found around temperature $T \approx t$. 
Since $J \ll t \ll U$ with $J=4t^2/U$ being 
the superexchange interaction between the neighboring spins, 
the plateau-like temperature dependence indicates 
the existence of the localized but thermally disordered spin $1/2$ at each site.  
For the smaller values of $U/t$, the plateau-like temperature dependence is hardly observed.

The specific heat $C(T)$ shows a two-peak structure. 
The high-temperature peak shifts towards the higher temperature 
with increasing $U/t$, indicating that the peak corresponds to the 
energy fluctuation due to the charge excitation which involves the energy scale of 
$\sim U-W$ for large $U$.  
Therefore, we refer to the temperature at which $C(T)$ exhibits the high-temperature peak as $T_{\rm charge}$,  
even in a small $U/t$ regime since the peak in a large $U/t$ regime 
is smoothly connected to that in a small $U/t$ regime with decreasing $U$. 
On the other hand, the position of the low-temperature peak 
moves non-monotonically with $U$ and the $U$ dependence is rather similar to that of 
$T_{\rm N}$ [see Fig.~\ref{phasediagram} and also Fig.~\ref{ctmap}(c)]. 
Indeed, for large $U/t$, the entropy almost reaches to 
the maximum entropy $\ln 2$ of a localized free spin at $T_{\rm dip}$, i.e., 
\begin{equation} 
  S(T_{\rm dip}) = \int_{0}^{T_{\rm dip}} \dd T \frac{C(T)}{T} \approx \ln 2, 
\end{equation} 
where $T_{\rm dip}$ is the temperature at which $C(T)$ takes the minimum between the two peaks. 
Therefore, the low-temperature peak of $C(T)$ corresponds 
to the energy fluctuation due to the spin excitation. 
We thus refer to the temperature at which $C(T)$ exhibits the low-temperature peak as $T_{\rm spin}$, 
even in a small $U/t$ regime since the peak in a large $U/t$ regime 
is smoothly connected to that in a small $U/t$ regime with decreasing $U$.

\subsubsection{Thermodynamic quantities in $(U,T)$ plane}

\begin{figure*}
  \begin{center}
    \includegraphics[width=0.95\textwidth]{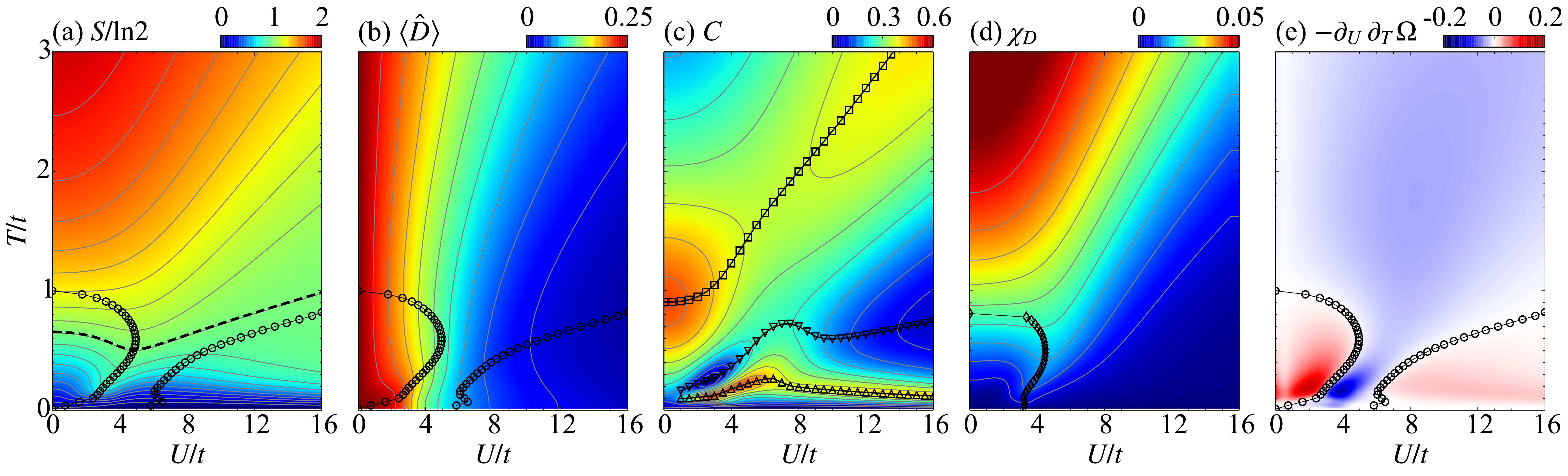}
    \caption{
      The contour plots of 
      (a) the entropy $S/\ln{2}$, 
      (b) the double occupancy $\langle \hat{D} \rangle$,
      (c) the specific heat $C$, 
      (d) the double-occupancy susceptibility $\chi_D$, and 
      (e) the mixed derivative $-\partial_U \partial_T \Omega$ 
      in the $(U,T)$ plane with a range of $0 \leqslant U/t \leqslant 16$ and $ 0 \leqslant T/t \leqslant 3$. 
      The results are obtained for the paramagnetic state with the cluster of $L_{\rm c} = 4 \times 2$ sites. 
      The black lines with open circles in (a), (b), and (e) indicate 
      the contours on which $\partial_U \partial_T \Omega = 0$. 
      The black dashed line in (a) indicates $S(U,T)=\ln 2$. 
      The black line with open triangles (squares) in (c) 
      indicates the peak of $C(U,T)$ at low (high) temperatures, 
      and the black line with inverted open triangles indicates 
      the dip of $C(U,T)$. 
      The black line with open diamonds in (d) indicates 
      the contour on which $\chi_D$ takes the maximum at finite $U$. 
      \label{ctmap}}
  \end{center}
\end{figure*}

Thermodynamic quantities in the $(U,T)$ plane is summarized in Fig.~\ref{ctmap}. 
Here, the results include 
the entropy $S$, 
the specific heat $C$, 
the mixed derivative $-\partial_U \partial_T \Omega$, 
the double occupancy $\langle \hat{D} \rangle$, and 
the double-occupancy susceptibility $\chi_D$,  
the latter two quantities being defined below. 
The increment of $U/t$ ($T/t$) is set to be 0.5 (0.01) and 
the derivatives are evaluated by quadratically fitting $\Omega(U,T)$ first.

\begin{figure}
  \begin{center}
    \includegraphics[width=1.0\columnwidth]{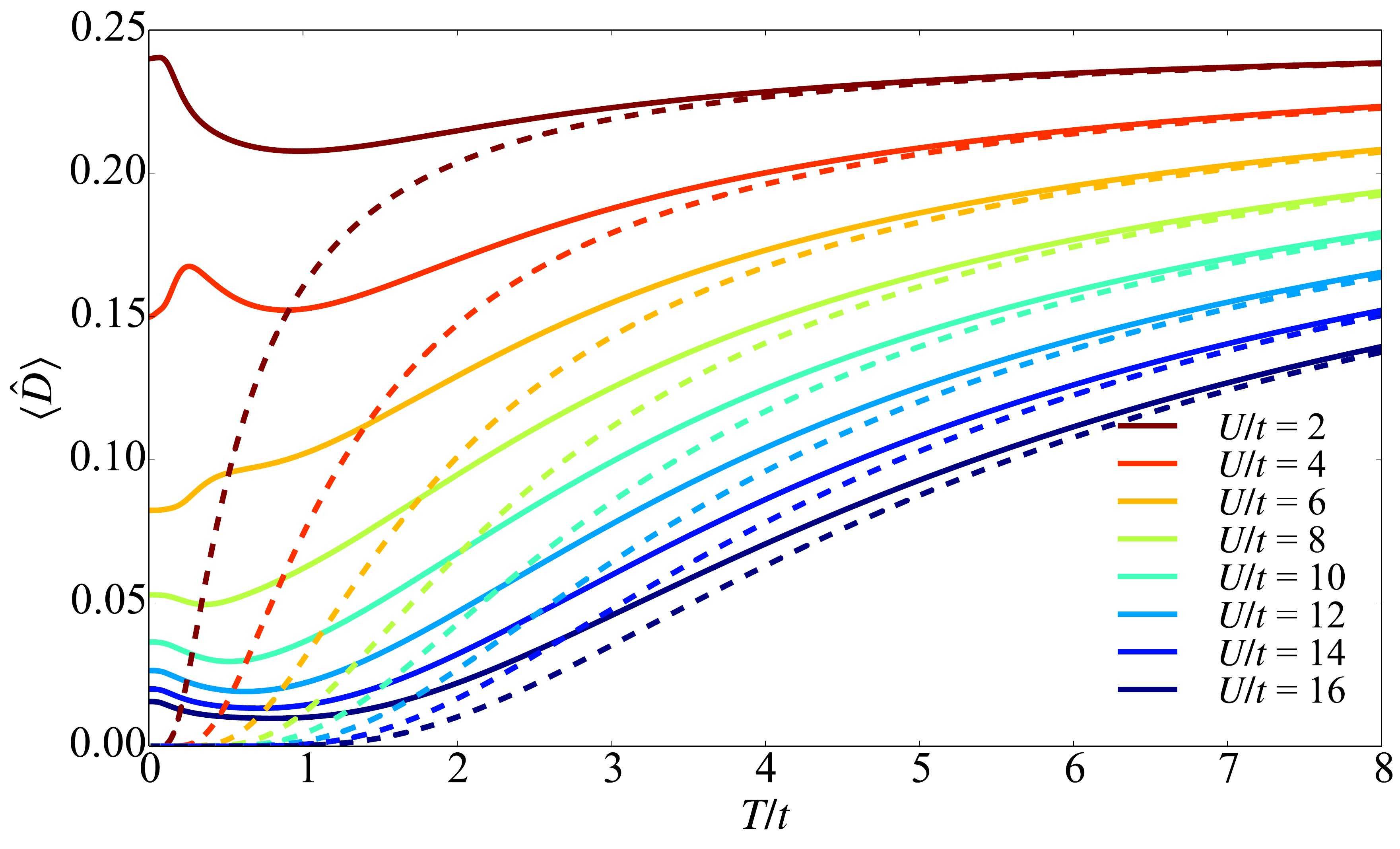}
    \caption{
      The double occupancy $\langle \hat{D} \rangle$ as a function of $T$ 
      for $U/t=2,4,6,8,10,12,14$, and $16$ (from top to bottom). 
      The results are for the paramagnetic state with the cluster of $L_{\rm c} = 4 \times 2$ sites. 
      The increment of $T/t$ is set to be 0.01. 
      For comparison, the double occupancy, 
      $\langle \hat{D} \rangle = 1/\left[2+2\exp{(U/2T)}\right]$,   
      of the single-site Hubbard model (i.e, in the atomic limit) 
      at half filling is also shown by dashed lines for the same values of  $U/t$. 
      \label{docc}}
  \end{center}
\end{figure}

The entropy is an increasing function of $T$ but not a monotonic function of $U$. 
Indeed, the entropy takes extrema $\partial_U S=0$ at certain $U$ values for a fixed temperature, 
as indicated by lines with open circles in Figs.~\ref{ctmap}(a), \ref{ctmap}(b), and \ref{ctmap}(e). 
The $U$ derivative of the entropy is related to the $T$ derivative 
of the double occupancy $\langle \hat{D} \rangle$ 
through the Maxwell relation, 
\begin{equation}
  \frac{\partial S}{\partial U}  
  = -\frac{\partial^2 \Omega}{\partial U \partial T} 
  = -\frac{\partial \langle \hat{D} \rangle}{\partial T}, 
\end{equation}
where $\hat{D}=\frac{1}{NL_{\rm c}}\sum_i^{NL_{\rm c}} \hat{D}_{i}$, 
$\hat{D}_i = \hat{n}_{i\up} \hat{n}_{i\dn}$, and $\langle \cdots \rangle$ denotes the thermal average. 
In the parameter regions surrounded by these lines, 
$\partial_U S>0$ or equivalently $\partial_T \langle \hat{D} \rangle<0$ [see Fig.~\ref{ctmap}(e)], except for $U=0$. 
This behavior has been observed previously 
in several approximate or unbiased methods~\cite{Georges1992,Werner2005,Paiva2010,Fuchs2011,Khatami2011,LeBlanc2013,Laubach2015,Misawa2016,Takai2016}.  
It has also been suggested that this can be utilized for the adiabatic cooling of cold atoms by tuning 
$U/t$~\cite{Werner2005}. 
Note that the $T$ dependence of the double occupancy is counterintuitive 
because the increase of $T$ is expected to increase the charge fluctuation 
and thus increase the double occupancy. 
Indeed, in the atomic limit, the double occupancy 
increases monotonically with increasing $T$, as shown in Fig.~\ref{docc}. 
Interestingly, we observe a non-monotonic 
behavior of $\partial_T \langle \hat{D} \rangle$ 
at low temperatures with increasing $U$: 
the sign of $\partial_T \langle \hat{D} \rangle$ is 
negative, positive, and negative again 
with increasing $U$ at a fixed temperature [see Fig.~\ref{ctmap}(e)].

We now discuss the counterintuitive sign 
of $\partial_U S = - \partial_T \langle \hat{D} \rangle >0$ 
for a small $U$ regime by considering 
the double-occupancy susceptibility $\chi_D$ 
defined as a dimensionless second derivative of $\Omega$ with respect to $U$ 
\begin{equation} 
  \chi_D 
  = -T \frac{\partial^2 \Omega}{\partial U^2} 
  = -T \frac{\partial \langle \hat{D} \rangle}{\partial U}. 
  \label{Dfluc}
\end{equation}
The result of $\chi_D(U,T)$ is shown in Fig.~\ref{ctmap}(d). 
It is found that the $(U,T)$ domain in which $\chi_D$ increases with $U$, 
surrounded by the line with open diamonds in Fig.~\ref{ctmap}(d), well agrees 
with the domain in which $\partial_U S > 0$ for the small $U/t$ regime. 
Since the local spin moment squared,   
$\hat{s}=\frac{1}{NL_{\rm c}}\sum_{i}\left(\hat{n}_{i\up} - \hat{n}_{i\dn} \right)^2 $, 
is related to $\hat{D}$ as 
$  \hat{s} = 1-2\hat{D} $ at half filling, 
the decrease of the double occupancy implies 
the larger fluctuation of the local spin moment. 
Therefore, the increase of the entropy as a function of $U$ can be assigned to 
the increase of the spin fluctuation due to the electron correlation. 
It should be noted that the $(U,T)$ domain where the spin fluctuation increases with $U$ obtained here 
qualitatively agrees with the domain where the spin-fluctuation theory 
is expected to be appropriate~\cite{Moriya2006} 
and also with the interaction region where the nonlinear sigma model finds the 
Slater-type antiferromagnet~\cite{Borejsza2004}.

For sufficiently large $U$ and 
at $T \sim J \ll U$, the energy scale of the thermodynamic quantities 
is expected to be determined by $J$, because the Hubbard model at half filling 
with large $U/t$ is effectively described by the Heisenberg model with 
the superexchange interaction $J$. 
Therefore, in this parameter region, the increase of $U$ results in the decrease of 
$J=4t^2/U$, which is the only energy scale of the Hamiltonian, and 
hence $\partial_U S >0$ is expected. 
Indeed, we can show that 
\begin{equation}
  \frac{\partial S(T/J)}{\partial U} = \frac{1}{U}(T/J) \frac{\partial S(T/J)}{\partial (T/J)} = \frac{1}{U} C(T/J) >0. 
\end{equation} 
On the other hand, 
for sufficiently large $U$ but at much higher temperatures, i.e., $T \gg J$, 
the spin correlation of the energy scale of $J$ is negligible 
and the system can be considered as a collection of Hubbard atoms. 
Therefore, in this parameter regime, $\partial_U S = -\partial_T \langle \hat{D} \rangle <0$ 
(see Fig.~\ref{docc}), 
as intuitively expected from the atomic limit of the Hubbard model.

\subsubsection{Crossover diagram}

\begin{figure}
  \begin{center}
    \includegraphics[width=0.8\columnwidth]{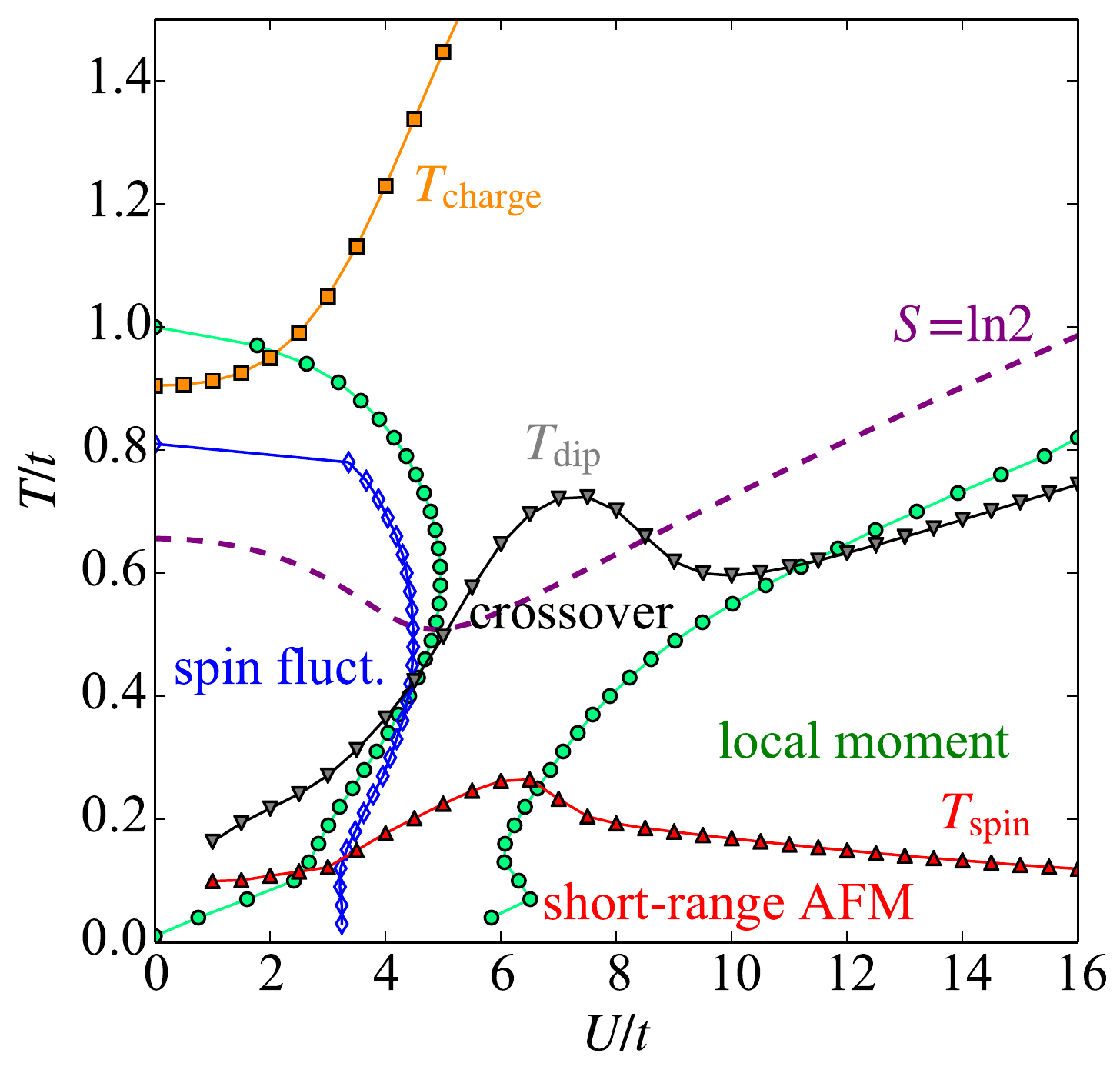}
    \caption{
      A finite-temperature crossover diagram for the half-filled Hubbard model in the paramagnetic state 
      obtained by the VCA with the cluster of $L_{\rm c}=4\times2$ sites. 
      $T_{\rm spin}$ (red line with solid triangles) denotes the temperature where $C(T)$ shows the low-temperature peak,  
      $T_{\rm charge}$ (orange line with solid squares) denotes the temperature where $C(T)$ shows 
      the high-temperature peak, and 
      $T_{\rm dip}$ (grey line with solid inverted triangles) denotes the temperature where $C(T)$ shows the dip 
      between $T_{\rm spin}$ and $T_{\rm charge}$.  
      The violet dashed line indicates $T$ and $U$, where $S(T,U)=\ln 2$. 
      The blue line with open diamonds denotes the value of $U$ on which $\chi_D$ 
      takes a maximum with increasing $U$ for a given $T$. 
      The green lines with circles denote the parameters across which 
      $\partial S/\partial U$ changes the sign. 
      The antiferromagnetic correlations are expected to develop below $T_{\rm spin}$. 
      The large spin fluctuations are expected in the region around the blue line with open diamonds. 
      Local moments are formed in the region below the green line with open triangles 
      for $U\gtrsim 6$. 
      \label{crossover}}
  \end{center}
\end{figure}

Figure~\ref{crossover} summarizes 
the finite-temperature crossover diagram 
of the half-filled Hubbard model in the $(U,T)$ plane, 
featuring the thermodynamic quantities. 
Here, $T_{\rm spin}$ and $T_{\rm charge}$ are 
the temperatures at which $C(T)$ takes the maximum at low and high temperatures, respectively,  
and $T_{\rm dip}$ is the temperature at which $C(T)$ takes the minimum between $T_{\rm spin}$ and $T_{\rm charge}$. 
The antiferromagnetic correlation is expected to develop below $T_{\rm spin}$. 
Above $T_{\rm spin}$,
the formation of local moments is expected for 
$U/t \gtrsim 10$ (referred to as a ``local moment'' region in Fig.~\ref{crossover}), 
while the large spin fluctuations without the formation of local moments are expected for 
$U/t \lesssim 4$ (referred to as a ``spin fluct.'' region in Fig.~\ref{crossover}). 
These parameter regions characterize the Mott-Heisenberg-type and the Slater-type 
antiferromangets, respectively. 
The parameter region between these two regions is referred to as a ``crossover'' region in Fig.~\ref{crossover}. 
From the high-temperature side, this crossover region can be signaled 
as a shallow dip of the specific heat $C(T)$ with relatively high $T_{\rm dip}$ 
and relatively high entropy $S(T_{\rm dip}) \geqslant \ln 2$ 
(note the crossing of the two lines $T_{\rm dip} $ and $S=\ln 2$ 
near the crossover region in Fig.~\ref{crossover}), 
which can also be measured in the ultracold-atom experiment~\cite{Cocchi2017}.

\section{Finite-temperature CDIA study of paramagnetic Mott metal-insulator transition}\label{cdia_results}

In this section, we investigate the paramagnetic Mott metal-insulator transition 
at finite temperatures using the CDIA with the exact-diagonalization cluster solver 
developed in Secs.~\ref{method} and ~\ref{sec.BL}. 
Our study in this section can be considered as a 
counterpart of the preceding CDMFT study~\cite{Park2008}, 
whore the phase diagram has been revisited in combination with various DMFT-related methods 
including the zero-temperature CDIA~\cite{Schafer2015} 
and also extended for doped cases with the CDMFT~\cite{Sordi2011}.  
Our study here can also be considered 
as a finite-temperature extension of the previous zero-temperature CDIA study~\cite{Balzer2009}, 
which calculated the zero-temperature metal-insulator phase diagram and also gave 
a $U$-$T$ phase diagram schematically based on their zero-temperature snalysis~\cite{Balzer2009}. 

In the following, we first review the formalism of the finite-temperature CDIA 
and show how the bath degrees of freedom should be treated 
in the grand-potential functional calculations. 
We then use the finite-temperature CDIA to calculate the $U$-$T$ phase diagram and the single-particle 
excitation spectra.

\subsection{Formalism of CDIA}

\subsubsection{Subtlety regarding the bath degrees of freedom}

\begin{figure*}
  \includegraphics[width=0.9\textwidth]{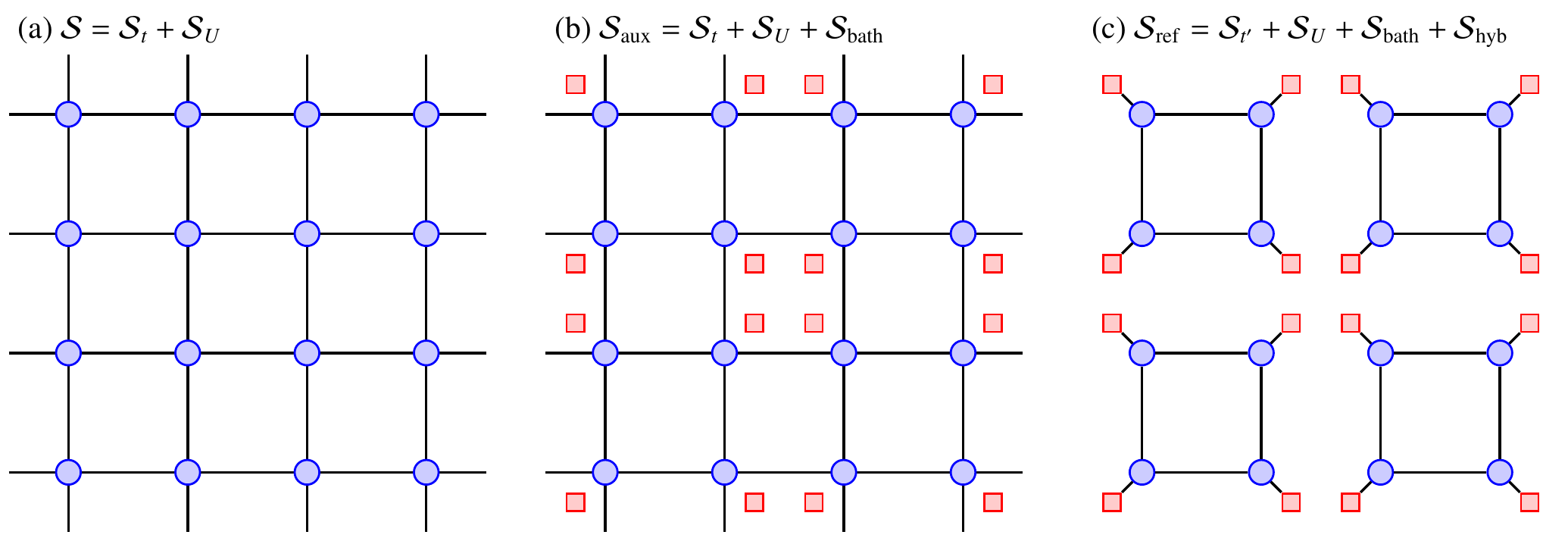}
  \caption{\label{CDIAsystems}
    Schematic figures of 
    (a) original, (b) auxiliary, and (c) reference systems
    considered in the CDIA. 
    (a) The original system consists of the correlated sites (blue circles).  
    (b) The auxiliary system consists of the correlated sites and bath orbitals (red squares) but they are decoupled. 
    (c) The reference system is composed of a collection of the disconnected small clusters, 
    each of which consists of the correlated sites and bath orbitals with hybridization 
    (solid lines between blue circles and red squares). 
    The corresponding actions of  
    the original system $\Sact$ in Eq.~(\ref{Sori}), 
    the auxiliary system $\Sact_{\rm aux}$ in Eq.~(\ref{Saux}), and 
    the reference system $\Sact_{\rm ref}$ in Eq.~(\ref{Sref}) are also indicated. 
  } 
\end{figure*}

As shown schematically in Fig.~\ref{CDIAsystems}(c), the reference system in the CDIA has the bath orbitals, 
which are absent in the original system of interest [Fig.~\ref{CDIAsystems}(a)]. 
Therefore, the degrees of freedom in the reference system differ from those in the original system. 
This causes a difficulty that 
$\Omega[\Sigma_{\rm r}] - \Omega_{\rm r}[\Sigma_{\rm r}]$ cannot be defined in the CDIA 
because the definition of the trace in Eq.~(\ref{eq.Tr}) for the reference system 
differs from that for the original system. 
This subtlety due to the presence of the bath degrees of freedom has been briefly mentioned 
in Sec.~VI~A of Ref.~\cite{Senechal2008}.

In order to address how the grand-potential functional of the original system should be 
calculated in the CDIA, we re-examine the formalism 
by considering the ratio between partition functions~\cite{Feynman-Hibbs,Feynman} 
of the original and the reference systems 
in the fermion-coherent-state path-integral formalism~\cite{Potthoff2006,Shanker1994,Imada1998,NO}. 
The strong-coupling expansion~\cite{Pairault1998,Pairault2000} 
in the lowest order~\cite{Senechal2002} is applied  
to derive an approximate grand potential relevant to the CDIA. 
Interpreting the result in terms of the SFT, we finally show 
how the grand-potential functional should be calculated in the CDIA. 

\subsubsection{Original system} 

The original system of interest is described by the following Hamiltonian:  
\begin{equation}
  \hat{H} = \hat{H}_t + \hat{H}_U, 
\end{equation}
where 
\begin{equation}
  \hat{H}_t = \sum_{i,j} \left(t_{ij} \hat{c}_i^\dag \hat{c}_j + {\rm H.c.} \right)
\end{equation} 
represents the single particle term with $i$ and $j$ being 
the generalized single-particle indices and $\hat{H}_U$ is the interaction term. 
We assume that $\hat{H}_{t}$ includes the chemical-potential term.  
The partition function $Z$ of the original system 
is given in a path-integral form as 
\begin{equation}
  Z = \int \mcal{D}[c^{\star} c] \e^{-\Sact}, 
\end{equation}
where 
\begin{equation}
  \Sact = \Sact_t + \Sact_U \label{Sori}
\end{equation}
is the action of the original system $\hat{H}$ with 
\begin{equation}
  \Sact_t=\int_0^\beta \dd \tau \left[\sum_{i} c_i^\star(\tau) \frac{\partial}{\partial \tau} c_i(\tau) + H_t(c^\star(\tau),c(\tau)) \right]
\end{equation}
and
\begin{equation} 
  \Sact_U=\int_0^\beta \dd \tau H_U(c^\star(\tau),c(\tau)). 
\end{equation}
Here, $c^\star_i(\tau)$ and $c_i(\tau)$ are the Grassmann fields at imaginary-time $\tau$, defined as 
the left and right eigenvalues of the fermion operators $\hat{c}^\dag_i$ and $\hat{c}_i$ 
with respect to the fermion-coherent state, respectively. 
Note that $H_U(c^\star(\tau),c(\tau))$ is obtained by 
normal ordering the interaction Hamiltonian $\hat{H}_U$ and also replacing 
$\hat{c}^\dag_i$ and $\hat{c}_i$ by $c^\star_i(\tau)$ and $c_i(\tau)$, respectively. 
The same applies for $H_t(c^\star(\tau),c(\tau))$. 
In the following, we refer to lattice sites of the system as correlated 
sites to distinguish from bath orbitals.

\subsubsection{Reference system}
As shown schematically in Fig.~\ref{CDIAsystems}(c), the reference system is composed of a collection of 
disconnected clusters (i.e., no hopping between clusters), each of which consists of the correlated sites and 
bath orbitals, and it is  
described by the following Hamiltonian: 
\begin{equation}
  \hat{H}_{\rm ref} = \hat{H}_{t'} + \hat{H}_U + \hat{H}_{\rm bath} + \hat{H}_{\rm hyb},  
  \label{HrefCDIA}
\end{equation}
where 
\begin{equation}
  \hat{H}_{\rm bath} = \sum_{k} \eps_{k} \hat{b}_k^\dag \hat{b}_k 
\end{equation}
is the bath Hamiltonian with $\hat{b}^\dag_k$ being the fermion creation operator 
for a bath of single-particle index $k$ and 
\begin{equation}
  \hat{H}_{\rm hyb} = \sum_{ik} \left(v_{ik} \hat{c}_i^\dag \hat{b}_k + {\rm H.c.} \right)
\end{equation}
represents the hybridization between the correlated sites 
and the bath orbitals. 
The partition function of the reference system can be written as 
\begin{equation}
  Z_{\rm ref} = \int \mcal{D}[c^{\star} c] \mcal{D}[b^{\star} b] \e^{-\Sact_{\rm ref}},
\end{equation}
where 
\begin{equation}
  \Sact_{\rm ref} = \Sact_{t'} + \Sact_U + \Sact_{\rm bath} + \Sact_{\rm hyb} \label{Sref}
\end{equation}
is the action of the reference system with
\begin{eqnarray}
  \Sact_{\rm bath} &=& 
  \int_0^\beta \dd \tau 
  \left[\sum_k b_k^\star(\tau) \frac{\partial}{\partial \tau} b_k(\tau)+ H_{\rm bath}(b^\star(\tau),b(\tau)) \right]
\end{eqnarray}
and 
\begin{eqnarray}
  \Sact_{\rm hyb} &=& \int_0^\beta \dd \tau H_{\rm hyb}(c^\star(\tau),c(\tau),b^\star(\tau),b(\tau)). 
\end{eqnarray} 
Here, $b^\star(\tau)$ and $b(\tau)$ are the Grassmann fields of the bath electrons at imaginary-time $\tau$.

\subsubsection{Auxiliary system}

Let us now introduce an auxiliary system defined by an action 
\begin{equation}
  \Sact_{\rm aux} = \Sact + \Sact_{\rm bath}. \label{Saux}
\end{equation}
The partition function of the auxiliary system is given as 
\begin{eqnarray}
  Z_{\rm aux} &=&  \int \mcal{D}[c^{\star}c]\mcal{D}[b^{\star}b] \e^{-\Sact_{\rm aux}} \\
  & = & Z\cdot Z_{\rm bath},  
\end{eqnarray}
where we introduced the partition function of the bath system 
\begin{equation}
  Z_{\rm bath} = \int \mcal{D}[b^{\star}b] \e^{-\Sact_{\rm bath}}.   
\end{equation}
Therefore, the ratio of the partition functions between the original and reference systems is 
\begin{equation}
  \frac{Z}{Z_{\rm ref}} = \frac{Z_{\rm aux}}{Z_{\rm ref}} \frac{1}{Z_{\rm bath}}. 
  \label{ratioZZ}
\end{equation}
Note that the auxiliary system 
consists of the original system and bath orbitals  
but they are decoupled, as schematically shown in Fig.~\ref{CDIAsystems}(b). 
Since the bath system does not contain the interaction terms, 
$Z_{\rm bath}$ can be readily evaluated numerically or even analytically. 
Therefore, in the following, we focus on the ratio $Z_{\rm aux}/Z_{\rm ref}$.  
The ratio $Z_{\rm aux}/Z_{\rm ref}$ can be treated within the path-integral formalism 
because the auxiliary system $\Sact_{\rm aux}(c^\star,c,b^\star,b)$ has the same degrees of freedom 
with the reference system $\Sact'(c^\star,c,b^\star,b)$.  
This is precisely the reason why we have introduced the auxiliary system~\cite{note_hyb}.

\subsubsection{Ratio of partition functions}

The ratio of the two partition functions can be written as 
\begin{eqnarray}
  \frac{Z_{\rm aux}}{Z_{\rm ref}} &=& 
  \left \langle \e^{-(\Sact_{\rm aux}-\Sact_{\rm ref})}\right \rangle', \label{average}
\end{eqnarray}
where 
\begin{equation}
  \langle \cdots \rangle'= \frac{1}{Z_{\rm ref}} {\int \mcal{D}[c^{\star}c]\mcal{D}[b^{\star}b] \cdots \e^{-\Sact_{\rm ref}}} 
  \label{def_average}
\end{equation}
denotes the expectation value with respect to the reference system. 
To simplify the notation, 
we now denote the Grassmann fields $c^\star$ and $b^\star$ ($c$ and $b$) 
simply as a single symbol $\gamma^\star$ ($\gamma$).  
The expectation value is thus shortly written as 
$\langle \cdots \rangle'= {\int \mcal{D}[\gamma^{\star}\gamma]\cdots \e^{-\Sact_{\rm ref}}}/Z_{\rm ref}$ 
and the action in the exponent of Eq.~(\ref{average}) can be written as 
\begin{eqnarray}
  \Sact_{\rm aux}-\Sact_{\rm ref} 
  &=&\int_0^\beta \dd \tau  \left[\sum_{ij} \gamma_i^\star(\tau) \left(t_{ij}-t'_{ij} - v_{ij}\right) \gamma_j(\tau) \right]   \notag \\
  &=&\int_0^\beta \dd \tau \int_0^\beta \dd \tau' \bs{\gamma}^\star(\tau) \bs{V}(\tau-\tau') \bs{\gamma}(\tau') , \notag \\
  &\equiv&\bs{\gamma}^\star \bs{V} \bs{\gamma}, \label{abb} 
\end{eqnarray}
where 
\begin{equation}
  [\bs{V}(\tau-\tau')]_{ij} = [\bs{V}]_{ij}\delta(\tau-\tau')=\left( t_{ij} - t'_{ij} - v_{ij}\right)\delta(\tau-\tau'), \label{VCDIA}
\end{equation}
$\bs{\gamma}^\star(\tau) =(\gamma_1^\star(\tau),\gamma_2^\star(\tau),\cdots )$, and 
$\bs{\gamma}(\tau) =(\gamma_1(\tau),\gamma_2(\tau),\cdots )^{\rm T}$. 
We also follow the convention that 
the integration over $\tau$ and $\tau'$ is implicitly assumed in Eq.~(\ref{abb}). 
Notice in Eq.~(\ref{VCDIA}) that 
the bath energy $\eps_k$ does not appear in $\bs{V}$ matrix 
because $\bs{V}$ represents the difference 
between the Hamiltonian of the auxiliary system and the Hamiltonian of the reference system. 
Equation~(\ref{VCDIA}) thus clarifies 
another subtlety of the CDIA regarding 
the bath energy discussed in Sec.~IV A of Ref.~\cite{Senechal2008}.

Although $\Sact_{\rm aux}-\Sact_{\rm ref}$ is quadratic in $\gamma^\star$ and $\gamma$, 
Eq.~(\ref{average}) cannot be evaluated in general because the interaction term of the cluster is exponentiated 
in the definition of average $\langle \cdots \rangle'$ in Eq.~(\ref{def_average}). 
To proceed the calculation further, we follow the previous studies~\cite{Pairault1998,Pairault2000,Senechal2002} 
by applying the Hubbard-Stratonovich transformation of the Grassmann variables~\cite{NO,Imada1998}
to the residual single-particle part 
$\e^{-(\Sact_{\rm aux}-\Sact_{\rm ref})}= \e^{-\bs{\gamma}^\star \bs{V} \bs{\gamma}}$, i.e., 
\begin{eqnarray}
  \frac{Z_{\rm aux}}{Z_{\rm ref}}
  &=&  \left \langle \e^{\bs{\gamma}^\star (-\bs{V}) \bs{\gamma}} \right \rangle' \notag \\
  &=& {\rm Det} (-\bs{V}) \int  {\cal D} [\psi^\star \psi] \e^{\bs{\psi}^\star \bs{V}^{-1} \bs{\psi} } 
  \left \langle \e^{ \left(\bs{\psi}^\star \bs{\gamma} + \bs{\gamma}^\star \bs{\psi} \right)} \right \rangle', 
  \label{Zaux}
\end{eqnarray}
where ${\rm Det}(\cdots)$ represents the functional determinant 
and is carried out over all the indices of the Grassmann fields, 
as explicitly shown later in Eq.~(\ref{Det}). 
We note that the auxiliary Grassmann fields, $\psi^\star$ and $\psi$ in Eq.~(\ref{Zaux}), 
introduced by the Hubbard-Stratonovich transformation 
play a key role in the so-called dual fermions approach, a recent extension of the DMFT~\cite{Rubtsov2008}. 
We also note that in the preceding studies~\cite{Pairault2000,Senechal2002,Adibi2016} 
the prefactor similar to that in Eq.~(\ref{Zaux}) is derived but 
the argument of the determinant is opposite. 
The sign of the argument in ${\rm Det}{(-\bs{V}})$ is crucial  
for the present study to obtain 
the final form of the approximate grand potential
[see the first line in Eq.~(\ref{Zaux/Z'})].

\subsubsection{Cumulant expansion}
In the right-hand side of Eq.~(\ref{Zaux}), the expectation value of the exponential 
can be written as the exponential of the cumulant average, 
\begin{equation}
  \left \langle \e^{ \left(\bs{\psi}^\star \bs{\gamma} + \bs{\gamma}^\star \bs{\psi} \right)} \right \rangle' 
  = \exp{
    \left[
      \sum_{n=1}^{\infty} \frac{1}{n!}\left \langle \left(\bs{\psi}^\star \bs{\gamma} + \bs{\gamma}^\star \bs{\psi}\right)^n \right \rangle'_{\rm c}
      \right]}, 
  \label{cumulant}
\end{equation} 
where $\langle \cdots \rangle'_{\rm c}$ denotes the cumulant average~\cite{Kubo1962}. 
For instance, the first three cumulant averages are given as 
$\langle A \rangle'_{\rm c} = \langle A \rangle'$, 
$\langle A^2 \rangle'_{\rm c} = \langle A^2 \rangle' - \langle A \rangle'^2$, and 
$\langle A^3 \rangle'_{\rm c} = \langle A^3 \rangle' -3 \langle A \rangle' \langle A^2 \rangle' + 2\langle A \rangle'^3 $. 
Note that odd cumulants are zero as they involve 
products of the odd numbers of the Grassmann fields $\gamma^\star$ and $\gamma$,  
while even cumulants are nonvanishing in general. 
The $n$th cumulants ($n$: even) can be expressed as 
\begin{eqnarray}
  &&\left \langle \left(\bs{\psi}^\star \bs{\gamma} + \bs{\gamma}^\star \bs{\psi}\right)^n \right \rangle'_{\rm c} \notag \\
  &=& \binom{n}{n/2} \int_0^\beta \prod_{k=1}^{n/2} \dd \tau_k \int_0^\beta \prod_{l=1}^{n/2} \dd \tau'_l
  \sum_{i_1, \cdots, i_{n/2}} \sum_{j_1, \cdots, j_{n/2}} \notag \\
  &\times& 
  \psi^\star_{i_1}(\tau_1) \cdots  \psi^\star_{i_{n/2}}(\tau_{n/2})  
  \psi_{j_{n/2}}(\tau'_{n/2}) \cdots \psi_{j_1}(\tau'_1)   \notag \\  
  &\times& 
  \left\langle 
  \gamma_{i_1}(\tau_1) \cdots \gamma_{i_{n/2}}(\tau_{n/2}) 
  \gamma_{j_{n/2}}^\star(\tau'_{n/2}) \cdots \gamma^\star_{j_1}(\tau'_{1}) 
  \right\rangle'_{\rm c}, 
  \label{cumulantn}
\end{eqnarray} 
where $\binom{n}{n/2}=n!/(n/2)!^2$ is the binomial coefficient. 
Notice that the $n$th cumulant involves $n/2$-body correlation functions. 
The cumulant expansion in Eq.~(\ref{cumulant}) with Eq.~(\ref{cumulantn}) allows 
one to systematically approximate the original system, 
depending on the selection of the reference system and the expansion order~\cite{Pairault1998,Pairault2000,Senechal2002,Adibi2016}.  
Note also that $n/2$ in Eq.~(\ref{cumulantn}) corresponds to 
the expansion order ``$R$'' in Refs.~\cite{Pairault1998,Pairault2000,Senechal2002,Adibi2016}.

\subsubsection{Lowest order approximation}

So far, no approximation has been made. 
Here, as in the CPT~\cite{Senechal2002}, 
we make an approximation by taking the cumulant expansion 
in Eq.~(\ref{cumulant}) only up to the lowest order ($n=2$).  
The exponent in Eq.~(\ref{cumulant}) for $n=2$ is given as 
\begin{eqnarray}
  \frac{1}{2!} 
  \left \langle \left(\bs{\psi}^\star \bs{\gamma} + \bs{\gamma}^\star \bs{\psi}\right)^2 \right \rangle'_{\rm c}
  &=& \int_0^\beta \dd \tau \int_0^\beta \dd \tau' \sum_{i}\sum_{j} \notag \\
  &\times& \psi^\star_i(\tau)  \psi_j(\tau')  
  \left\langle \gamma_i(\tau) \gamma^\star_j(\tau') \right\rangle', 
  \label{cumulant2}
\end{eqnarray} 
and Eq.~(\ref{cumulant}) is now approximated as  
\begin{equation}
  \left \langle \e^{ \left(\bs{\psi}^\star \bs{\gamma} + \bs{\gamma}^\star \bs{\psi} \right)} \right \rangle' 
  \approx \e^{-\bs{\psi}^\star \bs{G}_{\rm r} \bs{\psi}}, \label{cpt_approx} 
\end{equation} 
where $[\bs{G}_{\rm r}(\tau-\tau')]_{ij}=-\langle \gamma_i(\tau) \gamma_j^\star(\tau') \rangle'$ is 
the imaginary-time single-particle Green's function of the reference system $\hat{H}_{\rm ref}$
and can be evaluated numerically exactly. 
Note that the quadratic form $\bs{\psi}^\star \bs{G}_{\rm r} \bs{\psi}$ 
can be diagonalized with respect to the Matsubara frequency as 
\begin{eqnarray}
  \bs{\psi}^\star \bs{G}_{\rm r} \bs{\psi} 
  &=& \int_0^\beta \dd \tau \int_0^\beta \dd \tau' \sum_{ij} \psi^\star_i(\tau) [\bs{G}_{\rm r}(\tau-\tau')]_{ij} \psi_j(\tau') \notag \\
  &=& \sum_{\nu=-\infty}^\infty \sum_{ij} \psi^\star_i(\imag \w_\nu) [\bs{G}_{\rm r}(\imag \w_\nu)]_{ij} \psi_j(\imag \w_\nu), 
\end{eqnarray} 
where $\bs{G}_{\rm r}(\imag \w_\nu)$ is the Fourier transformation of $\bs{G}_{\rm r}(\tau-\tau')$, i.e., 
\begin{equation}
 \bs{G}_{\rm r}(\tau-\tau')=\frac{1}{\beta} \sum_{\nu=-\infty}^\infty \bs{G}_{\rm r}(\imag \w_\nu) \e^{-\imag \w_\nu (\tau-\tau')}, 
\end{equation}
and the Fourier transformations of the Grassmann fields 
\begin{eqnarray}
  \psi_i(\imag \w_\nu) &=&       \frac{1}{\sqrt{\beta}} \int_0^\beta \dd \tau \psi_i (\tau) \e^{\imag \w_\nu \tau},
  \label{gt1}
\end{eqnarray}
and
\begin{eqnarray}
  \psi_i^\star(\imag \w_\nu) &=& \frac{1}{\sqrt{\beta}} \int_0^\beta \dd \tau \psi_i^\star (\tau) \e^{-\imag \w_\nu \tau} 
  \label{gt2}
\end{eqnarray} 
are introduced. Since $\bs{V}$ is a static quantity, 
$\bs{\psi}^\star \bs{V}^{-1} \bs{\psi}$ is diagonalized 
either in the imaginary-time 
or Matsubara-frequency representation, i.e., 
%\begin{eqnarray} 
$
\bs{\psi}^\star \bs{V}^{-1} \bs{\psi} 
= \int_0^\beta \dd \tau \bs{\psi}^\star(\tau) \bs{V}^{-1} \bs{\psi}(\tau) 
= \sum_{\nu=-\infty}^{\infty} \bs{\psi}^\star(\imag \w_\nu) \bs{V}^{-1} \bs{\psi}(\imag \w_\nu), 
$
%\end{eqnarray} 
where $\beta^{-1}\int_0^\beta \e^{\imag (\w_\nu - \w_{\nu'}) \tau} =\delta_{\nu, \nu'} $ is used. 

Substituting the approximation~(\ref{cpt_approx}) into Eq.~(\ref{Zaux}) yields  
\begin{eqnarray}
  \frac{Z_{\rm aux}}{Z_{\rm ref}} 
  &\approx& {\rm Det} (-\bs{V}) \int {\cal D} [\psi^\star \psi] \e^{-\bs{\psi}^\star
    \left(\bs{G}_{\rm r} -\bs{V}^{-1} \right) \bs{\psi} } \notag \\
  &=&\prod_{\nu=-\infty}^\infty 
    \det{\left[\bs{I}-\bs{V}\bs{G}_{\rm r}(\imag \w_\nu) \right]}, 
  \label{Zaux/Z'}
\end{eqnarray} 
where all the Grassmann fields ($\gamma$, $\gamma^\star$, $\psi$, and $\psi^\star$) 
are assumed to be in the Matsubara-frequency representation 
and thus ${\rm Det}(\cdots)$ is given as 
\begin{equation}
  {\rm Det} (\cdots) = \prod_{\nu=-\infty}^\infty \det [\cdots], \label{Det} 
\end{equation} 
and $\det[\cdots]$ is the determinant with respect to the remaining single-particle indices. 
The Gaussian integral with respect to 
$\psi^\star$ and $\psi$ is performed in Eq.~(\ref{Zaux/Z'}). 
Note that the Jacobians for the Grassmann-variable transformation are not necessarily to be considered here 
because they cancel out between the numerator and the denominator of $Z_{\rm aux}/Z_{\rm ref}$.

\subsubsection{Grand-potential functional in CDIA}

Taking the logarithm of Eq.~(\ref{ratioZZ}) with 
the approximation in Eq.~(\ref{Zaux/Z'}) yields the grand potential $\Omega$ of the system as  
\begin{eqnarray}
  \Omega
  &\approx&
  \Omega_{\rm r}-\Omega_{\rm bath} \notag \\
  &-&\frac{1}{\beta} \sum_{\nu=-\infty}^{\infty} \sum_{\tilde{\mb{k}}} \ln \det \left[\bs{I}-\bs{V}(\tilde{\mb{k}})\bs{G}'(\imag \w_\nu)\right]
  \label{grand_cdia}
\end{eqnarray}
where $\Omega_{\rm r}$ is the grand potential of the reference system and 
\begin{equation}
  \Omega_{\rm bath} = -\frac{1}{\beta}\ln{Z_{\rm bath}}  
\end{equation} 
is the grand potential of the isolated bath system. 
In Eq.~(\ref{grand_cdia}), assuming the reference system being composed of the identical clusters and thus  
the translational symmetry of the superlattice of clusters, 
we decomposed the single-particle indices into 
the wavevector $\tilde{\mb{k}}$ belonging to the reduced Brillouin zone of the superlattice and 
the remaining indices for the cluster and bath orbitals which are considered in $\det[\cdots]$. 
$\bs{G}'(\imag \w_\nu)$ is the single-particle Green's function of a single cluster in the reference system. 

The grand-potential functional in the CDIA is obtained from the approximate grand potential in Eq.~(\ref{grand_cdia}). 
Since the auxiliary system and the reference system 
have the same degrees of freedom and the same interaction term, 
the VCA can be made between these two systems. 
Thus, the CDIA evaluates approximately the grand potential $\Omega$ of the original system as follows. 
First, apply the VCA to the auxiliary system defined in Eq.~(\ref{Saux}) 
with the reference system given in Eq.~(\ref{Sref}) 
[also see Figs.~\ref{CDIAsystems}(b) and \ref{CDIAsystems}(c)] 
to evaluate the grand-potential functional of the auxiliary system 
\begin{equation}
  \Omega_{\rm aux}[\Sigma_{{\rm r},\bs{\lambda}}] = \Omega_{\rm r}[\Sigma_{{\rm r},\bs{\lambda}}] - \frac{1}{\beta} {\rm Tr} \ln (\bs{I}-\bs{V}\bs{G}_{\rm r}), 
\end{equation} 
where $\Sigma_{{\rm r},\bs{\lambda}}$ is the self-energy of the reference system parametrized by the single-particle 
parameter $\bs{\lambda}$,
and find the stationary condition 
\begin{equation}
  \left.\frac{\partial \Omega_{\rm aux}[\Sigma_{{\rm r},\bs{\lambda}}]}{\partial \bs{\lambda}}\right|_{\bs{\lambda}=\bs{\lambda}^*} = \bs{0}
  \label{eq:sc}
\end{equation} 
to obtain the grand potentials $\Omega_{\rm aux}$ and $\Omega_{\rm bath}$ with the optimized 
single-particle parameters $\bs{\lambda}^*$. 
Next, subtract $\Omega_{\rm bath}$ from $\Omega_{\rm aux}$ to finally obtain the 
approximate grand-potential of the original system [see Figs.~\ref{CDIAsystems}(a) and \ref{CDIAsystems}(b)] 
\begin{equation}
  \Omega[\Sigma_{{\rm r},\bs{\lambda}^*}]  = \Omega_{\rm aux}[\Sigma_{{\rm r},\bs{\lambda}^*}] - \Omega_{\rm bath}. \label{CDIAomega} 
\end{equation}
Note that $\Omega_{\rm bath}= - T \ln Z_{\rm bath}$ has to be subtracted at finite temperatures in Eq.~(\ref{CDIAomega}), 
although this term is irrelevant at $T=0$.

\subsection{Application of finite-temperature CDIA} 

\subsubsection{Setting up}

Having formulated the finite-temperature CDIA, we now apply the method to 
examine the finite-temperature phase diagram and the single-particle excitations 
for the paramagnetic Mott metal-insulator transition of the 2D Hubbard model at half filling.  

For this purpose, here we consider the reference system composed of the clusters of $L_{\rm c}=2 \times 2$ 
correlated sites connected to four bath orbitals, as schematically shown in Fig.~\ref{CDIAsystems}(c).  
Because of bath orbitals, the CDIA can induce the density fluctuations within 
the correlated sites in the cluster, which is absent in the VCA.  
Each correlated site in the cluster is connected to a single bath orbital with 
the hybridization parameter $V'$ which is treated as a variational parameter to be optimized. 
The cluster Hamiltonian $\hat{H}'$ is thus given as 
\begin{equation}
  \hat{H}'=\hat{h} + \hat{h}_{\rm bath} + \hat{h}_{\rm hyb}, 
\end{equation}
where $\hat{h}$ is the single-band Hubbard Hamiltonian $\hat{H}$ in Eq.~(\ref{eq.ham}) defined on the 
correlated sites within the cluster 
under open boundary conditions, 
\begin{equation}
  \hat{h}_{\rm bath} = \sum_{k=1}^{4}\sum_{\s=\up,\dn} \eps_{k} \hat{b}_{k \s}^\dag \hat{b}_{k\s} \label{eq.Hbath}
\end{equation}
is the bath Hamiltonian with $\hat{b}_{k \s}^\dag$ being the electron creation operator with spin 
$\s\,(=\uparrow,\downarrow)$ at bath orbital $k$, and 
\begin{equation}
  \hat{h}_{\rm hyb} = V' \sum_{i=1}^{4}\sum_{k=1}^{4}\sum_{\s=\up,\dn} 
  \delta_{i,k}\left(\hat{c}_{i \s}^\dag \hat{b}_{k\s} + {\rm H. c.} \right) \label{eq.HV}
\end{equation}
represents the hybridization between the correlated sites 
and the bath orbitals. 
Here, $ \delta_{i,k}$ is the Kronecker delta. 
Since we consider the particle-hole symmetric case at half filling, 
the bath energy can be fixed at $\eps_{k}=0$  as in Eq.~(\ref{eq.Hbath}) 
even at finite temperatures. 
Therefore, the hybridization $V'$ in Eq.~(\ref{eq.HV}) 
is the only variational parameter to be optimized.

\subsubsection{Finite-temperature phase diagram} 

\begin{figure*}
  \begin{center}
    \includegraphics[width=0.9\textwidth]{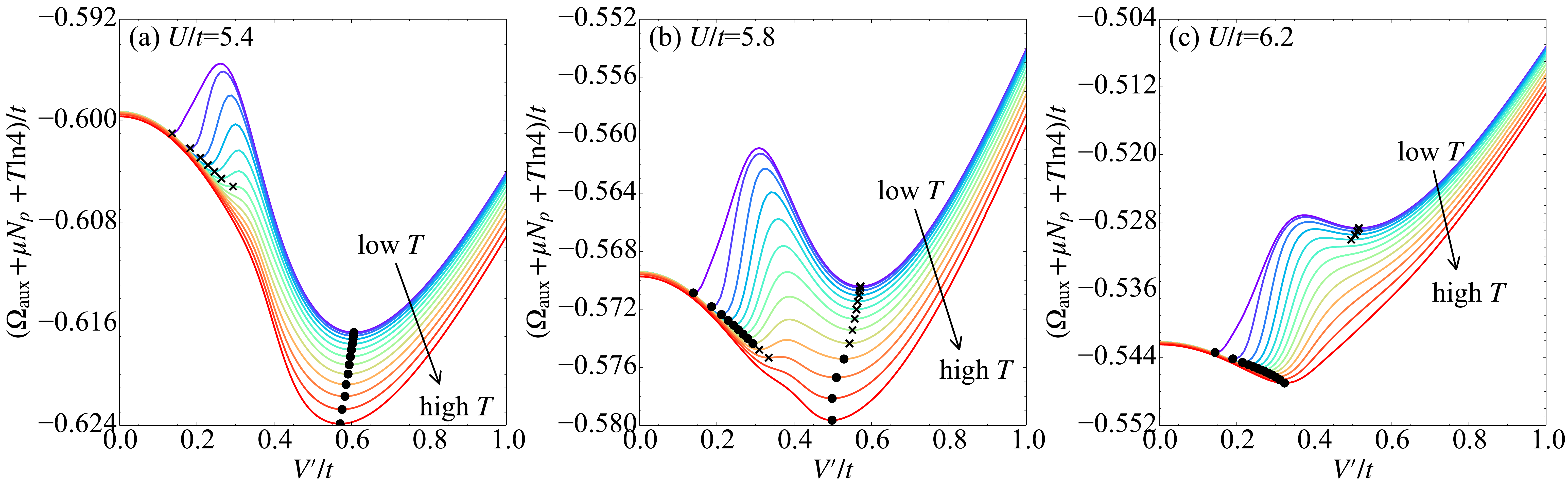}
    \caption{
      The grand-potential functional $\Omega_{\rm aux}(V)$ per site of the auxiliary system 
      as a function of the variational parameter $V'$ for 
      (a) $U/t=5.4$ in the metallic phase, 
      (b) $U/t=5.8$ in the vicinity of the Mott metal-insulator transition, and 
      (c) $U/t=6.2$ in the paramagnetic Mott insulating phase at temperatures 
      $T/t=0.001, 0.005, 0.01, 0.015, \cdots, 0.055$, and $0.06$ (from violet to red lines).
      The cluster composing $L_{\rm c}=2 \times 2$ correlated sites connected to four bath orbitals is used. 
      Dots (crosses) indicate the solutions of nonzero $V'$ with the lowest (second-lowest) grand potential 
      satisfying the stationary condition in Eq.~(\ref{eq:sc}).  
      \label{MITomega}} 
  \end{center}
\end{figure*}

Figure~\ref{MITomega} shows 
the $V'$ dependence of the grand-potential functional $\Omega_{\rm aux}(V')$ per site 
of the auxiliary system at different temperatures 
for three values of $U$ representative for 
the metallic phase ($U/t=5.4$), 
the vicinity of the Mott metal-insulator transition ($U/t=5.8$), and 
the Mott insulating phase ($U/t=6.2$).
The increment of $V'/t$ is set to be 0.01. 
For comparison with the zero-temperature results in Ref.~\cite{Balzer2009}, 
we plot $\Omega_{\rm aux}(V') + \mu N_p-\Omega_{\rm bath}/L_{\rm c}$, where 
$N_p\,(=1)$ is the particle number density with 
the chemical potential $\mu=U/2$ for the 
particle-hole symmetric case at half filling, and  
$\Omega_{\rm bath}=-T \ln W_{\rm bath} = -TL_{\rm c}\ln 4$ is the grand potential of the bath system 
($W_{\rm bath}=4^{L_{\rm c}}$ is the degeneracy of the bath system with $L_{\rm c}$ orbitals). 
The results in Fig.~\ref{MITomega} should be compared with the internal-energy functional 
calculated in Ref.~\cite{Balzer2009} at zero temperature. 
Note also that $\Omega_{\rm aux}(V') -\Omega_{\rm bath}$ 
is the grand-potential functional $\Omega(V')$ of the system for a given $V$ [see Eq.~(\ref{CDIAomega})].

The CDIA grand-potential functional depends sensitively on $T$ specially at low temperatures, 
as compared with the $T$ dependence of the VCA grand-potential functional shown in Fig.~\ref{grandpotential}. 
This is due to the fact that many low-lying excited states 
exist in $\hat{H}'$ for the CDIA because of the bath orbitals. 
At low temperatures, $\Omega_{\rm aux}(V')$ exhibits two minima at $V'^* \not = 0$. 
The minimum of the grand-potential functional with the smaller $V'^*$ corresponds to the insulating solution, 
while the larger $V'^*$ corresponds to the metallic one. 
The coexistence region in the $U$-$T$ phase diagram is thus identified as 
the parameter region in which $\Omega_{\rm aux}(V')$ shows the two minima at $V'^* \not =0$.

The metal-insulator transition takes place at a critical interaction strength $U_{\rm c}(T)$
where the metallic and insulating solutions have the same value of the grand potential 
for a given temperature $T$. 
The solution jumps from one to the other by varying $U$ across $U_{\rm c}(T)$. 
Therefore, the metal-insulator transition is discontinuous. 
The discontinuity of the transition persists down to the zero-temperature limit, thus in  
good agreement with the previous CDIA result at zero temperature~\cite{Balzer2009}. 
Similarly, the solution jumps from one to the other by varying $T$ across the transition temperature 
[see Fig.~\ref{MITomega}(b)]. 
It is also found in Fig.~\ref{MITomega} that 
one of the two solutions vanishes above a certain temperature, indicating that 
the metallic and insulating states become no longer distinguishable.

\begin{figure}
  \begin{center}
    \includegraphics[width=7.5cm]{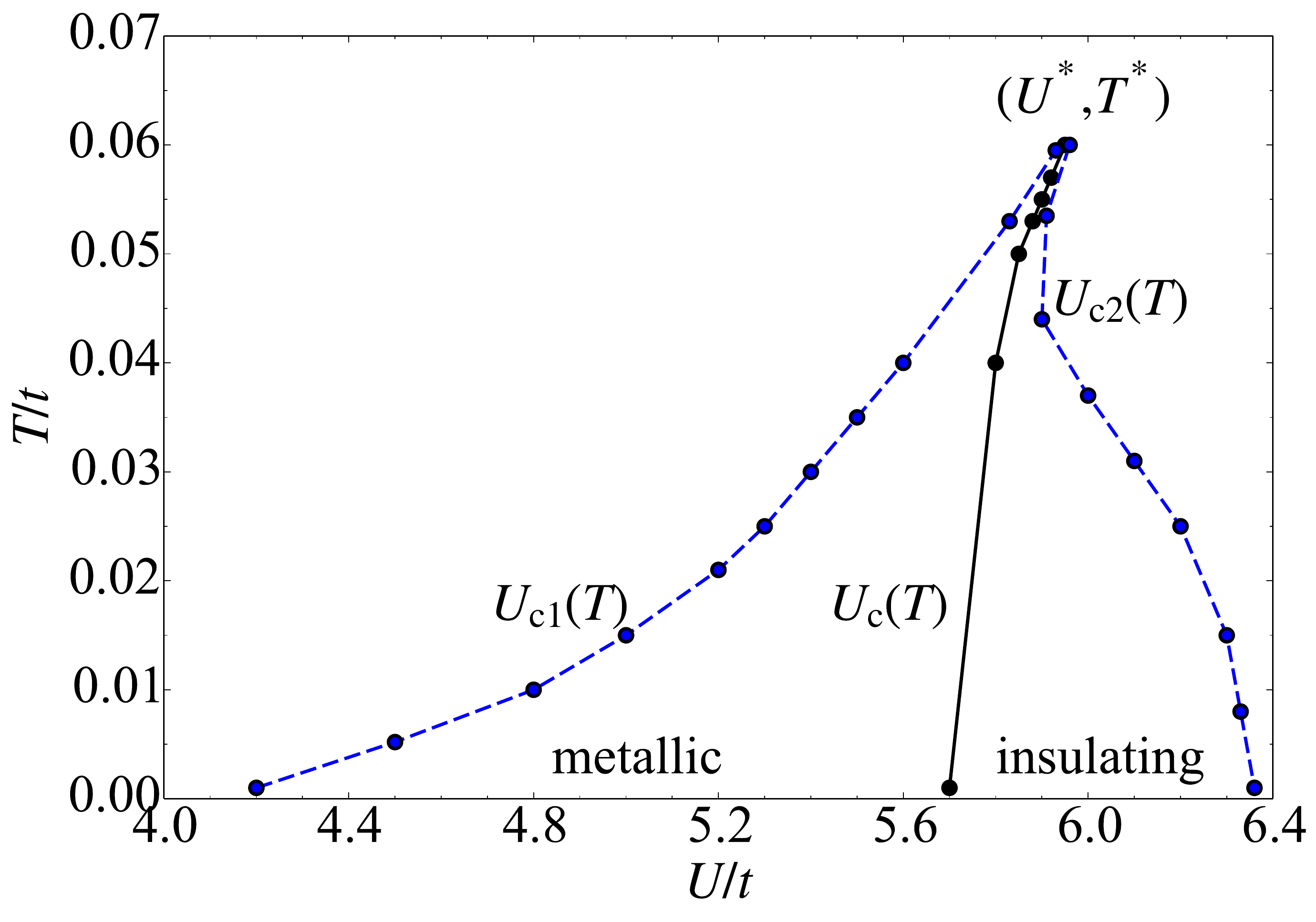}
    \caption{
      The finite-temperature phase diagram for the paramagnetic Mott metal-insulator transition 
      obtained by the finite-temperature CDIA. The cluster composing $L_{\rm c}=2\times 2$ correlated 
      sites connected to four bath orbitals 
      is used. The three lines with dots represent  
      $U_{{\rm c}1}(T)$, $U_{\rm c}(T)$, and $U_{{\rm c}2}(T)$, 
      which terminate at a critical point $(U^*/t,T^*/t)=(5.95,0.061)$. 
      Black dots denote the first-order transition boundary separating the metallic and insulating phases. 
      The metallic and insulating phases coexist in the region surrounded by blue dots.  
      \label{MITphase}}
  \end{center}
\end{figure}

Figure~\ref{MITphase} shows the finite-temperature phase diagram in the $U$-$T$ plane.  
The phase diagram contains the three boundaries, $U_{\rm c}(T)$, $U_{{\rm c}1}(T)$, and $U_{{\rm c}2}(T)$. 
$U_{\rm c}(T)$ is the critical $U$ for the metal-insulator transition at temperature $T$, 
while $U_{{\rm c}1}(T)$ and $U_{{\rm c}2}(T)$ bound the $(U,T)$ region 
in which the metallic and the insulating solutions coexist.  
The three boundaries terminate at a critical point $(U^*/t,T^*/t)\approx(5.95,0.061)$. 
The finite-temperature phase diagram expected in the zero-temperature CDIA study~\cite{Balzer2009}  
is in qualitative agreement with our result, but 
the $U_{\rm c2}(T)$ boundary is more complicated ``S-shape''-like in 
the region of $5.85 \lesssim U/t \lesssim 5.95$ at $0.04 \lesssim T/t \lesssim 0.061$ in Fig.~\ref{MITphase}.

\subsubsection{Single-particle excitations}

\begin{figure*}
  \begin{center}
    \includegraphics[width=.9\textwidth]{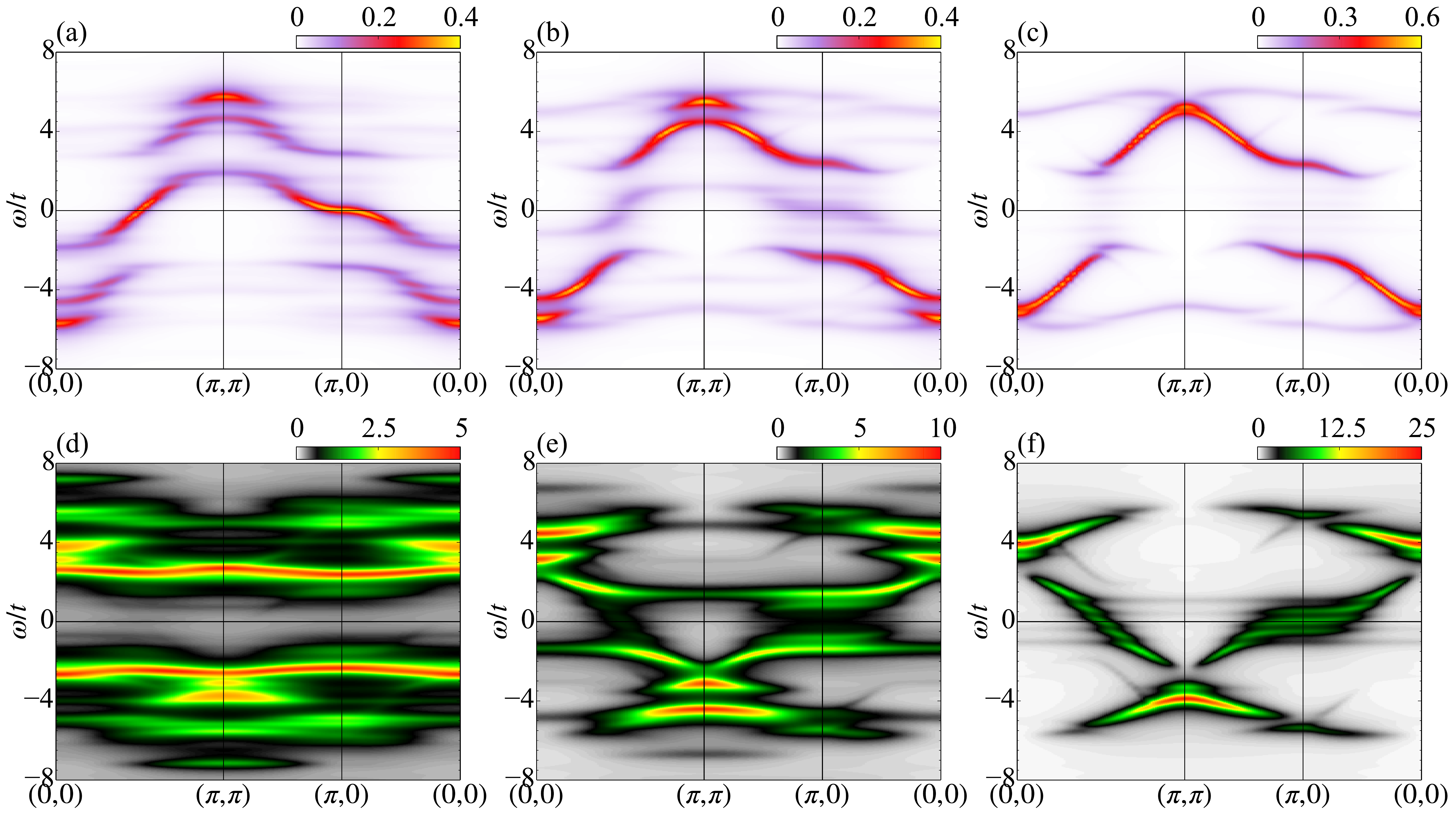}
    \caption{
      Single-particle excitation spectrum $\mcal{A}(\mb{k},\w)$ [(a)--(c)] 
      and the imaginary part $\mcal{S}(\mb{k},\w)$ of the self energy [(d)--(f)] 
      for the half-filled Hubbard model at 
      (a, d) $U/t=5.4$ and $T/t=0.02$ in the metallic phase, 
      (b, e) $U/t=5.95$ and $T/t=0.061$ at the critical point, and 
      (c, f) $U/t=6.2$ and $T/t=0.02$ in the paramagnetic Mott insulating phase. 
      The horizontal line at $\w=0$ denotes the Fermi level.
      The Lorentzian broadening of $\eta/t=0.2$ is used. 
      The CDIA is employed with the cluster of $L_{\rm c}=2\times2$
      correlated sites connected to four bath orbitals. 
      Note that different figures use different intensity scales as indicated in the color bars. 
      \label{spectra2}}
  \end{center}
\end{figure*}

Figure~\ref{spectra2} summarizes the single-particle excitation spectrum $\mcal{A}(\mb{k},\w)$ 
and the imaginary part of the self-energy $\mcal{S}(\mb{k},\w)$ 
for three representative sets of parameters at $(U/t,T/t)=(5.4, 0.02)$
in the metallic phase, 
$(5.95, 0.061)$ at the critical point, and 
$(6.2, 0.02)$ in the paramagnetic Mott insulating phase. 

As shown in Fig.~\ref{spectra2}(a), $\mcal{A}(\mb{k},\w)$ in the metallic phase exhibits 
a three-peak structure, i.e., 
the coherent quasiparticle dispersion near the Fermi level at $\w=0$, 
the upper Hubbard band around $\w/t \sim 4$, and 
the lower Hubbard band around $\w/t \sim -4$.   
Accordingly, $\mcal{S}(\mb{k},\w)$ shown in Fig.~\ref{spectra2}(d) 
does not have a finite spectral weight around the Fermi level but instead 
has a sizable spectral weight with a less-dispersive structure around $\w/t \sim \pm2.5$, 
separating the quasiparticle dispersion from the upper and the lower Hubbard bands in $\mcal{A}(\mb{k},\w)$.

In the insulating phase, 
the coherent quasiparticle dispersion is absent near the Fermi level but   
a tiny amount of spectral weight remains, as shown in Fig.~\ref{spectra2}(c). 
This tiny spectral weight is due to the finite hybridization 
between the correlated sites and the bath orbitals. 
On the other hand, the upper and lower Hubbard bands 
appearing around $2 \lesssim |\w/t| \lesssim 5$ acquire a sizable spectral weight. 
As shown in Fig.~\ref{spectra2}(f), 
$\mcal{S}(\mb{k},\w)$ around the Fermi level in $|\w/t|\alt2$ is more dispersive, 
as compared with that in the Hubbard-I approximation [see Fig.~\ref{spectra}(j)],  
and rather similar to that in the SDW mean-field 
theory [see Fig.~\ref{spectra}(f)]. This implies the presence of the antiferromagnetic fluctuations  
in the paramagnetic Mott insulating state. 

The results at the critical point are shown in Figs.~\ref{spectra2}(b) and \ref{spectra2}(e).  
Although there appears a finite single-particle excitation spectrum around the Fermi level 
as in the metallic state, 
it is no longer coherent and thus quasiparticles do not exist. 
In contrast, the upper and lower Hubbard bands can be clearly observed 
as in the insulating state. 
As shown in Figs.~\ref{spectra2}(e), 
$\mcal{S}(\mb{k},\w)$ exhibits a clear structure at $\w/t \sim \pm1.5$, which separates  
the low-energy incoherent excitations around the Fermi level 
and the upper and lower Hubbard bands in $\mcal{A}(\mb{k},\w)$. 
The similar structure is found in the metallic state at $\w/t \sim \pm2.5$ 
shown in Fig.~\ref{spectra2}(d).   
It is also noticed in Fig.~\ref{spectra2}(e) that 
$\mcal{S}(\mb{k},\w)$ shows 
incoherent spectra around $4\alt |\omega|\alt6$, similar to that in the insulating state, although 
the intensity is quite smaller. 
The overall feature of $\mcal{A}(\mb{k},\w)$ and $\mcal{S}(\mb{k},\w)$ at the critical point is thus 
characterized by the average over the metallic and insulating states.

\section{Summary and Discussion}\label{summary} 

A finite-temperature VCA algorithm suitable 
for the exact-diagonalization cluster solver has been formulated. 
The major difficulty of the current finite-temperature VCA is overcome 
by analyzing the analytical properties of logarithm of the complex determinant 
function which appears in the SFT grand-potential functional. 
Explicit formulas of the thermodynamic quantities in the SFT have been derived. 
These quantities include the 
grand potential, entropy, and specific heat. 
The block-Lanczos method has also been proposed to efficiently calculate 
the single-particle Green's function of the cluster.

The finite-temperature VCA developed here is applied to 
the single-band Hubbard model on the square lattice at half filling. 
We have obtained 
the finite-temperature phase diagram containing the paramagnetic and antiferromagnetic phases.  
Although we have found a finite N\'{e}el temperature, 
this is due to 
the mean-field like treatment of the spatial correlations 
beyond the size of clusters in the VCA. 
Moreover, we have examined the temperature dependence of the single-particle excitations 
and the results are compared with the Hubbard-I approximation and the SDW 
mean-field theory. 
In order to characterize the crossover between 
the weak-coupling Slater-type insulator and 
the strong-coupling Mott-type insulator 
in the $U$-$T$ plane, 
we have calculated various thermodynamic quantities such as 
the entropy $S$, the double occupancy $\langle \hat{D} \rangle$, the specific heat $C$, 
the double-occupancy susceptibility $\chi_D$, and the mixed derivative 
$\partial_U S = \partial_U \partial_T \Omega = - \partial_T \langle \hat{D} \rangle$ 
of the grand potential. 
The entropy and the specific heat 
show a kink and a jump at the N\'{e}el temperature, respectively, 
indicating that the antiferromagnetic transition is second ordered. 
We have also examined the third law of the thermodynamics within the VCA 
and shown rigorously that the third law of the thermodynamics is guaranteed 
if and only if the ground state of the cluster is unique.

Furthermore, we have extended the finite-temperature VCA scheme to the finite-temperature CDIA to 
investigate the finite-temperature paramagnetic Mott metal-insulator transition for the single-band 
Hubbard model on the square lattice at half filling. 
After formulating the finite-temperature CDIA, 
we have demonstrated that the systematic evaluation of the grand-potential functional as a function of 
the hybridization parameter $V'$ allows us to clearly identify  
the metallic phase, the Mott insulating phase, the coexisting region, and the crossover region 
in the $U$-$T$ phase diagram. 
We have shown that 
the first-order metal-insulator transition boundary $U_{\rm c}(T)$ is terminated 
at a critical point $(U^*/t,T^*/t)$. 
We have also calculated the single-particle excitation spectrum and found that the coherent 
quasiparticle dispersion 
exists near 
the Fermi level in the metallic phase, while only the upper and lower Hubbard bands have a sizable spectral weight   
in the Mott insulating phase. At the critical point, no quasiparticle dispersion crossing the Fermi level is 
found but the incoherent excitations with a small spectral weight 
are observed around the Fermi level, in addition to the upper and lower Hubbard bands.

The finite-temperature VCA scheme developed here is particularly suitable 
for low to intermediate temperatures because it has to truncate the high-energy excited states 
when the large clusters are employed. 
However, as demonstrated in Sec~\ref{results} and Sec.~\ref{cdia_results}, 
the method can be applied successfully to obtain the finite-temperature phase diagrams and 
examine the single-particle excitations across the transitions for the two-dimensional 
single-band Hubbard model at half filling.

The finite-temperature VCA can treat exactly the thermal and quantum fluctuations on an equal footing 
within the clusters. Therefore, it is highly interesting to apply the method to various strongly interacting 
fermions. The immediate application is to 
investigate the carrier-doped Mott insulator and an emergent $d$-wave 
superconductivity, and to elucidate the pseudogap phenomena in 
cuprates~\cite{Lee2006}.

One- or two-atom-thick-layer $^3$He atoms on graphite surface 
are also interesting strongly correlated spin-$1/2$ fermion systems~\cite{Fukuyama2008,Neumann2007}. 
Here, $^3$He atoms repel each other due to their hard core potentials with each $^3$He atom holding 
a nuclear-spin $1/2$. 
Experimental measurements of the thermodynamic quantities at low temperatures 
are valuable to reveal the ground-state and low-lying excitation properties of these systems.  
Theoretically, the nuclear magnetism of the monolayer $^3$He system has been studied by analyzing 
the Heisenberg or $t$-$J$ models on the triangular lattice with 
cyclic exchange interactions~\cite{Momoi1999,Momoi2006,Momoi2012,Fuseya2009,Seki2009}. 
The bilayer $^3$He system has also been studied theoretically using a periodic Anderson model on a 
stacked triangular lattices~\cite{Beach2011}, for which the finite-temperature VCA can also be adopted. 

Another interesting class of systems to which the finite-temperature VCA can be applied is 
organic frustrated Mott insulating materials~\cite{Powell2011,Zhou2016}. 
In these materials, various thermodynamic quantities are 
measured experimentally with controlling the electron correlation parameters~\cite{Kurosaki2005}, 
the degree of geometric frustration~\cite{Watanabe2012}, and 
the electron filling~\cite{Kawasugi2009,Kawasugi2011,Kawasugi2016,Sato_Kawasugi2016}. 
Recently, it has been observed in an organic Mott insulator ${\rm EtMe_{3}Sb[Pd(dmit)_{2}]_{2}}$~\cite{Itou2017}  
that the first-order nature of the correlation-induced metal-insulator transition 
is obscured by disorders and instead an intermediate region called 
an electronic Griffiths phase~\cite{Andrade2009} emerges. 
One possible way to treat the disorder effect based on the scheme developed here is 
the finite-temperature CDIA method based on 
the SFT formalism for disordered systems~\cite{Potthoff2007}. 

Finally, ultracold atoms now allow us to 
study not only the static quantities~\cite{Cocchi2016,Cocchi2017} 
but also the dynamical ones~\cite{Bohrdt2018,Brown2018}
of interacting fermions at finite temperatures. 
Although there are still experimental difficulties with lowering 
the temperature down to extremely low temperatures,  
they can reach to relatively low temperatures where 
the short-range correlations are important. 
There, the finite-temperature VCA scheme can be used,  
as a complement or an extension of the DMFT-like methods, 
to make comparison with the experiment for better understanding  
the finite-temperature properties of interacting fermions.  
%}

In order to reach the higher temperatures, 
stochastic sampling techniques for the many-body-state vectors in the Krylov subspace 
would be promising, instead of directly solving 
the eigenvalue problems of the large Hamiltonian matrix.  
These stochastic methods include the finite-temperature Lanczos method~\cite{Prelovsek,Jaklic1994,Jaklic2000,Kokalj2017}, 
the low-temperature Lanczos method~\cite{Aichhorn2003}, and 
the thermal-pure-quantum-state-based method~\cite{Sugiura2013}. 
A block extension of these stochastic methods 
would also be of technical interest to reduce the computational cost.
In particular, a block-Lanczos extension of 
the finite- and low-temperature Lanczos methods  
would be straightforward by following the description in Sec.~\ref{sec.BL}.

\section*{ACKNOWLEDGMENTS} 
The authors are grateful to 
Hiroshi Watanabe, Jure Kokalj, and Alexander Wei{\ss}e 
for helpful comments. 
The numerical computations have been done 
on HOKUSAI GreatWave supercomputer at RIKEN Advanced Center for 
Computing and Communication (ACCC) under
Projects No. G16029, No. G17032, and No. G18025. 
K.S. acknowledges support from the JSPS Overseas Research Fellowships. 
T.S. acknowledges the Simons Foundation for funding. 
This work was also supported by RIKEN Molecular Systems Project 
and RIKEN iTHES Project.

\appendix

\section{Another expression of thermodynamic quantities}\label{app.anotherOmega}

In this Appendix, we derive an analytical expression of 
the second term of the right hand side in Eq.~(\ref{eq.Omega-num}) 
and discuss briefly a possible application of the KPM~\cite{Weisse2006} to the VCA.  

\begin{figure}
  \begin{center}
    \includegraphics[width=1.0\columnwidth]{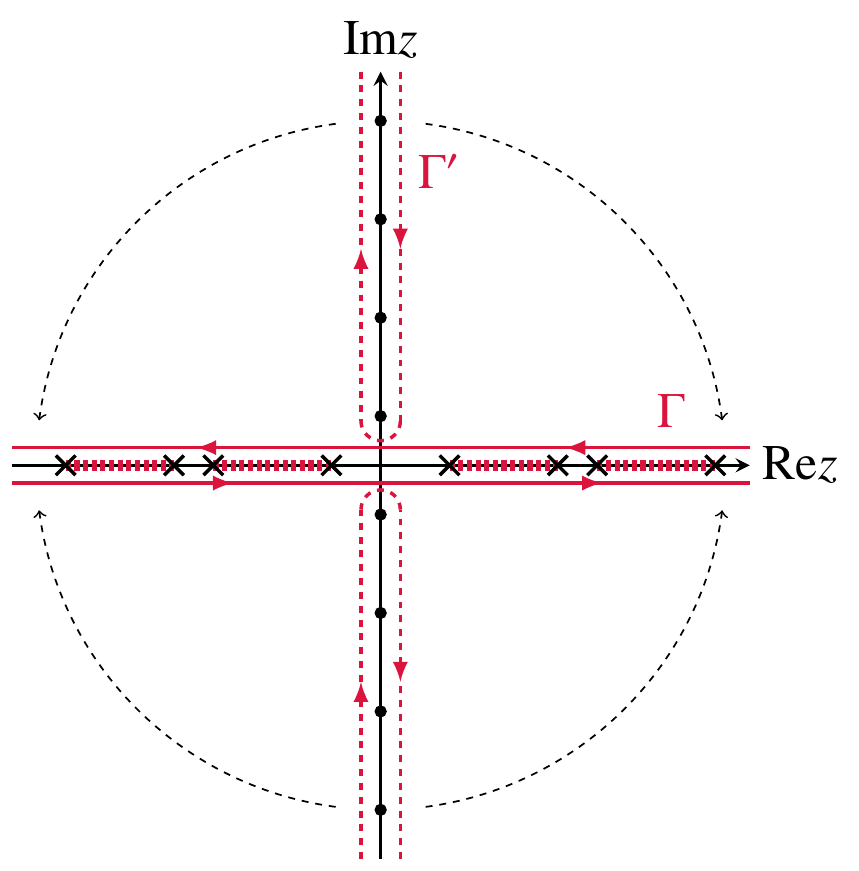}
    \caption{
      Contour $\Gamma'$ (dashed lines) encloses the fermionic Matsubara 
      frequencies (solid circles) in a clockwise manner. 
      Contour $\Gamma$ (solid lines) enclosing the real axis is obtained by deforming contour $\Gamma'$.  
      The branch cuts and the branch points 
      of $ \ln \det\left[\bs{I} - \bs{V}(\tilde{\mb{k}}) \bs{G}'(z) \right]$ are indicated by 
      magenta dotted lines and black crosses on the real axis, respectively (see also Fig.~\ref{contour} 
      and Fig.~\ref{fig.cuts}).  
      \label{fig.Gamma}}
  \end{center}
\end{figure}

Substituting Eq.~(\ref{eq.branch}) into the second term of 
the right-hand side in Eq.~(\ref{eq.Omega-num}) yields 
\begin{eqnarray}
  & &\Omegapersite - \frac{\Omega'}{L_{\rm c}} \notag \\
  &=&-\frac{1}{NL_{\rm c}} \sum_{\tilde{\mb{k}},p} \oint_{\Gamma'} \frac{\dd z}{2\pi \imag} n_{\rm F} (z) \ln \left( \frac{z - \w_{\tilde{\mb{k}},p}}{z - \w_{p}} \right)  \label{eq:nz_poles}\\
  &=& \frac{1}{NL_{\rm c}} \sum_{\tilde{\mb{k}},p} \oint_{\Gamma} \frac{\dd z}{2\pi \imag} n_{\rm F} (z) \left(\int_{\w_p}^{\w_{\tilde{\mb{k}},p}} \frac{\dd x}{z - x} \right)  \label{eq:nz_poles2}\\
  &=& \frac{1}{NL_{\rm c}} \sum_{\tilde{\mb{k}},p} \int_{\w_p}^{\w_{\tilde{\mb{k}},p}} \dd x n_{\rm F} (x) \label{eq.Omega_nf}  \label{eq:nz} \\
  &=&-\frac{1}{L_{\rm c}   \beta}\left[\frac{1}{N}\sum_{\tilde{\mb{k}},p} \ln \left(1 + \e^{-\beta \w_{\tilde{\mb{k}},p}}\right) - \sum_{p} \ln \left(1 + \e^{-\beta \w_{p}} \right)\right], 
  \label{eq.Omega-pole}
\end{eqnarray} 
where contour $\Gamma'$ in Eq.~(\ref{eq:nz_poles}) encloses all the poles of $n_{\rm F}(z)$ in a clockwise 
manner and can be deformed into contour $\Gamma$ since the integrand 
 $n_{\rm F}(z) \ln \det\left[\bs{I} - \bs{V}(\tilde{\mb{k}}) \bs{G}'(z) \right]$ 
is analytical in the complex region surrounded 
by contours $\Gamma'$ and $\Gamma$ (see Fig.~\ref{fig.Gamma}). 
We can therefore convert the contour integral into 
the real-valued integral of the Fermi-distribution function $n_{\rm F}(x)$
over $[\w_p,\  \w_{\tilde{\mb{k}},p}]$ in Eqs.~(\ref{eq:nz_poles2}) and (\ref{eq:nz}). 
The real-valued integral in Eq.~(\ref{eq:nz}) can be performed by noticing that 
$n_{\rm F}(x) = -\frac{1}{\beta} \frac{\dd}{\dd x} \ln (1+\e^{-\beta x})$, as shown in Eq.~(\ref{eq.Omega-pole}). 
The analytical expression of the grand-potential functional derived in Eq.~(\ref{eq.Omega-pole}) 
is formally similar to the grand potential for the ideal Fermi gas~\cite{Feynman}, and  
is identical to that in Refs.~\cite{Potthoff2003b} and \cite{Eder2008} obtained in different ways.

Differentiating Eq.~(\ref{eq.Omega-pole}) with respect to $T$ 
yields another expression for the entropy, i.e.,  
\begin{eqnarray}
  & & S - \frac{S'}{L_{\rm c}} = -\frac{\dd }{\dd T}
   \left(\Omega - \frac{\Omega'}{L_{\rm c}} \right) \notag \\
  &=& \frac{-1}{L_{\rm c}}
  \left[\frac{1}{N}\sum_{\tilde{\mb{k}},p} \left( 
    n_{\rm F}(-\w_{\tilde{\mb{k}},p}) \ln {n_{\rm F}(-\w_{\tilde{\mb{k}},p})} 
  + n_{\rm F}( \w_{\tilde{\mb{k}},p}) \ln {n_{\rm F}( \w_{\tilde{\mb{k}},p})} \right) \right. \notag \\
  &-& \left.
  \sum_{p} \left( 
    n_{\rm F}(-\w_{p}) \ln {n_{\rm F}(-\w_{p})} 
  + n_{\rm F}( \w_{p}) \ln {n_{\rm F}( \w_{p})} \right) \right], 
  \label{eq.Sc}
\end{eqnarray} 
where $n_{\rm F}(-\w)=1-n_{\rm F}(\w)$ and $\beta \w = \ln \left[n_{\rm F}(-\w)/n_{\rm F}(\w)\right]$ are
used. 
By definition, the internal energy is obtained as 
\begin{eqnarray}
  E - \frac{E'}{L_{\rm c}} &=& \left(\Omega - \frac{\Omega'}{L_{\rm c}} \right) + T \left( S - \frac{S'}{L_{\rm c}}  \right) \notag \\
&=& \frac{1}{L_{\rm c}} \left[ 
  \frac{1}{N} \sum_{\tilde{\mb{k}},p} \w_{\tilde{\mb{k}},p} n_{\rm F} (\w_{\tilde{\mb{k}},p}) 
  - \sum_{p} \w_{p} n_{\rm F} (\w_p)  
  \right]. \label{eq.Ec}
\end{eqnarray}
The specific heat is evaluated as the $T$ derivative of the entropy in Eq.~(\ref{eq.Sc}), i.e., 
\begin{eqnarray}
  C - \frac{C'}{L_{\rm c}} &=&  \frac{\beta^2}{L_{\rm c}} \left[ 
    \frac{1}{N} \sum_{\tilde{\mb{k}},p} \w_{\tilde{\mb{k}},p}^2 n_{\rm F} (\w_{\tilde{\mb{k}},p}) n_{\rm F} (- \w_{\tilde{\mb{k}},p}) \right. \notag \\
     &-& \left. \sum_{p} \w_{p}^2 n_{\rm F} (\w_p)  n_{\rm F} (-\w_p) 
    \right], \label{eq.Cc}
\end{eqnarray} 
where the temperature dependence of the variational parameter is ignored [see Eq.~(\ref{eq.CT})] and this is justified, 
e.g., in the paramagnetic state when the variational parameter is zero. 
Equations ~(\ref{eq.Omega-pole})--(\ref{eq.Cc}) show 
that the thermodynamic quantities $\Omega$, $S$, $E$, and $C$ within the VCA 
involve only the single-particle excitation energies $\w_{\tilde{\mb{k}},p}$ and $\w_{p}$, 
in addition to the corresponding quantities $\Omega'$, $S'$, $E'$, and $C'$ of the cluster, which 
can be calculated numerically exactly.

The expression of the grand-potential functional in Eq.~(\ref{eq.Omega_nf}) 
is remarkably simple because it is expressed solely by the integral of 
the real-valued Fermi-distribution function. 
This is further simplified in the zero-temperature limit, 
where the Fermi-distribution function is replaced by the step function, i.e., 
\begin{eqnarray}
  & &\lim_{T \to 0} 
  \frac{1}{NL_{\rm c}} \sum_{\tilde{\mb{k}},p} \int_{\w_p}^{\w_{\tilde{\mb{k}},p}} \dd x n_{\rm F} (x) \notag \\
  &=&\frac{1}{L_{\rm c}}\left[ 
    \frac{1}{N} \sum_{\tilde{\mb{k}},p}{\w_{\tilde{\mb{k}},p}} \Theta(-{\w_{\tilde{\mb{k}},p}})
    -\sum_{p} {\w_p} \Theta(-{\w_p}) 
               \right], 
  \label{eq.Omega_nfTzero}
\end{eqnarray} 
with $\Theta(x)=0 \ (1)$ for $x<0 \ (x>0)$. 
This can also be derived by taking the 
the zero-temperature limit directly in Eq.~(\ref{eq.Omega-pole})~\cite{Aichhorn2006_Qmat} 
and indeed agrees with the zero-temperature limit of the internal energy in Eq.~(\ref{eq.Ec}).

Recently, Wei{\ss}e has reported 
the Green-function-based Monte Carlo method for 
a double-exchange model with classical local spins~\cite{Weisse2009}, 
where the change of the effective action is calculated efficiently 
by the Chebyshev expansion of the Green's function based on the KPM~\cite{Weisse2006}.  
The KPM is an efficient method to calculate the dynamical correlation functions 
including the single-particle Green's function on the real-frequency axis. 
The similarity between Eq.~(\ref{eq.Omega-pole}) and the change of the effective action expressed in 
Refs.~\cite{Weisse2009} and \cite{Zhang2013} 
suggests that the KPM can also be used to calculate the grand-potential functional 
in the finite-temperature VCA. 
Indeed, the free-fermion-like formulas of the SFT thermodynamic functions in 
Eqs.~(\ref{eq.Omega-pole})--(\ref{eq.Cc}) 
suggest that if the sum of the $\delta$ functions 
\begin{equation}
  \tilde{\rho}(\w) = \frac{1}{N} \sum_{\tilde{\mb{k}},p} \left[\delta(\w - \w_{\tilde{\mb{k}},p}) - \delta(\w - \w_p)\right] 
  \label{countpoles}
\end{equation}
can be evaluated accurately by, e.g., the KPM, 
these thermodynamic quantities are obtained as 
\begin{eqnarray}
  \Omega-\frac{\Omega'}{L_{\rm c}} &=& -\frac{1}{\beta L_{\rm c}} \int_{-\infty}^{\infty} \dd \w \tilde{\rho}(\w) \ln \left(1+\e^{-\beta \w}\right),\\
  S-\frac{S'}{L_{\rm c}} &=& -\frac{1}{L_{\rm c}} \int_{-\infty}^{\infty} \dd \w \tilde{\rho}(\w) \notag \\
  &\times&\left[n_{\rm F}(-\w)\ln n_{\rm F}(-\w) + n_{\rm F}(\w) \ln n_{\rm F}(\w) \right] ,\\
  E-\frac{E'}{L_{\rm c}} &=& \frac{1}{L_{\rm c}} \int_{-\infty}^{\infty} \dd \w \tilde{\rho}(\w) \w n_{\rm F}(\w), 
\end{eqnarray}
and
\begin{equation}
  C-\frac{C'}{L_{\rm c}} = \frac{\beta^2}{L_{\rm c}} \int_{-\infty}^{\infty} \dd \w \tilde{\rho}(\w) \w^2 n_{\rm F}(\w) n_{\rm F}(-\w). 
\end{equation}
Note that $\tilde{\rho}(\w)$ in Eq.~(\ref{countpoles}) is {\it not} the difference 
of density of states between the original and reference systems.  
Instead, $\tilde{\rho}(\w)$ can be expressed as a sum of the logarithmic derivative of 
$\det(I-\bs{V}\bs{G}'(z))$ [see Eq.~(\ref{eq.branch})], i.e.,  
\begin{eqnarray}
  \tilde{\rho}(\w)
  &=&-\lim_{\eta \to 0} {\rm Im} \frac{1}{\pi N} \sum_{\tilde{\mb{k}},p} \left[\frac{1}{\w - \w_{\tilde{\mb{k}},p} + \imag \eta} - \frac{1}{\w - \w_p + \imag \eta}\right], \notag \\
  &=&-\lim_{\eta \to 0} {\rm Im} \frac{1}{\pi N} \sum_{\tilde{\mb{k}},p} \left[\frac{\partial}{\partial z} \ln \det \left(\bs{I}-\bs{V}\bs{G}'(z)\right) \right]_{z=\w+\imag \eta} \notag \\
  &=& \lim_{\eta \to 0} {\rm Im} \frac{1}{\pi N} \sum_{\tilde{\mb{k}},p} {\rm tr} \left[(\bs{I}-\bs{V}\bs{G}'(z))^{-1} \bs{V}\partial_z \bs{G}'(z) \right]_{z=\w+\imag \eta}.  
\end{eqnarray}
In this study, the block-Lanczos method is used to efficiently calculate the single-particle Green's function as 
described in Sec.~\ref{sec.BL} and the complex contour integral is employed for 
the thermodynamic quantities in Sec.~\ref{method}. 
However, it is highly interesting to explore the efficiency of the KPM for the finite-temperature VCA in the future.

\section{Another derivation of entropy}\label{app.ST}

In this Appendix, we show another derivation of 
the entropy given in Eq.~(\ref{SFTentropy}).

\subsection{Contour integrals involving the derivatives of the Fermi-distribution function}
First, 
we recall the formula for the Fermi-distribution function 
\begin{equation}
  n_{\rm F}(z) = \frac{1}{2} + \frac{1}{\beta} \sum_{\nu = -\infty}^\infty \frac{1}{\imag \w_\nu - z}, 
\end{equation}
where $\w_\nu=(2\nu+1)\pi/\beta$ and $\nu$ is integer~\cite{Luttinger1960}. 
The $n$th derivative of $n_{\rm F}(z)$ with respect to $z$ is thus given as 
\begin{equation}
  n_{\rm F}^{(n)}(z)  = \frac{n!}{\beta} \sum_{\nu = -\infty}^\infty \frac{1}{(\imag \w_\nu - z)^{n+1}}.  
\end{equation}
Note that $n_{\rm F}^{(n)}(z)$ has poles of $(n+1)$st order at each fermionic Matsubara frequency 
$i\w_\nu$.
From the Cauchy's integral formula, one can easily show that  
\begin{equation}
  g^{(n)}(z') =\frac{n!}{2 \pi \imag} \oint_{P} \dd z \frac{g(z)}{(z-z')^{n+1}}
  \label{eq:gn}
\end{equation}
for any regular function $g(z)$ in a complex $z$ domain 
containing a non-self-intersecting continuous loop $P$ which encloses $z'$. 
The contour integral in Eq.~(\ref{eq:gn}) is directed in a counter-clockwise manner along contour $P$. 
We thus finally obtain that 
\begin{equation}
  \frac{1}{2 \pi \imag} \oint_{\Gamma'} \dd z  n_{\rm F}^{(n)}(z) g(z) 
  = \frac{(-1)^n}{\beta} \sum_{\nu = -\infty}^\infty  g^{(n)}(\imag \w_\nu),  
  \label{eq.nf}
\end{equation}
where contour $\Gamma'$ is shown in Fig.~\ref{fig.Gamma} and $g(z)$ is assumed to be analytic on and inside 
contour $\Gamma'$. Note that contour $\Gamma'$ in Fig.~\ref{fig.Gamma} encloses the fermionic Matsubara 
frequencies in a clockwise manner.

\subsection{Derivation of Equation (\ref{SFTentropy})}

The frequency sum in the second term of the right hand side in Eq.~(\ref{eq.Omega-num}) can be evaluated using 
the contour integral, i.e., 
\begin{eqnarray}\label{eq:gng}
  \frac{1}{\beta}  \sum_{\nu=-\infty}^\infty  
  g(\imag \w_\nu)
  = \oint_{\Gamma'} \frac{\dd z}{2\pi \imag} n_{\rm F} (z) g(z), 
\end{eqnarray}
where $g(z)=N^{-1} \sum_{\tilde{\mb{k}}} \ln \det \left[ \bs{I}-\bs{V}(\tilde{\mb{k}}) \bs{G}'(z) \right]$ 
and is analytic on and inside contour $\Gamma'$ 
defined in Fig.~\ref{fig.Gamma}. 
The temperature derivative of the right hand side in Eq.~(\ref{eq:gng}) is
\begin{equation}\label{eq.dTlndet}
  \frac{1}{2 \pi \imag} \oint_{\Gamma'} \dd z 
  \left[ 
    \frac{\partial n_{\rm F}(z)}{\partial T}g(z) + n_{\rm F}(z) \frac{\partial g(z)}{\partial T}
    \right].
\end{equation}
Because of the relation
\begin{equation}
  \frac{\partial n_{\rm F}(z)}{\partial T} = -\beta z \frac{\partial n_{\rm F}(z)}{\partial z}
\end{equation}
and Eq.~(\ref{eq.nf}), Eq.~(\ref{eq.dTlndet}) can be written as 
\begin{equation}\label{T-derivative1}
  \frac{1}{\beta} \sum_{\nu = -\infty}^\infty 
  \left[\beta g(\imag \w_\nu) 
  + \imag \w_\nu \beta \left. \frac{\partial g(z)}{\partial z}\right|_{z = \imag \w_\nu} 
  + \left. \frac{\partial g(z)}{\partial T} \right|_{z = \imag \w_\nu}  \right],
\end{equation} 
which proves Eq.~(\ref{SFTentropy}). 

Note that the same result in Eq.~(\ref{T-derivative1}) can be obtained simply by taking into account the 
$T$ dependence of the Matsubara frequency when the $T$ derivative is performed on the left hand side in 
Eq.~(\ref{eq:gng}), 
which is equivalent to replacing the differential operator as
\begin{equation}\label{T-derivative2}
  \frac{\partial}{\partial T} \to 
  \frac{{\rm D}}{{\rm D} T} := \frac{\partial}{\partial T} + \frac{\partial \imag \w_\nu} {\partial T} \frac{\partial }{\partial (\imag \w_\nu)}.  
\end{equation} 
Similarly, we can obtain the second derivative of the grand-potential functional with respect to $T$ 
and the result is given in Eq.~(\ref{C-matrix}).

\section{Single-particle Green's function in the continued-fraction representation}
\label{app.dG}

In this Appendix, we describe 
how to calculate $\partial_z \bs{G}'(z)$ and $\partial_z^2 \bs{G}'(z)$ numerically 
using the coefficients appearing in the continued-fraction representation of $\bs{G}'(z)$ 
obtained by the standard Lanczos method with a single initial vector. 
The method described here corresponds to a direct calculation of 
the (selected) matrix element and its derivatives of the inversion of the tridiagonal matrix 
generated by the Lanczos iteration.  

Let us first consider the particle-addition part of the single-particle Green's function $G^{+}_{ij,s}(z)$ 
of the cluster given in Eq.~(\ref{eq.GeLehmann}). 
In order to evaluate $\partial_z G^{+}_{ij,s}(z)$ and $\partial_z^2 G^{+}_{ij,s}(z)$ 
using the standard Lanczos method, we define the following auxiliary single-particle Green's function:  
\begin{equation}
  {\mathcal X}_{ij,s}^{+}(z) = 
  \left \bra \Psi_s \left| \hat{x}_{ij} \left[z - \left(\hat{H}' - E_s \right) \right]^{-1} \hat{x}_{ij}^{\dag} \right| \Psi_s \right \ket,
\end{equation}
where 
\begin{equation}
  \hat{x}_{ij} = \hat{c}_{i} + \hat{c}_{j} 
\end{equation}
and the subscripts $i$ and $j\,(=1,2,\cdots,L)$ are the generalized single-particle indices including the site 
and spin indices (see Sec.~\ref{sec.BL}). 
$|\Psi_s\rangle$ is the $s$th eigenstate of the cluster Hamiltonian $\hat H'$ with the eigenvalue $E_s$.

The auxiliary single-particle Green's function ${\mathcal X}_{ij,s}^{+}(z)$ can be calculated from the tridiagonal 
matrix representation of 
$\hat{H}'$ obtained iteratively by the Lanczos method starting with the normalized initial vector 
\begin{equation}
|q_1\ket = \hat{x}_{ij}^{\dag} |\Psi_s \ket/B_0, 
\end{equation}
where 
\begin{equation}
  B_0^2= \bra \Psi_s |\hat{x}_{ij} \hat{x}_{ij}^{\dag} |\Psi_s \ket 
  \label{b02}
\end{equation}
is the static correlation function~\cite{Prelovsek,Fulde1995,Dagotto1994}.  
The continued-fraction representation of  ${\mathcal X}_{ij,s}^+ (z)$ reads 
\begin{equation}\label{eq.cfrac}
  {\mathcal X}_{ij,s}^{+}(z) = \frac{B_0^2}{z + E_s - A_1 - \cfrac{B_1^2}{z + E_s - A_2 - \cdots}},
\end{equation}
where $A_k$ and $B_k$ are respectively 
the diagonal and subdiagonal elements of the real-symmetric tridiagonal matrix obtained by 
the standard Lanczos method at $k$th iteration. 
Here, the procedure of the the standard Lanczos method can be obtained 
simply by setting the block size $L=1$ in Eqs.~(\ref{BL1})--(\ref{BL3}).   
In particular, the QR factorization of $\bs{X}_k$
in Eq.~(\ref{BL3}) is now merely the normalization of $\bs{X}_k$ and 
$B_k$ corresponds to the norm of $\bs{X}_k$. 
Once ${\mathcal X}_{ij,s}^{+}(z)$ is obtained after $M$ times of Lanczos iterations, 
the particle-addition part of the single-particle Green's function $G_{ij,s}^{+}(z)$ is easily evaluated as 
\begin{equation}
  G_{ij,s}^{+}(z) = \frac{1}{2} {\mathcal X}_{ij,s}^{+}(z) - \frac{1}{8} \left[ {\mathcal X}_{ii,s}^{+}(z) + {\mathcal X}_{jj,s}^{+}(z) \right].
  \label{eq:g-x}
\end{equation}

We now show how to evaluate $\partial_z {\mathcal X}_{ij,s}^{+}(z)$. 
For this purpose, it is important to notice that the continued fraction in Eq.~(\ref{eq.cfrac}) can be written 
as a rational function
\begin{equation}\label{eq.X}
  {\mathcal X}_{ij,s}^{+}(z)= -\frac{P_{M}(z)}{Q_M(z)},
\end{equation}
where the polynomials $P_k(z)$ and $Q_k(z)$ ($k = 1, 2,\cdots, M$) 
are given via the following recurrence formulas:
\begin{equation}
  P_k(z) = \tilde{A}_{k}(z) P_{k-1}(z) + \tilde{B}_{k-1} P_{k-2}(z) \label{eq.P}
\end{equation}
and
\begin{equation}
  Q_k(z) = \tilde{A}_{k}(z) Q_{k-1}(z) + \tilde{B}_{k-1} Q_{k-2}(z), \label{eq.Q}
\end{equation}
where $\tilde{A}_{k}(z) = z + E_s - A_{k}$ and $\tilde{B}_{k-1} = -B_{k-1}^2$ 
with $P_{-1} = 1$, $Q_{-1} = 0$, $P_{0} = 0$, and $Q_{0} = 1$~\cite{Beach2000,NR3,Liesen}. 
Differentiating Eq.~(\ref{eq.X}) with respect to $z$ yields 
\begin{equation}
 \partial_z {\mathcal X}_{ij,s}^{+}(z) = -\frac{{\mathcal X}_{ij,s}^{+}(z) \partial_z Q_M(z) + \partial_z P_M(z)}{Q_M(z)},
 \label{eq:dx}
\end{equation}
where $\partial_z P_k(z)$ and $\partial_z Q_k(z)$ are also given recursively as 
\begin{equation}
  \partial_z P_k(z) = P_{k-1}(z) + \tilde{A}_{k} \partial_z P_{k-1}(z) + \tilde{B}_{k-1} \partial_z P_{k-2}(z) 
\end{equation}
and
\begin{equation}
  \partial_z Q_k(z) = Q_{k-1}(z) + \tilde{A}_{k} \partial_z Q_{k-1}(z) + \tilde{B}_{k-1} \partial_z Q_{k-2}(z).
\end{equation}
Similarly, the second derivative of ${\mathcal X}_{ij,s}^{+}(z)$ with respect to $z$ is evaluated as
\begin{eqnarray}\label{eq:d2x}
  \partial_z^2 {\mathcal X}_{ij,s}^{+}(z) 
  = 
  -\frac{2\partial_z {\mathcal X}_{ij,s}^{+}(z) \partial_z Q_M(z) + {\mathcal X}_{ij,s}^{+}(z) \partial_z^2 Q_M(z)  + \partial_z^2 P_M(z)}{Q_M(z)} \nonumber \\
\end{eqnarray}
with the recurrence formulas 
\begin{equation}
  \partial_z^2 P_k(z) = 2 \partial_z P_{k-1}(z) + \tilde{A}_{k} \partial_z^2 P_{k-1}(z) + \tilde{B}_{k-1} \partial_z^2 P_{k-2}(z)
\end{equation}
and
\begin{equation} 
  \partial_z^2 Q_k(z) = 2 \partial_z Q_{k-1}(z) + \tilde{A}_{k} \partial_z^2 Q_{k-1}(z) + \tilde{B}_{k-1} \partial_z^2 Q_{k-2}(z).
\end{equation}
Using Eqs.~(\ref{eq:dx}) and (\ref{eq:d2x}), we can now easily evaluate $\partial_z G^{+}_{ij,s}(z)$ and 
$\partial_z^2 G^{+}_{ij,s}(z)$. 
The same procedure can be applied for the particle-removal part of the single-particle Green's function 
$G_{ij,s}^-(z)$ of the cluster given in Eq.~(\ref{eq.GhLehmann}) to evaluate $\partial_z G^{-}_{ij,s}(z)$ and 
$\partial_z^2 G^{-}_{ij,s}(z)$, and therefore we can calculate $\partial_z \bs{G}'(z)$  and $\partial_z^2 \bs{G}'(z)$.

Finally, we should note that $G_{ij,s}^{+}(z) = G_{ji,s}^{+}(z)$ is assumed in Eq.~(\ref{eq:g-x}). 
However, even if $G_{ij,s}^{+}(z) \not = G_{ji,s}^{+}(z)$, we can easily generalize the above derivation by 
introducing an additional auxiliary single-particle Green's function 
\begin{equation}
  {\mathcal Y}_{ij,s}^{+}(z) = 
  \left \bra \Psi_s \left| \hat{y}_{ij} \left[z - \left(\hat{H}' - E_s\right)\right]^{-1} \hat{y}_{ij}^{\dag} \right| \Psi_s \right \ket,
\end{equation}
where $\hat{y}_{ij} = \hat{c}_i - \imag \hat{c}_j$ 
and   $\hat{y}_{ij}^\dag = \hat{c}_i^\dag + \imag \hat{c}_j^\dag$.  
Indeed, using ${\mathcal X}_{ij,s}^+ (z)$ and ${\mathcal Y}_{ij,s}^{+}(z)$, 
the particle-addition part of the single-particle Green's function $G_{ij,s}^{+}(z)$ is obtained as 
\begin{eqnarray}
  G_{ij,s}^{+}(z) &=& \frac{1}{2}\mcal{X}_{ij,s}^{+}(z) 
  -   \frac{1}    {4} \left(\mcal{Y}_{ij,s}^{+}(z) + \mcal{Y}_{ji,s}^{+}(z)\right) \notag \\
  &-&     \frac{\imag}{4} \left(\mcal{Y}_{ij,s}^{+}(z) - \mcal{Y}_{ji,s}^{+}(z)\right).
\end{eqnarray}
A similar procedure was employed to evaluate the anomalous single-particle Green's function using 
the Lanczos method~\cite{Ohta1994}. 
Note that, similarly to Eq.~(\ref{eq.cfrac}), the auxiliary single-particle Green's function ${\mathcal Y}_{ij,s}^{+}(z)$ 
can be represented in the 
continued-fraction form but now with the initial 
Lanczos vector $|q_1\ket = \hat{y}_{ij}^{\dag} |\Psi_s \ket/B_0$ with 
$B_0^2= \bra \Psi_s |\hat{y}_{ij} \hat{y}_{ij}^{\dag} |\Psi_s \ket$.

\section{Benchmark results of the energy}\label{app.energy}
In this Appendix, we show
benchmark results of the energy of the half-filled Hubbard
model within the VCA at zero temperature.

Recently, an extensive numerical study 
on the two-dimensional Hubbard model has reported 
the energy and other static quantities such as
the expectation values of the double occupancy and 
the magnetization with several unbiased methods and approximate methods~\cite{LeBlanc2015}. 
Here, we show the finite-size scaling analysis of the ground-state energy for the two-dimensional
Hubbard model at half filling within the VCA. 
We note that a benchmark of the VCA for
the one-dimensional Hubbard model has already been reported in Ref.~\cite{Balzer2008}

Compared to the one-dimensional system~\cite{Balzer2008},
the finite-size scaling of the energy in two dimensions 
is more difficult because the finite-size effect 
due to the open-boundary conditions is more significant~\cite{Senechal2008}. 
Following Refs.~\cite{Senechal2008,Sahebsara2008}, 
we introduce a scaling factor $Q$ 
which is defined as the number of links connecting neighboring sites though the hopping within the cluster
divided by the total number of links 
of the original lattice within a unit cell of the superlattice of clusters.
Taking the cluster size in the thermodynamic limit corresponds to $Q \to 1$, or equivalently $1-Q \to 0$.
Note that $1-Q$ behaves similarly to $1/\sqrt{L_{\rm c}}$ 
for clusters whose aspect ratio is close to unity, and 
in particular these two quantities are identical for $L_{\rm c} = l \times l$, i.e., $1-Q=1/\sqrt{L_c}=1/l$, 
where $l$ is an integer (see the third and fourth columns of Table~\ref{tab.U4} or Table~\ref{tab.U8}). 
A nice property of $1-Q$, as compared to $1/\sqrt{L_{\rm c}}$, is 
that $1-Q$ can distinguish clusters with 
the same $L_{\rm c}$ but with the different shape 
because $Q$ takes into account the boundary effect. 
Moreover, $1-Q$ can even reverse the order of $1/\sqrt{L_{\rm c}}$ 
for some particular values of $L_{\rm c}$, e.g., 
for $L_{\rm c} = 3 \times 4$ and
$L_{\rm c} = 2 \times 8$ clusters.

\begin{figure}
  \begin{center}
    \includegraphics[width=1.0\columnwidth]{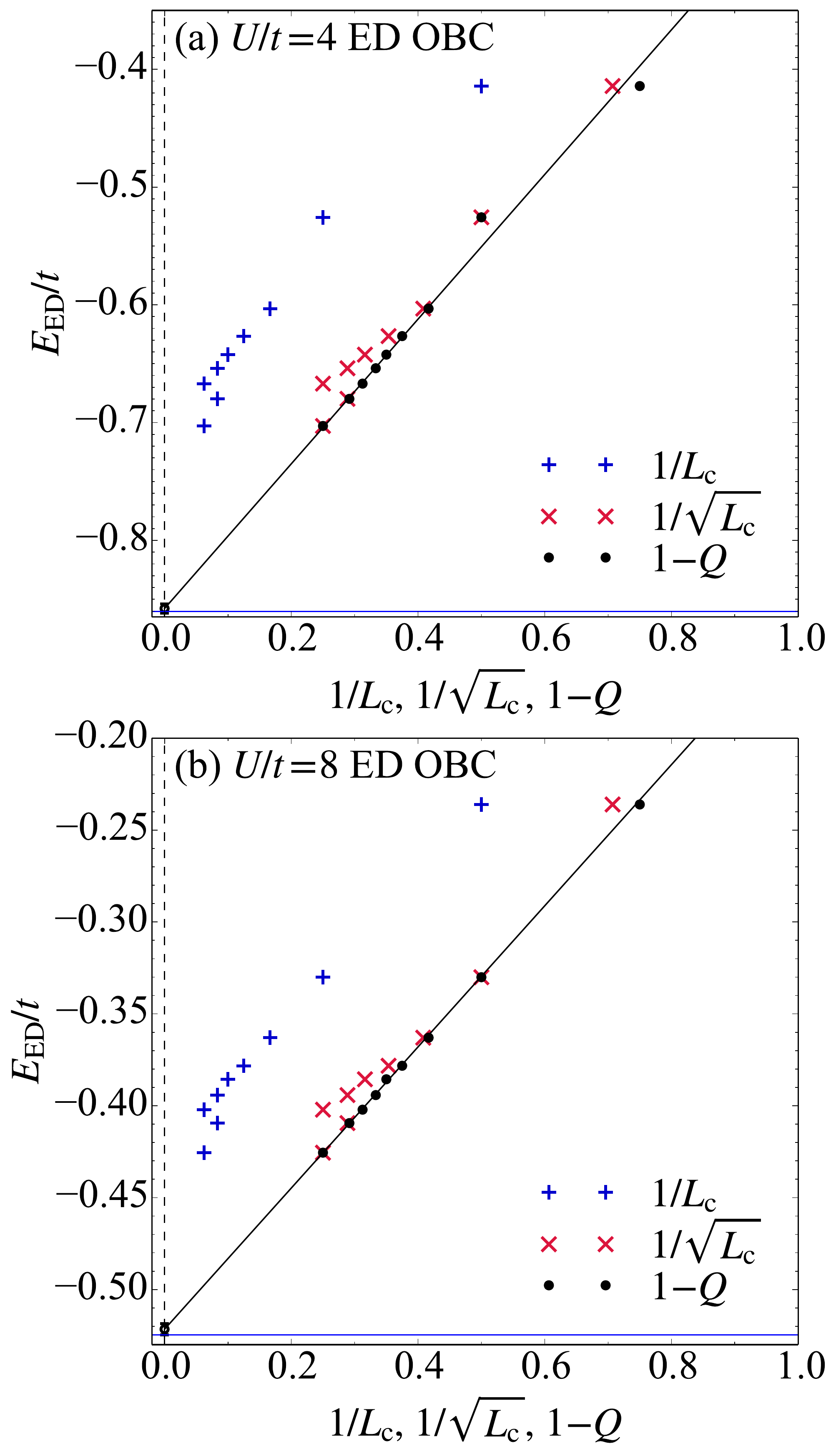}
    \caption{
      Ground-state energy $E_{\rm ED}$ per site for (a) $U/t=4$ and (b) $U/t=8$ at half filling 
      calculated by the exact-diagonalization method for different size of clusters under the
      open-boundary conditions. 
      The pluses, crosses, and dots represent the energies plotted with respect to
      $1/L_{\rm c}$, $1/\sqrt{L_{\rm c}}$, and $1-Q$, respectively.
      The blue horizontal line is the ground-state energy
      in the thermodynamic limit obtained by the auxiliary-field QMC method taken from Ref.~\cite{LeBlanc2015}. 
      \label{fig.edT0}}
  \end{center}
\end{figure}

Before showing the results obtained by the VCA, we first study 
the finite-size scaling of the exact ground-state energy 
of small clusters under open-boundary conditions, denoted as $E_{\rm ED}$. 
Tables~\ref{tab.U4} and \ref{tab.U8} show the energy per site at zero temperature 
for $U/t=4$ and $U/t=8$ with various clusters, respectively. 
Figures~\ref{fig.edT0}(a) and \ref{fig.edT0}(b) show $E_{\rm ED}$ 
as a function of $1/L_{\rm c}$, $1/\sqrt{L_{\rm c}}$, and $1-Q$  
for $U/t=4$ and $U/t=8$, respectively. 
As expected for small-sized and open-boundary clusters, 
the energy depends strongly on the size and the shape 
of the cluster.
No systematic dependence of the energy  
on $1/L_{\rm c}$ or $1/\sqrt{L_{\rm c}}$ can be found,
while the energy scales nicely with respect to $1-Q$. 
By a linear fit to the data with excluding the smallest 
three clusters, we obtain 
$\lim_{1-Q\to 0} E_{\rm ED}/t=-0.8579(40)$ for $U/t=4$ and 
$\lim_{1-Q\to 0} E_{\rm ED}/t=-0.5216(31)$ for $U/t=8$, where the numbers in the parentheses 
indicate the uncertainty due to the extrapolation in the last digits. Although the 
uncertainties are larger by an order of magnitude than those reported in Ref.~\cite{LeBlanc2015},
these extrapolated values are consistent with the ones in the literature~\cite{LeBlanc2015,Sorella2015,Qin2016,Karakuzu2018}. 
The reasonable extrapolation of the energy with $1-Q$ scaling 
is rather surprising because the clusters used are quite small. 

Next, we show the results for the VCA. 
Here, in addition to the variational magnetic field $h'$ defined in Eq.~(\ref{eq.v-parm}), 
we introduce a variational intra-cluster nearest-neighbor hopping parameter $\delta t'$ as
\begin{equation}\label{eq.dt-parm}
  \hat{H}_{\delta t'} =
  - \delta t' \sum_{\bra i,j \ket} \sum_\sigma \left(\hat{c}_{i\s}^{\dag} \hat{c}_{j\s} + \mathrm{H.c.}\right).  
\end{equation}
Tables~\ref{tab.U4} and \ref{tab.U8} show the ground-state energy per site 
for $U/t=4$ and $U/t=8$ with various clusters, respectively. 
We denote as $E(h'^*)$ the energy obtained by optimizing only $h'$, and
as $E(h'^*,\delta t'^*)$ the energy obtained by optimizing both $h'$ and $\delta t'$. 
The importance for optimizing $\delta t'$, especially for small $U/t$ regime,
has been reported in Refs.~\cite{Balzer2008,Laubach2014,Laubach2015,Rachel2015}.
Indeed, it is found in Tables~\ref{tab.U4} and \ref{tab.U8} that
the variation of $\delta t'$ provides the larger energy gain
for the smaller cluster and the smaller $U/t$.

\begin{figure}
  \begin{center}
    \includegraphics[width=1.0\columnwidth]{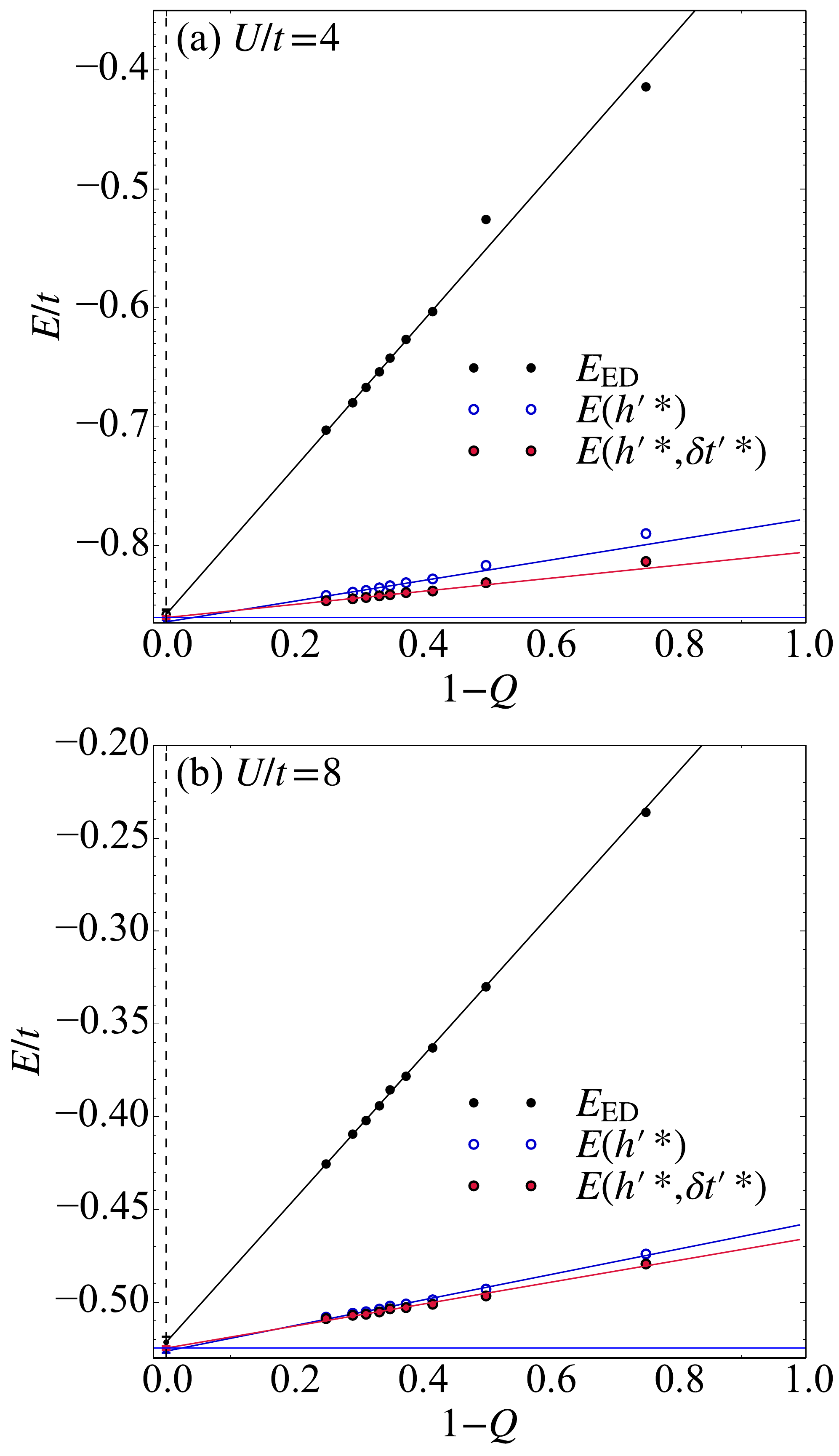}
    \caption{
      Ground-state energy $E$ per site for (a) $U/t=4$ and (b) $U/t=8$ at half filling.
      The empty and the filled circles denote $E(h'^*)$ and $E(h'^*,\delta t'^*)$, respectively.
      The dots are obtained by the exact-diagonalization method for the clusters under the open-boundary
      conditions, as shown also in Fig.~\ref{fig.edT0}.  
      The blue horizontal line is the ground-state energy
      in the thermodynamic limit obtained by the auxiliary-field QMC method taken from Ref.~\cite{LeBlanc2015}.
      \label{fig.energyT0}}
  \end{center}
\end{figure}

Figures~\ref{fig.energyT0}(a) and \ref{fig.energyT0}(b) show 
the finite-size scaling of the energy within the VCA for $U/t=4$ and $U/t=8$, respectively. 
Here, only the scaling factor $1-Q$ is employed since, as 
in the exact-diagonalization study, only $1-Q$ allows 
for a reasonable finite-size scaling.
The dependence of the energy on $1-Q$ obtained by the VCA
is weaker than that of $E_{\rm ED}$. 
By optimizing both $h'$ and $\delta t'$, the energies are extrapolated to
$\lim_{1-Q\to 0} E(h'^*,\delta t'^*)/t=-0.8605(10)$ for $U/t=4$ and
$\lim_{1-Q\to 0} E(h'^*,\delta t'^*)/t=-0.5247(9)$ for $U/t=8$, 
being consistent with those reported, although the uncertainties are still large. 
Here, the data corresponding to the smallest three clusters are excluded from
the linear fit for $U/t=4$, while all the data are included in the linear fit for $U/t=8$. 
With the same extrapolation scheme but by optimizing only $h'$,
the energies are extrapolated to the lower values
$\lim_{1-Q\to 0} E(h'^*)/t=-0.8641(17)$ for $U/t=4$ and
$\lim_{1-Q\to 0} E(h'^*)/t=-0.5263(8)$ for $U/t=8$, 
which are inconsistent with those reported in the literature. 
This might be because the optimization of $\delta t'$ reduces the
finite-size effect by providing the more energy gain
(compared to the one without optimizing $\delta t'$) 
for the smaller clusters and thereby decreases the slope of the energy with respect to $1-Q$. 

Finally, we remind that the VCA grand potential, which corresponds to the energy 
up to the constant shift $\mu=U/2$ in the zero-temperature limit at half filling, 
consists of the cluster term $\Omega'/L_{\rm c}$  and
the inter-cluster term $-{\rm Tr} \ln (\bs{I}-\bs{VG}')/\beta N L_{\rm c}$.  
Although the cluster term should contribute dominantly for large $L_{\rm c}$, 
it is not obvious whether the sum of the two can be scaled well with $1-Q$ 
for the small clusters studied here. 
Our results suggest that the scaling of the VCA energy with respect to
$1-Q$ is reasonable, at least, for $U/t=4$ and $U/t=8$ with
the cluster sizes studied here 
when the two variational parameters $h'$ and $\delta t'$
are optimized. 
The similar analysis could be useful also 
at finite temperatures and can be done straightforwardly.

\begin{table*}
  \caption{
    \label{tab.U4}
    Ground-state energy per site 
    of the Hubbard model on the square lattice with different sizes $L_{\rm c}$ of clusters 
    at half filling for $U/t=4$ calculated using 
    the exact diagonalization method and the VCA.
    $E_{\rm ED}$ is the exact ground-state energy of the cluster under the open-boundary conditions. 
    $E(h'^*)$ is obtained by optimizing one variational parameter $h'$ and 
    $E(h'^*,\delta t'^*)$ by optimizing two variational parameters $h'$ and $\delta t'$.
    The optimal variational parameters $h'^*$ and $(h'^*,\delta t'^*)$ are also shown.
  }
  \begin{tabular}{clllccccc}
    \hline
    \hline
    \multicolumn{9}{c}{$U/t=4$} \\
    \hline
    $L_{\rm c}$ & $1/L_{\rm c}$ & $1/\sqrt{L_{\rm c}}$ & $1-Q$ & $E_{\rm ED}/t$ & $E({h^\prime}^*)/t$  & $E({h^\prime}^*, {\delta t^\prime}^*)/t$  & ${h^\prime}^*/t$  & $({h^\prime}^*/t, {\delta t^\prime}^*/t)$  \\
    \hline
    $2\times1$& 0.5    & 0.7071 & 0.75   & $-$0.414214 & $-$0.789850 & $-$0.813333 & 0.352469 & (1.474002, 1.750708) \\
    $2\times2$& 0.25   & 0.5    & 0.5    & $-$0.525687 & $-$0.816607 & $-$0.831151 & 0.218936 & (0.361820, 0.918256) \\
    $2\times3$& 0.1667 & 0.4082 & 0.4167 & $-$0.603220 & $-$0.827991 & $-$0.838147 & 0.188001 & (0.391725, 0.577525) \\
    $2\times4$& 0.125  & 0.3536 & 0.375  & $-$0.626563 & $-$0.831151 & $-$0.839467 & 0.173734 & (0.336096, 0.496004) \\
    $10$&       0.1    & 0.3162 & 0.35   & $-$0.642306 & $-$0.833671 & $-$0.841210 & 0.151529 & (0.285646, 0.446568) \\
    $2\times6$& 0.0833 & 0.2887 & 0.3333 & $-$0.653863 & $-$0.835492 & $-$0.842223 & 0.155622 & (0.281362, 0.413290) \\
    $2\times8$& 0.0625 & 0.25   & 0.3125 & $-$0.666924 & $-$0.837667 & $-$0.843579 & 0.146469 & (0.255994, 0.372645) \\ 
    $3\times4$& 0.0833 & 0.2887 & 0.2917 & $-$0.679842 & $-$0.839112 & $-$0.844663 & 0.130976 & (0.227668, 0.351127) \\
    $4\times4$& 0.0625 & 0.25   & 0.25   & $-$0.702877 & $-$0.841893 & $-$0.846322 & 0.114508 & (0.179996, 0.301982) \\
    \hline
    \hline
  \end{tabular}
\end{table*}

\begin{table*}
  \caption{
    \label{tab.U8}
    Same as Table~\ref{tab.U4} but for $U/t=8$. 
  }
  \begin{tabular}{clllccccc}
    \hline
    \hline
    \multicolumn{9}{c}{$U/t=8$} \\
    \hline
    $L_{\rm c}$ & $1/L_{\rm c}$ & $1/\sqrt{L_{\rm c}}$ & $1-Q$ & $E_{\rm ED}/t$ & $E({h^\prime}^*)/t$  & $E({h^\prime}^*, {\delta t^\prime}^*)/t$  & ${h^\prime}^*/t$  & $({h^\prime}^*/t, {\delta t^\prime}^*/t)$  \\
    \hline
    $2\times1$& 0.5    & 0.7071 & 0.75    & $-$0.236068 & $-$0.474032 & $-$0.479400 & 0.292618 & (0.847075, 0.698321)\\
    $2\times2$& 0.25   & 0.5    & 0.5     & $-$0.330059 & $-$0.492911 & $-$0.496630 & 0.195523 & (0.334300, 0.345593)\\
    $2\times3$& 0.1667 & 0.4082 & 0.4167  & $-$0.362966 & $-$0.498704 & $-$0.501105 & 0.159398 & (0.242098, 0.229336)\\
    $2\times4$& 0.125  & 0.3536 & 0.375   & $-$0.378240 & $-$0.500951 & $-$0.502899 & 0.149863 & (0.215346, 0.195757)\\
    $10$&       0.1    & 0.3162 & 0.35    & $-$0.385580 & $-$0.502058 & $-$0.503551 & 0.134032 & (0.183029, 0.161724)\\
    $2\times6$& 0.0833 & 0.2887 & 0.3333  & $-$0.394188 & $-$0.503697 & $-$0.505236 & 0.135106 & (0.184715, 0.162582)\\
    $2\times8$& 0.0625 & 0.25   & 0.3125  & $-$0.402132 & $-$0.505126 & $-$0.506483 & 0.127109 & (0.169825, 0.147633)\\
    $3\times4$& 0.0833 & 0.2887 & 0.2917  & $-$0.409438 & $-$0.505953 & $-$0.507042 & 0.113759 & (0.147501, 0.128113)\\
    $4\times4$& 0.0625 & 0.25   & 0.25    & $-$0.425526 & $-$0.508044 & $-$0.508827 & 0.102302 & (0.126727, 0.103465)\\
    \hline
    \hline
  \end{tabular}
\end{table*}

%\bibliography{ftvca}

\begin{thebibliography}{204}%
\makeatletter
\providecommand \@ifxundefined [1]{%
 \@ifx{#1\undefined}
}%
\providecommand \@ifnum [1]{%
 \ifnum #1\expandafter \@firstoftwo
 \else \expandafter \@secondoftwo
 \fi
}%
\providecommand \@ifx [1]{%
 \ifx #1\expandafter \@firstoftwo
 \else \expandafter \@secondoftwo
 \fi
}%
\providecommand \natexlab [1]{#1}%
\providecommand \enquote  [1]{``#1''}%
\providecommand \bibnamefont  [1]{#1}%
\providecommand \bibfnamefont [1]{#1}%
\providecommand \citenamefont [1]{#1}%
\providecommand \href@noop [0]{\@secondoftwo}%
\providecommand \href [0]{\begingroup \@sanitize@url \@href}%
\providecommand \@href[1]{\@@startlink{#1}\@@href}%
\providecommand \@@href[1]{\endgroup#1\@@endlink}%
\providecommand \@sanitize@url [0]{\catcode `\\12\catcode `\$12\catcode
  `\&12\catcode `\#12\catcode `\^12\catcode `\_12\catcode `\%12\relax}%
\providecommand \@@startlink[1]{}%
\providecommand \@@endlink[0]{}%
\providecommand \url  [0]{\begingroup\@sanitize@url \@url }%
\providecommand \@url [1]{\endgroup\@href {#1}{\urlprefix }}%
\providecommand \urlprefix  [0]{URL }%
\providecommand \Eprint [0]{\href }%
\providecommand \doibase [0]{http://dx.doi.org/}%
\providecommand \selectlanguage [0]{\@gobble}%
\providecommand \bibinfo  [0]{\@secondoftwo}%
\providecommand \bibfield  [0]{\@secondoftwo}%
\providecommand \translation [1]{[#1]}%
\providecommand \BibitemOpen [0]{}%
\providecommand \bibitemStop [0]{}%
\providecommand \bibitemNoStop [0]{.\EOS\space}%
\providecommand \EOS [0]{\spacefactor3000\relax}%
\providecommand \BibitemShut  [1]{\csname bibitem#1\endcsname}%
\let\auto@bib@innerbib\@empty
%</preamble>
\bibitem [{\citenamefont {Matsubara}(1955)}]{Matsubara1955}%
  \BibitemOpen
  \bibfield  {author} {\bibinfo {author} {\bibfnamefont {T.}~\bibnamefont
  {Matsubara}},\ }\href
  {http://ptp.oxfordjournals.org/content/14/4/351.abstract} {\bibfield
  {journal} {\bibinfo  {journal} {Prog. Theor. Phys.}\ }\textbf {\bibinfo
  {volume} {14}},\ \bibinfo {pages} {351} (\bibinfo {year} {1955})}\BibitemShut
  {NoStop}%
\bibitem [{\citenamefont {Ezawa}\ \emph {et~al.}(1957)\citenamefont {Ezawa},
  \citenamefont {Tomozawa},\ and\ \citenamefont {Umezawa}}]{Ezawa1957}%
  \BibitemOpen
  \bibfield  {author} {\bibinfo {author} {\bibfnamefont {H.}~\bibnamefont
  {Ezawa}}, \bibinfo {author} {\bibfnamefont {Y.}~\bibnamefont {Tomozawa}}, \
  and\ \bibinfo {author} {\bibfnamefont {H.}~\bibnamefont {Umezawa}},\ }\href
  {http://link.springer.com/article/10.1007\%2FBF02903206} {\bibfield
  {journal} {\bibinfo  {journal} {Nuovo Cimento}\ }\textbf {\bibinfo {volume}
  {5}},\ \bibinfo {pages} {810} (\bibinfo {year} {1957})}\BibitemShut {NoStop}%
\bibitem [{\citenamefont {Abrikosov}\ \emph {et~al.}(1975)\citenamefont
  {Abrikosov}, \citenamefont {Gorkov},\ and\ \citenamefont
  {Dzyaloshinski}}]{AGD}%
  \BibitemOpen
  \bibfield  {author} {\bibinfo {author} {\bibfnamefont {A.~A.}\ \bibnamefont
  {Abrikosov}}, \bibinfo {author} {\bibfnamefont {L.~P.}\ \bibnamefont
  {Gorkov}}, \ and\ \bibinfo {author} {\bibfnamefont {I.~E.}\ \bibnamefont
  {Dzyaloshinski}},\ }\href@noop {} {\emph {\bibinfo {title} {Methods of
  Quantum Field Theory in Statistical Physics}}}\ (\bibinfo  {publisher}
  {Dover},\ \bibinfo {address} {New York},\ \bibinfo {year} {1975})\BibitemShut
  {NoStop}%
\bibitem [{\citenamefont {Fetter}\ and\ \citenamefont {Walecka}(2003)}]{FW}%
  \BibitemOpen
  \bibfield  {author} {\bibinfo {author} {\bibfnamefont {A.~L.}\ \bibnamefont
  {Fetter}}\ and\ \bibinfo {author} {\bibfnamefont {J.~D.}\ \bibnamefont
  {Walecka}},\ }\href@noop {} {\emph {\bibinfo {title} {Quantum Theory of Many
  Particle Physics}}}\ (\bibinfo  {publisher} {Dover},\ \bibinfo {address} {New
  York},\ \bibinfo {year} {2003})\BibitemShut {NoStop}%
\bibitem [{\citenamefont {Cabra}\ \emph {et~al.}(2012)\citenamefont {Cabra},
  \citenamefont {Honecker},\ and\ \citenamefont {(eds.)}}]{text}%
  \BibitemOpen
  \bibfield  {author} {\bibinfo {author} {\bibfnamefont {D.~C.}\ \bibnamefont
  {Cabra}}, \bibinfo {author} {\bibfnamefont {A.}~\bibnamefont {Honecker}}, \
  and\ \bibinfo {author} {\bibfnamefont {P.~P.}\ \bibnamefont {(eds.)}},\
  }\href@noop {} {\emph {\bibinfo {title} {Modern Theories of Many-Particle
  Systems in Condensed Matter Physics}}},\ Vol.\ \bibinfo {volume} {843}\
  (\bibinfo  {publisher} {Springer},\ \bibinfo {address} {Heidelberg},\
  \bibinfo {year} {2012})\BibitemShut {NoStop}%
\bibitem [{\citenamefont {Luttinger}\ and\ \citenamefont
  {Ward}(1960)}]{Luttinger1960}%
  \BibitemOpen
  \bibfield  {author} {\bibinfo {author} {\bibfnamefont {J.~M.}\ \bibnamefont
  {Luttinger}}\ and\ \bibinfo {author} {\bibfnamefont {J.~C.}\ \bibnamefont
  {Ward}},\ }\href {\doibase 10.1103/PhysRev.118.1417} {\bibfield  {journal}
  {\bibinfo  {journal} {Phys. Rev.}\ }\textbf {\bibinfo {volume} {118}},\
  \bibinfo {pages} {1417} (\bibinfo {year} {1960})}\BibitemShut {NoStop}%
\bibitem [{\citenamefont {Potthoff}(2006)}]{Potthoff2006}%
  \BibitemOpen
  \bibfield  {author} {\bibinfo {author} {\bibfnamefont {M.}~\bibnamefont
  {Potthoff}},\ }\href {http://dx.doi.org/10.5488/CMP.9.3.557} {\bibfield
  {journal} {\bibinfo  {journal} {Condens. Mat. Phys.}\ }\textbf {\bibinfo
  {volume} {9}},\ \bibinfo {pages} {557} (\bibinfo {year} {2006})}\BibitemShut
  {NoStop}%
\bibitem [{\citenamefont {Negele}\ and\ \citenamefont {Orland}(1988)}]{NO}%
  \BibitemOpen
  \bibfield  {author} {\bibinfo {author} {\bibfnamefont {J.~W.}\ \bibnamefont
  {Negele}}\ and\ \bibinfo {author} {\bibfnamefont {H.}~\bibnamefont
  {Orland}},\ }\href@noop {} {\emph {\bibinfo {title} {{Quantum Many-Particle
  Systems}}}}\ (\bibinfo  {publisher} {Addison-Wesley},\ \bibinfo {address}
  {New York},\ \bibinfo {year} {1988})\ Chap.\ \bibinfo {chapter} {1, 2 and
  5.}\BibitemShut {Stop}%
\bibitem [{\citenamefont {Potthoff}(2003{\natexlab{a}})}]{Potthoff2003a}%
  \BibitemOpen
  \bibfield  {author} {\bibinfo {author} {\bibfnamefont {M.}~\bibnamefont
  {Potthoff}},\ }\href
  {http://epjb.epj.org/articles/epjb/abs/2003/08/b03014/b03014.html} {\bibfield
   {journal} {\bibinfo  {journal} {Eur. Phys. J. B}\ }\textbf {\bibinfo
  {volume} {32}},\ \bibinfo {pages} {429} (\bibinfo {year}
  {2003}{\natexlab{a}})}\BibitemShut {NoStop}%
\bibitem [{\citenamefont {Potthoff}(2003{\natexlab{b}})}]{Potthoff2003b}%
  \BibitemOpen
  \bibfield  {author} {\bibinfo {author} {\bibfnamefont {M.}~\bibnamefont
  {Potthoff}},\ }\href
  {http://epjb.epj.org/articles/epjb/abs/2003/23/b03321/b03321.html} {\bibfield
   {journal} {\bibinfo  {journal} {Eur. Phys. J. B}\ }\textbf {\bibinfo
  {volume} {36}},\ \bibinfo {pages} {335} (\bibinfo {year}
  {2003}{\natexlab{b}})}\BibitemShut {NoStop}%
\bibitem [{\citenamefont {Potthoff}(2012)}]{Potthoff2012}%
  \BibitemOpen
  \bibfield  {author} {\bibinfo {author} {\bibfnamefont {M.}~\bibnamefont
  {Potthoff}},\ }\enquote {\bibinfo {title} {Self-energy-functional theory},}\
  in\ \href {\doibase 10.1007/978-3-642-21831-6_10} {\emph {\bibinfo
  {booktitle} {Strongly Correlated Systems: Theoretical Methods}}},\ \bibinfo
  {editor} {edited by\ \bibinfo {editor} {\bibfnamefont {A.}~\bibnamefont
  {Avella}}\ and\ \bibinfo {editor} {\bibfnamefont {F.}~\bibnamefont
  {Mancini}}}\ (\bibinfo  {publisher} {Springer Berlin Heidelberg},\ \bibinfo
  {address} {Berlin, Heidelberg},\ \bibinfo {year} {2012})\ pp.\ \bibinfo
  {pages} {303--339}\BibitemShut {NoStop}%
\bibitem [{\citenamefont {Maier}\ \emph
  {et~al.}(2005{\natexlab{a}})\citenamefont {Maier}, \citenamefont {Jarrell},
  \citenamefont {Pruschke},\ and\ \citenamefont {Hettler}}]{Maier2005_RMP}%
  \BibitemOpen
  \bibfield  {author} {\bibinfo {author} {\bibfnamefont {T.}~\bibnamefont
  {Maier}}, \bibinfo {author} {\bibfnamefont {M.}~\bibnamefont {Jarrell}},
  \bibinfo {author} {\bibfnamefont {T.}~\bibnamefont {Pruschke}}, \ and\
  \bibinfo {author} {\bibfnamefont {M.~H.}\ \bibnamefont {Hettler}},\ }\href
  {\doibase 10.1103/RevModPhys.77.1027} {\bibfield  {journal} {\bibinfo
  {journal} {Rev. Mod. Phys.}\ }\textbf {\bibinfo {volume} {77}},\ \bibinfo
  {pages} {1027} (\bibinfo {year} {2005}{\natexlab{a}})}\BibitemShut {NoStop}%
\bibitem [{\citenamefont {S\'{e}n\'{e}chal}()}]{Senechal2008}%
  \BibitemOpen
  \bibfield  {author} {\bibinfo {author} {\bibfnamefont {D.}~\bibnamefont
  {S\'{e}n\'{e}chal}},\ }\href {http://arxiv.org/abs/0806.2690} {\enquote
  {\bibinfo {title} {{An introduction to quantum cluster methods}},}\ }\Eprint
  {http://arxiv.org/abs/cond-mat/0806.2690} {arXiv:cond-mat/0806.2690}
  \BibitemShut {NoStop}%
\bibitem [{\citenamefont {Georges}\ \emph {et~al.}(1996)\citenamefont
  {Georges}, \citenamefont {Kotliar}, \citenamefont {Krauth},\ and\
  \citenamefont {Rozenberg}}]{Georges1996}%
  \BibitemOpen
  \bibfield  {author} {\bibinfo {author} {\bibfnamefont {A.}~\bibnamefont
  {Georges}}, \bibinfo {author} {\bibfnamefont {G.}~\bibnamefont {Kotliar}},
  \bibinfo {author} {\bibfnamefont {W.}~\bibnamefont {Krauth}}, \ and\ \bibinfo
  {author} {\bibfnamefont {M.~J.}\ \bibnamefont {Rozenberg}},\ }\href {\doibase
  10.1103/RevModPhys.68.13} {\bibfield  {journal} {\bibinfo  {journal} {Rev.
  Mod. Phys.}\ }\textbf {\bibinfo {volume} {68}},\ \bibinfo {pages} {13}
  (\bibinfo {year} {1996})}\BibitemShut {NoStop}%
\bibitem [{\citenamefont {Lichtenstein}\ and\ \citenamefont
  {Katsnelson}(2000)}]{Lichtenstein2000}%
  \BibitemOpen
  \bibfield  {author} {\bibinfo {author} {\bibfnamefont {A.~I.}\ \bibnamefont
  {Lichtenstein}}\ and\ \bibinfo {author} {\bibfnamefont {M.~I.}\ \bibnamefont
  {Katsnelson}},\ }\href {\doibase 10.1103/PhysRevB.62.R9283} {\bibfield
  {journal} {\bibinfo  {journal} {Phys. Rev. B}\ }\textbf {\bibinfo {volume}
  {62}},\ \bibinfo {pages} {R9283} (\bibinfo {year} {2000})}\BibitemShut
  {NoStop}%
\bibitem [{\citenamefont {Kotliar}\ \emph {et~al.}(2001)\citenamefont
  {Kotliar}, \citenamefont {Savrasov}, \citenamefont {P\'alsson},\ and\
  \citenamefont {Biroli}}]{Kotliar2001}%
  \BibitemOpen
  \bibfield  {author} {\bibinfo {author} {\bibfnamefont {G.}~\bibnamefont
  {Kotliar}}, \bibinfo {author} {\bibfnamefont {S.~Y.}\ \bibnamefont
  {Savrasov}}, \bibinfo {author} {\bibfnamefont {G.}~\bibnamefont {P\'alsson}},
  \ and\ \bibinfo {author} {\bibfnamefont {G.}~\bibnamefont {Biroli}},\ }\href
  {\doibase 10.1103/PhysRevLett.87.186401} {\bibfield  {journal} {\bibinfo
  {journal} {Phys. Rev. Lett.}\ }\textbf {\bibinfo {volume} {87}},\ \bibinfo
  {pages} {186401} (\bibinfo {year} {2001})}\BibitemShut {NoStop}%
\bibitem [{\citenamefont {Gros}\ and\ \citenamefont
  {Valent\'{\i}}(1993)}]{Gros1993}%
  \BibitemOpen
  \bibfield  {author} {\bibinfo {author} {\bibfnamefont {C.}~\bibnamefont
  {Gros}}\ and\ \bibinfo {author} {\bibfnamefont {R.}~\bibnamefont
  {Valent\'{\i}}},\ }\href {\doibase 10.1103/PhysRevB.48.418} {\bibfield
  {journal} {\bibinfo  {journal} {Phys. Rev. B}\ }\textbf {\bibinfo {volume}
  {48}},\ \bibinfo {pages} {418} (\bibinfo {year} {1993})}\BibitemShut
  {NoStop}%
\bibitem [{\citenamefont {S\'en\'echal}\ \emph {et~al.}(2000)\citenamefont
  {S\'en\'echal}, \citenamefont {Perez},\ and\ \citenamefont
  {Pioro-Ladri\`ere}}]{Senechal2000}%
  \BibitemOpen
  \bibfield  {author} {\bibinfo {author} {\bibfnamefont {D.}~\bibnamefont
  {S\'en\'echal}}, \bibinfo {author} {\bibfnamefont {D.}~\bibnamefont {Perez}},
  \ and\ \bibinfo {author} {\bibfnamefont {M.}~\bibnamefont
  {Pioro-Ladri\`ere}},\ }\href {\doibase 10.1103/PhysRevLett.84.522} {\bibfield
   {journal} {\bibinfo  {journal} {Phys. Rev. Lett.}\ }\textbf {\bibinfo
  {volume} {84}},\ \bibinfo {pages} {522} (\bibinfo {year} {2000})}\BibitemShut
  {NoStop}%
\bibitem [{\citenamefont {S\'en\'echal}\ \emph {et~al.}(2002)\citenamefont
  {S\'en\'echal}, \citenamefont {Perez},\ and\ \citenamefont
  {Plouffe}}]{Senechal2002}%
  \BibitemOpen
  \bibfield  {author} {\bibinfo {author} {\bibfnamefont {D.}~\bibnamefont
  {S\'en\'echal}}, \bibinfo {author} {\bibfnamefont {D.}~\bibnamefont {Perez}},
  \ and\ \bibinfo {author} {\bibfnamefont {D.}~\bibnamefont {Plouffe}},\ }\href
  {\doibase 10.1103/PhysRevB.66.075129} {\bibfield  {journal} {\bibinfo
  {journal} {Phys. Rev. B}\ }\textbf {\bibinfo {volume} {66}},\ \bibinfo
  {pages} {075129} (\bibinfo {year} {2002})}\BibitemShut {NoStop}%
\bibitem [{\citenamefont {S{\'e}n{\'e}chal}(2012)}]{Senechal2012}%
  \BibitemOpen
  \bibfield  {author} {\bibinfo {author} {\bibfnamefont {D.}~\bibnamefont
  {S{\'e}n{\'e}chal}},\ }\enquote {\bibinfo {title} {Cluster perturbation
  theory},}\ in\ \href {\doibase 10.1007/978-3-642-21831-6_8} {\emph {\bibinfo
  {booktitle} {Strongly Correlated Systems: Theoretical Methods}}},\ \bibinfo
  {editor} {edited by\ \bibinfo {editor} {\bibfnamefont {A.}~\bibnamefont
  {Avella}}\ and\ \bibinfo {editor} {\bibfnamefont {F.}~\bibnamefont
  {Mancini}}}\ (\bibinfo  {publisher} {Springer Berlin Heidelberg},\ \bibinfo
  {address} {Berlin, Heidelberg},\ \bibinfo {year} {2012})\ pp.\ \bibinfo
  {pages} {237--270}\BibitemShut {NoStop}%
\bibitem [{\citenamefont {Potthoff}\ \emph {et~al.}(2003)\citenamefont
  {Potthoff}, \citenamefont {Aichhorn},\ and\ \citenamefont
  {Dahnken}}]{Potthoff2003_PRL}%
  \BibitemOpen
  \bibfield  {author} {\bibinfo {author} {\bibfnamefont {M.}~\bibnamefont
  {Potthoff}}, \bibinfo {author} {\bibfnamefont {M.}~\bibnamefont {Aichhorn}},
  \ and\ \bibinfo {author} {\bibfnamefont {C.}~\bibnamefont {Dahnken}},\ }\href
  {\doibase 10.1103/PhysRevLett.91.206402} {\bibfield  {journal} {\bibinfo
  {journal} {Phys. Rev. Lett.}\ }\textbf {\bibinfo {volume} {91}},\ \bibinfo
  {pages} {206402} (\bibinfo {year} {2003})}\BibitemShut {NoStop}%
\bibitem [{\citenamefont {Balzer}\ \emph {et~al.}(2009)\citenamefont {Balzer},
  \citenamefont {Kyung}, \citenamefont {S\'{e}n\'{e}chal}, \citenamefont
  {Tremblay},\ and\ \citenamefont {Potthoff}}]{Balzer2009}%
  \BibitemOpen
  \bibfield  {author} {\bibinfo {author} {\bibfnamefont {M.}~\bibnamefont
  {Balzer}}, \bibinfo {author} {\bibfnamefont {B.}~\bibnamefont {Kyung}},
  \bibinfo {author} {\bibfnamefont {D.}~\bibnamefont {S\'{e}n\'{e}chal}},
  \bibinfo {author} {\bibfnamefont {A.-M.~S.}\ \bibnamefont {Tremblay}}, \ and\
  \bibinfo {author} {\bibfnamefont {M.}~\bibnamefont {Potthoff}},\ }\href
  {http://iopscience.iop.org/0295-5075/85/1/17002} {\bibfield  {journal}
  {\bibinfo  {journal} {Europhys. Lett.}\ }\textbf {\bibinfo {volume} {85}},\
  \bibinfo {pages} {17002} (\bibinfo {year} {2009})}\BibitemShut {NoStop}%
\bibitem [{\citenamefont {Balzer}\ \emph {et~al.}(2010)\citenamefont {Balzer},
  \citenamefont {Hanke},\ and\ \citenamefont {Potthoff}}]{Balzer2010cdia}%
  \BibitemOpen
  \bibfield  {author} {\bibinfo {author} {\bibfnamefont {M.}~\bibnamefont
  {Balzer}}, \bibinfo {author} {\bibfnamefont {W.}~\bibnamefont {Hanke}}, \
  and\ \bibinfo {author} {\bibfnamefont {M.}~\bibnamefont {Potthoff}},\ }\href
  {\doibase 10.1103/PhysRevB.81.144516} {\bibfield  {journal} {\bibinfo
  {journal} {Phys. Rev. B}\ }\textbf {\bibinfo {volume} {81}},\ \bibinfo
  {pages} {144516} (\bibinfo {year} {2010})}\BibitemShut {NoStop}%
\bibitem [{\citenamefont {S\'en\'echal}(2010)}]{Senechal2010}%
  \BibitemOpen
  \bibfield  {author} {\bibinfo {author} {\bibfnamefont {D.}~\bibnamefont
  {S\'en\'echal}},\ }\href {\doibase 10.1103/PhysRevB.81.235125} {\bibfield
  {journal} {\bibinfo  {journal} {Phys. Rev. B}\ }\textbf {\bibinfo {volume}
  {81}},\ \bibinfo {pages} {235125} (\bibinfo {year} {2010})}\BibitemShut
  {NoStop}%
\bibitem [{\citenamefont {Dahnken}\ \emph {et~al.}(2004)\citenamefont
  {Dahnken}, \citenamefont {Aichhorn}, \citenamefont {Hanke}, \citenamefont
  {Arrigoni},\ and\ \citenamefont {Potthoff}}]{Dahnken2004}%
  \BibitemOpen
  \bibfield  {author} {\bibinfo {author} {\bibfnamefont {C.}~\bibnamefont
  {Dahnken}}, \bibinfo {author} {\bibfnamefont {M.}~\bibnamefont {Aichhorn}},
  \bibinfo {author} {\bibfnamefont {W.}~\bibnamefont {Hanke}}, \bibinfo
  {author} {\bibfnamefont {E.}~\bibnamefont {Arrigoni}}, \ and\ \bibinfo
  {author} {\bibfnamefont {M.}~\bibnamefont {Potthoff}},\ }\href {\doibase
  10.1103/PhysRevB.70.245110} {\bibfield  {journal} {\bibinfo  {journal} {Phys.
  Rev. B}\ }\textbf {\bibinfo {volume} {70}},\ \bibinfo {pages} {245110}
  (\bibinfo {year} {2004})}\BibitemShut {NoStop}%
\bibitem [{\citenamefont {Tong}(2005)}]{Tong2005}%
  \BibitemOpen
  \bibfield  {author} {\bibinfo {author} {\bibfnamefont {N.-H.}\ \bibnamefont
  {Tong}},\ }\href {\doibase 10.1103/PhysRevB.72.115104} {\bibfield  {journal}
  {\bibinfo  {journal} {Phys. Rev. B}\ }\textbf {\bibinfo {volume} {72}},\
  \bibinfo {pages} {115104} (\bibinfo {year} {2005})}\BibitemShut {NoStop}%
\bibitem [{\citenamefont {Koller}\ and\ \citenamefont
  {Dupuis}(2006)}]{Koller2006}%
  \BibitemOpen
  \bibfield  {author} {\bibinfo {author} {\bibfnamefont {W.}~\bibnamefont
  {Koller}}\ and\ \bibinfo {author} {\bibfnamefont {N.}~\bibnamefont
  {Dupuis}},\ }\href {http://iopscience.iop.org/0953-8984/18/41/019} {\bibfield
   {journal} {\bibinfo  {journal} {J. Phys.: Condens. Matter}\ }\textbf
  {\bibinfo {volume} {18}},\ \bibinfo {pages} {9525} (\bibinfo {year}
  {2006})}\BibitemShut {NoStop}%
\bibitem [{\citenamefont {Knap}\ \emph {et~al.}(2010)\citenamefont {Knap},
  \citenamefont {Arrigoni},\ and\ \citenamefont {von~der Linden}}]{Knap2010}%
  \BibitemOpen
  \bibfield  {author} {\bibinfo {author} {\bibfnamefont {M.}~\bibnamefont
  {Knap}}, \bibinfo {author} {\bibfnamefont {E.}~\bibnamefont {Arrigoni}}, \
  and\ \bibinfo {author} {\bibfnamefont {W.}~\bibnamefont {von~der Linden}},\
  }\href {\doibase 10.1103/PhysRevB.81.024301} {\bibfield  {journal} {\bibinfo
  {journal} {Phys. Rev. B}\ }\textbf {\bibinfo {volume} {81}},\ \bibinfo
  {pages} {024301} (\bibinfo {year} {2010})}\BibitemShut {NoStop}%
\bibitem [{\citenamefont {Knap}\ \emph {et~al.}(2011)\citenamefont {Knap},
  \citenamefont {Arrigoni},\ and\ \citenamefont {von~der Linden}}]{Knap2011}%
  \BibitemOpen
  \bibfield  {author} {\bibinfo {author} {\bibfnamefont {M.}~\bibnamefont
  {Knap}}, \bibinfo {author} {\bibfnamefont {E.}~\bibnamefont {Arrigoni}}, \
  and\ \bibinfo {author} {\bibfnamefont {W.}~\bibnamefont {von~der Linden}},\
  }\href {\doibase 10.1103/PhysRevB.83.134507} {\bibfield  {journal} {\bibinfo
  {journal} {Phys. Rev. B}\ }\textbf {\bibinfo {volume} {83}},\ \bibinfo
  {pages} {134507} (\bibinfo {year} {2011})}\BibitemShut {NoStop}%
\bibitem [{\citenamefont {Arrigoni}\ \emph {et~al.}(2011)\citenamefont
  {Arrigoni}, \citenamefont {Knap},\ and\ \citenamefont {von~der
  Linden}}]{Arrigoni2011}%
  \BibitemOpen
  \bibfield  {author} {\bibinfo {author} {\bibfnamefont {E.}~\bibnamefont
  {Arrigoni}}, \bibinfo {author} {\bibfnamefont {M.}~\bibnamefont {Knap}}, \
  and\ \bibinfo {author} {\bibfnamefont {W.}~\bibnamefont {von~der Linden}},\
  }\href {\doibase 10.1103/PhysRevB.84.014535} {\bibfield  {journal} {\bibinfo
  {journal} {Phys. Rev. B}\ }\textbf {\bibinfo {volume} {84}},\ \bibinfo
  {pages} {014535} (\bibinfo {year} {2011})}\BibitemShut {NoStop}%
\bibitem [{\citenamefont {Ejima}\ \emph {et~al.}(2012)\citenamefont {Ejima},
  \citenamefont {Fehske}, \citenamefont {Gebhard}, \citenamefont
  {zu~M\"unster}, \citenamefont {Knap}, \citenamefont {Arrigoni},\ and\
  \citenamefont {von~der Linden}}]{Ejima2012}%
  \BibitemOpen
  \bibfield  {author} {\bibinfo {author} {\bibfnamefont {S.}~\bibnamefont
  {Ejima}}, \bibinfo {author} {\bibfnamefont {H.}~\bibnamefont {Fehske}},
  \bibinfo {author} {\bibfnamefont {F.}~\bibnamefont {Gebhard}}, \bibinfo
  {author} {\bibfnamefont {K.}~\bibnamefont {zu~M\"unster}}, \bibinfo {author}
  {\bibfnamefont {M.}~\bibnamefont {Knap}}, \bibinfo {author} {\bibfnamefont
  {E.}~\bibnamefont {Arrigoni}}, \ and\ \bibinfo {author} {\bibfnamefont
  {W.}~\bibnamefont {von~der Linden}},\ }\href {\doibase
  10.1103/PhysRevA.85.053644} {\bibfield  {journal} {\bibinfo  {journal} {Phys.
  Rev. A}\ }\textbf {\bibinfo {volume} {85}},\ \bibinfo {pages} {053644}
  (\bibinfo {year} {2012})}\BibitemShut {NoStop}%
\bibitem [{\citenamefont {Filor}\ and\ \citenamefont
  {Pruschke}(2010)}]{Filor2010}%
  \BibitemOpen
  \bibfield  {author} {\bibinfo {author} {\bibfnamefont {S.}~\bibnamefont
  {Filor}}\ and\ \bibinfo {author} {\bibfnamefont {T.}~\bibnamefont
  {Pruschke}},\ }\href {http://iopscience.iop.org/1742-6596/200/2/022007}
  {\bibfield  {journal} {\bibinfo  {journal} {J. Phys.: Conf. Ser.}\ }\textbf
  {\bibinfo {volume} {200}},\ \bibinfo {pages} {022007} (\bibinfo {year}
  {2010})}\BibitemShut {NoStop}%
\bibitem [{\citenamefont {Filor}\ and\ \citenamefont
  {Pruschke}(2014)}]{Filor2014}%
  \BibitemOpen
  \bibfield  {author} {\bibinfo {author} {\bibfnamefont {S.}~\bibnamefont
  {Filor}}\ and\ \bibinfo {author} {\bibfnamefont {T.}~\bibnamefont
  {Pruschke}},\ }\href {http://iopscience.iop.org/1367-2630/16/6/063059/}
  {\bibfield  {journal} {\bibinfo  {journal} {New. J. Phys.}\ }\textbf
  {\bibinfo {volume} {16}},\ \bibinfo {pages} {065039} (\bibinfo {year}
  {2014})}\BibitemShut {NoStop}%
\bibitem [{\citenamefont {Hofmann}\ \emph {et~al.}(2013)\citenamefont
  {Hofmann}, \citenamefont {Eckstein}, \citenamefont {Arrigoni},\ and\
  \citenamefont {Potthoff}}]{Hofmann2013}%
  \BibitemOpen
  \bibfield  {author} {\bibinfo {author} {\bibfnamefont {F.}~\bibnamefont
  {Hofmann}}, \bibinfo {author} {\bibfnamefont {M.}~\bibnamefont {Eckstein}},
  \bibinfo {author} {\bibfnamefont {E.}~\bibnamefont {Arrigoni}}, \ and\
  \bibinfo {author} {\bibfnamefont {M.}~\bibnamefont {Potthoff}},\ }\href
  {\doibase 10.1103/PhysRevB.88.165124} {\bibfield  {journal} {\bibinfo
  {journal} {Phys. Rev. B}\ }\textbf {\bibinfo {volume} {88}},\ \bibinfo
  {pages} {165124} (\bibinfo {year} {2013})}\BibitemShut {NoStop}%
\bibitem [{\citenamefont {Hofmann}\ \emph {et~al.}(2016)\citenamefont
  {Hofmann}, \citenamefont {Eckstein},\ and\ \citenamefont
  {Potthoff}}]{Hofmann2016}%
  \BibitemOpen
  \bibfield  {author} {\bibinfo {author} {\bibfnamefont {F.}~\bibnamefont
  {Hofmann}}, \bibinfo {author} {\bibfnamefont {M.}~\bibnamefont {Eckstein}}, \
  and\ \bibinfo {author} {\bibfnamefont {M.}~\bibnamefont {Potthoff}},\ }\href
  {\doibase 10.1103/PhysRevB.93.235104} {\bibfield  {journal} {\bibinfo
  {journal} {Phys. Rev. B}\ }\textbf {\bibinfo {volume} {93}},\ \bibinfo
  {pages} {235104} (\bibinfo {year} {2016})}\BibitemShut {NoStop}%
\bibitem [{\citenamefont {Asadzadeh}\ \emph {et~al.}(2016)\citenamefont
  {Asadzadeh}, \citenamefont {Fabrizio},\ and\ \citenamefont
  {Arrigoni}}]{Asadzadeh2016}%
  \BibitemOpen
  \bibfield  {author} {\bibinfo {author} {\bibfnamefont {M.~Z.}\ \bibnamefont
  {Asadzadeh}}, \bibinfo {author} {\bibfnamefont {M.}~\bibnamefont {Fabrizio}},
  \ and\ \bibinfo {author} {\bibfnamefont {E.}~\bibnamefont {Arrigoni}},\
  }\href {\doibase 10.1103/PhysRevB.94.205146} {\bibfield  {journal} {\bibinfo
  {journal} {Phys. Rev. B}\ }\textbf {\bibinfo {volume} {94}},\ \bibinfo
  {pages} {205146} (\bibinfo {year} {2016})}\BibitemShut {NoStop}%
\bibitem [{\citenamefont {Kosugi}\ \emph {et~al.}(2018)\citenamefont {Kosugi},
  \citenamefont {Nishi}, \citenamefont {Furukawa},\ and\ \citenamefont
  {Matsushita}}]{Kosugi2018}%
  \BibitemOpen
  \bibfield  {author} {\bibinfo {author} {\bibfnamefont {T.}~\bibnamefont
  {Kosugi}}, \bibinfo {author} {\bibfnamefont {H.}~\bibnamefont {Nishi}},
  \bibinfo {author} {\bibfnamefont {Y.}~\bibnamefont {Furukawa}}, \ and\
  \bibinfo {author} {\bibfnamefont {Y.-i.}\ \bibnamefont {Matsushita}},\ }\href
  {\doibase 10.1063/1.5029535} {\bibfield  {journal} {\bibinfo  {journal} {The
  Journal of Chemical Physics}\ }\textbf {\bibinfo {volume} {148}},\ \bibinfo
  {pages} {224103} (\bibinfo {year} {2018})}\BibitemShut {NoStop}%
\bibitem [{\citenamefont {Po\v{z}gaj\v{c}i\'{c}}()}]{Pozgajcic2004}%
  \BibitemOpen
  \bibfield  {author} {\bibinfo {author} {\bibfnamefont {K.}~\bibnamefont
  {Po\v{z}gaj\v{c}i\'{c}}},\ }\href {http://arxiv.org/abs/cond-mat/0407172}
  {\enquote {\bibinfo {title} {{Quantitative aspects of the dynamical impurity
  approach}},}\ }\Eprint {http://arxiv.org/abs/cond-mat/0407172}
  {arXiv:cond-mat/0407172} \BibitemShut {NoStop}%
\bibitem [{\citenamefont {Inaba}\ \emph
  {et~al.}(2005{\natexlab{a}})\citenamefont {Inaba}, \citenamefont {Koga},
  \citenamefont {Suga},\ and\ \citenamefont {Kawakami}}]{Inaba2005_PRB}%
  \BibitemOpen
  \bibfield  {author} {\bibinfo {author} {\bibfnamefont {K.}~\bibnamefont
  {Inaba}}, \bibinfo {author} {\bibfnamefont {A.}~\bibnamefont {Koga}},
  \bibinfo {author} {\bibfnamefont {S.}~\bibnamefont {Suga}}, \ and\ \bibinfo
  {author} {\bibfnamefont {N.}~\bibnamefont {Kawakami}},\ }\href {\doibase
  10.1103/PhysRevB.72.085112} {\bibfield  {journal} {\bibinfo  {journal} {Phys.
  Rev. B}\ }\textbf {\bibinfo {volume} {72}},\ \bibinfo {pages} {085112}
  (\bibinfo {year} {2005}{\natexlab{a}})}\BibitemShut {NoStop}%
\bibitem [{\citenamefont {Inaba}\ \emph
  {et~al.}(2005{\natexlab{b}})\citenamefont {Inaba}, \citenamefont {Koga},
  \citenamefont {Suga},\ and\ \citenamefont {Kawakami}}]{Inaba2005_JPSJ}%
  \BibitemOpen
  \bibfield  {author} {\bibinfo {author} {\bibfnamefont {K.}~\bibnamefont
  {Inaba}}, \bibinfo {author} {\bibfnamefont {A.}~\bibnamefont {Koga}},
  \bibinfo {author} {\bibfnamefont {S.}~\bibnamefont {Suga}}, \ and\ \bibinfo
  {author} {\bibfnamefont {N.}~\bibnamefont {Kawakami}},\ }\href
  {http://dx.doi.org/10.1143/JPSJ.74.2393} {\bibfield  {journal} {\bibinfo
  {journal} {J. Phys. Soc. Jpn.}\ }\textbf {\bibinfo {volume} {74}},\ \bibinfo
  {pages} {2393} (\bibinfo {year} {2005}{\natexlab{b}})}\BibitemShut {NoStop}%
\bibitem [{\citenamefont {Eckstein}\ \emph {et~al.}(2007)\citenamefont
  {Eckstein}, \citenamefont {Kollar}, \citenamefont {Potthoff},\ and\
  \citenamefont {Vollhardt}}]{Eckstein2007}%
  \BibitemOpen
  \bibfield  {author} {\bibinfo {author} {\bibfnamefont {M.}~\bibnamefont
  {Eckstein}}, \bibinfo {author} {\bibfnamefont {M.}~\bibnamefont {Kollar}},
  \bibinfo {author} {\bibfnamefont {M.}~\bibnamefont {Potthoff}}, \ and\
  \bibinfo {author} {\bibfnamefont {D.}~\bibnamefont {Vollhardt}},\ }\href
  {\doibase 10.1103/PhysRevB.75.125103} {\bibfield  {journal} {\bibinfo
  {journal} {Phys. Rev. B}\ }\textbf {\bibinfo {volume} {75}},\ \bibinfo
  {pages} {125103} (\bibinfo {year} {2007})}\BibitemShut {NoStop}%
\bibitem [{\citenamefont {Eder}(2007)}]{Eder2007}%
  \BibitemOpen
  \bibfield  {author} {\bibinfo {author} {\bibfnamefont {R.}~\bibnamefont
  {Eder}},\ }\href
  {http://link.springer.com/article/10.1007\%2Fs10909-007-9310-4} {\bibfield
  {journal} {\bibinfo  {journal} {J. Low Temp. Phys.}\ }\textbf {\bibinfo
  {volume} {147}},\ \bibinfo {pages} {295} (\bibinfo {year}
  {2007})}\BibitemShut {NoStop}%
\bibitem [{\citenamefont {Eder}(2008)}]{Eder2008}%
  \BibitemOpen
  \bibfield  {author} {\bibinfo {author} {\bibfnamefont {R.}~\bibnamefont
  {Eder}},\ }\href {\doibase 10.1103/PhysRevB.78.115111} {\bibfield  {journal}
  {\bibinfo  {journal} {Phys. Rev. B}\ }\textbf {\bibinfo {volume} {78}},\
  \bibinfo {pages} {115111} (\bibinfo {year} {2008})}\BibitemShut {NoStop}%
\bibitem [{\citenamefont {Li}\ \emph {et~al.}(2009)\citenamefont {Li},
  \citenamefont {Hanke}, \citenamefont {Rubtsov}, \citenamefont {B\"ase},\ and\
  \citenamefont {Potthoff}}]{Li2009}%
  \BibitemOpen
  \bibfield  {author} {\bibinfo {author} {\bibfnamefont {G.}~\bibnamefont
  {Li}}, \bibinfo {author} {\bibfnamefont {W.}~\bibnamefont {Hanke}}, \bibinfo
  {author} {\bibfnamefont {A.~N.}\ \bibnamefont {Rubtsov}}, \bibinfo {author}
  {\bibfnamefont {S.}~\bibnamefont {B\"ase}}, \ and\ \bibinfo {author}
  {\bibfnamefont {M.}~\bibnamefont {Potthoff}},\ }\href {\doibase
  10.1103/PhysRevB.80.195118} {\bibfield  {journal} {\bibinfo  {journal} {Phys.
  Rev. B}\ }\textbf {\bibinfo {volume} {80}},\ \bibinfo {pages} {195118}
  (\bibinfo {year} {2009})}\BibitemShut {NoStop}%
\bibitem [{\citenamefont {Seki}\ \emph {et~al.}(2014)\citenamefont {Seki},
  \citenamefont {Wakisaka}, \citenamefont {Kaneko}, \citenamefont {Toriyama},
  \citenamefont {Konishi}, \citenamefont {Sudayama}, \citenamefont {Saini},
  \citenamefont {Arita}, \citenamefont {Namatame}, \citenamefont {Taniguchi},
  \citenamefont {Katayama}, \citenamefont {Nohara}, \citenamefont {Takagi},
  \citenamefont {Mizokawa},\ and\ \citenamefont {Ohta}}]{Seki2014}%
  \BibitemOpen
  \bibfield  {author} {\bibinfo {author} {\bibfnamefont {K.}~\bibnamefont
  {Seki}}, \bibinfo {author} {\bibfnamefont {Y.}~\bibnamefont {Wakisaka}},
  \bibinfo {author} {\bibfnamefont {T.}~\bibnamefont {Kaneko}}, \bibinfo
  {author} {\bibfnamefont {T.}~\bibnamefont {Toriyama}}, \bibinfo {author}
  {\bibfnamefont {T.}~\bibnamefont {Konishi}}, \bibinfo {author} {\bibfnamefont
  {T.}~\bibnamefont {Sudayama}}, \bibinfo {author} {\bibfnamefont {N.~L.}\
  \bibnamefont {Saini}}, \bibinfo {author} {\bibfnamefont {M.}~\bibnamefont
  {Arita}}, \bibinfo {author} {\bibfnamefont {H.}~\bibnamefont {Namatame}},
  \bibinfo {author} {\bibfnamefont {M.}~\bibnamefont {Taniguchi}}, \bibinfo
  {author} {\bibfnamefont {N.}~\bibnamefont {Katayama}}, \bibinfo {author}
  {\bibfnamefont {M.}~\bibnamefont {Nohara}}, \bibinfo {author} {\bibfnamefont
  {H.}~\bibnamefont {Takagi}}, \bibinfo {author} {\bibfnamefont
  {T.}~\bibnamefont {Mizokawa}}, \ and\ \bibinfo {author} {\bibfnamefont
  {Y.}~\bibnamefont {Ohta}},\ }\href {\doibase 10.1103/PhysRevB.90.155116}
  {\bibfield  {journal} {\bibinfo  {journal} {Phys. Rev. B}\ }\textbf {\bibinfo
  {volume} {90}},\ \bibinfo {pages} {155116} (\bibinfo {year}
  {2014})}\BibitemShut {NoStop}%
\bibitem [{\citenamefont {Eder}(2015)}]{Eder2015}%
  \BibitemOpen
  \bibfield  {author} {\bibinfo {author} {\bibfnamefont {R.}~\bibnamefont
  {Eder}},\ }\href {\doibase 10.1103/PhysRevB.91.245146} {\bibfield  {journal}
  {\bibinfo  {journal} {Phys. Rev. B}\ }\textbf {\bibinfo {volume} {91}},\
  \bibinfo {pages} {245146} (\bibinfo {year} {2015})}\BibitemShut {NoStop}%
\bibitem [{\citenamefont {Seki}\ \emph {et~al.}(2016)\citenamefont {Seki},
  \citenamefont {Shirakawa}, \citenamefont {Zhang}, \citenamefont {Li},\ and\
  \citenamefont {Yunoki}}]{Seki2016}%
  \BibitemOpen
  \bibfield  {author} {\bibinfo {author} {\bibfnamefont {K.}~\bibnamefont
  {Seki}}, \bibinfo {author} {\bibfnamefont {T.}~\bibnamefont {Shirakawa}},
  \bibinfo {author} {\bibfnamefont {Q.}~\bibnamefont {Zhang}}, \bibinfo
  {author} {\bibfnamefont {T.}~\bibnamefont {Li}}, \ and\ \bibinfo {author}
  {\bibfnamefont {S.}~\bibnamefont {Yunoki}},\ }\href {\doibase
  10.1103/PhysRevB.93.155419} {\bibfield  {journal} {\bibinfo  {journal} {Phys.
  Rev. B}\ }\textbf {\bibinfo {volume} {93}},\ \bibinfo {pages} {155419}
  (\bibinfo {year} {2016})}\BibitemShut {NoStop}%
\bibitem [{\citenamefont {Wei{\ss}e}\ and\ \citenamefont
  {Fehske}(2008)}]{Weisse}%
  \BibitemOpen
  \bibfield  {author} {\bibinfo {author} {\bibfnamefont {A.}~\bibnamefont
  {Wei{\ss}e}}\ and\ \bibinfo {author} {\bibfnamefont {H.}~\bibnamefont
  {Fehske}},\ }\enquote {\bibinfo {title} {Exact diagonalization techniques},}\
  in\ \href {\doibase 10.1007/978-3-540-74686-7_18} {\emph {\bibinfo
  {booktitle} {Computational Many-Particle Physics}}},\ Vol.\ \bibinfo {volume}
  {739},\ \bibinfo {editor} {edited by\ \bibinfo {editor} {\bibfnamefont
  {H.}~\bibnamefont {Fehske}}, \bibinfo {editor} {\bibfnamefont
  {R.}~\bibnamefont {Schneider}}, \ and\ \bibinfo {editor} {\bibfnamefont
  {A.}~\bibnamefont {Wei{\ss}e}}}\ (\bibinfo  {publisher} {Springer Berlin
  Heidelberg},\ \bibinfo {address} {Berlin, Heidelberg},\ \bibinfo {year}
  {2008})\ pp.\ \bibinfo {pages} {529--544}\BibitemShut {NoStop}%
\bibitem [{\citenamefont {Prelov{\v{s}}ek}\ and\ \citenamefont
  {Bon{\v{c}}a}(2013)}]{Prelovsek}%
  \BibitemOpen
  \bibfield  {author} {\bibinfo {author} {\bibfnamefont {P.}~\bibnamefont
  {Prelov{\v{s}}ek}}\ and\ \bibinfo {author} {\bibfnamefont {J.}~\bibnamefont
  {Bon{\v{c}}a}},\ }\enquote {\bibinfo {title} {Ground state and finite
  temperature lanczos methods},}\ in\ \href {\doibase
  10.1007/978-3-642-35106-8_1} {\emph {\bibinfo {booktitle} {Strongly
  Correlated Systems: Numerical Methods}}},\ \bibinfo {editor} {edited by\
  \bibinfo {editor} {\bibfnamefont {A.}~\bibnamefont {Avella}}\ and\ \bibinfo
  {editor} {\bibfnamefont {F.}~\bibnamefont {Mancini}}}\ (\bibinfo  {publisher}
  {Springer Berlin Heidelberg},\ \bibinfo {address} {Berlin, Heidelberg},\
  \bibinfo {year} {2013})\ pp.\ \bibinfo {pages} {1--30}\BibitemShut {NoStop}%
\bibitem [{\citenamefont {Kim}\ \emph {et~al.}(2008)\citenamefont {Kim},
  \citenamefont {Jin}, \citenamefont {Moon}, \citenamefont {Kim}, \citenamefont
  {Park}, \citenamefont {Leem}, \citenamefont {Yu}, \citenamefont {Noh},
  \citenamefont {Kim}, \citenamefont {Oh}, \citenamefont {Park}, \citenamefont
  {Durairaj}, \citenamefont {Cao},\ and\ \citenamefont {Rotenberg}}]{Kim2008}%
  \BibitemOpen
  \bibfield  {author} {\bibinfo {author} {\bibfnamefont {B.~J.}\ \bibnamefont
  {Kim}}, \bibinfo {author} {\bibfnamefont {H.}~\bibnamefont {Jin}}, \bibinfo
  {author} {\bibfnamefont {S.~J.}\ \bibnamefont {Moon}}, \bibinfo {author}
  {\bibfnamefont {J.-Y.}\ \bibnamefont {Kim}}, \bibinfo {author} {\bibfnamefont
  {B.-G.}\ \bibnamefont {Park}}, \bibinfo {author} {\bibfnamefont {C.~S.}\
  \bibnamefont {Leem}}, \bibinfo {author} {\bibfnamefont {J.}~\bibnamefont
  {Yu}}, \bibinfo {author} {\bibfnamefont {T.~W.}\ \bibnamefont {Noh}},
  \bibinfo {author} {\bibfnamefont {C.}~\bibnamefont {Kim}}, \bibinfo {author}
  {\bibfnamefont {S.-J.}\ \bibnamefont {Oh}}, \bibinfo {author} {\bibfnamefont
  {J.-H.}\ \bibnamefont {Park}}, \bibinfo {author} {\bibfnamefont
  {V.}~\bibnamefont {Durairaj}}, \bibinfo {author} {\bibfnamefont
  {G.}~\bibnamefont {Cao}}, \ and\ \bibinfo {author} {\bibfnamefont
  {E.}~\bibnamefont {Rotenberg}},\ }\href {\doibase
  10.1103/PhysRevLett.101.076402} {\bibfield  {journal} {\bibinfo  {journal}
  {Phys. Rev. Lett.}\ }\textbf {\bibinfo {volume} {101}},\ \bibinfo {pages}
  {076402} (\bibinfo {year} {2008})}\BibitemShut {NoStop}%
\bibitem [{\citenamefont {Kim}\ \emph {et~al.}(2012)\citenamefont {Kim},
  \citenamefont {Casa}, \citenamefont {Upton}, \citenamefont {Gog},
  \citenamefont {Kim}, \citenamefont {Mitchell}, \citenamefont {van
  Veenendaal}, \citenamefont {Daghofer}, \citenamefont {van~den Brink},
  \citenamefont {Khaliullin},\ and\ \citenamefont {Kim}}]{Kim2012}%
  \BibitemOpen
  \bibfield  {author} {\bibinfo {author} {\bibfnamefont {J.}~\bibnamefont
  {Kim}}, \bibinfo {author} {\bibfnamefont {D.}~\bibnamefont {Casa}}, \bibinfo
  {author} {\bibfnamefont {M.~H.}\ \bibnamefont {Upton}}, \bibinfo {author}
  {\bibfnamefont {T.}~\bibnamefont {Gog}}, \bibinfo {author} {\bibfnamefont
  {Y.-J.}\ \bibnamefont {Kim}}, \bibinfo {author} {\bibfnamefont {J.~F.}\
  \bibnamefont {Mitchell}}, \bibinfo {author} {\bibfnamefont {M.}~\bibnamefont
  {van Veenendaal}}, \bibinfo {author} {\bibfnamefont {M.}~\bibnamefont
  {Daghofer}}, \bibinfo {author} {\bibfnamefont {J.}~\bibnamefont {van~den
  Brink}}, \bibinfo {author} {\bibfnamefont {G.}~\bibnamefont {Khaliullin}}, \
  and\ \bibinfo {author} {\bibfnamefont {B.~J.}\ \bibnamefont {Kim}},\ }\href
  {\doibase 10.1103/PhysRevLett.108.177003} {\bibfield  {journal} {\bibinfo
  {journal} {Phys. Rev. Lett.}\ }\textbf {\bibinfo {volume} {108}},\ \bibinfo
  {pages} {177003} (\bibinfo {year} {2012})}\BibitemShut {NoStop}%
\bibitem [{\citenamefont {Jackeli}\ and\ \citenamefont
  {Khaliullin}(2009)}]{Jackeli2009}%
  \BibitemOpen
  \bibfield  {author} {\bibinfo {author} {\bibfnamefont {G.}~\bibnamefont
  {Jackeli}}\ and\ \bibinfo {author} {\bibfnamefont {G.}~\bibnamefont
  {Khaliullin}},\ }\href {\doibase 10.1103/PhysRevLett.102.017205} {\bibfield
  {journal} {\bibinfo  {journal} {Phys. Rev. Lett.}\ }\textbf {\bibinfo
  {volume} {102}},\ \bibinfo {pages} {017205} (\bibinfo {year}
  {2009})}\BibitemShut {NoStop}%
\bibitem [{\citenamefont {Jin}\ \emph {et~al.}(2009)\citenamefont {Jin},
  \citenamefont {Jeong}, \citenamefont {Ozaki},\ and\ \citenamefont
  {Yu}}]{Jin2009}%
  \BibitemOpen
  \bibfield  {author} {\bibinfo {author} {\bibfnamefont {H.}~\bibnamefont
  {Jin}}, \bibinfo {author} {\bibfnamefont {H.}~\bibnamefont {Jeong}}, \bibinfo
  {author} {\bibfnamefont {T.}~\bibnamefont {Ozaki}}, \ and\ \bibinfo {author}
  {\bibfnamefont {J.}~\bibnamefont {Yu}},\ }\href {\doibase
  10.1103/PhysRevB.80.075112} {\bibfield  {journal} {\bibinfo  {journal} {Phys.
  Rev. B}\ }\textbf {\bibinfo {volume} {80}},\ \bibinfo {pages} {075112}
  (\bibinfo {year} {2009})}\BibitemShut {NoStop}%
\bibitem [{\citenamefont {Watanabe}\ \emph {et~al.}(2010)\citenamefont
  {Watanabe}, \citenamefont {Shirakawa},\ and\ \citenamefont
  {Yunoki}}]{Watanabe2010}%
  \BibitemOpen
  \bibfield  {author} {\bibinfo {author} {\bibfnamefont {H.}~\bibnamefont
  {Watanabe}}, \bibinfo {author} {\bibfnamefont {T.}~\bibnamefont {Shirakawa}},
  \ and\ \bibinfo {author} {\bibfnamefont {S.}~\bibnamefont {Yunoki}},\ }\href
  {\doibase 10.1103/PhysRevLett.105.216410} {\bibfield  {journal} {\bibinfo
  {journal} {Phys. Rev. Lett.}\ }\textbf {\bibinfo {volume} {105}},\ \bibinfo
  {pages} {216410} (\bibinfo {year} {2010})}\BibitemShut {NoStop}%
\bibitem [{\citenamefont {Shirakawa}\ \emph {et~al.}(2011)\citenamefont
  {Shirakawa}, \citenamefont {Watanabe},\ and\ \citenamefont
  {Yunoki}}]{Shirakawa2011}%
  \BibitemOpen
  \bibfield  {author} {\bibinfo {author} {\bibfnamefont {T.}~\bibnamefont
  {Shirakawa}}, \bibinfo {author} {\bibfnamefont {H.}~\bibnamefont {Watanabe}},
  \ and\ \bibinfo {author} {\bibfnamefont {S.}~\bibnamefont {Yunoki}},\ }\href
  {http://dx.doi.org/10.1143/JPSJS.80SB.SB010} {\bibfield  {journal} {\bibinfo
  {journal} {J. Phys. Soc. Jpn.}\ }\textbf {\bibinfo {volume} {80}},\ \bibinfo
  {pages} {SB010} (\bibinfo {year} {2011})}\BibitemShut {NoStop}%
\bibitem [{\citenamefont {Watanabe}\ \emph {et~al.}(2011)\citenamefont
  {Watanabe}, \citenamefont {Shirakawa},\ and\ \citenamefont
  {Yunoki}}]{Watanabe2011}%
  \BibitemOpen
  \bibfield  {author} {\bibinfo {author} {\bibfnamefont {H.}~\bibnamefont
  {Watanabe}}, \bibinfo {author} {\bibfnamefont {T.}~\bibnamefont {Shirakawa}},
  \ and\ \bibinfo {author} {\bibfnamefont {S.}~\bibnamefont {Yunoki}},\ }\href
  {http://dx.doi.org/10.1143/JPSJS.80SB.SB006} {\bibfield  {journal} {\bibinfo
  {journal} {J. Phys. Soc. Jpn.}\ }\textbf {\bibinfo {volume} {80}},\ \bibinfo
  {pages} {SB006} (\bibinfo {year} {2011})}\BibitemShut {NoStop}%
\bibitem [{\citenamefont {Sato}\ \emph {et~al.}(2015)\citenamefont {Sato},
  \citenamefont {Shirakawa},\ and\ \citenamefont {Yunoki}}]{Sato2015}%
  \BibitemOpen
  \bibfield  {author} {\bibinfo {author} {\bibfnamefont {T.}~\bibnamefont
  {Sato}}, \bibinfo {author} {\bibfnamefont {T.}~\bibnamefont {Shirakawa}}, \
  and\ \bibinfo {author} {\bibfnamefont {S.}~\bibnamefont {Yunoki}},\ }\href
  {\doibase 10.1103/PhysRevB.91.125122} {\bibfield  {journal} {\bibinfo
  {journal} {Phys. Rev. B}\ }\textbf {\bibinfo {volume} {91}},\ \bibinfo
  {pages} {125122} (\bibinfo {year} {2015})}\BibitemShut {NoStop}%
\bibitem [{\citenamefont {Wang}\ and\ \citenamefont
  {Senthil}(2011)}]{Wang2011}%
  \BibitemOpen
  \bibfield  {author} {\bibinfo {author} {\bibfnamefont {F.}~\bibnamefont
  {Wang}}\ and\ \bibinfo {author} {\bibfnamefont {T.}~\bibnamefont {Senthil}},\
  }\href {\doibase 10.1103/PhysRevLett.106.136402} {\bibfield  {journal}
  {\bibinfo  {journal} {Phys. Rev. Lett.}\ }\textbf {\bibinfo {volume} {106}},\
  \bibinfo {pages} {136402} (\bibinfo {year} {2011})}\BibitemShut {NoStop}%
\bibitem [{\citenamefont {Watanabe}\ \emph {et~al.}(2013)\citenamefont
  {Watanabe}, \citenamefont {Shirakawa},\ and\ \citenamefont
  {Yunoki}}]{Watanabe2013}%
  \BibitemOpen
  \bibfield  {author} {\bibinfo {author} {\bibfnamefont {H.}~\bibnamefont
  {Watanabe}}, \bibinfo {author} {\bibfnamefont {T.}~\bibnamefont {Shirakawa}},
  \ and\ \bibinfo {author} {\bibfnamefont {S.}~\bibnamefont {Yunoki}},\ }\href
  {\doibase 10.1103/PhysRevLett.110.027002} {\bibfield  {journal} {\bibinfo
  {journal} {Phys. Rev. Lett.}\ }\textbf {\bibinfo {volume} {110}},\ \bibinfo
  {pages} {027002} (\bibinfo {year} {2013})}\BibitemShut {NoStop}%
\bibitem [{\citenamefont {Yang}\ \emph {et~al.}(2014)\citenamefont {Yang},
  \citenamefont {Wang}, \citenamefont {Liu}, \citenamefont {Chen},
  \citenamefont {Dai},\ and\ \citenamefont {Wang}}]{Yang2014}%
  \BibitemOpen
  \bibfield  {author} {\bibinfo {author} {\bibfnamefont {Y.}~\bibnamefont
  {Yang}}, \bibinfo {author} {\bibfnamefont {W.-S.}\ \bibnamefont {Wang}},
  \bibinfo {author} {\bibfnamefont {J.-G.}\ \bibnamefont {Liu}}, \bibinfo
  {author} {\bibfnamefont {H.}~\bibnamefont {Chen}}, \bibinfo {author}
  {\bibfnamefont {J.-H.}\ \bibnamefont {Dai}}, \ and\ \bibinfo {author}
  {\bibfnamefont {Q.-H.}\ \bibnamefont {Wang}},\ }\href {\doibase
  10.1103/PhysRevB.89.094518} {\bibfield  {journal} {\bibinfo  {journal} {Phys.
  Rev. B}\ }\textbf {\bibinfo {volume} {89}},\ \bibinfo {pages} {094518}
  (\bibinfo {year} {2014})}\BibitemShut {NoStop}%
\bibitem [{\citenamefont {Meng}\ \emph {et~al.}(2014)\citenamefont {Meng},
  \citenamefont {Kim},\ and\ \citenamefont {Kee}}]{Meng2014}%
  \BibitemOpen
  \bibfield  {author} {\bibinfo {author} {\bibfnamefont {Z.~Y.}\ \bibnamefont
  {Meng}}, \bibinfo {author} {\bibfnamefont {Y.~B.}\ \bibnamefont {Kim}}, \
  and\ \bibinfo {author} {\bibfnamefont {H.-Y.}\ \bibnamefont {Kee}},\ }\href
  {\doibase 10.1103/PhysRevLett.113.177003} {\bibfield  {journal} {\bibinfo
  {journal} {Phys. Rev. Lett.}\ }\textbf {\bibinfo {volume} {113}},\ \bibinfo
  {pages} {177003} (\bibinfo {year} {2014})}\BibitemShut {NoStop}%
\bibitem [{\citenamefont {Kim}\ \emph {et~al.}(2014)\citenamefont {Kim},
  \citenamefont {Krupin}, \citenamefont {Denlinger}, \citenamefont {Bostwick},
  \citenamefont {Rotenberg}, \citenamefont {Zhao}, \citenamefont {Mitchell},
  \citenamefont {Allen},\ and\ \citenamefont {Kim}}]{Kim2014}%
  \BibitemOpen
  \bibfield  {author} {\bibinfo {author} {\bibfnamefont {Y.~K.}\ \bibnamefont
  {Kim}}, \bibinfo {author} {\bibfnamefont {O.}~\bibnamefont {Krupin}},
  \bibinfo {author} {\bibfnamefont {J.~D.}\ \bibnamefont {Denlinger}}, \bibinfo
  {author} {\bibfnamefont {A.}~\bibnamefont {Bostwick}}, \bibinfo {author}
  {\bibfnamefont {E.}~\bibnamefont {Rotenberg}}, \bibinfo {author}
  {\bibfnamefont {Q.}~\bibnamefont {Zhao}}, \bibinfo {author} {\bibfnamefont
  {J.~F.}\ \bibnamefont {Mitchell}}, \bibinfo {author} {\bibfnamefont {J.~W.}\
  \bibnamefont {Allen}}, \ and\ \bibinfo {author} {\bibfnamefont {B.~J.}\
  \bibnamefont {Kim}},\ }\href {\doibase 10.1126/science.1251151} {\bibfield
  {journal} {\bibinfo  {journal} {Science}\ }\textbf {\bibinfo {volume}
  {345}},\ \bibinfo {pages} {187} (\bibinfo {year} {2014})}\BibitemShut
  {NoStop}%
\bibitem [{\citenamefont {de~la Torre}\ \emph {et~al.}(2015)\citenamefont
  {de~la Torre}, \citenamefont {McKeown~Walker}, \citenamefont {Bruno},
  \citenamefont {Ricc\'o}, \citenamefont {Wang}, \citenamefont
  {Gutierrez~Lezama}, \citenamefont {Scheerer}, \citenamefont {Giriat},
  \citenamefont {Jaccard}, \citenamefont {Berthod}, \citenamefont {Kim},
  \citenamefont {Hoesch}, \citenamefont {Hunter}, \citenamefont {Perry},
  \citenamefont {Tamai},\ and\ \citenamefont {Baumberger}}]{Torre2015}%
  \BibitemOpen
  \bibfield  {author} {\bibinfo {author} {\bibfnamefont {A.}~\bibnamefont
  {de~la Torre}}, \bibinfo {author} {\bibfnamefont {S.}~\bibnamefont
  {McKeown~Walker}}, \bibinfo {author} {\bibfnamefont {F.~Y.}\ \bibnamefont
  {Bruno}}, \bibinfo {author} {\bibfnamefont {S.}~\bibnamefont {Ricc\'o}},
  \bibinfo {author} {\bibfnamefont {Z.}~\bibnamefont {Wang}}, \bibinfo {author}
  {\bibfnamefont {I.}~\bibnamefont {Gutierrez~Lezama}}, \bibinfo {author}
  {\bibfnamefont {G.}~\bibnamefont {Scheerer}}, \bibinfo {author}
  {\bibfnamefont {G.}~\bibnamefont {Giriat}}, \bibinfo {author} {\bibfnamefont
  {D.}~\bibnamefont {Jaccard}}, \bibinfo {author} {\bibfnamefont
  {C.}~\bibnamefont {Berthod}}, \bibinfo {author} {\bibfnamefont {T.~K.}\
  \bibnamefont {Kim}}, \bibinfo {author} {\bibfnamefont {M.}~\bibnamefont
  {Hoesch}}, \bibinfo {author} {\bibfnamefont {E.~C.}\ \bibnamefont {Hunter}},
  \bibinfo {author} {\bibfnamefont {R.~S.}\ \bibnamefont {Perry}}, \bibinfo
  {author} {\bibfnamefont {A.}~\bibnamefont {Tamai}}, \ and\ \bibinfo {author}
  {\bibfnamefont {F.}~\bibnamefont {Baumberger}},\ }\href {\doibase
  10.1103/PhysRevLett.115.176402} {\bibfield  {journal} {\bibinfo  {journal}
  {Phys. Rev. Lett.}\ }\textbf {\bibinfo {volume} {115}},\ \bibinfo {pages}
  {176402} (\bibinfo {year} {2015})}\BibitemShut {NoStop}%
\bibitem [{\citenamefont {Yan}\ \emph {et~al.}(2015)\citenamefont {Yan},
  \citenamefont {Ren}, \citenamefont {Xu}, \citenamefont {Xie}, \citenamefont
  {Tao}, \citenamefont {Choi}, \citenamefont {Lee}, \citenamefont {Choi},
  \citenamefont {Zhang},\ and\ \citenamefont {Feng}}]{Yan2015}%
  \BibitemOpen
  \bibfield  {author} {\bibinfo {author} {\bibfnamefont {Y.~J.}\ \bibnamefont
  {Yan}}, \bibinfo {author} {\bibfnamefont {M.~Q.}\ \bibnamefont {Ren}},
  \bibinfo {author} {\bibfnamefont {H.~C.}\ \bibnamefont {Xu}}, \bibinfo
  {author} {\bibfnamefont {B.~P.}\ \bibnamefont {Xie}}, \bibinfo {author}
  {\bibfnamefont {R.}~\bibnamefont {Tao}}, \bibinfo {author} {\bibfnamefont
  {H.~Y.}\ \bibnamefont {Choi}}, \bibinfo {author} {\bibfnamefont
  {N.}~\bibnamefont {Lee}}, \bibinfo {author} {\bibfnamefont {Y.~J.}\
  \bibnamefont {Choi}}, \bibinfo {author} {\bibfnamefont {T.}~\bibnamefont
  {Zhang}}, \ and\ \bibinfo {author} {\bibfnamefont {D.~L.}\ \bibnamefont
  {Feng}},\ }\href {\doibase 10.1103/PhysRevX.5.041018} {\bibfield  {journal}
  {\bibinfo  {journal} {Phys. Rev. X}\ }\textbf {\bibinfo {volume} {5}},\
  \bibinfo {pages} {041018} (\bibinfo {year} {2015})}\BibitemShut {NoStop}%
\bibitem [{\citenamefont {Liu}\ \emph {et~al.}(2015)\citenamefont {Liu},
  \citenamefont {Yu}, \citenamefont {Jia}, \citenamefont {Zhao}, \citenamefont
  {Weng}, \citenamefont {Peng}, \citenamefont {Chen}, \citenamefont {Xie},
  \citenamefont {Mou}, \citenamefont {He}, \citenamefont {Liu}, \citenamefont
  {Feng}, \citenamefont {Yi}, \citenamefont {Zhao}, \citenamefont {Liu},
  \citenamefont {He}, \citenamefont {Dong}, \citenamefont {Zhang},
  \citenamefont {Xu}, \citenamefont {Chen}, \citenamefont {Cao}, \citenamefont
  {Dai}, \citenamefont {Fang},\ and\ \citenamefont {Zhou}}]{Liu2015}%
  \BibitemOpen
  \bibfield  {author} {\bibinfo {author} {\bibfnamefont {Y.}~\bibnamefont
  {Liu}}, \bibinfo {author} {\bibfnamefont {L.}~\bibnamefont {Yu}}, \bibinfo
  {author} {\bibfnamefont {X.}~\bibnamefont {Jia}}, \bibinfo {author}
  {\bibfnamefont {J.}~\bibnamefont {Zhao}}, \bibinfo {author} {\bibfnamefont
  {H.}~\bibnamefont {Weng}}, \bibinfo {author} {\bibfnamefont {Y.}~\bibnamefont
  {Peng}}, \bibinfo {author} {\bibfnamefont {C.}~\bibnamefont {Chen}}, \bibinfo
  {author} {\bibfnamefont {Z.}~\bibnamefont {Xie}}, \bibinfo {author}
  {\bibfnamefont {D.}~\bibnamefont {Mou}}, \bibinfo {author} {\bibfnamefont
  {J.}~\bibnamefont {He}}, \bibinfo {author} {\bibfnamefont {X.}~\bibnamefont
  {Liu}}, \bibinfo {author} {\bibfnamefont {Y.}~\bibnamefont {Feng}}, \bibinfo
  {author} {\bibfnamefont {H.}~\bibnamefont {Yi}}, \bibinfo {author}
  {\bibfnamefont {L.}~\bibnamefont {Zhao}}, \bibinfo {author} {\bibfnamefont
  {G.}~\bibnamefont {Liu}}, \bibinfo {author} {\bibfnamefont {S.}~\bibnamefont
  {He}}, \bibinfo {author} {\bibfnamefont {X.}~\bibnamefont {Dong}}, \bibinfo
  {author} {\bibfnamefont {J.}~\bibnamefont {Zhang}}, \bibinfo {author}
  {\bibfnamefont {Z.}~\bibnamefont {Xu}}, \bibinfo {author} {\bibfnamefont
  {C.}~\bibnamefont {Chen}}, \bibinfo {author} {\bibfnamefont {G.}~\bibnamefont
  {Cao}}, \bibinfo {author} {\bibfnamefont {X.}~\bibnamefont {Dai}}, \bibinfo
  {author} {\bibfnamefont {Z.}~\bibnamefont {Fang}}, \ and\ \bibinfo {author}
  {\bibfnamefont {X.~J.}\ \bibnamefont {Zhou}},\ }\href
  {http://dx.doi.org/10.1038/srep13036} {\bibfield  {journal} {\bibinfo
  {journal} {Scientific Reports}\ }\textbf {\bibinfo {volume} {5}},\ \bibinfo
  {pages} {13036} (\bibinfo {year} {2015})}\BibitemShut {NoStop}%
\bibitem [{\citenamefont {Zhao}\ \emph {et~al.}(2016)\citenamefont {Zhao},
  \citenamefont {Torchinsky}, \citenamefont {Chu}, \citenamefont {Ivanov},
  \citenamefont {Lifshitz}, \citenamefont {Flint}, \citenamefont {Qi},
  \citenamefont {Cao},\ and\ \citenamefont {Hsieh}}]{Zhao16}%
  \BibitemOpen
  \bibfield  {author} {\bibinfo {author} {\bibfnamefont {L.}~\bibnamefont
  {Zhao}}, \bibinfo {author} {\bibfnamefont {D.~H.}\ \bibnamefont
  {Torchinsky}}, \bibinfo {author} {\bibfnamefont {H.}~\bibnamefont {Chu}},
  \bibinfo {author} {\bibfnamefont {V.}~\bibnamefont {Ivanov}}, \bibinfo
  {author} {\bibfnamefont {R.}~\bibnamefont {Lifshitz}}, \bibinfo {author}
  {\bibfnamefont {R.}~\bibnamefont {Flint}}, \bibinfo {author} {\bibfnamefont
  {T.}~\bibnamefont {Qi}}, \bibinfo {author} {\bibfnamefont {G.}~\bibnamefont
  {Cao}}, \ and\ \bibinfo {author} {\bibfnamefont {D.}~\bibnamefont {Hsieh}},\
  }\href {https://www.nature.com/articles/nphys3517#ref27} {\bibfield
  {journal} {\bibinfo  {journal} {Nature Physics}\ }\textbf {\bibinfo {volume}
  {12}},\ \bibinfo {pages} {32} (\bibinfo {year} {2016})}\BibitemShut {NoStop}%
\bibitem [{\citenamefont {Kim}\ \emph {et~al.}(2016)\citenamefont {Kim},
  \citenamefont {Sung}, \citenamefont {Denlinger},\ and\ \citenamefont
  {Kim}}]{Kim2016}%
  \BibitemOpen
  \bibfield  {author} {\bibinfo {author} {\bibfnamefont {Y.~K.}\ \bibnamefont
  {Kim}}, \bibinfo {author} {\bibfnamefont {N.~H.}\ \bibnamefont {Sung}},
  \bibinfo {author} {\bibfnamefont {J.~D.}\ \bibnamefont {Denlinger}}, \ and\
  \bibinfo {author} {\bibfnamefont {B.~J.}\ \bibnamefont {Kim}},\ }\href
  {http://www.nature.com/nphys/journal/v12/n1/abs/nphys3503.html} {\bibfield
  {journal} {\bibinfo  {journal} {Nature Physics}\ }\textbf {\bibinfo {volume}
  {12}},\ \bibinfo {pages} {37} (\bibinfo {year} {2016})}\BibitemShut {NoStop}%
\bibitem [{\citenamefont {Battisti}\ \emph {et~al.}(2017)\citenamefont
  {Battisti}, \citenamefont {Bastiaans}, \citenamefont {Fedoseev},
  \citenamefont {Torre}, \citenamefont {Iliopoulos}, \citenamefont {Tamai},
  \citenamefont {Hunter}, \citenamefont {Perry}, \citenamefont {Zaanen},
  \citenamefont {Baumberger},\ and\ \citenamefont {Allan}}]{Battisti17}%
  \BibitemOpen
  \bibfield  {author} {\bibinfo {author} {\bibfnamefont {I.}~\bibnamefont
  {Battisti}}, \bibinfo {author} {\bibfnamefont {K.~M.}\ \bibnamefont
  {Bastiaans}}, \bibinfo {author} {\bibfnamefont {V.}~\bibnamefont {Fedoseev}},
  \bibinfo {author} {\bibfnamefont {A.~d.~l.}\ \bibnamefont {Torre}}, \bibinfo
  {author} {\bibfnamefont {N.}~\bibnamefont {Iliopoulos}}, \bibinfo {author}
  {\bibfnamefont {A.}~\bibnamefont {Tamai}}, \bibinfo {author} {\bibfnamefont
  {E.~C.}\ \bibnamefont {Hunter}}, \bibinfo {author} {\bibfnamefont {R.~S.}\
  \bibnamefont {Perry}}, \bibinfo {author} {\bibfnamefont {J.}~\bibnamefont
  {Zaanen}}, \bibinfo {author} {\bibfnamefont {F.}~\bibnamefont {Baumberger}},
  \ and\ \bibinfo {author} {\bibfnamefont {M.~P.}\ \bibnamefont {Allan}},\
  }\href {https://www.nature.com/articles/nphys3894} {\bibfield  {journal}
  {\bibinfo  {journal} {Nature Physics}\ }\textbf {\bibinfo {volume} {13}},\
  \bibinfo {pages} {21} (\bibinfo {year} {2017})}\BibitemShut {NoStop}%
\bibitem [{\citenamefont {Seo}\ \emph {et~al.}(2017)\citenamefont {Seo},
  \citenamefont {Ahn}, \citenamefont {Song}, \citenamefont {Chen},
  \citenamefont {Wilson},\ and\ \citenamefont {Moon}}]{Seo17}%
  \BibitemOpen
  \bibfield  {author} {\bibinfo {author} {\bibfnamefont {J.~H.}\ \bibnamefont
  {Seo}}, \bibinfo {author} {\bibfnamefont {G.~H.}\ \bibnamefont {Ahn}},
  \bibinfo {author} {\bibfnamefont {S.~J.}\ \bibnamefont {Song}}, \bibinfo
  {author} {\bibfnamefont {X.}~\bibnamefont {Chen}}, \bibinfo {author}
  {\bibfnamefont {S.~D.}\ \bibnamefont {Wilson}}, \ and\ \bibinfo {author}
  {\bibfnamefont {S.~J.}\ \bibnamefont {Moon}},\ }\href
  {https://www.nature.com/articles/s41598-017-10725-z} {\bibfield  {journal}
  {\bibinfo  {journal} {Scientific Reports}\ }\textbf {\bibinfo {volume} {7}},\
  \bibinfo {pages} {10494} (\bibinfo {year} {2017})}\BibitemShut {NoStop}%
\bibitem [{\citenamefont {Terashima}\ \emph {et~al.}(2017)\citenamefont
  {Terashima}, \citenamefont {Sunagawa}, \citenamefont {Fujiwara},
  \citenamefont {Fukura}, \citenamefont {Fujii}, \citenamefont {Okada},
  \citenamefont {Horigane}, \citenamefont {Kobayashi}, \citenamefont {Horie},
  \citenamefont {Akimitsu}, \citenamefont {Golias}, \citenamefont {Marchenko},
  \citenamefont {Varykhalov}, \citenamefont {Saini}, \citenamefont {Wakita},
  \citenamefont {Muraoka},\ and\ \citenamefont {Yokoya}}]{Terashima17}%
  \BibitemOpen
  \bibfield  {author} {\bibinfo {author} {\bibfnamefont {K.}~\bibnamefont
  {Terashima}}, \bibinfo {author} {\bibfnamefont {M.}~\bibnamefont {Sunagawa}},
  \bibinfo {author} {\bibfnamefont {H.}~\bibnamefont {Fujiwara}}, \bibinfo
  {author} {\bibfnamefont {T.}~\bibnamefont {Fukura}}, \bibinfo {author}
  {\bibfnamefont {M.}~\bibnamefont {Fujii}}, \bibinfo {author} {\bibfnamefont
  {K.}~\bibnamefont {Okada}}, \bibinfo {author} {\bibfnamefont
  {K.}~\bibnamefont {Horigane}}, \bibinfo {author} {\bibfnamefont
  {K.}~\bibnamefont {Kobayashi}}, \bibinfo {author} {\bibfnamefont
  {R.}~\bibnamefont {Horie}}, \bibinfo {author} {\bibfnamefont
  {J.}~\bibnamefont {Akimitsu}}, \bibinfo {author} {\bibfnamefont
  {E.}~\bibnamefont {Golias}}, \bibinfo {author} {\bibfnamefont
  {D.}~\bibnamefont {Marchenko}}, \bibinfo {author} {\bibfnamefont
  {A.}~\bibnamefont {Varykhalov}}, \bibinfo {author} {\bibfnamefont {N.~L.}\
  \bibnamefont {Saini}}, \bibinfo {author} {\bibfnamefont {T.}~\bibnamefont
  {Wakita}}, \bibinfo {author} {\bibfnamefont {Y.}~\bibnamefont {Muraoka}}, \
  and\ \bibinfo {author} {\bibfnamefont {T.}~\bibnamefont {Yokoya}},\ }\href
  {\doibase 10.1103/PhysRevB.96.041106} {\bibfield  {journal} {\bibinfo
  {journal} {Phys. Rev. B}\ }\textbf {\bibinfo {volume} {96}},\ \bibinfo
  {pages} {041106(R)} (\bibinfo {year} {2017})}\BibitemShut {NoStop}%
\bibitem [{\citenamefont {Chen}\ \emph {et~al.}(2018)\citenamefont {Chen},
  \citenamefont {Schmehr}, \citenamefont {Islam}, \citenamefont {Porter},
  \citenamefont {Zoghlin}, \citenamefont {Finkelstein}, \citenamefont {Ruff},\
  and\ \citenamefont {Wilson}}]{Chen18}%
  \BibitemOpen
  \bibfield  {author} {\bibinfo {author} {\bibfnamefont {X.}~\bibnamefont
  {Chen}}, \bibinfo {author} {\bibfnamefont {J.~L.}\ \bibnamefont {Schmehr}},
  \bibinfo {author} {\bibfnamefont {Z.}~\bibnamefont {Islam}}, \bibinfo
  {author} {\bibfnamefont {Z.}~\bibnamefont {Porter}}, \bibinfo {author}
  {\bibfnamefont {E.}~\bibnamefont {Zoghlin}}, \bibinfo {author} {\bibfnamefont
  {K.}~\bibnamefont {Finkelstein}}, \bibinfo {author} {\bibfnamefont
  {J.~P.~C.}\ \bibnamefont {Ruff}}, \ and\ \bibinfo {author} {\bibfnamefont
  {S.~D.}\ \bibnamefont {Wilson}},\ }\href
  {https://www.nature.com/articles/s41467-017-02647-1} {\bibfield  {journal}
  {\bibinfo  {journal} {Nature Communications}\ }\textbf {\bibinfo {volume}
  {9}},\ \bibinfo {pages} {103} (\bibinfo {year} {2018})}\BibitemShut {NoStop}%
\bibitem [{\citenamefont {Arita}\ \emph {et~al.}(2012)\citenamefont {Arita},
  \citenamefont {Kune\ifmmode~\check{s}\else \v{s}\fi{}}, \citenamefont
  {Kozhevnikov}, \citenamefont {Eguiluz},\ and\ \citenamefont
  {Imada}}]{Arita2012}%
  \BibitemOpen
  \bibfield  {author} {\bibinfo {author} {\bibfnamefont {R.}~\bibnamefont
  {Arita}}, \bibinfo {author} {\bibfnamefont {J.}~\bibnamefont
  {Kune\ifmmode~\check{s}\else \v{s}\fi{}}}, \bibinfo {author} {\bibfnamefont
  {A.~V.}\ \bibnamefont {Kozhevnikov}}, \bibinfo {author} {\bibfnamefont
  {A.~G.}\ \bibnamefont {Eguiluz}}, \ and\ \bibinfo {author} {\bibfnamefont
  {M.}~\bibnamefont {Imada}},\ }\href {\doibase 10.1103/PhysRevLett.108.086403}
  {\bibfield  {journal} {\bibinfo  {journal} {Phys. Rev. Lett.}\ }\textbf
  {\bibinfo {volume} {108}},\ \bibinfo {pages} {086403} (\bibinfo {year}
  {2012})}\BibitemShut {NoStop}%
\bibitem [{\citenamefont {Watanabe}\ \emph {et~al.}(2014)\citenamefont
  {Watanabe}, \citenamefont {Shirakawa},\ and\ \citenamefont
  {Yunoki}}]{Watanabe2014}%
  \BibitemOpen
  \bibfield  {author} {\bibinfo {author} {\bibfnamefont {H.}~\bibnamefont
  {Watanabe}}, \bibinfo {author} {\bibfnamefont {T.}~\bibnamefont {Shirakawa}},
  \ and\ \bibinfo {author} {\bibfnamefont {S.}~\bibnamefont {Yunoki}},\ }\href
  {\doibase 10.1103/PhysRevB.89.165115} {\bibfield  {journal} {\bibinfo
  {journal} {Phys. Rev. B}\ }\textbf {\bibinfo {volume} {89}},\ \bibinfo
  {pages} {165115} (\bibinfo {year} {2014})}\BibitemShut {NoStop}%
\bibitem [{\citenamefont {Calder}\ \emph {et~al.}(2012)\citenamefont {Calder},
  \citenamefont {Garlea}, \citenamefont {McMorrow}, \citenamefont {Lumsden},
  \citenamefont {Stone}, \citenamefont {Lang}, \citenamefont {Kim},
  \citenamefont {Schlueter}, \citenamefont {Shi}, \citenamefont {Yamaura},
  \citenamefont {Sun}, \citenamefont {Tsujimoto},\ and\ \citenamefont
  {Christianson}}]{Calder2012}%
  \BibitemOpen
  \bibfield  {author} {\bibinfo {author} {\bibfnamefont {S.}~\bibnamefont
  {Calder}}, \bibinfo {author} {\bibfnamefont {V.~O.}\ \bibnamefont {Garlea}},
  \bibinfo {author} {\bibfnamefont {D.~F.}\ \bibnamefont {McMorrow}}, \bibinfo
  {author} {\bibfnamefont {M.~D.}\ \bibnamefont {Lumsden}}, \bibinfo {author}
  {\bibfnamefont {M.~B.}\ \bibnamefont {Stone}}, \bibinfo {author}
  {\bibfnamefont {J.~C.}\ \bibnamefont {Lang}}, \bibinfo {author}
  {\bibfnamefont {J.-W.}\ \bibnamefont {Kim}}, \bibinfo {author} {\bibfnamefont
  {J.~A.}\ \bibnamefont {Schlueter}}, \bibinfo {author} {\bibfnamefont {Y.~G.}\
  \bibnamefont {Shi}}, \bibinfo {author} {\bibfnamefont {K.}~\bibnamefont
  {Yamaura}}, \bibinfo {author} {\bibfnamefont {Y.~S.}\ \bibnamefont {Sun}},
  \bibinfo {author} {\bibfnamefont {Y.}~\bibnamefont {Tsujimoto}}, \ and\
  \bibinfo {author} {\bibfnamefont {A.~D.}\ \bibnamefont {Christianson}},\
  }\href {\doibase 10.1103/PhysRevLett.108.257209} {\bibfield  {journal}
  {\bibinfo  {journal} {Phys. Rev. Lett.}\ }\textbf {\bibinfo {volume} {108}},\
  \bibinfo {pages} {257209} (\bibinfo {year} {2012})}\BibitemShut {NoStop}%
\bibitem [{\citenamefont {Cui}\ \emph {et~al.}(2016)\citenamefont {Cui},
  \citenamefont {Cheng}, \citenamefont {Fan}, \citenamefont {Taylor},
  \citenamefont {Calder}, \citenamefont {McGuire}, \citenamefont {Yan},
  \citenamefont {Meyers}, \citenamefont {Li}, \citenamefont {Cai},
  \citenamefont {Jiao}, \citenamefont {Choi}, \citenamefont {Haskel},
  \citenamefont {Gotou}, \citenamefont {Uwatoko}, \citenamefont {Chakhalian},
  \citenamefont {Christianson}, \citenamefont {Yunoki}, \citenamefont
  {Goodenough},\ and\ \citenamefont {Zhou}}]{Cui2016}%
  \BibitemOpen
  \bibfield  {author} {\bibinfo {author} {\bibfnamefont {Q.}~\bibnamefont
  {Cui}}, \bibinfo {author} {\bibfnamefont {J.-G.}\ \bibnamefont {Cheng}},
  \bibinfo {author} {\bibfnamefont {W.}~\bibnamefont {Fan}}, \bibinfo {author}
  {\bibfnamefont {A.~E.}\ \bibnamefont {Taylor}}, \bibinfo {author}
  {\bibfnamefont {S.}~\bibnamefont {Calder}}, \bibinfo {author} {\bibfnamefont
  {M.~A.}\ \bibnamefont {McGuire}}, \bibinfo {author} {\bibfnamefont {J.-Q.}\
  \bibnamefont {Yan}}, \bibinfo {author} {\bibfnamefont {D.}~\bibnamefont
  {Meyers}}, \bibinfo {author} {\bibfnamefont {X.}~\bibnamefont {Li}}, \bibinfo
  {author} {\bibfnamefont {Y.~Q.}\ \bibnamefont {Cai}}, \bibinfo {author}
  {\bibfnamefont {Y.~Y.}\ \bibnamefont {Jiao}}, \bibinfo {author}
  {\bibfnamefont {Y.}~\bibnamefont {Choi}}, \bibinfo {author} {\bibfnamefont
  {D.}~\bibnamefont {Haskel}}, \bibinfo {author} {\bibfnamefont
  {H.}~\bibnamefont {Gotou}}, \bibinfo {author} {\bibfnamefont
  {Y.}~\bibnamefont {Uwatoko}}, \bibinfo {author} {\bibfnamefont
  {J.}~\bibnamefont {Chakhalian}}, \bibinfo {author} {\bibfnamefont {A.~D.}\
  \bibnamefont {Christianson}}, \bibinfo {author} {\bibfnamefont
  {S.}~\bibnamefont {Yunoki}}, \bibinfo {author} {\bibfnamefont {J.~B.}\
  \bibnamefont {Goodenough}}, \ and\ \bibinfo {author} {\bibfnamefont {J.-S.}\
  \bibnamefont {Zhou}},\ }\href {\doibase 10.1103/PhysRevLett.117.176603}
  {\bibfield  {journal} {\bibinfo  {journal} {Phys. Rev. Lett.}\ }\textbf
  {\bibinfo {volume} {117}},\ \bibinfo {pages} {176603} (\bibinfo {year}
  {2016})}\BibitemShut {NoStop}%
\bibitem [{\citenamefont {Chikara}\ \emph {et~al.}(2009)\citenamefont
  {Chikara}, \citenamefont {Korneta}, \citenamefont {Crummett}, \citenamefont
  {DeLong}, \citenamefont {Schlottmann},\ and\ \citenamefont
  {Cao}}]{Chikara2009}%
  \BibitemOpen
  \bibfield  {author} {\bibinfo {author} {\bibfnamefont {S.}~\bibnamefont
  {Chikara}}, \bibinfo {author} {\bibfnamefont {O.}~\bibnamefont {Korneta}},
  \bibinfo {author} {\bibfnamefont {W.~P.}\ \bibnamefont {Crummett}}, \bibinfo
  {author} {\bibfnamefont {L.~E.}\ \bibnamefont {DeLong}}, \bibinfo {author}
  {\bibfnamefont {P.}~\bibnamefont {Schlottmann}}, \ and\ \bibinfo {author}
  {\bibfnamefont {G.}~\bibnamefont {Cao}},\ }\href {\doibase
  10.1103/PhysRevB.80.140407} {\bibfield  {journal} {\bibinfo  {journal} {Phys.
  Rev. B}\ }\textbf {\bibinfo {volume} {80}},\ \bibinfo {pages} {140407}
  (\bibinfo {year} {2009})}\BibitemShut {NoStop}%
\bibitem [{\citenamefont {Li}\ \emph {et~al.}(2013)\citenamefont {Li},
  \citenamefont {Cao}, \citenamefont {Okamoto}, \citenamefont {Yi},
  \citenamefont {Lin}, \citenamefont {Sales}, \citenamefont {Yan},
  \citenamefont {Arita}, \citenamefont {Kunes}, \citenamefont {Kozhevnikov},
  \citenamefont {Eguiluz}, \citenamefont {Imada}, \citenamefont {Gai},
  \citenamefont {Pan},\ and\ \citenamefont {Mandrus}}]{Li2013}%
  \BibitemOpen
  \bibfield  {author} {\bibinfo {author} {\bibfnamefont {Q.}~\bibnamefont
  {Li}}, \bibinfo {author} {\bibfnamefont {G.}~\bibnamefont {Cao}}, \bibinfo
  {author} {\bibfnamefont {S.}~\bibnamefont {Okamoto}}, \bibinfo {author}
  {\bibfnamefont {J.}~\bibnamefont {Yi}}, \bibinfo {author} {\bibfnamefont
  {W.}~\bibnamefont {Lin}}, \bibinfo {author} {\bibfnamefont {B.~C.}\
  \bibnamefont {Sales}}, \bibinfo {author} {\bibfnamefont {J.}~\bibnamefont
  {Yan}}, \bibinfo {author} {\bibfnamefont {R.}~\bibnamefont {Arita}}, \bibinfo
  {author} {\bibfnamefont {J.}~\bibnamefont {Kunes}}, \bibinfo {author}
  {\bibfnamefont {A.~V.}\ \bibnamefont {Kozhevnikov}}, \bibinfo {author}
  {\bibfnamefont {A.~G.}\ \bibnamefont {Eguiluz}}, \bibinfo {author}
  {\bibfnamefont {M.}~\bibnamefont {Imada}}, \bibinfo {author} {\bibfnamefont
  {Z.}~\bibnamefont {Gai}}, \bibinfo {author} {\bibfnamefont {M.}~\bibnamefont
  {Pan}}, \ and\ \bibinfo {author} {\bibfnamefont {D.~G.}\ \bibnamefont
  {Mandrus}},\ }\href {http://dx.doi.org/10.1038/srep03073} {\bibfield
  {journal} {\bibinfo  {journal} {Scientific Reports}\ }\textbf {\bibinfo
  {volume} {3}},\ \bibinfo {pages} {3073} (\bibinfo {year} {2013})}\BibitemShut
  {NoStop}%
\bibitem [{\citenamefont {Nakayama}\ \emph {et~al.}(2016)\citenamefont
  {Nakayama}, \citenamefont {Kondo}, \citenamefont {Tian}, \citenamefont
  {Ishikawa}, \citenamefont {Halim}, \citenamefont {Bareille}, \citenamefont
  {Malaeb}, \citenamefont {Kuroda}, \citenamefont {Tomita}, \citenamefont
  {Ideta}, \citenamefont {Tanaka}, \citenamefont {Matsunami}, \citenamefont
  {Kimura}, \citenamefont {Inami}, \citenamefont {Ono}, \citenamefont
  {Kumigashira}, \citenamefont {Balents}, \citenamefont {Nakatsuji},\ and\
  \citenamefont {Shin}}]{Nakayama2016}%
  \BibitemOpen
  \bibfield  {author} {\bibinfo {author} {\bibfnamefont {M.}~\bibnamefont
  {Nakayama}}, \bibinfo {author} {\bibfnamefont {T.}~\bibnamefont {Kondo}},
  \bibinfo {author} {\bibfnamefont {Z.}~\bibnamefont {Tian}}, \bibinfo {author}
  {\bibfnamefont {J.~J.}\ \bibnamefont {Ishikawa}}, \bibinfo {author}
  {\bibfnamefont {M.}~\bibnamefont {Halim}}, \bibinfo {author} {\bibfnamefont
  {C.}~\bibnamefont {Bareille}}, \bibinfo {author} {\bibfnamefont
  {W.}~\bibnamefont {Malaeb}}, \bibinfo {author} {\bibfnamefont
  {K.}~\bibnamefont {Kuroda}}, \bibinfo {author} {\bibfnamefont
  {T.}~\bibnamefont {Tomita}}, \bibinfo {author} {\bibfnamefont
  {S.}~\bibnamefont {Ideta}}, \bibinfo {author} {\bibfnamefont
  {K.}~\bibnamefont {Tanaka}}, \bibinfo {author} {\bibfnamefont
  {M.}~\bibnamefont {Matsunami}}, \bibinfo {author} {\bibfnamefont
  {S.}~\bibnamefont {Kimura}}, \bibinfo {author} {\bibfnamefont
  {N.}~\bibnamefont {Inami}}, \bibinfo {author} {\bibfnamefont
  {K.}~\bibnamefont {Ono}}, \bibinfo {author} {\bibfnamefont {H.}~\bibnamefont
  {Kumigashira}}, \bibinfo {author} {\bibfnamefont {L.}~\bibnamefont
  {Balents}}, \bibinfo {author} {\bibfnamefont {S.}~\bibnamefont {Nakatsuji}},
  \ and\ \bibinfo {author} {\bibfnamefont {S.}~\bibnamefont {Shin}},\ }\href
  {\doibase 10.1103/PhysRevLett.117.056403} {\bibfield  {journal} {\bibinfo
  {journal} {Phys. Rev. Lett.}\ }\textbf {\bibinfo {volume} {117}},\ \bibinfo
  {pages} {056403} (\bibinfo {year} {2016})}\BibitemShut {NoStop}%
\bibitem [{\citenamefont {Aichhorn}\ \emph
  {et~al.}(2006{\natexlab{a}})\citenamefont {Aichhorn}, \citenamefont
  {Arrigoni}, \citenamefont {Potthoff},\ and\ \citenamefont
  {Hanke}}]{Aichhorn2006_Qmat}%
  \BibitemOpen
  \bibfield  {author} {\bibinfo {author} {\bibfnamefont {M.}~\bibnamefont
  {Aichhorn}}, \bibinfo {author} {\bibfnamefont {E.}~\bibnamefont {Arrigoni}},
  \bibinfo {author} {\bibfnamefont {M.}~\bibnamefont {Potthoff}}, \ and\
  \bibinfo {author} {\bibfnamefont {W.}~\bibnamefont {Hanke}},\ }\href
  {\doibase 10.1103/PhysRevB.74.235117} {\bibfield  {journal} {\bibinfo
  {journal} {Phys. Rev. B}\ }\textbf {\bibinfo {volume} {74}},\ \bibinfo
  {pages} {235117} (\bibinfo {year} {2006}{\natexlab{a}})}\BibitemShut
  {NoStop}%
\bibitem [{\citenamefont {Aichhorn}\ \emph {et~al.}(2003)\citenamefont
  {Aichhorn}, \citenamefont {Daghofer}, \citenamefont {Evertz},\ and\
  \citenamefont {von~der Linden}}]{Aichhorn2003}%
  \BibitemOpen
  \bibfield  {author} {\bibinfo {author} {\bibfnamefont {M.}~\bibnamefont
  {Aichhorn}}, \bibinfo {author} {\bibfnamefont {M.}~\bibnamefont {Daghofer}},
  \bibinfo {author} {\bibfnamefont {H.~G.}\ \bibnamefont {Evertz}}, \ and\
  \bibinfo {author} {\bibfnamefont {W.}~\bibnamefont {von~der Linden}},\ }\href
  {\doibase 10.1103/PhysRevB.67.161103} {\bibfield  {journal} {\bibinfo
  {journal} {Phys. Rev. B}\ }\textbf {\bibinfo {volume} {67}},\ \bibinfo
  {pages} {161103} (\bibinfo {year} {2003})}\BibitemShut {NoStop}%
\bibitem [{\citenamefont {Perroni}\ \emph {et~al.}(2007)\citenamefont
  {Perroni}, \citenamefont {Ishida},\ and\ \citenamefont
  {Liebsch}}]{Perroni2007}%
  \BibitemOpen
  \bibfield  {author} {\bibinfo {author} {\bibfnamefont {C.~A.}\ \bibnamefont
  {Perroni}}, \bibinfo {author} {\bibfnamefont {H.}~\bibnamefont {Ishida}}, \
  and\ \bibinfo {author} {\bibfnamefont {A.}~\bibnamefont {Liebsch}},\ }\href
  {\doibase 10.1103/PhysRevB.75.045125} {\bibfield  {journal} {\bibinfo
  {journal} {Phys. Rev. B}\ }\textbf {\bibinfo {volume} {75}},\ \bibinfo
  {pages} {045125} (\bibinfo {year} {2007})}\BibitemShut {NoStop}%
\bibitem [{\citenamefont {Capone}\ \emph {et~al.}(2007)\citenamefont {Capone},
  \citenamefont {de' Medici},\ and\ \citenamefont {Georges}}]{Capone2007}%
  \BibitemOpen
  \bibfield  {author} {\bibinfo {author} {\bibfnamefont {M.}~\bibnamefont
  {Capone}}, \bibinfo {author} {\bibfnamefont {L.}~\bibnamefont {de' Medici}},
  \ and\ \bibinfo {author} {\bibfnamefont {A.}~\bibnamefont {Georges}},\ }\href
  {\doibase 10.1103/PhysRevB.76.245116} {\bibfield  {journal} {\bibinfo
  {journal} {Phys. Rev. B}\ }\textbf {\bibinfo {volume} {76}},\ \bibinfo
  {pages} {245116} (\bibinfo {year} {2007})}\BibitemShut {NoStop}%
\bibitem [{\citenamefont {Liebsch}\ and\ \citenamefont
  {Ishida}(2012)}]{Liebsch2012}%
  \BibitemOpen
  \bibfield  {author} {\bibinfo {author} {\bibfnamefont {A.}~\bibnamefont
  {Liebsch}}\ and\ \bibinfo {author} {\bibfnamefont {H.}~\bibnamefont
  {Ishida}},\ }\href {http://iopscience.iop.org/0953-8984/24/5/053201}
  {\bibfield  {journal} {\bibinfo  {journal} {J. Phys.: Condens. Matter}\
  }\textbf {\bibinfo {volume} {24}},\ \bibinfo {pages} {053201} (\bibinfo
  {year} {2012})}\BibitemShut {NoStop}%
\bibitem [{\citenamefont {Wildberger}\ \emph {et~al.}(1995)\citenamefont
  {Wildberger}, \citenamefont {Lang}, \citenamefont {Zeller},\ and\
  \citenamefont {Dederichs}}]{Wildberger1995}%
  \BibitemOpen
  \bibfield  {author} {\bibinfo {author} {\bibfnamefont {K.}~\bibnamefont
  {Wildberger}}, \bibinfo {author} {\bibfnamefont {P.}~\bibnamefont {Lang}},
  \bibinfo {author} {\bibfnamefont {R.}~\bibnamefont {Zeller}}, \ and\ \bibinfo
  {author} {\bibfnamefont {P.~H.}\ \bibnamefont {Dederichs}},\ }\href {\doibase
  10.1103/PhysRevB.52.11502} {\bibfield  {journal} {\bibinfo  {journal} {Phys.
  Rev. B}\ }\textbf {\bibinfo {volume} {52}},\ \bibinfo {pages} {11502}
  (\bibinfo {year} {1995})}\BibitemShut {NoStop}%
\bibitem [{\citenamefont {Lu}\ and\ \citenamefont {Arrigoni}(2009)}]{Lu2009}%
  \BibitemOpen
  \bibfield  {author} {\bibinfo {author} {\bibfnamefont {X.}~\bibnamefont
  {Lu}}\ and\ \bibinfo {author} {\bibfnamefont {E.}~\bibnamefont {Arrigoni}},\
  }\href {\doibase 10.1103/PhysRevB.79.245109} {\bibfield  {journal} {\bibinfo
  {journal} {Phys. Rev. B}\ }\textbf {\bibinfo {volume} {79}},\ \bibinfo
  {pages} {245109} (\bibinfo {year} {2009})}\BibitemShut {NoStop}%
\bibitem [{\citenamefont {Seki}\ and\ \citenamefont {Yunoki}(2017)}]{Seki2017}%
  \BibitemOpen
  \bibfield  {author} {\bibinfo {author} {\bibfnamefont {K.}~\bibnamefont
  {Seki}}\ and\ \bibinfo {author} {\bibfnamefont {S.}~\bibnamefont {Yunoki}},\
  }\href {\doibase 10.1103/PhysRevB.96.085124} {\bibfield  {journal} {\bibinfo
  {journal} {Phys. Rev. B}\ }\textbf {\bibinfo {volume} {96}},\ \bibinfo
  {pages} {085124} (\bibinfo {year} {2017})}\BibitemShut {NoStop}%
\bibitem [{\citenamefont {Dzyaloshinskii}(2003)}]{Dzyaloshinskii2003}%
  \BibitemOpen
  \bibfield  {author} {\bibinfo {author} {\bibfnamefont {I.}~\bibnamefont
  {Dzyaloshinskii}},\ }\href {\doibase 10.1103/PhysRevB.68.085113} {\bibfield
  {journal} {\bibinfo  {journal} {Phys. Rev. B}\ }\textbf {\bibinfo {volume}
  {68}},\ \bibinfo {pages} {085113} (\bibinfo {year} {2003})}\BibitemShut
  {NoStop}%
\bibitem [{\citenamefont {Seki}\ and\ \citenamefont
  {Yunoki}(2016)}]{Seki2016_tetra}%
  \BibitemOpen
  \bibfield  {author} {\bibinfo {author} {\bibfnamefont {K.}~\bibnamefont
  {Seki}}\ and\ \bibinfo {author} {\bibfnamefont {S.}~\bibnamefont {Yunoki}},\
  }\href {\doibase 10.1103/PhysRevB.93.245115} {\bibfield  {journal} {\bibinfo
  {journal} {Phys. Rev. B}\ }\textbf {\bibinfo {volume} {93}},\ \bibinfo
  {pages} {245115} (\bibinfo {year} {2016})}\BibitemShut {NoStop}%
\bibitem [{\citenamefont {Ahlfors}(1979)}]{Ahlfors}%
  \BibitemOpen
  \bibfield  {author} {\bibinfo {author} {\bibfnamefont {L.~V.}\ \bibnamefont
  {Ahlfors}},\ }\href@noop {} {\emph {\bibinfo {title} {Complex analysis}}},\
  \bibinfo {edition} {3rd}\ ed.\ (\bibinfo  {publisher} {McGraw-Hill},\
  \bibinfo {address} {New York},\ \bibinfo {year} {1979})\ Chap.\ \bibinfo
  {chapter} {1 and 2}\BibitemShut {NoStop}%
\bibitem [{\citenamefont {Wei\ss{}e}\ \emph {et~al.}(2006)\citenamefont
  {Wei\ss{}e}, \citenamefont {Wellein}, \citenamefont {Alvermann},\ and\
  \citenamefont {Fehske}}]{Weisse2006}%
  \BibitemOpen
  \bibfield  {author} {\bibinfo {author} {\bibfnamefont {A.}~\bibnamefont
  {Wei\ss{}e}}, \bibinfo {author} {\bibfnamefont {G.}~\bibnamefont {Wellein}},
  \bibinfo {author} {\bibfnamefont {A.}~\bibnamefont {Alvermann}}, \ and\
  \bibinfo {author} {\bibfnamefont {H.}~\bibnamefont {Fehske}},\ }\href
  {\doibase 10.1103/RevModPhys.78.275} {\bibfield  {journal} {\bibinfo
  {journal} {Rev. Mod. Phys.}\ }\textbf {\bibinfo {volume} {78}},\ \bibinfo
  {pages} {275} (\bibinfo {year} {2006})}\BibitemShut {NoStop}%
\bibitem [{\citenamefont {Eder}(2010)}]{Eder2010chi}%
  \BibitemOpen
  \bibfield  {author} {\bibinfo {author} {\bibfnamefont {R.}~\bibnamefont
  {Eder}},\ }\href {\doibase 10.1103/PhysRevB.81.035101} {\bibfield  {journal}
  {\bibinfo  {journal} {Phys. Rev. B}\ }\textbf {\bibinfo {volume} {81}},\
  \bibinfo {pages} {035101} (\bibinfo {year} {2010})}\BibitemShut {NoStop}%
\bibitem [{\citenamefont {Balzer}\ and\ \citenamefont
  {Potthoff}(2010)}]{Balzer2010}%
  \BibitemOpen
  \bibfield  {author} {\bibinfo {author} {\bibfnamefont {M.}~\bibnamefont
  {Balzer}}\ and\ \bibinfo {author} {\bibfnamefont {M.}~\bibnamefont
  {Potthoff}},\ }\href {\doibase 10.1103/PhysRevB.82.174441} {\bibfield
  {journal} {\bibinfo  {journal} {Phys. Rev. B}\ }\textbf {\bibinfo {volume}
  {82}},\ \bibinfo {pages} {174441} (\bibinfo {year} {2010})}\BibitemShut
  {NoStop}%
\bibitem [{\citenamefont {Aichhorn}\ \emph
  {et~al.}(2006{\natexlab{b}})\citenamefont {Aichhorn}, \citenamefont
  {Arrigoni}, \citenamefont {Potthoff},\ and\ \citenamefont
  {Hanke}}]{Aichhorn2006_muvar}%
  \BibitemOpen
  \bibfield  {author} {\bibinfo {author} {\bibfnamefont {M.}~\bibnamefont
  {Aichhorn}}, \bibinfo {author} {\bibfnamefont {E.}~\bibnamefont {Arrigoni}},
  \bibinfo {author} {\bibfnamefont {M.}~\bibnamefont {Potthoff}}, \ and\
  \bibinfo {author} {\bibfnamefont {W.}~\bibnamefont {Hanke}},\ }\href
  {\doibase 10.1103/PhysRevB.74.024508} {\bibfield  {journal} {\bibinfo
  {journal} {Phys. Rev. B}\ }\textbf {\bibinfo {volume} {74}},\ \bibinfo
  {pages} {024508} (\bibinfo {year} {2006}{\natexlab{b}})}\BibitemShut
  {NoStop}%
\bibitem [{\citenamefont {Sakai}\ \emph {et~al.}(2012)\citenamefont {Sakai},
  \citenamefont {Sangiovanni}, \citenamefont {Civelli}, \citenamefont {Motome},
  \citenamefont {Held},\ and\ \citenamefont {Imada}}]{Sakai2012}%
  \BibitemOpen
  \bibfield  {author} {\bibinfo {author} {\bibfnamefont {S.}~\bibnamefont
  {Sakai}}, \bibinfo {author} {\bibfnamefont {G.}~\bibnamefont {Sangiovanni}},
  \bibinfo {author} {\bibfnamefont {M.}~\bibnamefont {Civelli}}, \bibinfo
  {author} {\bibfnamefont {Y.}~\bibnamefont {Motome}}, \bibinfo {author}
  {\bibfnamefont {K.}~\bibnamefont {Held}}, \ and\ \bibinfo {author}
  {\bibfnamefont {M.}~\bibnamefont {Imada}},\ }\href {\doibase
  10.1103/PhysRevB.85.035102} {\bibfield  {journal} {\bibinfo  {journal} {Phys.
  Rev. B}\ }\textbf {\bibinfo {volume} {85}},\ \bibinfo {pages} {035102}
  (\bibinfo {year} {2012})}\BibitemShut {NoStop}%
\bibitem [{\citenamefont {Kawasugi}\ \emph {et~al.}(2016)\citenamefont
  {Kawasugi}, \citenamefont {Seki}, \citenamefont {Edagawa}, \citenamefont
  {Sato}, \citenamefont {Pu}, \citenamefont {Takenobu}, \citenamefont {Yunoki},
  \citenamefont {Yamamoto},\ and\ \citenamefont {Kato}}]{Kawasugi2016}%
  \BibitemOpen
  \bibfield  {author} {\bibinfo {author} {\bibfnamefont {Y.}~\bibnamefont
  {Kawasugi}}, \bibinfo {author} {\bibfnamefont {K.}~\bibnamefont {Seki}},
  \bibinfo {author} {\bibfnamefont {Y.}~\bibnamefont {Edagawa}}, \bibinfo
  {author} {\bibfnamefont {Y.}~\bibnamefont {Sato}}, \bibinfo {author}
  {\bibfnamefont {J.}~\bibnamefont {Pu}}, \bibinfo {author} {\bibfnamefont
  {T.}~\bibnamefont {Takenobu}}, \bibinfo {author} {\bibfnamefont
  {S.}~\bibnamefont {Yunoki}}, \bibinfo {author} {\bibfnamefont {H.~M.}\
  \bibnamefont {Yamamoto}}, \ and\ \bibinfo {author} {\bibfnamefont
  {R.}~\bibnamefont {Kato}},\ }\href {http://dx.doi.org/10.1038/ncomms12356}
  {\bibfield  {journal} {\bibinfo  {journal} {Nature Communications}\ }\textbf
  {\bibinfo {volume} {7}},\ \bibinfo {pages} {12356} (\bibinfo {year}
  {2016})}\BibitemShut {NoStop}%
\bibitem [{\citenamefont {Preuss}\ \emph {et~al.}(1995)\citenamefont {Preuss},
  \citenamefont {Hanke},\ and\ \citenamefont {von~der Linden}}]{Preuss1995}%
  \BibitemOpen
  \bibfield  {author} {\bibinfo {author} {\bibfnamefont {R.}~\bibnamefont
  {Preuss}}, \bibinfo {author} {\bibfnamefont {W.}~\bibnamefont {Hanke}}, \
  and\ \bibinfo {author} {\bibfnamefont {W.}~\bibnamefont {von~der Linden}},\
  }\href {\doibase 10.1103/PhysRevLett.75.1344} {\bibfield  {journal} {\bibinfo
   {journal} {Phys. Rev. Lett.}\ }\textbf {\bibinfo {volume} {75}},\ \bibinfo
  {pages} {1344} (\bibinfo {year} {1995})}\BibitemShut {NoStop}%
\bibitem [{\citenamefont {Preuss}\ \emph {et~al.}(1997)\citenamefont {Preuss},
  \citenamefont {Hanke}, \citenamefont {Gr\"ober},\ and\ \citenamefont
  {Evertz}}]{Preuss1997}%
  \BibitemOpen
  \bibfield  {author} {\bibinfo {author} {\bibfnamefont {R.}~\bibnamefont
  {Preuss}}, \bibinfo {author} {\bibfnamefont {W.}~\bibnamefont {Hanke}},
  \bibinfo {author} {\bibfnamefont {C.}~\bibnamefont {Gr\"ober}}, \ and\
  \bibinfo {author} {\bibfnamefont {H.~G.}\ \bibnamefont {Evertz}},\ }\href
  {\doibase 10.1103/PhysRevLett.79.1122} {\bibfield  {journal} {\bibinfo
  {journal} {Phys. Rev. Lett.}\ }\textbf {\bibinfo {volume} {79}},\ \bibinfo
  {pages} {1122} (\bibinfo {year} {1997})}\BibitemShut {NoStop}%
\bibitem [{\citenamefont {Gr\"ober}\ \emph {et~al.}(2000)\citenamefont
  {Gr\"ober}, \citenamefont {Eder},\ and\ \citenamefont {Hanke}}]{Groeber2000}%
  \BibitemOpen
  \bibfield  {author} {\bibinfo {author} {\bibfnamefont {C.}~\bibnamefont
  {Gr\"ober}}, \bibinfo {author} {\bibfnamefont {R.}~\bibnamefont {Eder}}, \
  and\ \bibinfo {author} {\bibfnamefont {W.}~\bibnamefont {Hanke}},\ }\href
  {\doibase 10.1103/PhysRevB.62.4336} {\bibfield  {journal} {\bibinfo
  {journal} {Phys. Rev. B}\ }\textbf {\bibinfo {volume} {62}},\ \bibinfo
  {pages} {4336} (\bibinfo {year} {2000})}\BibitemShut {NoStop}%
\bibitem [{\citenamefont {Hubbard}(1963)}]{Hubbard1963}%
  \BibitemOpen
  \bibfield  {author} {\bibinfo {author} {\bibfnamefont {J.}~\bibnamefont
  {Hubbard}},\ }\href
  {http://rspa.royalsocietypublishing.org/content/276/1365/238} {\bibfield
  {journal} {\bibinfo  {journal} {Proc. Roy. Soc. London A}\ }\textbf {\bibinfo
  {volume} {276}},\ \bibinfo {pages} {238} (\bibinfo {year}
  {1963})}\BibitemShut {NoStop}%
\bibitem [{\citenamefont {Gebhard}(1997)}]{Gebhard1997}%
  \BibitemOpen
  \bibfield  {author} {\bibinfo {author} {\bibfnamefont {F.}~\bibnamefont
  {Gebhard}},\ }\href@noop {} {\emph {\bibinfo {title} {{The Mott
  Metal-Insulator Transition}}}},\ Vol.\ \bibinfo {volume} {137}\ (\bibinfo
  {publisher} {Springer},\ \bibinfo {address} {Berlin},\ \bibinfo {year}
  {1997})\ Chap.~\bibinfo {chapter} {3}\BibitemShut {NoStop}%
\bibitem [{not()}]{note1}%
  \BibitemOpen
  \href@noop {} {}\bibinfo {note} {See Appendix E of Ref.~\cite{Senechal2008}
  for estimation of the computational cost in evaluating the grand-potential
  functional by the $Q$-matrix method at zero temperature}\BibitemShut
  {NoStop}%
\bibitem [{\citenamefont {Dallaire-Demers}\ and\ \citenamefont
  {Wilhelm}(2016)}]{Dallaire-Demers2016}%
  \BibitemOpen
  \bibfield  {author} {\bibinfo {author} {\bibfnamefont {P.-L.}\ \bibnamefont
  {Dallaire-Demers}}\ and\ \bibinfo {author} {\bibfnamefont {F.~K.}\
  \bibnamefont {Wilhelm}},\ }\href {\doibase 10.1103/PhysRevA.93.032303}
  {\bibfield  {journal} {\bibinfo  {journal} {Phys. Rev. A}\ }\textbf {\bibinfo
  {volume} {93}},\ \bibinfo {pages} {032303} (\bibinfo {year}
  {2016})}\BibitemShut {NoStop}%
\bibitem [{lap()}]{lapack}%
  \BibitemOpen
  \href@noop {} {}\bibinfo {note}
  {\href{http://www.netlib.org/lapack}{http://www.netlib.org/lapack}}\BibitemShut
  {NoStop}%
\bibitem [{\citenamefont {Freund}(2000)}]{Freund}%
  \BibitemOpen
  \bibfield  {author} {\bibinfo {author} {\bibfnamefont {R.}~\bibnamefont
  {Freund}},\ }\enquote {\bibinfo {title} {{Band Lanczos Method}},}\ in\ \href
  {http://www.netlib.org/utk/people/JackDongarra/etemplates} {\emph {\bibinfo
  {booktitle} {Templates for the solution of algebraic eigenvalue problems: A
  practical guide}}},\ \bibinfo {editor} {edited by\ \bibinfo {editor}
  {\bibfnamefont {Z.}~\bibnamefont {Bai}}, \bibinfo {editor} {\bibfnamefont
  {J.}~\bibnamefont {Demmel}}, \bibinfo {editor} {\bibfnamefont
  {J.}~\bibnamefont {Dongarra}}, \bibinfo {editor} {\bibfnamefont
  {A.}~\bibnamefont {Ruhe}}, \ and\ \bibinfo {editor} {\bibfnamefont
  {H.}~\bibnamefont {van~der Vorst}}}\ (\bibinfo  {publisher} {SIAM},\ \bibinfo
  {address} {Philadelphia},\ \bibinfo {year} {2000})\ Chap.\ \bibinfo {chapter}
  {4.6}\BibitemShut {NoStop}%
\bibitem [{Not()}]{Note_pointgroup}%
  \BibitemOpen
  \href@noop {} {}\bibinfo {note} {The index $i$ can be even a label for the
  symmetry-adapted molecular orbital index when the point-group symmetry of the
  cluster is adopted~\cite{Liebsch2012}. Indeed, we have utilized the
  point-group symmetry of the cluster which only has the one-dimensional
  irreducible representations, such as $C_{2}$, $C_{2v}$, or $C_{4}$, for
  solving the eigenvalue problem and calculating the single-particle Green's
  functions of the cluster, whenever it is available. In that case, the
  block-Lanczos procedure has to be applied for each irreducible
  representation.}\BibitemShut {Stop}%
\bibitem [{\citenamefont {Fulde}(1995)}]{Fulde1995}%
  \BibitemOpen
  \bibfield  {author} {\bibinfo {author} {\bibfnamefont {P.}~\bibnamefont
  {Fulde}},\ }\href@noop {} {\emph {\bibinfo {title} {{Electron Correlations in
  Molecules and Solids}}}}\ (\bibinfo  {publisher} {Springer},\ \bibinfo
  {address} {Berlin},\ \bibinfo {year} {1995})\ \bibinfo {note} {{Appendix
  M}}\BibitemShut {NoStop}%
\bibitem [{\citenamefont {Dagotto}(1994)}]{Dagotto1994}%
  \BibitemOpen
  \bibfield  {author} {\bibinfo {author} {\bibfnamefont {E.}~\bibnamefont
  {Dagotto}},\ }\href {\doibase 10.1103/RevModPhys.66.763} {\bibfield
  {journal} {\bibinfo  {journal} {Rev. Mod. Phys.}\ }\textbf {\bibinfo {volume}
  {66}},\ \bibinfo {pages} {763} (\bibinfo {year} {1994})}\BibitemShut
  {NoStop}%
\bibitem [{\citenamefont {Balzer}\ \emph {et~al.}(2012)\citenamefont {Balzer},
  \citenamefont {Gdaniec},\ and\ \citenamefont {Potthoff}}]{Balzer2011}%
  \BibitemOpen
  \bibfield  {author} {\bibinfo {author} {\bibfnamefont {M.}~\bibnamefont
  {Balzer}}, \bibinfo {author} {\bibfnamefont {N.}~\bibnamefont {Gdaniec}}, \
  and\ \bibinfo {author} {\bibfnamefont {M.}~\bibnamefont {Potthoff}},\ }\href
  {http://dx.doi.org/10.1088/0953-8984/24/3/035603} {\bibfield  {journal}
  {\bibinfo  {journal} {J. Phys.: Cond. Mat.}\ }\textbf {\bibinfo {volume}
  {24}},\ \bibinfo {pages} {035603} (\bibinfo {year} {2012})}\BibitemShut
  {NoStop}%
\bibitem [{\citenamefont {Jakli\v{c}}\ and\ \citenamefont
  {Prelov\v{s}ek}(2000)}]{Jaklic2000}%
  \BibitemOpen
  \bibfield  {author} {\bibinfo {author} {\bibfnamefont {J.}~\bibnamefont
  {Jakli\v{c}}}\ and\ \bibinfo {author} {\bibfnamefont {P.}~\bibnamefont
  {Prelov\v{s}ek}},\ }\href {http://dx.doi.org/10.1080/000187300243381}
  {\bibfield  {journal} {\bibinfo  {journal} {Adv. Phys.}\ }\textbf {\bibinfo
  {volume} {49}},\ \bibinfo {pages} {1} (\bibinfo {year} {2000})}\BibitemShut
  {NoStop}%
\bibitem [{\citenamefont {Jafari}(2008)}]{Jafari2008}%
  \BibitemOpen
  \bibfield  {author} {\bibinfo {author} {\bibfnamefont {S.~A.}\ \bibnamefont
  {Jafari}},\ }\href
  {http://ijpr.iut.ac.ir/browse.php?a_code=A-10-1-279&slc_lang=en&sid=1&sw=}
  {\bibfield  {journal} {\bibinfo  {journal} {Iranian J. Phys. Res.}\ }\textbf
  {\bibinfo {volume} {8}},\ \bibinfo {pages} {113} (\bibinfo {year}
  {2008})}\BibitemShut {NoStop}%
\bibitem [{\citenamefont {Chatelin}(1988)}]{Chatelin}%
  \BibitemOpen
  \bibfield  {author} {\bibinfo {author} {\bibfnamefont {F.}~\bibnamefont
  {Chatelin}},\ }\href@noop {} {\emph {\bibinfo {title} {{Valeurs propres de
  matrices}}}}\ (\bibinfo  {publisher} {Masson},\ \bibinfo {address} {Paris},\
  \bibinfo {year} {1988})\ \bibinfo {note} {{translation by M. Iri and Y. Iri
  (Springer, Tokyo, 2003)}}\BibitemShut {NoStop}%
\bibitem [{\citenamefont {Yamamoto}\ \emph {et~al.}(2014)\citenamefont
  {Yamamoto}, \citenamefont {Nakatsukasa}, \citenamefont {Yanagisawa},\ and\
  \citenamefont {Fukuya}}]{Yamamoto2014}%
  \BibitemOpen
  \bibfield  {author} {\bibinfo {author} {\bibfnamefont {Y.}~\bibnamefont
  {Yamamoto}}, \bibinfo {author} {\bibfnamefont {Y.}~\bibnamefont
  {Nakatsukasa}}, \bibinfo {author} {\bibfnamefont {Y.}~\bibnamefont
  {Yanagisawa}}, \ and\ \bibinfo {author} {\bibfnamefont {T.}~\bibnamefont
  {Fukuya}},\ }\href {http://www.keisu.t.u-tokyo.ac.jp/research/techrep/y2014}
  {\bibfield  {journal} {\bibinfo  {journal} {Mathematical Engineering
  Technical Reports}\ }\textbf {\bibinfo {volume} {36}} (\bibinfo {year}
  {2014})}\BibitemShut {NoStop}%
\bibitem [{\citenamefont {Fukuya}\ \emph {et~al.}(2014)\citenamefont {Fukuya},
  \citenamefont {Nakatsukasa}, \citenamefont {Yanagisawa},\ and\ \citenamefont
  {Yamamoto}}]{Fukuya2014}%
  \BibitemOpen
  \bibfield  {author} {\bibinfo {author} {\bibfnamefont {T.}~\bibnamefont
  {Fukuya}}, \bibinfo {author} {\bibfnamefont {Y.}~\bibnamefont {Nakatsukasa}},
  \bibinfo {author} {\bibfnamefont {Y.}~\bibnamefont {Yanagisawa}}, \ and\
  \bibinfo {author} {\bibfnamefont {Y.}~\bibnamefont {Yamamoto}},\ }\href
  {http://www.keisu.t.u-tokyo.ac.jp/research/techrep/y2014} {\bibfield
  {journal} {\bibinfo  {journal} {Mathematical Engineering Technical Reports}\
  }\textbf {\bibinfo {volume} {37}} (\bibinfo {year} {2014})}\BibitemShut
  {NoStop}%
\bibitem [{\citenamefont {Luttinger}(1961)}]{Luttinger1961}%
  \BibitemOpen
  \bibfield  {author} {\bibinfo {author} {\bibfnamefont {J.~M.}\ \bibnamefont
  {Luttinger}},\ }\href {\doibase 10.1103/PhysRev.121.942} {\bibfield
  {journal} {\bibinfo  {journal} {Phys. Rev.}\ }\textbf {\bibinfo {volume}
  {121}},\ \bibinfo {pages} {942} (\bibinfo {year} {1961})}\BibitemShut
  {NoStop}%
\bibitem [{\citenamefont {Harris}\ and\ \citenamefont
  {Lange}(1967)}]{Harris1967}%
  \BibitemOpen
  \bibfield  {author} {\bibinfo {author} {\bibfnamefont {A.~B.}\ \bibnamefont
  {Harris}}\ and\ \bibinfo {author} {\bibfnamefont {R.~V.}\ \bibnamefont
  {Lange}},\ }\href {\doibase 10.1103/PhysRev.157.295} {\bibfield  {journal}
  {\bibinfo  {journal} {Phys. Rev.}\ }\textbf {\bibinfo {volume} {157}},\
  \bibinfo {pages} {295} (\bibinfo {year} {1967})}\BibitemShut {NoStop}%
\bibitem [{\citenamefont {Seki}\ \emph {et~al.}(2011)\citenamefont {Seki},
  \citenamefont {Eder},\ and\ \citenamefont {Ohta}}]{Seki2011}%
  \BibitemOpen
  \bibfield  {author} {\bibinfo {author} {\bibfnamefont {K.}~\bibnamefont
  {Seki}}, \bibinfo {author} {\bibfnamefont {R.}~\bibnamefont {Eder}}, \ and\
  \bibinfo {author} {\bibfnamefont {Y.}~\bibnamefont {Ohta}},\ }\href {\doibase
  10.1103/PhysRevB.84.245106} {\bibfield  {journal} {\bibinfo  {journal} {Phys.
  Rev. B}\ }\textbf {\bibinfo {volume} {84}},\ \bibinfo {pages} {245106}
  (\bibinfo {year} {2011})}\BibitemShut {NoStop}%
\bibitem [{\citenamefont {Shirakawa}\ and\ \citenamefont
  {Yunoki}(2014)}]{Shirakawa2014}%
  \BibitemOpen
  \bibfield  {author} {\bibinfo {author} {\bibfnamefont {T.}~\bibnamefont
  {Shirakawa}}\ and\ \bibinfo {author} {\bibfnamefont {S.}~\bibnamefont
  {Yunoki}},\ }\href {\doibase 10.1103/PhysRevB.90.195109} {\bibfield
  {journal} {\bibinfo  {journal} {Phys. Rev. B}\ }\textbf {\bibinfo {volume}
  {90}},\ \bibinfo {pages} {195109} (\bibinfo {year} {2014})}\BibitemShut
  {NoStop}%
\bibitem [{\citenamefont {Mermin}\ and\ \citenamefont
  {Wagner}(1966)}]{Mermin1966}%
  \BibitemOpen
  \bibfield  {author} {\bibinfo {author} {\bibfnamefont {N.~D.}\ \bibnamefont
  {Mermin}}\ and\ \bibinfo {author} {\bibfnamefont {H.}~\bibnamefont
  {Wagner}},\ }\href {\doibase 10.1103/PhysRevLett.17.1133} {\bibfield
  {journal} {\bibinfo  {journal} {Phys. Rev. Lett.}\ }\textbf {\bibinfo
  {volume} {17}},\ \bibinfo {pages} {1133} (\bibinfo {year}
  {1966})}\BibitemShut {NoStop}%
\bibitem [{\citenamefont {Maier}\ \emph
  {et~al.}(2005{\natexlab{b}})\citenamefont {Maier}, \citenamefont {Jarrell},
  \citenamefont {Schulthess}, \citenamefont {Kent},\ and\ \citenamefont
  {White}}]{Maier2005}%
  \BibitemOpen
  \bibfield  {author} {\bibinfo {author} {\bibfnamefont {T.~A.}\ \bibnamefont
  {Maier}}, \bibinfo {author} {\bibfnamefont {M.}~\bibnamefont {Jarrell}},
  \bibinfo {author} {\bibfnamefont {T.~C.}\ \bibnamefont {Schulthess}},
  \bibinfo {author} {\bibfnamefont {P.~R.~C.}\ \bibnamefont {Kent}}, \ and\
  \bibinfo {author} {\bibfnamefont {J.~B.}\ \bibnamefont {White}},\ }\href
  {\doibase 10.1103/PhysRevLett.95.237001} {\bibfield  {journal} {\bibinfo
  {journal} {Phys. Rev. Lett.}\ }\textbf {\bibinfo {volume} {95}},\ \bibinfo
  {pages} {237001} (\bibinfo {year} {2005}{\natexlab{b}})}\BibitemShut
  {NoStop}%
\bibitem [{\citenamefont {Sato}\ and\ \citenamefont
  {Tsunetsugu}(2016)}]{Sato2016}%
  \BibitemOpen
  \bibfield  {author} {\bibinfo {author} {\bibfnamefont {T.}~\bibnamefont
  {Sato}}\ and\ \bibinfo {author} {\bibfnamefont {H.}~\bibnamefont
  {Tsunetsugu}},\ }\href {\doibase 10.1103/PhysRevB.94.085110} {\bibfield
  {journal} {\bibinfo  {journal} {Phys. Rev. B}\ }\textbf {\bibinfo {volume}
  {94}},\ \bibinfo {pages} {085110} (\bibinfo {year} {2016})}\BibitemShut
  {NoStop}%
\bibitem [{\citenamefont {Rozenberg}\ \emph {et~al.}(1994)\citenamefont
  {Rozenberg}, \citenamefont {Kotliar},\ and\ \citenamefont
  {Zhang}}]{Rosenberg1994}%
  \BibitemOpen
  \bibfield  {author} {\bibinfo {author} {\bibfnamefont {M.~J.}\ \bibnamefont
  {Rozenberg}}, \bibinfo {author} {\bibfnamefont {G.}~\bibnamefont {Kotliar}},
  \ and\ \bibinfo {author} {\bibfnamefont {X.~Y.}\ \bibnamefont {Zhang}},\
  }\href {\doibase 10.1103/PhysRevB.49.10181} {\bibfield  {journal} {\bibinfo
  {journal} {Phys. Rev. B}\ }\textbf {\bibinfo {volume} {49}},\ \bibinfo
  {pages} {10181} (\bibinfo {year} {1994})}\BibitemShut {NoStop}%
\bibitem [{\citenamefont {Ortloff}\ \emph {et~al.}(2007)\citenamefont
  {Ortloff}, \citenamefont {Balzer},\ and\ \citenamefont
  {Potthoff}}]{Ortloff2007}%
  \BibitemOpen
  \bibfield  {author} {\bibinfo {author} {\bibfnamefont {J.}~\bibnamefont
  {Ortloff}}, \bibinfo {author} {\bibfnamefont {M.}~\bibnamefont {Balzer}}, \
  and\ \bibinfo {author} {\bibfnamefont {M.}~\bibnamefont {Potthoff}},\ }\href
  {http://link.springer.com/article/10.1140\%2Fepjb\%2Fe2007-00203-7}
  {\bibfield  {journal} {\bibinfo  {journal} {Eur. Phys. J. B}\ }\textbf
  {\bibinfo {volume} {58}},\ \bibinfo {pages} {37} (\bibinfo {year}
  {2007})}\BibitemShut {NoStop}%
\bibitem [{\citenamefont {Eder}\ \emph {et~al.}(2011)\citenamefont {Eder},
  \citenamefont {Seki},\ and\ \citenamefont {Ohta}}]{Eder2011}%
  \BibitemOpen
  \bibfield  {author} {\bibinfo {author} {\bibfnamefont {R.}~\bibnamefont
  {Eder}}, \bibinfo {author} {\bibfnamefont {K.}~\bibnamefont {Seki}}, \ and\
  \bibinfo {author} {\bibfnamefont {Y.}~\bibnamefont {Ohta}},\ }\href {\doibase
  10.1103/PhysRevB.83.205137} {\bibfield  {journal} {\bibinfo  {journal} {Phys.
  Rev. B}\ }\textbf {\bibinfo {volume} {83}},\ \bibinfo {pages} {205137}
  (\bibinfo {year} {2011})}\BibitemShut {NoStop}%
\bibitem [{\citenamefont {Sakai}\ \emph
  {et~al.}(2016{\natexlab{a}})\citenamefont {Sakai}, \citenamefont {Civelli},\
  and\ \citenamefont {Imada}}]{Sakai2014}%
  \BibitemOpen
  \bibfield  {author} {\bibinfo {author} {\bibfnamefont {S.}~\bibnamefont
  {Sakai}}, \bibinfo {author} {\bibfnamefont {M.}~\bibnamefont {Civelli}}, \
  and\ \bibinfo {author} {\bibfnamefont {M.}~\bibnamefont {Imada}},\ }\href
  {\doibase 10.1103/PhysRevLett.116.057003} {\bibfield  {journal} {\bibinfo
  {journal} {Phys. Rev. Lett.}\ }\textbf {\bibinfo {volume} {116}},\ \bibinfo
  {pages} {057003} (\bibinfo {year} {2016}{\natexlab{a}})}\BibitemShut
  {NoStop}%
\bibitem [{\citenamefont {Sakai}\ \emph
  {et~al.}(2016{\natexlab{b}})\citenamefont {Sakai}, \citenamefont {Civelli},\
  and\ \citenamefont {Imada}}]{Sakai2016}%
  \BibitemOpen
  \bibfield  {author} {\bibinfo {author} {\bibfnamefont {S.}~\bibnamefont
  {Sakai}}, \bibinfo {author} {\bibfnamefont {M.}~\bibnamefont {Civelli}}, \
  and\ \bibinfo {author} {\bibfnamefont {M.}~\bibnamefont {Imada}},\ }\href
  {\doibase 10.1103/PhysRevB.94.115130} {\bibfield  {journal} {\bibinfo
  {journal} {Phys. Rev. B}\ }\textbf {\bibinfo {volume} {94}},\ \bibinfo
  {pages} {115130} (\bibinfo {year} {2016}{\natexlab{b}})}\BibitemShut
  {NoStop}%
\bibitem [{\citenamefont {Fulde}(2012)}]{Fulde2012}%
  \BibitemOpen
  \bibfield  {author} {\bibinfo {author} {\bibfnamefont {P.}~\bibnamefont
  {Fulde}},\ }\href@noop {} {\emph {\bibinfo {title} {{Correlated Electrons in
  Quantum Matter}}}}\ (\bibinfo  {publisher} {World Scientific},\ \bibinfo
  {address} {Singapore},\ \bibinfo {year} {2012})\ Chap.~\bibinfo {chapter}
  {10}\BibitemShut {NoStop}%
\bibitem [{\citenamefont {Imada}\ \emph {et~al.}()\citenamefont {Imada},
  \citenamefont {Yamaji}, \citenamefont {Sakai},\ and\ \citenamefont
  {Motome}}]{Imada2011}%
  \BibitemOpen
  \bibfield  {author} {\bibinfo {author} {\bibfnamefont {M.}~\bibnamefont
  {Imada}}, \bibinfo {author} {\bibfnamefont {Y.}~\bibnamefont {Yamaji}},
  \bibinfo {author} {\bibfnamefont {S.}~\bibnamefont {Sakai}}, \ and\ \bibinfo
  {author} {\bibfnamefont {Y.}~\bibnamefont {Motome}},\ }\href {\doibase
  10.1002/andp.201100028} {\bibfield  {journal} {\bibinfo  {journal} {Annalen
  der Physik}\ }\textbf {\bibinfo {volume} {523}},\ \bibinfo {pages}
  {629}}\BibitemShut {NoStop}%
\bibitem [{\citenamefont {Eder}\ and\ \citenamefont {Becker}(1990)}]{Eder1990}%
  \BibitemOpen
  \bibfield  {author} {\bibinfo {author} {\bibfnamefont {R.}~\bibnamefont
  {Eder}}\ and\ \bibinfo {author} {\bibfnamefont {K.~W.}\ \bibnamefont
  {Becker}},\ }\href {https://doi.org/10.1007/BF01307839} {\bibfield  {journal}
  {\bibinfo  {journal} {Z. Phys. B}\ }\textbf {\bibinfo {volume} {78}},\
  \bibinfo {pages} {219} (\bibinfo {year} {1990})}\BibitemShut {NoStop}%
\bibitem [{\citenamefont {Trugman}(1988)}]{Trugman1988}%
  \BibitemOpen
  \bibfield  {author} {\bibinfo {author} {\bibfnamefont {S.~A.}\ \bibnamefont
  {Trugman}},\ }\href {\doibase 10.1103/PhysRevB.37.1597} {\bibfield  {journal}
  {\bibinfo  {journal} {Phys. Rev. B}\ }\textbf {\bibinfo {volume} {37}},\
  \bibinfo {pages} {1597} (\bibinfo {year} {1988})}\BibitemShut {NoStop}%
\bibitem [{\citenamefont {Eder}\ \emph {et~al.}(2010)\citenamefont {Eder},
  \citenamefont {Wr\'obel},\ and\ \citenamefont {Ohta}}]{Eder2010}%
  \BibitemOpen
  \bibfield  {author} {\bibinfo {author} {\bibfnamefont {R.}~\bibnamefont
  {Eder}}, \bibinfo {author} {\bibfnamefont {P.}~\bibnamefont {Wr\'obel}}, \
  and\ \bibinfo {author} {\bibfnamefont {Y.}~\bibnamefont {Ohta}},\ }\href
  {\doibase 10.1103/PhysRevB.82.155109} {\bibfield  {journal} {\bibinfo
  {journal} {Phys. Rev. B}\ }\textbf {\bibinfo {volume} {82}},\ \bibinfo
  {pages} {155109} (\bibinfo {year} {2010})}\BibitemShut {NoStop}%
\bibitem [{\citenamefont {Eder}\ and\ \citenamefont {Ohta}(1994)}]{Eder1994}%
  \BibitemOpen
  \bibfield  {author} {\bibinfo {author} {\bibfnamefont {R.}~\bibnamefont
  {Eder}}\ and\ \bibinfo {author} {\bibfnamefont {Y.}~\bibnamefont {Ohta}},\
  }\href {\doibase 10.1103/PhysRevB.50.10043} {\bibfield  {journal} {\bibinfo
  {journal} {Phys. Rev. B}\ }\textbf {\bibinfo {volume} {50}},\ \bibinfo
  {pages} {10043} (\bibinfo {year} {1994})}\BibitemShut {NoStop}%
\bibitem [{\citenamefont {Schrieffer}\ \emph {et~al.}(1989)\citenamefont
  {Schrieffer}, \citenamefont {Wen},\ and\ \citenamefont
  {Zhang}}]{Schrieffer1989}%
  \BibitemOpen
  \bibfield  {author} {\bibinfo {author} {\bibfnamefont {J.~R.}\ \bibnamefont
  {Schrieffer}}, \bibinfo {author} {\bibfnamefont {X.~G.}\ \bibnamefont {Wen}},
  \ and\ \bibinfo {author} {\bibfnamefont {S.~C.}\ \bibnamefont {Zhang}},\
  }\href {\doibase 10.1103/PhysRevB.39.11663} {\bibfield  {journal} {\bibinfo
  {journal} {Phys. Rev. B}\ }\textbf {\bibinfo {volume} {39}},\ \bibinfo
  {pages} {11663} (\bibinfo {year} {1989})}\BibitemShut {NoStop}%
\bibitem [{\citenamefont {Koch}\ \emph {et~al.}(2008)\citenamefont {Koch},
  \citenamefont {Sangiovanni},\ and\ \citenamefont {Gunnarsson}}]{Koch2008}%
  \BibitemOpen
  \bibfield  {author} {\bibinfo {author} {\bibfnamefont {E.}~\bibnamefont
  {Koch}}, \bibinfo {author} {\bibfnamefont {G.}~\bibnamefont {Sangiovanni}}, \
  and\ \bibinfo {author} {\bibfnamefont {O.}~\bibnamefont {Gunnarsson}},\
  }\href {\doibase 10.1103/PhysRevB.78.115102} {\bibfield  {journal} {\bibinfo
  {journal} {Phys. Rev. B}\ }\textbf {\bibinfo {volume} {78}},\ \bibinfo
  {pages} {115102} (\bibinfo {year} {2008})}\BibitemShut {NoStop}%
\bibitem [{\citenamefont {Kusunose}(2006)}]{Kusunose2006}%
  \BibitemOpen
  \bibfield  {author} {\bibinfo {author} {\bibfnamefont {H.}~\bibnamefont
  {Kusunose}},\ }\href {http://dx.doi.org/10.1143/JPSJ.75.054713} {\bibfield
  {journal} {\bibinfo  {journal} {J. Phys. Soc. Jpn.}\ }\textbf {\bibinfo
  {volume} {75}},\ \bibinfo {pages} {054713} (\bibinfo {year}
  {2006})}\BibitemShut {NoStop}%
\bibitem [{\citenamefont {Nozi\`{e}res}(1998)}]{Nozieres1998}%
  \BibitemOpen
  \bibfield  {author} {\bibinfo {author} {\bibfnamefont {P.}~\bibnamefont
  {Nozi\`{e}res}},\ }\href {http://dx.doi.org/10.1007/s100510050571} {\bibfield
   {journal} {\bibinfo  {journal} {Eur. Phys. J. B}\ }\textbf {\bibinfo
  {volume} {6}},\ \bibinfo {pages} {447} (\bibinfo {year} {1998})}\BibitemShut
  {NoStop}%
\bibitem [{\citenamefont {Kotliar}(1999)}]{Kotliar1999}%
  \BibitemOpen
  \bibfield  {author} {\bibinfo {author} {\bibfnamefont {G.}~\bibnamefont
  {Kotliar}},\ }\href {http://dx.doi.org/10.1007/s100510050914} {\bibfield
  {journal} {\bibinfo  {journal} {Eur. Phys. J. B}\ }\textbf {\bibinfo {volume}
  {11}},\ \bibinfo {pages} {27} (\bibinfo {year} {1999})}\BibitemShut {NoStop}%
\bibitem [{\citenamefont {Rost}\ \emph {et~al.}(2012)\citenamefont {Rost},
  \citenamefont {Gorelik}, \citenamefont {Assaad},\ and\ \citenamefont
  {Bl\"umer}}]{Rost2012}%
  \BibitemOpen
  \bibfield  {author} {\bibinfo {author} {\bibfnamefont {D.}~\bibnamefont
  {Rost}}, \bibinfo {author} {\bibfnamefont {E.~V.}\ \bibnamefont {Gorelik}},
  \bibinfo {author} {\bibfnamefont {F.}~\bibnamefont {Assaad}}, \ and\ \bibinfo
  {author} {\bibfnamefont {N.}~\bibnamefont {Bl\"umer}},\ }\href {\doibase
  10.1103/PhysRevB.86.155109} {\bibfield  {journal} {\bibinfo  {journal} {Phys.
  Rev. B}\ }\textbf {\bibinfo {volume} {86}},\ \bibinfo {pages} {155109}
  (\bibinfo {year} {2012})}\BibitemShut {NoStop}%
\bibitem [{\citenamefont {Tocchio}\ \emph {et~al.}(2016)\citenamefont
  {Tocchio}, \citenamefont {Becca},\ and\ \citenamefont
  {Sorella}}]{Tocchio2016}%
  \BibitemOpen
  \bibfield  {author} {\bibinfo {author} {\bibfnamefont {L.~F.}\ \bibnamefont
  {Tocchio}}, \bibinfo {author} {\bibfnamefont {F.}~\bibnamefont {Becca}}, \
  and\ \bibinfo {author} {\bibfnamefont {S.}~\bibnamefont {Sorella}},\ }\href
  {\doibase 10.1103/PhysRevB.94.195126} {\bibfield  {journal} {\bibinfo
  {journal} {Phys. Rev. B}\ }\textbf {\bibinfo {volume} {94}},\ \bibinfo
  {pages} {195126} (\bibinfo {year} {2016})}\BibitemShut {NoStop}%
\bibitem [{\citenamefont {Fratino}\ \emph {et~al.}(2017)\citenamefont
  {Fratino}, \citenamefont {S\'emon}, \citenamefont {Charlebois}, \citenamefont
  {Sordi},\ and\ \citenamefont {Tremblay}}]{Fratino2017}%
  \BibitemOpen
  \bibfield  {author} {\bibinfo {author} {\bibfnamefont {L.}~\bibnamefont
  {Fratino}}, \bibinfo {author} {\bibfnamefont {P.}~\bibnamefont {S\'emon}},
  \bibinfo {author} {\bibfnamefont {M.}~\bibnamefont {Charlebois}}, \bibinfo
  {author} {\bibfnamefont {G.}~\bibnamefont {Sordi}}, \ and\ \bibinfo {author}
  {\bibfnamefont {A.-M.~S.}\ \bibnamefont {Tremblay}},\ }\href {\doibase
  10.1103/PhysRevB.95.235109} {\bibfield  {journal} {\bibinfo  {journal} {Phys.
  Rev. B}\ }\textbf {\bibinfo {volume} {95}},\ \bibinfo {pages} {235109}
  (\bibinfo {year} {2017})}\BibitemShut {NoStop}%
\bibitem [{\citenamefont {Karolak}\ \emph {et~al.}(2015)\citenamefont
  {Karolak}, \citenamefont {Edelmann},\ and\ \citenamefont
  {Sangiovanni}}]{Karolak2015}%
  \BibitemOpen
  \bibfield  {author} {\bibinfo {author} {\bibfnamefont {M.}~\bibnamefont
  {Karolak}}, \bibinfo {author} {\bibfnamefont {M.}~\bibnamefont {Edelmann}}, \
  and\ \bibinfo {author} {\bibfnamefont {G.}~\bibnamefont {Sangiovanni}},\
  }\href {\doibase 10.1103/PhysRevB.91.075108} {\bibfield  {journal} {\bibinfo
  {journal} {Phys. Rev. B}\ }\textbf {\bibinfo {volume} {91}},\ \bibinfo
  {pages} {075108} (\bibinfo {year} {2015})}\BibitemShut {NoStop}%
\bibitem [{\citenamefont {Cocchi}\ \emph {et~al.}(2016)\citenamefont {Cocchi},
  \citenamefont {Miller}, \citenamefont {Drewes}, \citenamefont {Koschorreck},
  \citenamefont {Pertot}, \citenamefont {Brennecke},\ and\ \citenamefont
  {K\"ohl}}]{Cocchi2016}%
  \BibitemOpen
  \bibfield  {author} {\bibinfo {author} {\bibfnamefont {E.}~\bibnamefont
  {Cocchi}}, \bibinfo {author} {\bibfnamefont {L.~A.}\ \bibnamefont {Miller}},
  \bibinfo {author} {\bibfnamefont {J.~H.}\ \bibnamefont {Drewes}}, \bibinfo
  {author} {\bibfnamefont {M.}~\bibnamefont {Koschorreck}}, \bibinfo {author}
  {\bibfnamefont {D.}~\bibnamefont {Pertot}}, \bibinfo {author} {\bibfnamefont
  {F.}~\bibnamefont {Brennecke}}, \ and\ \bibinfo {author} {\bibfnamefont
  {M.}~\bibnamefont {K\"ohl}},\ }\href {\doibase
  10.1103/PhysRevLett.116.175301} {\bibfield  {journal} {\bibinfo  {journal}
  {Phys. Rev. Lett.}\ }\textbf {\bibinfo {volume} {116}},\ \bibinfo {pages}
  {175301} (\bibinfo {year} {2016})}\BibitemShut {NoStop}%
\bibitem [{\citenamefont {Cocchi}\ \emph {et~al.}(2017)\citenamefont {Cocchi},
  \citenamefont {Miller}, \citenamefont {Drewes}, \citenamefont {Chan},
  \citenamefont {Pertot}, \citenamefont {Brennecke},\ and\ \citenamefont
  {K\"ohl}}]{Cocchi2017}%
  \BibitemOpen
  \bibfield  {author} {\bibinfo {author} {\bibfnamefont {E.}~\bibnamefont
  {Cocchi}}, \bibinfo {author} {\bibfnamefont {L.~A.}\ \bibnamefont {Miller}},
  \bibinfo {author} {\bibfnamefont {J.~H.}\ \bibnamefont {Drewes}}, \bibinfo
  {author} {\bibfnamefont {C.~F.}\ \bibnamefont {Chan}}, \bibinfo {author}
  {\bibfnamefont {D.}~\bibnamefont {Pertot}}, \bibinfo {author} {\bibfnamefont
  {F.}~\bibnamefont {Brennecke}}, \ and\ \bibinfo {author} {\bibfnamefont
  {M.}~\bibnamefont {K\"ohl}},\ }\href {\doibase 10.1103/PhysRevX.7.031025}
  {\bibfield  {journal} {\bibinfo  {journal} {Phys. Rev. X}\ }\textbf {\bibinfo
  {volume} {7}},\ \bibinfo {pages} {031025} (\bibinfo {year}
  {2017})}\BibitemShut {NoStop}%
\bibitem [{\citenamefont {Georges}\ and\ \citenamefont
  {Krauth}(1992)}]{Georges1992}%
  \BibitemOpen
  \bibfield  {author} {\bibinfo {author} {\bibfnamefont {A.}~\bibnamefont
  {Georges}}\ and\ \bibinfo {author} {\bibfnamefont {W.}~\bibnamefont
  {Krauth}},\ }\href {\doibase 10.1103/PhysRevLett.69.1240} {\bibfield
  {journal} {\bibinfo  {journal} {Phys. Rev. Lett.}\ }\textbf {\bibinfo
  {volume} {69}},\ \bibinfo {pages} {1240} (\bibinfo {year}
  {1992})}\BibitemShut {NoStop}%
\bibitem [{\citenamefont {Werner}\ \emph {et~al.}(2005)\citenamefont {Werner},
  \citenamefont {Parcollet}, \citenamefont {Georges},\ and\ \citenamefont
  {Hassan}}]{Werner2005}%
  \BibitemOpen
  \bibfield  {author} {\bibinfo {author} {\bibfnamefont {F.}~\bibnamefont
  {Werner}}, \bibinfo {author} {\bibfnamefont {O.}~\bibnamefont {Parcollet}},
  \bibinfo {author} {\bibfnamefont {A.}~\bibnamefont {Georges}}, \ and\
  \bibinfo {author} {\bibfnamefont {S.~R.}\ \bibnamefont {Hassan}},\ }\href
  {\doibase 10.1103/PhysRevLett.95.056401} {\bibfield  {journal} {\bibinfo
  {journal} {Phys. Rev. Lett.}\ }\textbf {\bibinfo {volume} {95}},\ \bibinfo
  {pages} {056401} (\bibinfo {year} {2005})}\BibitemShut {NoStop}%
\bibitem [{\citenamefont {Paiva}\ \emph {et~al.}(2010)\citenamefont {Paiva},
  \citenamefont {Scalettar}, \citenamefont {Randeria},\ and\ \citenamefont
  {Trivedi}}]{Paiva2010}%
  \BibitemOpen
  \bibfield  {author} {\bibinfo {author} {\bibfnamefont {T.}~\bibnamefont
  {Paiva}}, \bibinfo {author} {\bibfnamefont {R.}~\bibnamefont {Scalettar}},
  \bibinfo {author} {\bibfnamefont {M.}~\bibnamefont {Randeria}}, \ and\
  \bibinfo {author} {\bibfnamefont {N.}~\bibnamefont {Trivedi}},\ }\href
  {\doibase 10.1103/PhysRevLett.104.066406} {\bibfield  {journal} {\bibinfo
  {journal} {Phys. Rev. Lett.}\ }\textbf {\bibinfo {volume} {104}},\ \bibinfo
  {pages} {066406} (\bibinfo {year} {2010})}\BibitemShut {NoStop}%
\bibitem [{\citenamefont {Fuchs}\ \emph {et~al.}(2011)\citenamefont {Fuchs},
  \citenamefont {Gull}, \citenamefont {Pollet}, \citenamefont {Burovski},
  \citenamefont {Kozik}, \citenamefont {Pruschke},\ and\ \citenamefont
  {Troyer}}]{Fuchs2011}%
  \BibitemOpen
  \bibfield  {author} {\bibinfo {author} {\bibfnamefont {S.}~\bibnamefont
  {Fuchs}}, \bibinfo {author} {\bibfnamefont {E.}~\bibnamefont {Gull}},
  \bibinfo {author} {\bibfnamefont {L.}~\bibnamefont {Pollet}}, \bibinfo
  {author} {\bibfnamefont {E.}~\bibnamefont {Burovski}}, \bibinfo {author}
  {\bibfnamefont {E.}~\bibnamefont {Kozik}}, \bibinfo {author} {\bibfnamefont
  {T.}~\bibnamefont {Pruschke}}, \ and\ \bibinfo {author} {\bibfnamefont
  {M.}~\bibnamefont {Troyer}},\ }\href {\doibase
  10.1103/PhysRevLett.106.030401} {\bibfield  {journal} {\bibinfo  {journal}
  {Phys. Rev. Lett.}\ }\textbf {\bibinfo {volume} {106}},\ \bibinfo {pages}
  {030401} (\bibinfo {year} {2011})}\BibitemShut {NoStop}%
\bibitem [{\citenamefont {Khatami}\ and\ \citenamefont
  {Rigol}(2011)}]{Khatami2011}%
  \BibitemOpen
  \bibfield  {author} {\bibinfo {author} {\bibfnamefont {E.}~\bibnamefont
  {Khatami}}\ and\ \bibinfo {author} {\bibfnamefont {M.}~\bibnamefont
  {Rigol}},\ }\href {\doibase 10.1103/PhysRevA.84.053611} {\bibfield  {journal}
  {\bibinfo  {journal} {Phys. Rev. A}\ }\textbf {\bibinfo {volume} {84}},\
  \bibinfo {pages} {053611} (\bibinfo {year} {2011})}\BibitemShut {NoStop}%
\bibitem [{\citenamefont {LeBlanc}\ and\ \citenamefont
  {Gull}(2013)}]{LeBlanc2013}%
  \BibitemOpen
  \bibfield  {author} {\bibinfo {author} {\bibfnamefont {J.~P.~F.}\
  \bibnamefont {LeBlanc}}\ and\ \bibinfo {author} {\bibfnamefont
  {E.}~\bibnamefont {Gull}},\ }\href {\doibase 10.1103/PhysRevB.88.155108}
  {\bibfield  {journal} {\bibinfo  {journal} {Phys. Rev. B}\ }\textbf {\bibinfo
  {volume} {88}},\ \bibinfo {pages} {155108} (\bibinfo {year}
  {2013})}\BibitemShut {NoStop}%
\bibitem [{\citenamefont {Laubach}\ \emph {et~al.}(2015)\citenamefont
  {Laubach}, \citenamefont {Thomale}, \citenamefont {Platt}, \citenamefont
  {Hanke},\ and\ \citenamefont {Li}}]{Laubach2015}%
  \BibitemOpen
  \bibfield  {author} {\bibinfo {author} {\bibfnamefont {M.}~\bibnamefont
  {Laubach}}, \bibinfo {author} {\bibfnamefont {R.}~\bibnamefont {Thomale}},
  \bibinfo {author} {\bibfnamefont {C.}~\bibnamefont {Platt}}, \bibinfo
  {author} {\bibfnamefont {W.}~\bibnamefont {Hanke}}, \ and\ \bibinfo {author}
  {\bibfnamefont {G.}~\bibnamefont {Li}},\ }\href {\doibase
  10.1103/PhysRevB.91.245125} {\bibfield  {journal} {\bibinfo  {journal} {Phys.
  Rev. B}\ }\textbf {\bibinfo {volume} {91}},\ \bibinfo {pages} {245125}
  (\bibinfo {year} {2015})}\BibitemShut {NoStop}%
\bibitem [{\citenamefont {Misawa}\ and\ \citenamefont
  {Yamaji}(2018)}]{Misawa2016}%
  \BibitemOpen
  \bibfield  {author} {\bibinfo {author} {\bibfnamefont {T.}~\bibnamefont
  {Misawa}}\ and\ \bibinfo {author} {\bibfnamefont {Y.}~\bibnamefont
  {Yamaji}},\ }\href {https://doi.org/10.7566/JPSJ.87.023707} {\bibfield
  {journal} {\bibinfo  {journal} {J. Phys. Soc. Jpn.}\ }\textbf {\bibinfo
  {volume} {87}},\ \bibinfo {pages} {023707} (\bibinfo {year}
  {2018})}\BibitemShut {NoStop}%
\bibitem [{\citenamefont {Takai}\ \emph {et~al.}(2016)\citenamefont {Takai},
  \citenamefont {Ido}, \citenamefont {Misawa}, \citenamefont {Yamaji},\ and\
  \citenamefont {Imada}}]{Takai2016}%
  \BibitemOpen
  \bibfield  {author} {\bibinfo {author} {\bibfnamefont {K.}~\bibnamefont
  {Takai}}, \bibinfo {author} {\bibfnamefont {K.}~\bibnamefont {Ido}}, \bibinfo
  {author} {\bibfnamefont {T.}~\bibnamefont {Misawa}}, \bibinfo {author}
  {\bibfnamefont {Y.}~\bibnamefont {Yamaji}}, \ and\ \bibinfo {author}
  {\bibfnamefont {M.}~\bibnamefont {Imada}},\ }\href
  {http://dx.doi.org/10.7566/JPSJ.85.034601} {\bibfield  {journal} {\bibinfo
  {journal} {J. Phys. Soc. Jpn.}\ }\textbf {\bibinfo {volume} {85}},\ \bibinfo
  {pages} {034601} (\bibinfo {year} {2016})}\BibitemShut {NoStop}%
\bibitem [{\citenamefont {Moriya}(2006)}]{Moriya2006}%
  \BibitemOpen
  \bibfield  {author} {\bibinfo {author} {\bibfnamefont {T.}~\bibnamefont
  {Moriya}},\ }\href {https://www.ncbi.nlm.nih.gov/pmc/articles/PMC4322922/}
  {\bibfield  {journal} {\bibinfo  {journal} {Proc. Jpn. Acad., Ser. B}\
  }\textbf {\bibinfo {volume} {82}} (\bibinfo {year} {2006})}\BibitemShut
  {NoStop}%
\bibitem [{\citenamefont {Borejsza}\ and\ \citenamefont
  {Dupuis}(2004)}]{Borejsza2004}%
  \BibitemOpen
  \bibfield  {author} {\bibinfo {author} {\bibfnamefont {K.}~\bibnamefont
  {Borejsza}}\ and\ \bibinfo {author} {\bibfnamefont {N.}~\bibnamefont
  {Dupuis}},\ }\href {\doibase 10.1103/PhysRevB.69.085119} {\bibfield
  {journal} {\bibinfo  {journal} {Phys. Rev. B}\ }\textbf {\bibinfo {volume}
  {69}},\ \bibinfo {pages} {085119} (\bibinfo {year} {2004})}\BibitemShut
  {NoStop}%
\bibitem [{\citenamefont {Park}\ \emph {et~al.}(2008)\citenamefont {Park},
  \citenamefont {Haule},\ and\ \citenamefont {Kotliar}}]{Park2008}%
  \BibitemOpen
  \bibfield  {author} {\bibinfo {author} {\bibfnamefont {H.}~\bibnamefont
  {Park}}, \bibinfo {author} {\bibfnamefont {K.}~\bibnamefont {Haule}}, \ and\
  \bibinfo {author} {\bibfnamefont {G.}~\bibnamefont {Kotliar}},\ }\href
  {\doibase 10.1103/PhysRevLett.101.186403} {\bibfield  {journal} {\bibinfo
  {journal} {Phys. Rev. Lett.}\ }\textbf {\bibinfo {volume} {101}},\ \bibinfo
  {pages} {186403} (\bibinfo {year} {2008})}\BibitemShut {NoStop}%
\bibitem [{\citenamefont {Sch\"afer}\ \emph {et~al.}(2015)\citenamefont
  {Sch\"afer}, \citenamefont {Geles}, \citenamefont {Rost}, \citenamefont
  {Rohringer}, \citenamefont {Arrigoni}, \citenamefont {Held}, \citenamefont
  {Bl\"umer}, \citenamefont {Aichhorn},\ and\ \citenamefont
  {Toschi}}]{Schafer2015}%
  \BibitemOpen
  \bibfield  {author} {\bibinfo {author} {\bibfnamefont {T.}~\bibnamefont
  {Sch\"afer}}, \bibinfo {author} {\bibfnamefont {F.}~\bibnamefont {Geles}},
  \bibinfo {author} {\bibfnamefont {D.}~\bibnamefont {Rost}}, \bibinfo {author}
  {\bibfnamefont {G.}~\bibnamefont {Rohringer}}, \bibinfo {author}
  {\bibfnamefont {E.}~\bibnamefont {Arrigoni}}, \bibinfo {author}
  {\bibfnamefont {K.}~\bibnamefont {Held}}, \bibinfo {author} {\bibfnamefont
  {N.}~\bibnamefont {Bl\"umer}}, \bibinfo {author} {\bibfnamefont
  {M.}~\bibnamefont {Aichhorn}}, \ and\ \bibinfo {author} {\bibfnamefont
  {A.}~\bibnamefont {Toschi}},\ }\href {\doibase 10.1103/PhysRevB.91.125109}
  {\bibfield  {journal} {\bibinfo  {journal} {Phys. Rev. B}\ }\textbf {\bibinfo
  {volume} {91}},\ \bibinfo {pages} {125109} (\bibinfo {year}
  {2015})}\BibitemShut {NoStop}%
\bibitem [{\citenamefont {Sordi}\ \emph {et~al.}(2011)\citenamefont {Sordi},
  \citenamefont {Haule},\ and\ \citenamefont {Tremblay}}]{Sordi2011}%
  \BibitemOpen
  \bibfield  {author} {\bibinfo {author} {\bibfnamefont {G.}~\bibnamefont
  {Sordi}}, \bibinfo {author} {\bibfnamefont {K.}~\bibnamefont {Haule}}, \ and\
  \bibinfo {author} {\bibfnamefont {A.-M.~S.}\ \bibnamefont {Tremblay}},\
  }\href {\doibase 10.1103/PhysRevB.84.075161} {\bibfield  {journal} {\bibinfo
  {journal} {Phys. Rev. B}\ }\textbf {\bibinfo {volume} {84}},\ \bibinfo
  {pages} {075161} (\bibinfo {year} {2011})}\BibitemShut {NoStop}%
\bibitem [{\citenamefont {Feynman}\ and\ \citenamefont
  {Hibbs}(1965)}]{Feynman-Hibbs}%
  \BibitemOpen
  \bibfield  {author} {\bibinfo {author} {\bibfnamefont {R.~P.}\ \bibnamefont
  {Feynman}}\ and\ \bibinfo {author} {\bibfnamefont {A.~R.}\ \bibnamefont
  {Hibbs}},\ }\href@noop {} {\emph {\bibinfo {title} {{Quantum Mechanics and
  Path Integrals}}}}\ (\bibinfo  {publisher} {McGraw-Hill},\ \bibinfo {address}
  {New York},\ \bibinfo {year} {1965})\ Chap.~\bibinfo {chapter}
  {11}\BibitemShut {NoStop}%
\bibitem [{\citenamefont {Feynman}(1972)}]{Feynman}%
  \BibitemOpen
  \bibfield  {author} {\bibinfo {author} {\bibfnamefont {R.~P.}\ \bibnamefont
  {Feynman}},\ }\href@noop {} {\emph {\bibinfo {title} {{Statistical
  Mechanics}}}}\ (\bibinfo  {publisher} {Westview Press},\ \bibinfo {address}
  {Boulder},\ \bibinfo {year} {1972})\ Chap.\ \bibinfo {chapter} {1 and
  3}\BibitemShut {NoStop}%
\bibitem [{\citenamefont {Shankar}(1994)}]{Shanker1994}%
  \BibitemOpen
  \bibfield  {author} {\bibinfo {author} {\bibfnamefont {R.}~\bibnamefont
  {Shankar}},\ }\href {\doibase 10.1103/RevModPhys.66.129} {\bibfield
  {journal} {\bibinfo  {journal} {Rev. Mod. Phys.}\ }\textbf {\bibinfo {volume}
  {66}},\ \bibinfo {pages} {129} (\bibinfo {year} {1994})}\BibitemShut
  {NoStop}%
\bibitem [{\citenamefont {Imada}\ \emph {et~al.}(1998)\citenamefont {Imada},
  \citenamefont {Fujimori},\ and\ \citenamefont {Tokura}}]{Imada1998}%
  \BibitemOpen
  \bibfield  {author} {\bibinfo {author} {\bibfnamefont {M.}~\bibnamefont
  {Imada}}, \bibinfo {author} {\bibfnamefont {A.}~\bibnamefont {Fujimori}}, \
  and\ \bibinfo {author} {\bibfnamefont {Y.}~\bibnamefont {Tokura}},\ }\href
  {\doibase 10.1103/RevModPhys.70.1039} {\bibfield  {journal} {\bibinfo
  {journal} {Rev. Mod. Phys.}\ }\textbf {\bibinfo {volume} {70}},\ \bibinfo
  {pages} {1039} (\bibinfo {year} {1998})}\BibitemShut {NoStop}%
\bibitem [{\citenamefont {Pairault}\ \emph {et~al.}(1998)\citenamefont
  {Pairault}, \citenamefont {S\'en\'echal},\ and\ \citenamefont
  {Tremblay}}]{Pairault1998}%
  \BibitemOpen
  \bibfield  {author} {\bibinfo {author} {\bibfnamefont {S.}~\bibnamefont
  {Pairault}}, \bibinfo {author} {\bibfnamefont {D.}~\bibnamefont
  {S\'en\'echal}}, \ and\ \bibinfo {author} {\bibfnamefont {A.-M.~S.}\
  \bibnamefont {Tremblay}},\ }\href {\doibase 10.1103/PhysRevLett.80.5389}
  {\bibfield  {journal} {\bibinfo  {journal} {Phys. Rev. Lett.}\ }\textbf
  {\bibinfo {volume} {80}},\ \bibinfo {pages} {5389} (\bibinfo {year}
  {1998})}\BibitemShut {NoStop}%
\bibitem [{\citenamefont {Pairault}\ \emph {et~al.}(2000)\citenamefont
  {Pairault}, \citenamefont {S\'{e}n\'{e}chal}, ,\ and\ \citenamefont
  {Tremblay}}]{Pairault2000}%
  \BibitemOpen
  \bibfield  {author} {\bibinfo {author} {\bibfnamefont {S.}~\bibnamefont
  {Pairault}}, \bibinfo {author} {\bibfnamefont {D.}~\bibnamefont
  {S\'{e}n\'{e}chal}}, , \ and\ \bibinfo {author} {\bibfnamefont {A.-M.~S.}\
  \bibnamefont {Tremblay}},\ }\href {http://dx.doi.org/10.1007/s100510070253}
  {\bibfield  {journal} {\bibinfo  {journal} {Eur. Phys. J. B}\ }\textbf
  {\bibinfo {volume} {16}},\ \bibinfo {pages} {85} (\bibinfo {year}
  {2000})}\BibitemShut {NoStop}%
\bibitem [{not()}]{note_hyb}%
  \BibitemOpen
  \href@noop {} {}\bibinfo {note} {Instead of introducing $S_{\rm
  aux}(c^\star,c,b^\star,b)$, it is also possible to proceed the following
  discussion by considering the action $S'(c^\star,c)$ which is obtained by
  integrating out the bath fermions $b^\star$ and $b$ from the action
  $S'(c^\star,c,b^\star,b)$ of the reference system, i.e., $S'(c^\star,c)=-\ln
  \int \mcal{D}[b^\star b]\exp\left[-S'(c^\star, c, b^\star, b)\right]$.
  Indeed, the degrees of freedom of $S'(c^\star,c)$ are the same as those of
  the original system $S(c^\star,c)$. The integration over $b^\star$ and $b$
  results in the so-called hybridization function which enters in the
  coefficient of the quadratic term of $c^\star$ and $c$ in $S'(c^\star,c)$.
  The formalism along this line would be more suitable for the
  self-energy-funcitonal approach to the
  CDMFT~\cite{Senechal2010}.}\BibitemShut {Stop}%
\bibitem [{\citenamefont {Rubtsov}\ \emph {et~al.}(2008)\citenamefont
  {Rubtsov}, \citenamefont {Katsnelson},\ and\ \citenamefont
  {Lichtenstein}}]{Rubtsov2008}%
  \BibitemOpen
  \bibfield  {author} {\bibinfo {author} {\bibfnamefont {A.~N.}\ \bibnamefont
  {Rubtsov}}, \bibinfo {author} {\bibfnamefont {M.~I.}\ \bibnamefont
  {Katsnelson}}, \ and\ \bibinfo {author} {\bibfnamefont {A.~I.}\ \bibnamefont
  {Lichtenstein}},\ }\href {\doibase 10.1103/PhysRevB.77.033101} {\bibfield
  {journal} {\bibinfo  {journal} {Phys. Rev. B}\ }\textbf {\bibinfo {volume}
  {77}},\ \bibinfo {pages} {033101} (\bibinfo {year} {2008})}\BibitemShut
  {NoStop}%
\bibitem [{\citenamefont {Adibi}\ and\ \citenamefont
  {Jafari}(2016)}]{Adibi2016}%
  \BibitemOpen
  \bibfield  {author} {\bibinfo {author} {\bibfnamefont {E.}~\bibnamefont
  {Adibi}}\ and\ \bibinfo {author} {\bibfnamefont {S.~A.}\ \bibnamefont
  {Jafari}},\ }\href {\doibase 10.1103/PhysRevB.93.075122} {\bibfield
  {journal} {\bibinfo  {journal} {Phys. Rev. B}\ }\textbf {\bibinfo {volume}
  {93}},\ \bibinfo {pages} {075122} (\bibinfo {year} {2016})}\BibitemShut
  {NoStop}%
\bibitem [{\citenamefont {Kubo}(1962)}]{Kubo1962}%
  \BibitemOpen
  \bibfield  {author} {\bibinfo {author} {\bibfnamefont {R.}~\bibnamefont
  {Kubo}},\ }\href {http://dx.doi.org/10.1143/JPSJ.17.1100} {\bibfield
  {journal} {\bibinfo  {journal} {J. Phys. Soc. Jpn.}\ }\textbf {\bibinfo
  {volume} {17}},\ \bibinfo {pages} {1100} (\bibinfo {year}
  {1962})}\BibitemShut {NoStop}%
\bibitem [{\citenamefont {Lee}\ \emph {et~al.}(2006)\citenamefont {Lee},
  \citenamefont {Nagaosa},\ and\ \citenamefont {Wen}}]{Lee2006}%
  \BibitemOpen
  \bibfield  {author} {\bibinfo {author} {\bibfnamefont {P.~A.}\ \bibnamefont
  {Lee}}, \bibinfo {author} {\bibfnamefont {N.}~\bibnamefont {Nagaosa}}, \ and\
  \bibinfo {author} {\bibfnamefont {X.-G.}\ \bibnamefont {Wen}},\ }\href
  {\doibase 10.1103/RevModPhys.78.17} {\bibfield  {journal} {\bibinfo
  {journal} {Rev. Mod. Phys.}\ }\textbf {\bibinfo {volume} {78}},\ \bibinfo
  {pages} {17} (\bibinfo {year} {2006})}\BibitemShut {NoStop}%
\bibitem [{\citenamefont {Fukuyama}(2008)}]{Fukuyama2008}%
  \BibitemOpen
  \bibfield  {author} {\bibinfo {author} {\bibfnamefont {H.}~\bibnamefont
  {Fukuyama}},\ }\href {http://dx.doi.org/10.1143/JPSJ.77.111013} {\bibfield
  {journal} {\bibinfo  {journal} {J. Phys. Soc. Jpn.}\ }\textbf {\bibinfo
  {volume} {77}},\ \bibinfo {pages} {111013} (\bibinfo {year}
  {2008})}\BibitemShut {NoStop}%
\bibitem [{\citenamefont {Neumann}\ \emph {et~al.}(2007)\citenamefont
  {Neumann}, \citenamefont {Ny{\'e}ki}, \citenamefont {Cowan},\ and\
  \citenamefont {Saunders}}]{Neumann2007}%
  \BibitemOpen
  \bibfield  {author} {\bibinfo {author} {\bibfnamefont {M.}~\bibnamefont
  {Neumann}}, \bibinfo {author} {\bibfnamefont {J.}~\bibnamefont {Ny{\'e}ki}},
  \bibinfo {author} {\bibfnamefont {B.}~\bibnamefont {Cowan}}, \ and\ \bibinfo
  {author} {\bibfnamefont {J.}~\bibnamefont {Saunders}},\ }\href {\doibase
  10.1126/science.1143607} {\bibfield  {journal} {\bibinfo  {journal}
  {Science}\ }\textbf {\bibinfo {volume} {317}},\ \bibinfo {pages} {1356}
  (\bibinfo {year} {2007})}\BibitemShut {NoStop}%
\bibitem [{\citenamefont {Momoi}\ \emph {et~al.}(1999)\citenamefont {Momoi},
  \citenamefont {Sakamoto},\ and\ \citenamefont {Kubo}}]{Momoi1999}%
  \BibitemOpen
  \bibfield  {author} {\bibinfo {author} {\bibfnamefont {T.}~\bibnamefont
  {Momoi}}, \bibinfo {author} {\bibfnamefont {H.}~\bibnamefont {Sakamoto}}, \
  and\ \bibinfo {author} {\bibfnamefont {K.}~\bibnamefont {Kubo}},\ }\href
  {\doibase 10.1103/PhysRevB.59.9491} {\bibfield  {journal} {\bibinfo
  {journal} {Phys. Rev. B}\ }\textbf {\bibinfo {volume} {59}},\ \bibinfo
  {pages} {9491} (\bibinfo {year} {1999})}\BibitemShut {NoStop}%
\bibitem [{\citenamefont {Momoi}\ \emph {et~al.}(2006)\citenamefont {Momoi},
  \citenamefont {Sindzingre},\ and\ \citenamefont {Shannon}}]{Momoi2006}%
  \BibitemOpen
  \bibfield  {author} {\bibinfo {author} {\bibfnamefont {T.}~\bibnamefont
  {Momoi}}, \bibinfo {author} {\bibfnamefont {P.}~\bibnamefont {Sindzingre}}, \
  and\ \bibinfo {author} {\bibfnamefont {N.}~\bibnamefont {Shannon}},\ }\href
  {\doibase 10.1103/PhysRevLett.97.257204} {\bibfield  {journal} {\bibinfo
  {journal} {Phys. Rev. Lett.}\ }\textbf {\bibinfo {volume} {97}},\ \bibinfo
  {pages} {257204} (\bibinfo {year} {2006})}\BibitemShut {NoStop}%
\bibitem [{\citenamefont {Momoi}\ \emph {et~al.}(2012)\citenamefont {Momoi},
  \citenamefont {Sindzingre},\ and\ \citenamefont {Kubo}}]{Momoi2012}%
  \BibitemOpen
  \bibfield  {author} {\bibinfo {author} {\bibfnamefont {T.}~\bibnamefont
  {Momoi}}, \bibinfo {author} {\bibfnamefont {P.}~\bibnamefont {Sindzingre}}, \
  and\ \bibinfo {author} {\bibfnamefont {K.}~\bibnamefont {Kubo}},\ }\href
  {\doibase 10.1103/PhysRevLett.108.057206} {\bibfield  {journal} {\bibinfo
  {journal} {Phys. Rev. Lett.}\ }\textbf {\bibinfo {volume} {108}},\ \bibinfo
  {pages} {057206} (\bibinfo {year} {2012})}\BibitemShut {NoStop}%
\bibitem [{\citenamefont {Fuseya}\ and\ \citenamefont
  {Ogata}(2009)}]{Fuseya2009}%
  \BibitemOpen
  \bibfield  {author} {\bibinfo {author} {\bibfnamefont {Y.}~\bibnamefont
  {Fuseya}}\ and\ \bibinfo {author} {\bibfnamefont {M.}~\bibnamefont {Ogata}},\
  }\href {\doibase 10.1143/JPSJ.78.013601} {\bibfield  {journal} {\bibinfo
  {journal} {J. Phys. Soc. Jpn.}\ }\textbf {\bibinfo {volume} {78}},\ \bibinfo
  {pages} {013601} (\bibinfo {year} {2009})}\BibitemShut {NoStop}%
\bibitem [{\citenamefont {Seki}\ \emph {et~al.}(2009)\citenamefont {Seki},
  \citenamefont {Shirakawa},\ and\ \citenamefont {Ohta}}]{Seki2009}%
  \BibitemOpen
  \bibfield  {author} {\bibinfo {author} {\bibfnamefont {K.}~\bibnamefont
  {Seki}}, \bibinfo {author} {\bibfnamefont {T.}~\bibnamefont {Shirakawa}}, \
  and\ \bibinfo {author} {\bibfnamefont {Y.}~\bibnamefont {Ohta}},\ }\href
  {\doibase 10.1103/PhysRevB.79.024303} {\bibfield  {journal} {\bibinfo
  {journal} {Phys. Rev. B}\ }\textbf {\bibinfo {volume} {79}},\ \bibinfo
  {pages} {024303} (\bibinfo {year} {2009})}\BibitemShut {NoStop}%
\bibitem [{\citenamefont {Beach}\ and\ \citenamefont
  {Assaad}(2011)}]{Beach2011}%
  \BibitemOpen
  \bibfield  {author} {\bibinfo {author} {\bibfnamefont {K.~S.~D.}\
  \bibnamefont {Beach}}\ and\ \bibinfo {author} {\bibfnamefont {F.~F.}\
  \bibnamefont {Assaad}},\ }\href {\doibase 10.1103/PhysRevB.83.045103}
  {\bibfield  {journal} {\bibinfo  {journal} {Phys. Rev. B}\ }\textbf {\bibinfo
  {volume} {83}},\ \bibinfo {pages} {045103} (\bibinfo {year}
  {2011})}\BibitemShut {NoStop}%
\bibitem [{\citenamefont {Powell}\ and\ \citenamefont
  {McKenzie}(2011)}]{Powell2011}%
  \BibitemOpen
  \bibfield  {author} {\bibinfo {author} {\bibfnamefont {B.~J.}\ \bibnamefont
  {Powell}}\ and\ \bibinfo {author} {\bibfnamefont {R.~H.}\ \bibnamefont
  {McKenzie}},\ }\href {http://iopscience.iop.org/0034-4885/74/5/056501/}
  {\bibfield  {journal} {\bibinfo  {journal} {Rep. Prog. Phys.}\ }\textbf
  {\bibinfo {volume} {74}},\ \bibinfo {pages} {056501} (\bibinfo {year}
  {2011})}\BibitemShut {NoStop}%
\bibitem [{\citenamefont {Zhou}\ \emph {et~al.}(2017)\citenamefont {Zhou},
  \citenamefont {Kanoda},\ and\ \citenamefont {Ng}}]{Zhou2016}%
  \BibitemOpen
  \bibfield  {author} {\bibinfo {author} {\bibfnamefont {Y.}~\bibnamefont
  {Zhou}}, \bibinfo {author} {\bibfnamefont {K.}~\bibnamefont {Kanoda}}, \ and\
  \bibinfo {author} {\bibfnamefont {T.-K.}\ \bibnamefont {Ng}},\ }\href
  {\doibase 10.1103/RevModPhys.89.025003} {\bibfield  {journal} {\bibinfo
  {journal} {Rev. Mod. Phys.}\ }\textbf {\bibinfo {volume} {89}},\ \bibinfo
  {pages} {025003} (\bibinfo {year} {2017})}\BibitemShut {NoStop}%
\bibitem [{\citenamefont {Kurosaki}\ \emph {et~al.}(2005)\citenamefont
  {Kurosaki}, \citenamefont {Shimizu}, \citenamefont {Miyagawa}, \citenamefont
  {Kanoda},\ and\ \citenamefont {Saito}}]{Kurosaki2005}%
  \BibitemOpen
  \bibfield  {author} {\bibinfo {author} {\bibfnamefont {Y.}~\bibnamefont
  {Kurosaki}}, \bibinfo {author} {\bibfnamefont {Y.}~\bibnamefont {Shimizu}},
  \bibinfo {author} {\bibfnamefont {K.}~\bibnamefont {Miyagawa}}, \bibinfo
  {author} {\bibfnamefont {K.}~\bibnamefont {Kanoda}}, \ and\ \bibinfo {author}
  {\bibfnamefont {G.}~\bibnamefont {Saito}},\ }\href {\doibase
  10.1103/PhysRevLett.95.177001} {\bibfield  {journal} {\bibinfo  {journal}
  {Phys. Rev. Lett.}\ }\textbf {\bibinfo {volume} {95}},\ \bibinfo {pages}
  {177001} (\bibinfo {year} {2005})}\BibitemShut {NoStop}%
\bibitem [{\citenamefont {Watanabe}\ \emph {et~al.}(2012)\citenamefont
  {Watanabe}, \citenamefont {Yamashita}, \citenamefont {Tonegawa},
  \citenamefont {Oshima}, \citenamefont {Yamamoto}, \citenamefont {Kato},
  \citenamefont {Sheikin}, \citenamefont {Behnia}, \citenamefont {Terashima},
  \citenamefont {Uji}, \citenamefont {Shibauchi},\ and\ \citenamefont
  {Matsuda}}]{Watanabe2012}%
  \BibitemOpen
  \bibfield  {author} {\bibinfo {author} {\bibfnamefont {D.}~\bibnamefont
  {Watanabe}}, \bibinfo {author} {\bibfnamefont {M.}~\bibnamefont {Yamashita}},
  \bibinfo {author} {\bibfnamefont {S.}~\bibnamefont {Tonegawa}}, \bibinfo
  {author} {\bibfnamefont {Y.}~\bibnamefont {Oshima}}, \bibinfo {author}
  {\bibfnamefont {M.}~\bibnamefont {Yamamoto}, \bibfnamefont {H}}, \bibinfo
  {author} {\bibfnamefont {R.}~\bibnamefont {Kato}}, \bibinfo {author}
  {\bibfnamefont {I.}~\bibnamefont {Sheikin}}, \bibinfo {author} {\bibfnamefont
  {K.}~\bibnamefont {Behnia}}, \bibinfo {author} {\bibfnamefont
  {T.}~\bibnamefont {Terashima}}, \bibinfo {author} {\bibfnamefont
  {S.}~\bibnamefont {Uji}}, \bibinfo {author} {\bibfnamefont {T.}~\bibnamefont
  {Shibauchi}}, \ and\ \bibinfo {author} {\bibfnamefont {Y.}~\bibnamefont
  {Matsuda}},\ }\href
  {http://www.nature.com/ncomms/journal/v3/n9/full/ncomms2082.html} {\bibfield
  {journal} {\bibinfo  {journal} {Nature Communications}\ }\textbf {\bibinfo
  {volume} {3}},\ \bibinfo {pages} {1090} (\bibinfo {year} {2012})}\BibitemShut
  {NoStop}%
\bibitem [{\citenamefont {Kawasugi}\ \emph {et~al.}(2009)\citenamefont
  {Kawasugi}, \citenamefont {Yamamoto}, \citenamefont {Tajima}, \citenamefont
  {Fukunaga}, \citenamefont {Tsukagoshi},\ and\ \citenamefont
  {Kato}}]{Kawasugi2009}%
  \BibitemOpen
  \bibfield  {author} {\bibinfo {author} {\bibfnamefont {Y.}~\bibnamefont
  {Kawasugi}}, \bibinfo {author} {\bibfnamefont {H.~M.}\ \bibnamefont
  {Yamamoto}}, \bibinfo {author} {\bibfnamefont {N.}~\bibnamefont {Tajima}},
  \bibinfo {author} {\bibfnamefont {T.}~\bibnamefont {Fukunaga}}, \bibinfo
  {author} {\bibfnamefont {K.}~\bibnamefont {Tsukagoshi}}, \ and\ \bibinfo
  {author} {\bibfnamefont {R.}~\bibnamefont {Kato}},\ }\href {\doibase
  10.1103/PhysRevLett.103.116801} {\bibfield  {journal} {\bibinfo  {journal}
  {Phys. Rev. Lett.}\ }\textbf {\bibinfo {volume} {103}},\ \bibinfo {pages}
  {116801} (\bibinfo {year} {2009})}\BibitemShut {NoStop}%
\bibitem [{\citenamefont {Kawasugi}\ \emph {et~al.}(2011)\citenamefont
  {Kawasugi}, \citenamefont {Yamamoto}, \citenamefont {Tajima}, \citenamefont
  {Fukunaga}, \citenamefont {Tsukagoshi},\ and\ \citenamefont
  {Kato}}]{Kawasugi2011}%
  \BibitemOpen
  \bibfield  {author} {\bibinfo {author} {\bibfnamefont {Y.}~\bibnamefont
  {Kawasugi}}, \bibinfo {author} {\bibfnamefont {H.~M.}\ \bibnamefont
  {Yamamoto}}, \bibinfo {author} {\bibfnamefont {N.}~\bibnamefont {Tajima}},
  \bibinfo {author} {\bibfnamefont {T.}~\bibnamefont {Fukunaga}}, \bibinfo
  {author} {\bibfnamefont {K.}~\bibnamefont {Tsukagoshi}}, \ and\ \bibinfo
  {author} {\bibfnamefont {R.}~\bibnamefont {Kato}},\ }\href {\doibase
  10.1103/PhysRevB.84.125129} {\bibfield  {journal} {\bibinfo  {journal} {Phys.
  Rev. B}\ }\textbf {\bibinfo {volume} {84}},\ \bibinfo {pages} {125129}
  (\bibinfo {year} {2011})}\BibitemShut {NoStop}%
\bibitem [{\citenamefont {Sato}\ \emph {et~al.}(2017)\citenamefont {Sato},
  \citenamefont {Kawasugi}, \citenamefont {Suda}, \citenamefont {Yamamoto},\
  and\ \citenamefont {Kato}}]{Sato_Kawasugi2016}%
  \BibitemOpen
  \bibfield  {author} {\bibinfo {author} {\bibfnamefont {Y.}~\bibnamefont
  {Sato}}, \bibinfo {author} {\bibfnamefont {Y.}~\bibnamefont {Kawasugi}},
  \bibinfo {author} {\bibfnamefont {M.}~\bibnamefont {Suda}}, \bibinfo {author}
  {\bibfnamefont {H.~M.}\ \bibnamefont {Yamamoto}}, \ and\ \bibinfo {author}
  {\bibfnamefont {R.}~\bibnamefont {Kato}},\ }\href {\doibase
  10.1021/acs.nanolett.6b03817} {\bibfield  {journal} {\bibinfo  {journal}
  {Nano Letters}\ }\textbf {\bibinfo {volume} {17}},\ \bibinfo {pages} {708}
  (\bibinfo {year} {2017})}\BibitemShut {NoStop}%
\bibitem [{\citenamefont {Itou}\ \emph {et~al.}(2017)\citenamefont {Itou},
  \citenamefont {Watanabe}, \citenamefont {Maegawa}, \citenamefont {Tajima},
  \citenamefont {Tajima}, \citenamefont {Kubo}, \citenamefont {Kato},\ and\
  \citenamefont {Kanoda}}]{Itou2017}%
  \BibitemOpen
  \bibfield  {author} {\bibinfo {author} {\bibfnamefont {T.}~\bibnamefont
  {Itou}}, \bibinfo {author} {\bibfnamefont {E.}~\bibnamefont {Watanabe}},
  \bibinfo {author} {\bibfnamefont {S.}~\bibnamefont {Maegawa}}, \bibinfo
  {author} {\bibfnamefont {A.}~\bibnamefont {Tajima}}, \bibinfo {author}
  {\bibfnamefont {N.}~\bibnamefont {Tajima}}, \bibinfo {author} {\bibfnamefont
  {K.}~\bibnamefont {Kubo}}, \bibinfo {author} {\bibfnamefont {R.}~\bibnamefont
  {Kato}}, \ and\ \bibinfo {author} {\bibfnamefont {K.}~\bibnamefont
  {Kanoda}},\ }\href {https://doi.org/10.1126/sciadv.1601594} {\bibfield
  {journal} {\bibinfo  {journal} {Science Advances}\ }\textbf {\bibinfo
  {volume} {3}},\ \bibinfo {pages} {e1601594} (\bibinfo {year}
  {2017})}\BibitemShut {NoStop}%
\bibitem [{\citenamefont {Andrade}\ \emph {et~al.}(2009)\citenamefont
  {Andrade}, \citenamefont {Miranda},\ and\ \citenamefont
  {Dobrosavljevi\ifmmode~\acute{c}\else \'{c}\fi{}}}]{Andrade2009}%
  \BibitemOpen
  \bibfield  {author} {\bibinfo {author} {\bibfnamefont {E.~C.}\ \bibnamefont
  {Andrade}}, \bibinfo {author} {\bibfnamefont {E.}~\bibnamefont {Miranda}}, \
  and\ \bibinfo {author} {\bibfnamefont {V.}~\bibnamefont
  {Dobrosavljevi\ifmmode~\acute{c}\else \'{c}\fi{}}},\ }\href {\doibase
  10.1103/PhysRevLett.102.206403} {\bibfield  {journal} {\bibinfo  {journal}
  {Phys. Rev. Lett.}\ }\textbf {\bibinfo {volume} {102}},\ \bibinfo {pages}
  {206403} (\bibinfo {year} {2009})}\BibitemShut {NoStop}%
\bibitem [{\citenamefont {Potthoff}\ and\ \citenamefont
  {Balzer}(2007)}]{Potthoff2007}%
  \BibitemOpen
  \bibfield  {author} {\bibinfo {author} {\bibfnamefont {M.}~\bibnamefont
  {Potthoff}}\ and\ \bibinfo {author} {\bibfnamefont {M.}~\bibnamefont
  {Balzer}},\ }\href {\doibase 10.1103/PhysRevB.75.125112} {\bibfield
  {journal} {\bibinfo  {journal} {Phys. Rev. B}\ }\textbf {\bibinfo {volume}
  {75}},\ \bibinfo {pages} {125112} (\bibinfo {year} {2007})}\BibitemShut
  {NoStop}%
\bibitem [{\citenamefont {Bohrdt}\ \emph {et~al.}(2018)\citenamefont {Bohrdt},
  \citenamefont {Greif}, \citenamefont {Demler}, \citenamefont {Knap},\ and\
  \citenamefont {Grusdt}}]{Bohrdt2018}%
  \BibitemOpen
  \bibfield  {author} {\bibinfo {author} {\bibfnamefont {A.}~\bibnamefont
  {Bohrdt}}, \bibinfo {author} {\bibfnamefont {D.}~\bibnamefont {Greif}},
  \bibinfo {author} {\bibfnamefont {E.}~\bibnamefont {Demler}}, \bibinfo
  {author} {\bibfnamefont {M.}~\bibnamefont {Knap}}, \ and\ \bibinfo {author}
  {\bibfnamefont {F.}~\bibnamefont {Grusdt}},\ }\href {\doibase
  10.1103/PhysRevB.97.125117} {\bibfield  {journal} {\bibinfo  {journal} {Phys.
  Rev. B}\ }\textbf {\bibinfo {volume} {97}},\ \bibinfo {pages} {125117}
  (\bibinfo {year} {2018})}\BibitemShut {NoStop}%
\bibitem [{\citenamefont {Brown}\ \emph {et~al.}()\citenamefont {Brown},
  \citenamefont {Mitra}, \citenamefont {Guardado-Sanchez}, \citenamefont
  {Nourafkan}, \citenamefont {Reymbaut}, \citenamefont {Bergeron},
  \citenamefont {Tremblay}, \citenamefont {Kokalj}, \citenamefont {Huse},
  \citenamefont {Schau},\ and\ \citenamefont {Bakr}}]{Brown2018}%
  \BibitemOpen
  \bibfield  {author} {\bibinfo {author} {\bibfnamefont {P.~T.}\ \bibnamefont
  {Brown}}, \bibinfo {author} {\bibfnamefont {D.}~\bibnamefont {Mitra}},
  \bibinfo {author} {\bibfnamefont {E.}~\bibnamefont {Guardado-Sanchez}},
  \bibinfo {author} {\bibfnamefont {R.}~\bibnamefont {Nourafkan}}, \bibinfo
  {author} {\bibfnamefont {A.}~\bibnamefont {Reymbaut}}, \bibinfo {author}
  {\bibfnamefont {S.}~\bibnamefont {Bergeron}}, \bibinfo {author}
  {\bibfnamefont {A.-M.~S.}\ \bibnamefont {Tremblay}}, \bibinfo {author}
  {\bibfnamefont {J.}~\bibnamefont {Kokalj}}, \bibinfo {author} {\bibfnamefont
  {D.~A.}\ \bibnamefont {Huse}}, \bibinfo {author} {\bibfnamefont
  {P.}~\bibnamefont {Schau}}, \ and\ \bibinfo {author} {\bibfnamefont {W.~S.}\
  \bibnamefont {Bakr}},\ }\href {https://arxiv.org/abs/1802.09456} {\enquote
  {\bibinfo {title} {{Bad metallic transport in a cold atom Fermi-Hubbard
  system}},}\ }\Eprint {http://arxiv.org/abs/cond-mat/1802.09456}
  {arXiv:cond-mat/1802.09456} \BibitemShut {NoStop}%
\bibitem [{\citenamefont {Jakli\ifmmode~\check{c}\else \v{c}\fi{}}\ and\
  \citenamefont {Prelov\ifmmode~\check{s}\else
  \v{s}\fi{}ek}(1994)}]{Jaklic1994}%
  \BibitemOpen
  \bibfield  {author} {\bibinfo {author} {\bibfnamefont {J.}~\bibnamefont
  {Jakli\ifmmode~\check{c}\else \v{c}\fi{}}}\ and\ \bibinfo {author}
  {\bibfnamefont {P.}~\bibnamefont {Prelov\ifmmode~\check{s}\else
  \v{s}\fi{}ek}},\ }\href {\doibase 10.1103/PhysRevB.49.5065} {\bibfield
  {journal} {\bibinfo  {journal} {Phys. Rev. B}\ }\textbf {\bibinfo {volume}
  {49}},\ \bibinfo {pages} {5065} (\bibinfo {year} {1994})}\BibitemShut
  {NoStop}%
\bibitem [{\citenamefont {Kokalj}(2017)}]{Kokalj2017}%
  \BibitemOpen
  \bibfield  {author} {\bibinfo {author} {\bibfnamefont {J.}~\bibnamefont
  {Kokalj}},\ }\href {\doibase 10.1103/PhysRevB.95.041110} {\bibfield
  {journal} {\bibinfo  {journal} {Phys. Rev. B}\ }\textbf {\bibinfo {volume}
  {95}},\ \bibinfo {pages} {041110} (\bibinfo {year} {2017})}\BibitemShut
  {NoStop}%
\bibitem [{\citenamefont {Sugiura}\ and\ \citenamefont
  {Shimizu}(2013)}]{Sugiura2013}%
  \BibitemOpen
  \bibfield  {author} {\bibinfo {author} {\bibfnamefont {S.}~\bibnamefont
  {Sugiura}}\ and\ \bibinfo {author} {\bibfnamefont {A.}~\bibnamefont
  {Shimizu}},\ }\href {\doibase 10.1103/PhysRevLett.111.010401} {\bibfield
  {journal} {\bibinfo  {journal} {Phys. Rev. Lett.}\ }\textbf {\bibinfo
  {volume} {111}},\ \bibinfo {pages} {010401} (\bibinfo {year}
  {2013})}\BibitemShut {NoStop}%
\bibitem [{\citenamefont {Wei\ss{}e}(2009)}]{Weisse2009}%
  \BibitemOpen
  \bibfield  {author} {\bibinfo {author} {\bibfnamefont {A.}~\bibnamefont
  {Wei\ss{}e}},\ }\href {\doibase 10.1103/PhysRevLett.102.150604} {\bibfield
  {journal} {\bibinfo  {journal} {Phys. Rev. Lett.}\ }\textbf {\bibinfo
  {volume} {102}},\ \bibinfo {pages} {150604} (\bibinfo {year}
  {2009})}\BibitemShut {NoStop}%
\bibitem [{\citenamefont {Zhang}\ \emph {et~al.}(2013)\citenamefont {Zhang},
  \citenamefont {Yamagiwa},\ and\ \citenamefont {Yunoki}}]{Zhang2013}%
  \BibitemOpen
  \bibfield  {author} {\bibinfo {author} {\bibfnamefont {S.}~\bibnamefont
  {Zhang}}, \bibinfo {author} {\bibfnamefont {S.}~\bibnamefont {Yamagiwa}}, \
  and\ \bibinfo {author} {\bibfnamefont {S.}~\bibnamefont {Yunoki}},\ }\href
  {http://iopscience.iop.org/1742-6596/454/1/012049/} {\bibfield  {journal}
  {\bibinfo  {journal} {J. Phys.: Conf. Ser.}\ }\textbf {\bibinfo {volume}
  {454}},\ \bibinfo {pages} {012049} (\bibinfo {year} {2013})}\BibitemShut
  {NoStop}%
\bibitem [{\citenamefont {Beach}\ \emph {et~al.}(2000)\citenamefont {Beach},
  \citenamefont {Gooding},\ and\ \citenamefont {Marsiglio}}]{Beach2000}%
  \BibitemOpen
  \bibfield  {author} {\bibinfo {author} {\bibfnamefont {K.~S.~D.}\
  \bibnamefont {Beach}}, \bibinfo {author} {\bibfnamefont {R.~J.}\ \bibnamefont
  {Gooding}}, \ and\ \bibinfo {author} {\bibfnamefont {F.}~\bibnamefont
  {Marsiglio}},\ }\href {\doibase 10.1103/PhysRevB.61.5147} {\bibfield
  {journal} {\bibinfo  {journal} {Phys. Rev. B}\ }\textbf {\bibinfo {volume}
  {61}},\ \bibinfo {pages} {5147} (\bibinfo {year} {2000})}\BibitemShut
  {NoStop}%
\bibitem [{\citenamefont {Press}\ \emph {et~al.}(2007)\citenamefont {Press},
  \citenamefont {Teukolsky}, \citenamefont {Vetterling},\ and\ \citenamefont
  {Flannery}}]{NR3}%
  \BibitemOpen
  \bibfield  {author} {\bibinfo {author} {\bibfnamefont {W.~H.}\ \bibnamefont
  {Press}}, \bibinfo {author} {\bibfnamefont {S.~A.}\ \bibnamefont
  {Teukolsky}}, \bibinfo {author} {\bibfnamefont {W.~T.}\ \bibnamefont
  {Vetterling}}, \ and\ \bibinfo {author} {\bibfnamefont {B.~P.}\ \bibnamefont
  {Flannery}},\ }\href@noop {} {\emph {\bibinfo {title} {Numerical Recipes: the
  art of scientific computing}}},\ \bibinfo {edition} {3rd}\ ed.\ (\bibinfo
  {publisher} {Cambridge University Press},\ \bibinfo {address} {Cambridge},\
  \bibinfo {year} {2007})\ Chap.~\bibinfo {chapter} {5}\BibitemShut {NoStop}%
\bibitem [{\citenamefont {Liesen}\ and\ \citenamefont
  {Strako\v{s}}(2013)}]{Liesen}%
  \BibitemOpen
  \bibfield  {author} {\bibinfo {author} {\bibfnamefont {J.}~\bibnamefont
  {Liesen}}\ and\ \bibinfo {author} {\bibfnamefont {Z.}~\bibnamefont
  {Strako\v{s}}},\ }\href@noop {} {\emph {\bibinfo {title} {{Krylov Subspace
  Methods}}}}\ (\bibinfo  {publisher} {Oxford University Press},\ \bibinfo
  {address} {Oxford},\ \bibinfo {year} {2013})\ Chap.~\bibinfo {chapter}
  {3}\BibitemShut {NoStop}%
\bibitem [{\citenamefont {Ohta}\ \emph {et~al.}(1994)\citenamefont {Ohta},
  \citenamefont {Shimozato}, \citenamefont {Eder},\ and\ \citenamefont
  {Maekawa}}]{Ohta1994}%
  \BibitemOpen
  \bibfield  {author} {\bibinfo {author} {\bibfnamefont {Y.}~\bibnamefont
  {Ohta}}, \bibinfo {author} {\bibfnamefont {T.}~\bibnamefont {Shimozato}},
  \bibinfo {author} {\bibfnamefont {R.}~\bibnamefont {Eder}}, \ and\ \bibinfo
  {author} {\bibfnamefont {S.}~\bibnamefont {Maekawa}},\ }\href {\doibase
  10.1103/PhysRevLett.73.324} {\bibfield  {journal} {\bibinfo  {journal} {Phys.
  Rev. Lett.}\ }\textbf {\bibinfo {volume} {73}},\ \bibinfo {pages} {324}
  (\bibinfo {year} {1994})}\BibitemShut {NoStop}%
\bibitem [{\citenamefont {LeBlanc}\ \emph {et~al.}(2015)\citenamefont
  {LeBlanc}, \citenamefont {Antipov}, \citenamefont {Becca}, \citenamefont
  {Bulik}, \citenamefont {Chan}, \citenamefont {Chung}, \citenamefont {Deng},
  \citenamefont {Ferrero}, \citenamefont {Henderson}, \citenamefont
  {Jim\'enez-Hoyos}, \citenamefont {Kozik}, \citenamefont {Liu}, \citenamefont
  {Millis}, \citenamefont {Prokof'ev}, \citenamefont {Qin}, \citenamefont
  {Scuseria}, \citenamefont {Shi}, \citenamefont {Svistunov}, \citenamefont
  {Tocchio}, \citenamefont {Tupitsyn}, \citenamefont {White}, \citenamefont
  {Zhang}, \citenamefont {Zheng}, \citenamefont {Zhu},\ and\ \citenamefont
  {Gull}}]{LeBlanc2015}%
  \BibitemOpen
  \bibfield  {author} {\bibinfo {author} {\bibfnamefont {J.~P.~F.}\
  \bibnamefont {LeBlanc}}, \bibinfo {author} {\bibfnamefont {A.~E.}\
  \bibnamefont {Antipov}}, \bibinfo {author} {\bibfnamefont {F.}~\bibnamefont
  {Becca}}, \bibinfo {author} {\bibfnamefont {I.~W.}\ \bibnamefont {Bulik}},
  \bibinfo {author} {\bibfnamefont {G.~K.-L.}\ \bibnamefont {Chan}}, \bibinfo
  {author} {\bibfnamefont {C.-M.}\ \bibnamefont {Chung}}, \bibinfo {author}
  {\bibfnamefont {Y.}~\bibnamefont {Deng}}, \bibinfo {author} {\bibfnamefont
  {M.}~\bibnamefont {Ferrero}}, \bibinfo {author} {\bibfnamefont {T.~M.}\
  \bibnamefont {Henderson}}, \bibinfo {author} {\bibfnamefont {C.~A.}\
  \bibnamefont {Jim\'enez-Hoyos}}, \bibinfo {author} {\bibfnamefont
  {E.}~\bibnamefont {Kozik}}, \bibinfo {author} {\bibfnamefont {X.-W.}\
  \bibnamefont {Liu}}, \bibinfo {author} {\bibfnamefont {A.~J.}\ \bibnamefont
  {Millis}}, \bibinfo {author} {\bibfnamefont {N.~V.}\ \bibnamefont
  {Prokof'ev}}, \bibinfo {author} {\bibfnamefont {M.}~\bibnamefont {Qin}},
  \bibinfo {author} {\bibfnamefont {G.~E.}\ \bibnamefont {Scuseria}}, \bibinfo
  {author} {\bibfnamefont {H.}~\bibnamefont {Shi}}, \bibinfo {author}
  {\bibfnamefont {B.~V.}\ \bibnamefont {Svistunov}}, \bibinfo {author}
  {\bibfnamefont {L.~F.}\ \bibnamefont {Tocchio}}, \bibinfo {author}
  {\bibfnamefont {I.~S.}\ \bibnamefont {Tupitsyn}}, \bibinfo {author}
  {\bibfnamefont {S.~R.}\ \bibnamefont {White}}, \bibinfo {author}
  {\bibfnamefont {S.}~\bibnamefont {Zhang}}, \bibinfo {author} {\bibfnamefont
  {B.-X.}\ \bibnamefont {Zheng}}, \bibinfo {author} {\bibfnamefont
  {Z.}~\bibnamefont {Zhu}}, \ and\ \bibinfo {author} {\bibfnamefont
  {E.}~\bibnamefont {Gull}} (\bibinfo {collaboration} {Simons Collaboration on
  the Many-Electron Problem}),\ }\href {\doibase 10.1103/PhysRevX.5.041041}
  {\bibfield  {journal} {\bibinfo  {journal} {Phys. Rev. X}\ }\textbf {\bibinfo
  {volume} {5}},\ \bibinfo {pages} {041041} (\bibinfo {year}
  {2015})}\BibitemShut {NoStop}%
\bibitem [{\citenamefont {Balzer}\ \emph {et~al.}(2008)\citenamefont {Balzer},
  \citenamefont {Hanke},\ and\ \citenamefont {Potthoff}}]{Balzer2008}%
  \BibitemOpen
  \bibfield  {author} {\bibinfo {author} {\bibfnamefont {M.}~\bibnamefont
  {Balzer}}, \bibinfo {author} {\bibfnamefont {W.}~\bibnamefont {Hanke}}, \
  and\ \bibinfo {author} {\bibfnamefont {M.}~\bibnamefont {Potthoff}},\ }\href
  {\doibase 10.1103/PhysRevB.77.045133} {\bibfield  {journal} {\bibinfo
  {journal} {Phys. Rev. B}\ }\textbf {\bibinfo {volume} {77}},\ \bibinfo
  {pages} {045133} (\bibinfo {year} {2008})}\BibitemShut {NoStop}%
\bibitem [{\citenamefont {Sahebsara}\ and\ \citenamefont
  {S\'en\'echal}(2008)}]{Sahebsara2008}%
  \BibitemOpen
  \bibfield  {author} {\bibinfo {author} {\bibfnamefont {P.}~\bibnamefont
  {Sahebsara}}\ and\ \bibinfo {author} {\bibfnamefont {D.}~\bibnamefont
  {S\'en\'echal}},\ }\href {\doibase 10.1103/PhysRevLett.100.136402} {\bibfield
   {journal} {\bibinfo  {journal} {Phys. Rev. Lett.}\ }\textbf {\bibinfo
  {volume} {100}},\ \bibinfo {pages} {136402} (\bibinfo {year}
  {2008})}\BibitemShut {NoStop}%
\bibitem [{\citenamefont {Sorella}(2015)}]{Sorella2015}%
  \BibitemOpen
  \bibfield  {author} {\bibinfo {author} {\bibfnamefont {S.}~\bibnamefont
  {Sorella}},\ }\href {\doibase 10.1103/PhysRevB.91.241116} {\bibfield
  {journal} {\bibinfo  {journal} {Phys. Rev. B}\ }\textbf {\bibinfo {volume}
  {91}},\ \bibinfo {pages} {241116} (\bibinfo {year} {2015})}\BibitemShut
  {NoStop}%
\bibitem [{\citenamefont {Qin}\ \emph {et~al.}(2016)\citenamefont {Qin},
  \citenamefont {Shi},\ and\ \citenamefont {Zhang}}]{Qin2016}%
  \BibitemOpen
  \bibfield  {author} {\bibinfo {author} {\bibfnamefont {M.}~\bibnamefont
  {Qin}}, \bibinfo {author} {\bibfnamefont {H.}~\bibnamefont {Shi}}, \ and\
  \bibinfo {author} {\bibfnamefont {S.}~\bibnamefont {Zhang}},\ }\href
  {\doibase 10.1103/PhysRevB.94.085103} {\bibfield  {journal} {\bibinfo
  {journal} {Phys. Rev. B}\ }\textbf {\bibinfo {volume} {94}},\ \bibinfo
  {pages} {085103} (\bibinfo {year} {2016})}\BibitemShut {NoStop}%
\bibitem [{\citenamefont {Karakuzu}\ \emph {et~al.}(2018)\citenamefont
  {Karakuzu}, \citenamefont {Seki},\ and\ \citenamefont
  {Sorella}}]{Karakuzu2018}%
  \BibitemOpen
  \bibfield  {author} {\bibinfo {author} {\bibfnamefont {S.}~\bibnamefont
  {Karakuzu}}, \bibinfo {author} {\bibfnamefont {K.}~\bibnamefont {Seki}}, \
  and\ \bibinfo {author} {\bibfnamefont {S.}~\bibnamefont {Sorella}},\ }\href
  {\doibase 10.1103/PhysRevB.98.075156} {\bibfield  {journal} {\bibinfo
  {journal} {Phys. Rev. B}\ }\textbf {\bibinfo {volume} {98}},\ \bibinfo
  {pages} {075156} (\bibinfo {year} {2018})}\BibitemShut {NoStop}%
\bibitem [{\citenamefont {Laubach}\ \emph {et~al.}(2014)\citenamefont
  {Laubach}, \citenamefont {Reuther}, \citenamefont {Thomale},\ and\
  \citenamefont {Rachel}}]{Laubach2014}%
  \BibitemOpen
  \bibfield  {author} {\bibinfo {author} {\bibfnamefont {M.}~\bibnamefont
  {Laubach}}, \bibinfo {author} {\bibfnamefont {J.}~\bibnamefont {Reuther}},
  \bibinfo {author} {\bibfnamefont {R.}~\bibnamefont {Thomale}}, \ and\
  \bibinfo {author} {\bibfnamefont {S.}~\bibnamefont {Rachel}},\ }\href
  {\doibase 10.1103/PhysRevB.90.165136} {\bibfield  {journal} {\bibinfo
  {journal} {Phys. Rev. B}\ }\textbf {\bibinfo {volume} {90}},\ \bibinfo
  {pages} {165136} (\bibinfo {year} {2014})}\BibitemShut {NoStop}%
\bibitem [{\citenamefont {Rachel}\ \emph {et~al.}(2015)\citenamefont {Rachel},
  \citenamefont {Laubach}, \citenamefont {Reuther},\ and\ \citenamefont
  {Thomale}}]{Rachel2015}%
  \BibitemOpen
  \bibfield  {author} {\bibinfo {author} {\bibfnamefont {S.}~\bibnamefont
  {Rachel}}, \bibinfo {author} {\bibfnamefont {M.}~\bibnamefont {Laubach}},
  \bibinfo {author} {\bibfnamefont {J.}~\bibnamefont {Reuther}}, \ and\
  \bibinfo {author} {\bibfnamefont {R.}~\bibnamefont {Thomale}},\ }\href
  {\doibase 10.1103/PhysRevLett.114.167201} {\bibfield  {journal} {\bibinfo
  {journal} {Phys. Rev. Lett.}\ }\textbf {\bibinfo {volume} {114}},\ \bibinfo
  {pages} {167201} (\bibinfo {year} {2015})}\BibitemShut {NoStop}%
\end{thebibliography}

\end{document}